# Role of stress/strain in tailoring the magnetic and transport properties of magnetic thin films and multilayers

*A Thesis Submitted to*

**Devi Ahilya Vishwavidyalaya, Indore**

*for the Degree of*

**Doctor of Philosophy**

**in Physics**

**Faculty of science**

*by*

**Arun Singh Dev**

*Supervised by*

**Dr. Dileep Kumar**

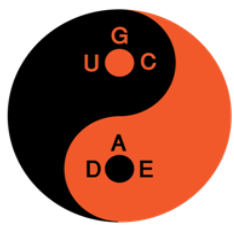

**In-situ thin film laboratory,**

**UGC-DAE Consortium for Scientific Research,**

**University Campus, Khandwa Road,**

**Indore-452001, Madhya Pradesh, India.**

**September 2022**

# CERTIFICATE OF THE SUPERVISOR

This is to certify that the work entitled "**Role of stress/strain in tailoring the magnetic and transport properties of magnetic thin films and multilayers**" is a piece of research work done by **Mr. Arun Singh Dev** under my supervision and guidance for the degree of **Doctor of Philosophy** of **Devi Ahilya Vishwavidyalaya (DAVV),** Indore, (M.P.) India. The candidate has put in an attendance of more than 200 days with me.

To the best of my knowledge and belief, the thesis:

(1) embodies the work of the candidate himself,

(2) has duly been completed,

(3) fulfils the requirement of the ordinance relating to the Ph. D. degree of the university and

(4) is up to the standard both in respect of contents and language for the being referred to the examiner.

Signature of the supervisor
(**Dr. Dileep Kumar**)

Forwarded

Signature of Head U.T.D./Principle

(**Dr. Vasant G. Sathe**)

# DECLARATION BY THE CANDIDATE

I declare that the thesis entitled "**Role of stress/strain in tailoring the magnetic and transport properties of magnetic thin films and multilayers**" is my own work conducted under the supervision of **Dr. Dileep Kumar**, Scientist -F, **UGC DAE CSR, Indore**, approved by Research Degree Committee. I have put in more than 200 days of attendance with the supervisor at the center.

I further declare that, to the best of my knowledge, the thesis does not contain any part of any work which has been submitted for the award of any degree either in this University or in any other University/Deemed University without proper citation.

Besides this-

- I have successfully completed the coursework as per UGC regulation 2009 norms.
- I have also given a pre-Ph.D. presentation and successfully incorporated the changes suggested on the basis of feedback and comments received.
- I have also published more than one research paper in an ISSN/referred journal from the research work of the thesis and have produced evidence of the same in the form of reprints.

Signature of the supervisor                                    Signature of the Candidate

**(Dr. Dileep Kumar)**                                              **(Arun Singh Dev)**

Forwarded

Signature of Head U.T.D./Principle

**(Dr. Vasant G. Sathe)**

*Dedicated to*

*Maa Bhagwati*

# Acknowledgements

*First and foremost, praises and thanks to God for his showers of blessings throughout my research work to complete the research successfully.*

*I would like to express my sincere gratitude to my supervisor* **Dr. Dileep Kumar** *for his continuous support during my Ph.D work, for his motivating, knowledgeful discussions and for his patience. His dynamism, vision and sincerity will remain ever inspiring to me. I learned the methodology to carry out the research and to present the outcome as clearly as possible from him. His guidance helped me all the time of research and writing of this thesis. It was a great experience to work and study under his guidance.*

*I gratefully acknowledge* **Prof. Amlan J. Pal,** *the Director,* **Dr. Vasant G. Sathe,** *the Centre Director, Dr.* **V. Siriguri,** *former Director,* **Dr. D. M. Phase**, *former Centre Director,* **Dr. A. K. Sinha**, *Former Director,* **Dr. Alok Banerjee** *and* **Dr. V. Ganesan,** *former Centre-Director, UGC-DAE CSR India, for their constant support and for providing me rich experimental facilities to carry out my Ph.D. research work. I am sincerely thankful to* **Prof. Ajay Gupta** *former Centre-Director, for mentoring,* **Dr. V. R. Reddy** *and* **Dr. Mukul Gupta** *for valuable suggestions and fruitful discussions during my Ph.D. work.*

*My sincere thanks are due to all scientists and researchers of UGC-DAE CSR, who helped me during different measurements.* **Dr. N. P. Lalla** *and* **Er. Preeti Mahajan** *are acknowledged for TEM measurements and fruitful discussions. I am thankful to* **Dr. V. R. Reddy**, **Mr. Anil Gome**, **Ms. Zaineb Hussain, Mr. Manik Kuila, Mr. Deepak** *for MOKE, Kerr microscopy, X-ray reflectivity, and GI-XRD measurements. I also sincerely thank* **Dr. Mukul Gupta**, **Mr. Layanta Behera**, **Ms. Jagriti Dwivedi, Ms. Nidhi Pandey, Mr. Prabhat Kumar** *and* **Ms. Niti dalal** *for sample preparation, SIMS and XRD measurements. I am thankful to* **Dr. D. M. Phase** *and specially to* **Dr. Satish Potdar** *for sample preparation and lab related fruitful discussions. I am also thankful to* **Dr V. Ganeshan** *and* **Mr. Mohan K. Gangrade** *for AFM measurements.* **Er. P.**

*Sarvanan* and his team are acknowledged for providing liquid nitrogen for performing temperature-dependent magnetic and structural studies.

I sincerely acknowledge **Dr. Pooja Gupta** (RRCAT, Indore) for MOKE measurements, guidance and support. I am thankful to **Dr. K. K. PANDEY** and **Dr. V. Srihari** (BARC, Mumbai) at BL-11 beamline, Indus-2, RRCAT for 2D-ADXRD measurement and Thanks to **Dr. Pragya Tiwari**, Synchrotron Utilization Division, RRCAT, for EDX measurements. I am thankful to **Dr. P. Nageswararao** for the AFM measurements at RRCAT.

**Prof. Stephan V. Roth, Dr. Matthias Schwartzkopf, Dr. Pallavi Pandit, Ms. Marie Betker (P03, DESY, Germany)** are thanked for help in GISAXS and GIWAXS measurements, useful discussions and insightful comments. I sincerely thank the Department of Science and Technology (DST), India, for providing travel support through the Jawaharlal Nehru Centre for Advanced Scientific Research (JNCASR),Bangalore, under India-German collaboration, for carrying out measurement at P03, PETRA III, DESY, Hamburg, Germany.

I would like to thank **Dr. V. Sathe, Dr. S. R. Barman, Dr. R. Rawat** and **Dr. D. K. Shukla** for fruitful discussions and for providing me a cheerful environment throughout my Ph.D. work.

I express my whole-heartedly thanks to **Dr. Sadhana Singh, Dr. Dhananjay Tiwari, Anup Kumar Bera, Md. Shahid Jamal, Avinash Khanderao, Sonia Kaushik, Sharanjeet Singh** and **Manisha Priyadarsini** who have been constant source of support throughout my Ph.D. work. I also thank them for their nice company and support and for maintaining the friendly environment in the lab. I express my sincere thanks to **Dr. Vaishali Pathak, Asst. Prof. Gagan Sharma**, **Dr. Pramod Vishwakarma** for providing constant technical support and guidance. I am blessed to have friends like **Tarachand, Vipin Kumar Singh, Rudra Prasad Jeena, Gyanendra Panchal, Mohit, Sumesh, Pooja, Prakash, Kaushik, Monika, Mukesh, Satyendra, Shailesh, Yogesh,**

*Hemant, Suman, Vinay, Vivek, Aakash, Megha, Anita, Indu, Sonali, Rakhul* and *Kali Prasanna Mondal* who were always there for me whenever I needed them and have devoted their precious time for me. I specially thank **Mr. Mohammad Balal (Ballu)** for standing beside me in each and every situation. I would like to thank **all the seniors, batchmates and junior research students** for their valuable support in one way or another during my entire research work. I specially thank my roommates **Shuvam, Pramod, Sajal** and **Bodhoday** for providing a wonderful environment. Sincere thanks are also due to the administrative personnel's and engineer staff, mainly **Er. Sanjay Singh Thakur** and **Er. Bhushan Jain** and the technical staff of UGC-DAE CSR, Indore, for rendering their help directly or indirectly.

I would like to thank the **Sports committee** of UGC-DAE CSR, Indore, for providing me the opportunity to explore extra-curricular activities. I sincerely thank **sub-inspector Avinash Gautam** and **Er. Rohit Pancholi** for being my fully motivated gym partner in Indore city. I would specially thank my friend **flight engineer Kamlesh Karki (Indian Air Force)** for always motivating me to achieve my goals.

Last but not least, I am extremely grateful to my Parents: **Shri. Prem Singh Dev**, **Smt. Nanda Devi,** my uncle: **Late. Ganga Singh Dev** and Aunty: **Smt. Devaki Devi** for their love, prayers, caring and sacrifices for educating and preparing me for my future. I am extremely thankful to my Joint family members, brothers: **Er. Dinkar Singh Dev, Mr. Kavindra Singh Dev, Asst. Prof. Laxman Singh Dev, Er. Pankaj Singh Dev;** Sisters: **Ms. Vimla Dev, Ms. Kanti Dev, Ms. Kastura Dev, Ms. Prema Dev;** Sister-in-law**: Smt. Kavita Dev;** nephew: **Bhawesh** and nice: **Bhawya** for being present with me during each step of this journey and making it wonderful.

*Arun Singh Dev*

# Preface

Magnetic anisotropy is a fundamental property of magnetic materials which determines the alignment of the spins along the preferential direction, called the easy axis of magnetisation. From the technological viewpoint, it is an important property of 2D magnetic materials, which often don't a has counterparts in the bulk material. Thin films, which possess spin alignment perpendicular to the film plane, i.e. perpendicular magnetic anisotropy (PMA), have been used in magnetic recording media such as hard disk drives (HDDs), whereas in-plane uniaxial magnetic anisotropy (UMA) is becoming increasingly important for applications in the rapidly developing field of spintronics. The key sources of magnetic anisotropy in a magnetic thin film are its (i) crystal structure, (ii) shape and (iii) strain. Anisotropy that arises from periodic lattice arrangement (crystal structure) is often realised in the epitaxial magnetic thin films such as Fe, Co, FeCo and SrRuO3 thin films deposited on substrates such as GaAs (001), MgO (100) and SrTiO3 (001). Here, spin-orbit interaction (SOC) is resulting preferred magnetisation direction with respect to the crystallographic structure of materials and giving rise to magnetocrystalline anisotropy (MCA). In epitaxial thin films, large PMA and in-plane are reported in several studies, which arises due to stress developed via lattice mismatch between substrate and thin film.

In the case of polycrystalline and amorphous thin films, MCA is not expected due to the random orientation of grains. Therefore, understanding the origin of UMA is generally difficult and can't be understood in terms of crystal orientation. The origin of UMA in polycrystalline films is often related to the preparation conditions and substrate properties. In some studies, the direction of deposition is found to have a definite relationship with the direction of the easy and hard axis of UMA in such a magnetic thin film. In a recent study by T. Kuschel et al., Co films of different thicknesses are deposited on glass substrates, where ex-situ investigation on a series of samples revealed that the preparation conditions like temperature, deposition rate, obliqueness of deposition, strain in the substrate and possible texturing, are not responsible for the origin of UMA. Although the substrate shape is reported to be responsible for the rotation of the anisotropy direction, the origin of UMA is not clearly understood.

On the other hand, in an in-situ thickness-dependent study of polycrystalline Co film on Si substrate, the appearance of UMA for thickness greater than 10 nm is attributed to the long-range internal stress developed during the film deposition. In beam-sputteredered nanocrystalline Co thin films, UMA exhibits a non-monotonic behaviour with thermal annealing, which is understood in terms of the variations in the internal stress, surface roughness, and grain structure. Similarly, in the case of ion beam-sputtered amorphous FINEMET film, stress is known to play an important role in the origin of UMA. Here, the internal stress in films is expected to couple with the magnetostriction, thereby giving rise to UMA. The origin of such stress in ion beam sputtered film is suggested to be related to the anisotropic ejection of the sputtered atoms from the target material. On the other hand, A. T. Hindmarch et al. have shown that UMA in amorphous CoFeB thin films is consistent with a bond-orientation anisotropy, which propagates the interface-induced UMA through the thickness of the interface.

A considerable amount of work has been done with an aim to understand the origin of UMA in polycrystalline and amorphous thin films, and most of the studies propose the stress (either growth induced or extrinsic) as a possible cause of the magnetic anisotropy present in the films. Still, the direct relationship between intrinsic stress and the magnetic anisotropy developed during the growth of the film is absent. In fact, several studies on stress development in thin-film have been carried out, where stress is found to vary significantly with film thickness. Also, the oblique angle deposition (OAD) technique can induce the UMA in such films. It is also a point of investigation that, in OAD, stress plays a role or not inducing UMA in such polycrystalline thin films, as this is not addressed in the literature to date.

In the present thesis, a detailed investigation of the origin of magnetic anisotropy in polycrystalline thin films is performed both in-situ and ex situ. During the in-situ process, UMA and stress measurements are performed in identical deposition conditions together with the transport measurements. Ex-situ measurements such as grazing incidence x-ray diffraction (GIXRD), reflection high energy electron diffraction (RHEED), transmission electron microscopy (TEM), x-ray reflectivity (XRR), atomic force microscopy (AFM), grazing-incidence small-angle x-ray scattering (GISAXS) and grazing incidence wide-angle x-ray scattering (GIWAXS) are performed for thorough study. We have divided our study into different sections to address the following points in the field of magnetic anisotropy thin film.

- Origin of magnetic anisotropy in polycrystalline and amorphous thin films.
- Internal stress during thin film growth and how it is related to magnetic anisotropy.

- Use of stress/strain in tailoring the magnetic anisotropy in polycrystalline thin films to get magnetic anisotropy according to the desired functionality.
- Role of stress in columnar growth to enhance the stress the induced magnetic anisotropy.
- Possibilities to control interface quality and magnetic properties in oblique angle deposited metal film on a thin organic layer to prepare high-quality organic spin-valve structures.

Co and FeCo polycrystalline thin films are selected as model systems for the present study to fulfil the objectives as mentioned earlier. In addition, a mini vacuum chamber, UHV heaters, and a mini evaporator for organic materials are designed and developed to perform detailed in-situ experiments at synchrotron radiation sources (Indus-2, RRCAT Indore and PETRA-III, DESY, Hamburg) and in the lab. The work in the present thesis is presented by dividing it into eight chapters, including one appendix. A brief summary of the individual chapter and an appendix are given in the following sections.

**Chapter 1:** The first chapter of the thesis presents a brief introduction to magnetic anisotropy and its importance in technological devices. A detailed review of the cause of UMA in crystalline, polycrystalline and amorphous thin films has also been provided. It also gives a brief description of the importance of an in-situ investigation to study UMA. Other deposition methods like OAD to get magnetic anisotropy have also been mentioned briefly. At the end of the chapter motivation and objective of the present thesis has been presented.

**Chapter 2:** A brief description of a versatile high vacuum chamber used for in-situ synthesis and characterisation has been provided. This chamber is equipped with a multi-beam optical stress sensor (MOSS) and magneto-optic Kerr effect (MOKE) measurements with simultaneous deposition and in-situ characterisations. A detailed description of the MOSS technique is also mentioned in this chapter. A brief description of a versatile UHV chamber used for in-situ synthesis and characterisation of ultra-thin films and multilayer structures is also included. Chamber is equipped with a facility for thin film deposition using electron beam evaporation and MOKE, XRR, RHEED and magnetoresistance (MR) techniques for in-situ magnetic, structural and transport characterisations. To perform in-situ x-ray scattering experiments, the whole set-up is attached with a lab x-ray source coupled with multilayer optics, which provides a beam of sufficient collimation and intensity. All the measurements can be done simultaneously during the deposition of the film or during annealing, thus making it possible to study the evolution of magnetic, transport and structural properties with film thickness under UHV conditions. The sample is attached to a 5-axis manipulator with an option to cool and heat the sample from 50 K to 1200 K. This chapter also includes a brief description of some of the ex-situ technique which has been used as complementary technique to the in-situ study, such as ion-beam sputtering (IBS) for deposition whereas x-ray diffraction (XRD), GIXRD, TEM, AFM, GISAXS and GIWAXS etc. for characterisation.

**Chapter 3:** This chapter is based on design and development. It describes a vacuum-compatible portable mini chamber's design and performance for temperature-dependent GISAXS and GIWAXS studies of thin films and multilayer structures. The water-cooled body of the chamber allows sample annealing up to 900 K using ultra-high vacuum compatible (UHV) pyrolytic boron nitride heater, thus making it possible to study the temperature-dependent evolution of structure and morphology of two-dimensional nanostructured materials. Due to its lightweight and small size, the chamber is portable and can be accommodated at synchrotron facilities worldwide. A systematic illustration of the versatility of the chamber has been demonstrated at beamline P03, PETRA-III, DESY, Hamburg, Germany. Temperature-dependent GISAXS and GIWAXS measurements were performed on oblique angle deposited Co/Ag multilayer structure, which jointly revealed that the surface diffusion in Co columns in Co/Ag multilayer enhances by increasing temperature from RT to ~ 573 K. This results in a morphology change from columnar tilted structure to densely packed morphological isotropic multilayer.

Also, this chapter includes the thin film heater developed in the lab for sample heating for various temperature ranges (320 °C and 500 °C, respectively). An organic material evaporator is also included, which is developed in-house, having a temperature range above 1000 °C.

**Chapter 4:** A detailed in-situ investigation to understand the origin of magnetic anisotropy in polycrystalline thin films, is presented in this chapter. For the case study, a cobalt thin-film system is selected as a model system due to its advantage of easy growth and high magnetocrystalline anisotropy. The growth of e-beam evaporated Co films on Si (001)/SiO2 substrate has been studied using real-time MOSS, four-probe resistivity (FPR) and in-situ MOKE measurements in-situ to understand the origin of magnetic anisotropy and correlate with the internal stress evolved during the film growth. The film grows via Volmer–Weber mechanism, where islands grow larger to impinge with other islands and eventually coalesce into a continuous film at around 10 nm thickness. The development of tensile stress in the film is found to be associated with the island coalescence process. The film exhibits a well-defined UMA, which varies synchronously with internal tensile stress developed during growth, suggesting that the observed UMA originates from the minimisation of magneto-elastic and morphological anisotropic energies in the presence of internal stress. In contrast to the ex-situ measurements, where thickness dependence involves growing a series of samples of varying film thicknesses, in the present study, by continuously monitoring real-time stress and magnetic hysteresis of a growing Co film in-situ, a complete thickness dependence curve is obtained in a single sample.

In the second part, to further understand and confirm the role of the stress on the origin of UMA, Co film is grown on Si/SiO2/Ag (10 nm) structure, where the Ag buffer layer was grown in the form of isolated islands. FPR measurements are used to confirm the island nature of the Ag film. Stress-thickness vs thickness plot obtained during Co growth on Ag islands clearly shows no significant change in the stress up to 15 nm Co thickness. In contrast to Co on Si/SiO2 sample, Co on Ag islands is magnetically isotropic. The absence of UMA in Co film having Ag underlayer in the form of islands is understood in terms of random short-range stress in the film caused by random pinning by the Ag islands.

Thus, the presented real-time measurements in this chapter show that in the case of polycrystalline films, stress plays a major role in defining magnetic anisotropy.

**Chapter 5:** In this chapter, the effect of externally applied stresses has been studied on polycrystalline Co films deposited on intentionally curved Si substrates. Tensile and compressive stresses of varying strengths were induced in the films by relieving the curvature. The strength of applied stress is correlated with the induced magnetic anisotropy and domain magnetisation during the magnetisation reversal process. It has been found that, within the elastic limit, the presence of stress leads to an in-plane magnetic anisotropy in the film, and its strength increases with increasing stress. The easy axis of magnetisation in the films is found to be parallel/ transverse to the compressive /tensile stresses, respectively. The origin of magnetic anisotropy in stressed films is understood in terms of magneto- elastic coupling, where the stress tries to align the magnetic moments in order to minimise the magneto-elastic as well as anisotropy energy. Tensile stress is also found to be responsible for the surface smoothening of the films, which is attributed to the movement of the atoms associated with the applied stress. Thus, this chapter provides a possible way to tailor the magnetic anisotropy and its direction in polycrystalline and amorphous films using controlled external stress.

**Chapter 6:** In this chapter, Iron-Cobalt (FeCo) columnar, multi-layered structure is prepared by depositing several thin FeCo layers by varying the angle between the surface normal and the evaporation direction as 0° (normal) and 60° (oblique), alternatively. The objective is to induce UMA in the FeCo layered structure by combining the shape and strain anisotropy contributions. Otherwise, FeCo films are magnetically isotropic in nature due to the absence of crystalline magnetic anisotropy. The correlation of the evolution of magnetic properties with that of morphology and structure of the multilayer was established by *in situ* X-ray scattering and MOKE measurements performed at room temperature and at elevated temperatures up to 450 °C. The strong shape anisotropy and compressive stress of nanocolumns in alternative FeCo layers resulted in a well-defined UMA with the easy axis of magnetisation along the projection of the tilted nanocolumns in the film plane. Stress in the film provides minimisation of magnetoelastic energy along the in-plane column direction, which couples with the columnar shape anisotropy energies and results in the preferential orientation of the easy magnetic axis along the OAD direction in the film plane. Drastic reduction in the in-plane UMA after annealing at 450 °C is attributed to the merging of columns and removal of stresses after heat treatment. It is demonstrated that the present method could be used to control and induce UMA in the FeCo layered structures. It provides a promising option for fabricating a pure and high in-plane UMA in films of increased thickness. The present chapter opens a new pathway to producing multilayer structures using single material and thus may have significant implications for future technological devices.

**Chapter 7:** Compared to the previous chapter, though this chapter is also based on OAD, but brings us some exciting and important results.

A systematic study of Co based multilayer structure [$Co_{oblique}$(4.4 nm)/$Co_{normal}$ (4.2 nm)] - 10 bilayers, where each alternative Co layer is deposited at oblique angle of 60°, has been done with an aim to induce high in-plane UMA and to enhance the thermal stability. Multilayer is grown in a UHV chamber using e-beam evaporation and characterised in-situ using MOKE and RHEED for their temperature-dependent magnetic and structural properties. Well-defined, strong UMA with the easy axis of magnetisation along the projection of the tilted nanocolumns in the film plane is observed. It is found to be present in the multilayer even after annealing at 500°C. To correlate the evolution of UMA with that of morphology and structure of the film, temperature-dependent GISAXS measurements were performed in-situ at a synchrotron radiation source, PETRA III, DESY, Germany. Observation of strong in-plane magnetic anisotropy in this multilayer is attributed to the combination of shape and MCA due to orientated/textured columnar growth at OAD. Crystalline texture in the film minimises spin-orbital coupling energy along the column direction, which couples with the columnar shape anisotropy energies and results in preferential orientation of the easy magnetic axis along the OAD direction in the film plane. Reduction in UMA after annealing is attributed to the merging of columns and removal of shape anisotropy after heat treatment.

**Chapter 8:** The overall conclusion of the research work carried out in this thesis has been given in this chapter. Further, the future scope of the studies presented in this thesis is also discussed in this chapter. Our further plans are - to further reduce the diffusion at the interfaces of organic spin-valve structures, tailoring magnetic anisotropy by controlling and aligning stress together with the structure of the material, to increase the high-temperature stability of magnetic anisotropy in OAD thin films by adequately choosing a nonmagnetic layer.

**Appendix:** Appendix describes our preliminary work on ferromagnetic/organic semiconductor thin films interfaces (Co/Alq3 and Co/Alq3/Co structures), where the OAD technique is used to achieve a good interface with the possibility of tuning the magnetic properties of the organic spin-valve systems.

# Contents

## Chapter 7: Oblique angle deposited (OAD) Nano-columnar layered structure for magnetic anisotropy    179-202

## Chapter 8: Conclusions and future Scope    203-210

## Appendix A: Spin valve structure -based on oblique angle deposition    211-227

This chapter presents a brief introduction to thin films and multilayers together with different types of magnetic anisotropies present. Then, a detailed description of the magnetic anisotropy phenomenon in polycrystal and amorphous thin films, including an understanding of its origin and applications, has been included. At last, the aim and scope of the present thesis have been discussed.

## 1.1  Magnetic nanostructure materials

Since its discovery in early times, magnetic nanostructure materials such as thin films and multilayers have found industrial applications with a rapid pace in optics and decorative purposes. This area has become one of the most widely and extensively studied research fields in magnetism as they possess remarkable properties that have both technological as well as fundamental significance [1,2,3 ,4]. Such kind of changes in properties often can't be found in the respective bulk counterpart as the thin films have reduced size and dimensionality, which provides a large surface-to-volume ratio [1,3,4]. In addition, their thickness can be varied to achieve various properties, and artificial structures can be produced that can't be prepared in a simple deposition process [2,3]. Such modifications have led to the discovery of various technology-changing phenomena such as perpendicular magnetic anisotropy (PMA) [1,3], increased magnetic moment at the interface [1,3], giant magneto-resistance (GMR) [5,6], exchange bias (EB) [7,8], exchange spring [9,10 ], tunnel magneto-resistance (TMR) [11], etc. However, a detailed understanding of magnetic behaviour is necessary for the creation and modification/tuning of such artificially created structures to achieve desired technological application. Because of these facts, we discuss various magnetic anisotropies and coupling in thin-film nanostructures.

## 1.2  Magnetic anisotropies

Magnetic anisotropy is one of the most important properties of thin films and multilayers. It has a wide field of applications. For example, when we talk about the applications of soft magnets, it includes static, low and high-frequency applications. Static applications include electromagnets, while  Low-frequency applications include inductors, transformers, motors, Magnetic amplifiers, generators, etc. High-frequency applications include chokes, high-frequency inductors and high-frequency transformers, pulse transformers and Microwave applications. Other very important applications are magnetic recording media, sensors and actuators.

Magnetic anisotropy is the direction dependence of any kind of magnetic property due to the difference of energy for the magnetic moment alignment along an easy and hard axis. Magnetic anisotropy determines the easy axis of magnetisation, which makes it a very important property in the field of magnetic thin films. The main source of magnetic anisotropies is either spin-orbit interaction or dipolar interaction. The type of magnetic anisotropy depends on the type of magnetic interaction and hence can be classified into the following four categories:

## 1.2.1 Magneto-crystalline anisotropy

Magneto-crystalline anisotropy (MCA), the intrinsic property of the thin film, is determined by the spin-orbit interaction of electrons in the magnetic material [12]. The spatial arrangement of the electron orbitals is strongly connected to the material's crystallographic structure. Here, electron spin couples with its orbit, which then couples to the lattice of the respective crystal. Resistance is observed from the lattice during the rotation of spins. Consequently, the interaction forces the spins to align along well-defined crystallographic axes. This interaction results in the easy and hard axis of magnetisation (preferential alignments of the spins) in the crystal structure of the magnetic material. The difference in energy for moment alignment along an easy and a hard axis or, in another way, the energy needed to overcome the coupling to rotate the spins away from the easy axis is known as magneto-crystalline anisotropy energy. The foldedness in the magnetic anisotropy is due to the symmetry of the lattice structure.

For a cubic crystal system, magneto-crystalline anisotropy energy is expressed in terms of $\theta$, which is the angle between the direction of the cubic axis and magnetisation [12] and expressed as

$$E = K_1\left(\alpha_x^2\,\alpha_y^2 + \alpha_y^2\,\alpha_z^2 + \alpha_z^2\,\alpha_x^2\right) + K_2\left(\alpha_x^2\,\alpha_y^2\,\alpha_z^2\right) + \dots \tag{1.1}$$

where $K_1$, $K_2$,…, and $\alpha_i$ etc., are the anisotropy constants and direction cosine of magnetisation with respect to crystal axes, respectively.

## 1.2.2 Shape anisotropy

Shape anisotropy is caused due to long-range dipolar interactions, which depend on the shape of a magnetised material. When a sample is magnetised, a magnetised body produces magnetic charges or poles at its surface. Surface charge distribution, in isolation, is itself a source of a magnetic field and produces a stray field. It then creates a field inside the sample known as the demagnetising field ($H_d$), which acts opposite the magnetisation. It is defined as $\overrightarrow{H_d} = NM$, where N is the demagnetisation factor and M is the magnetisation.

For a non-spherical material, e.g. long needle-shaped grain, The demagnetising field will be less along the long axis and high along the short axes. This results in the easy axis of magnetisation along the long axis while the short axis acts as the hard axis. While a sphere, on the other hand, possesses no shape anisotropy. The magnitude of shape anisotropy depends on the value of saturation magnetisation.

The equation can give shape anisotropy per unit volume as follow [13]

$$E = \frac{1}{2}\mu_0 M_s^2 \cos^2\theta \tag{1.2}$$

here $M_S$ and $\theta$ are the saturation magnetisation of the film and the angle of magnetisation with respect to the surface normal, respectively. $E$ is minimum for $\theta = 90^\circ$, i.e. when the moments lie parallel to the film plane.

### 1.2.3    Stress anisotropy

spin-orbit interaction is also responsible for magneto-elastic anisotropy (MEA), similar to MCA. MEA results from stress/strain in the crystal lattice. Strain causes the change in the distance between magnetic moments, which then changes the strength of the interaction.

For epitaxial thin films, stress generally results from lattice mismatch or the different thermal coefficients of the thin-film layer with the substrate or other layer. For an elastically isotropic medium, magneto-elastic energy is written as [12]

$$E = -\frac{3}{2}\lambda\sigma\cos^2\theta \tag{1.3}$$

here $\lambda$, $\sigma$ and $\theta$ are the magnetostriction constant, stress present in the film and the angle between the direction of stress and magnetisation, respectively.

In the case of polycrystalline and amorphous thin films, long-range atomic ordering is absent; hence the origin of MEA is often related to the substrate properties and preparation conditions. However, long-range stresses generally develop during thin film deposition and are attributed to causing the MEA in such thin films.

### 1.2.4    Surface/interface anisotropy

Interface or surface anisotropy is actually related to the symmetry breaking or reduced symmetry at the interfaces and surfaces of the thin films/multilayer. This, in turn, causes the effective anisotropy constant as a combination of volume ($K_V$) and surface anisotropy ($K_S$) as follows

$$K_{eff} = K_V + \frac{2K_S}{t} \tag{1.4}$$

here, t represents the thickness of the film. For bulk, the volume term dominates the MCA. For thin films, the surface anisotropy term is dominant as the surface-to-volume ratio increases as we go from bulk to thin film. As surface anisotropy overcomes the shape anisotropy, the thin-film system turns into perpendicular magnetisation.

## 1.3    Magnetic anisotropy in polycrystalline and amorphous thin films

In the field of magnetic thin films/multilayers, magnetic dipolar interactions and spin-orbit interactions are the two main causes responsible for in-plane uniaxial magnetic anisotropy (UMA). In the case of single crystals/epitaxial thin films, spin-orbit interaction (SOI) produces UMA, and that's why it is often observed in the epitaxial ferromagnetic thin films and is turned as MCA [14]. Grains are randomly oriented for the polycrystalline and amorphous films, hence due to the absence of long-range atomic ordering, magneto crystalline anisotropy is absent and therefore, the observed anisotropy can't be the magneto-crystalline in nature. In various studies, UMA has also been obtained in polycrystalline as well as amorphous films [15,16]. The origin of UMA in such films is often correlated to the deposition/preparation conditions and substrate properties, where long-range stresses develop in the film during deposition, and the presence of UMA is attributed to these long-range stress. Also, the direction of deposition defines the direction of the easy and hard axis of UMA in magnetic thin films [15]. The anisotropic ejection of the sputtered atoms from the target material relates to the long-range stress for the sputtered films [15,16]. When we pass through the literature, we find that the stress, present in the films, plays a significant role in the origin of UMA. But in amorphous and polycrystalline thin films, the origin of magnetic anisotropy and its correlation with the stress developed in the film during deposition is not clear. In literature, there are several studies where we find the combined investigation of the formation of intrinsic stresses and the evolution of surface morphology during the growth of the films and these intrinsic stresses are clearly correlated with the surface morphology [15,17]. But the proper correlation of the intrinsic stresses with magnetic anisotropy during the growth of the film is missing. The understanding of magnetic properties and their proper correlation with film properties like stress and morphology of the thin films are necessary for understanding magnetic anisotropy. It will also allow for modifying/tailoring the magnetic properties in polycrystalline and amorphous thin films.

**A brief review of the work already done in the field:**

The thin metallic films' properties, e.g., mechanical or any other physical properties, are strongly influenced by stress as excessive stress may approach the plastic-deformation limit. Excessive stresses may result in adhesion failure, and stress relaxation may cause alteration in the properties of the thin film during its service life [18]. More clearly, excessive stress can reduce the effectiveness of the thin film properties with time. In extreme cases, when the stresses are high enough, they are even found to be responsible for surface roughening, film cracking, peeling off of the films from the substrates or mass transport [19]. Depending on the deposition process, thin metallic films can have either tensile or compressive stresses. Typically, thin films deposited through evaporation possess tensile stress, mainly caused by grain boundaries of the thin films. However, when the growing film is bombarded with energetic particles during the deposition process, Compressive stresses may arise [18].

Stresses also cause a change in physical properties such as magnetic properties and mechanical hardness. Therefore, controlling the level and type of stress (compressive or tensile) during the deposition process is extremely important. It is also required to understand the compatible stress-relaxation processes that ensure that stress remains almost constant during the service life of the film with the specific environment (temperature, pressure, humidity etc.) [18]. Various research groups have recently shown considerable interest in inducing tunable UMA in different polycrystalline and amorphous thin films by intentionally applying controlled stress on the film-substrate system [20,21,22]. It is highly needed to obtain desired magnetic anisotropy. Below are some of the important studies done in this area:

- In 2016, R. Asai, et al. [23] deposited Ni thin films on a flexible polyethene naphthalate substrate and created a tensile strain of the order of a few percent which then switched the magnetic easy axis reversibly by 90° by the application of the applied stress. The easy axis was found to be orthogonal to the applied stress. The saturation magnetic field (in-plane) or the UMA energy changed linearly with respect to the applied tensile strain up to a strain of 2.3%. Also, They found a large difference up to ~ 0.3 T in the saturation magnetic field, which means a change of ~$7 \times 10^4$ J/m$^3$ in the magnetic anisotropy energy.

- F. Zighem, et al. (2013) [24] also deposited the polycrystalline Ni thin film deposited on the flexible substrate. They applied the voltage to study voltage-induced magnetic anisotropy quantitatively. The film/substrate system was attached to a piezoelectric actuator to induce anisotropy. They controlled the anisotropy through the applied elastic strains by the compliant elastic behavior of the substrate. They also found the non-linear variation of the resonance field as a function of the applied voltage, and they also reproduced it by taking into account the nonlinear-hysteretic variations of the induced in-plane strains.

- F. Zighem, et al. (2014) [25], deposited FeCuNbSiB (Finemet) film on flexible Kapton and studied using the micro-strip ferromagnetic resonance technique. They showed that the flexible substrate allows a good transfer of elastic strains produced by a piezoelectric actuator. They also controlled the magnetic anisotropy of the film by applying relatively small actuator voltages.

- A. Brandlmaier, et al. (2007) [26], manipulated the magnetic anisotropy of a crystalline $Fe_3O_4$ film in-situ with the application of tunable stress applied through a piezoelectric actuator system.

- K. An, et al. (2016) [27], demonstrated that charge current can effectively control the spin precession resonance in an $Al_2O_3$/CoFeB/Ta heterostructure. Numerical simulations concluded that the anisotropic stress produced by Joule heating plays an important role in controlling $H_k$. The results in this study provided new insights into the magnetic properties manipulation through current and have wide implications for spintronic devices.

- In 2014, D. Kumar, et al. [28] performed an in situ surface magneto-optical Kerr effect (MOKE) study for the ion-beam-sputtered amorphous $Fe_{73.9}Cu_{0.9}Nb_{3.1}Si_{13.2}B_{8.9}$ (FINEMET) ultra-thin films with film growth, ranging from a fraction of an nm to a few tens of nm. The magnetic dead layer was found to be absent at the substrate/FINEMET film interface, which means the absence of any intermixing. In this study, the attribution for the origin of UMA in the as-prepared stage is given to some long-range stresses generated in the film, which relieves near the onset temperature for nanocrystallisation.

- In 2016, a group of researchers, D. Chien, et al. [29] found that in the case of perpendicular magnetic tunnel junction (MTJ), in comparison to the current-controlled magnetisation switching, voltage- or electric field-induced magnetisation switching reduces the writing energy of the memory cell, which in turn results in increased memory density. In that work, an ultrathin PZT film with a high dielectric constant was deposited into the tunnelling oxide layer to enhance the voltage-controlled magnetic anisotropy (VCMA) effect. These fabricated MTJs with MgO/PZT/MgO barrier demonstrated a VCMA coefficient, which was ~40% higher than the control sample MTJs with a MgO barrier.

- A research group of S. J. Kernion, et al. (2012) [30], developed induced anisotropies, $K_u$, as huge as $1.89 \times 10^4$ $J/m^3$, for Co-rich nanocomposite alloys through strain annealing. These alloy systems have large negative magnetostrictive coefficients in Crystalline phases, which leads to an anisotropy field per unit stress twice as compared to those developed in FINEMET.

- S. Saranu, et al. (2011) [31], used magneto-elastic interactions to modify the total effective magnetic anisotropy of the nanoparticles. Large biaxial stress was applied on the nanoparticles embedded in a non-magnetic film, and this, in turn, demonstrated significant modification in the magnetic properties. In this study, stress-induced magnetic effects is presented through experimental evidence, which is based on temperature-dependent magnetisation curves of superparamagnetic Fe particles.

- In 2013, Y. Dusch, et al. [32] presented the implementation of a magnetoelectric memory. They used a ferroelectric relaxor substrate for the voltage-driven writing method. While applying a voltage across electrodes, this ferroelectric relaxor substrate was used to create anisotropic stress in the magnetic element. Applied voltage-controlled the effective anisotropy of the magnetic element because of the inverse magnetostrictive effect, and This voltage was used to switch magnetisation from one state to the other.

- In 2012, Z.K. Wang, et al. [33]deposited FeCo thin films using electrochemical deposition on the curving substrates. After the substrates are flattened, The compressive stress was produced in the as-deposited films. It was found that stress induces magnetic anisotropy, and the resonance frequency increases with the increase of substrate curvature. The induced magnetic easy axis was found to be perpendicular to the compressive stress direction.

- In 2016, X. Qiao, et al. [34] used flexible polyimide (PI) substrates to investigate the magnetic properties of amorphous $Co_{40}Fe_{40}B_{20}$ (CoFeB) thin films. What they have found is a uniaxial magnetic anisotropy in the CoFeB film due to the magnetostrictive effect because the flexible substrates were flattened from a convex state after fabrication. This research group also found the thickness of film and substrate changes the anisotropy.

**Noteworthy contributions in the field:**

➢ In 1997, H. Deng, et al.[35], represented a new method to investigate internal stress effects on magnetic properties of thin films. A series of oxidised silicon (111) substrates were already prestressed to different degrees. They deposited FeTaN film on such substrates and found that anisotropic stresses with different amplitude transfer after releasing the prestressed film. They found that the induced stress anisotropy strongly affects the anisotropy field ($H_k$), saturation magnetostriction ($\lambda_s$), initial permeability ($\mu_i$) as well as the orientation of the easy axis of FeTaN thin films.

➢ D. Hunter, et al. (2011) [36] found giant magnetostriction in $Co_{1-x}Fe_x$ alloy. This group observed that structural and chemical heterogeneity and also the interaction of coexisting phases can lead to extraordinary behaviours in oxides, as found in relaxor ferroelectrics and piezoelectric materials at morphotropic phase boundaries. However, such phenomena are very rare in metallic alloys. Effective magnetostriction $\lambda_{eff}$ as large as 260 p.p.m. has been achieved at a low saturation field of ~10 mT by tuning the structural heterogeneity in textured $Co_{1-x}Fe_x$ thin films. It is observed through microstructural analyses of $Co_{1-x}Fe_x$ films that maximal magnetostriction occurs for the compositions near the (fcc + bcc)/bcc phase boundary and originates from an equilibrium Co-rich fcc phase embedded in a Fe-rich bcc matrix.

➤ In 2013, X. Zhang, et al., [37], used $Ir_{20}Mn_{80}$ as the antiferromagnetic layer and magnetostrictive $Fe_{81}Co_{19}$ alloy as the ferromagnetic layer on polyethylene terephthalate substrates to fabricated flexible exchange biased heterostructures. With the magnetic field parallel to the pinning direction, under a compressive strain, a drastic decrease in the exchange bias field was observed, but under a tensile strain, only a slight decrease was shown. Based on the Stoner-Wohlfarth model calculation, they suggested that the distributions of both antiferromagnetic and ferromagnetic anisotropies are the keys to induce the mechanically tunable exchange bias.

➤ In 2013, G. Dai, et al. [22] proposed a method to induce a uniaxial magnetic anisotropy in magnetostrictive $Fe_{81}Ga_{19}$ films deposited on flexible PET substrates by curving the substrate prior to deposition. Changes in the shape of PET substrates induce internal stress on the $Fe_{81}Ga_{19}$ films, exhibiting a significant uniaxial magnetic anisotropy. The easy axis was observed along the tensile stress direction, and the increase in the internal tensile stress resulted in the increase in the coercive field along the easy axis.

➤ W. Karboul-Trojet, et al., (2012) [38], studied the effect of external stress on the domain configuration of a NiFe film. The interesting thing in this study was that this film was "obliquely" deposited on a compliant polyimide substrate.

## 1.4    Methodology used during the tenure of the research work:

The present research work is proposed with an aim to the studied role of stress/strain in tailoring the magnetic and transport properties of polycrystalline and amorphous thin films and multilayers. Therefore, the stress evolution during film growth is examined using the real-time curvature-based multi-beam optical stress sensor (MOSS) technique, whereas in-situ MOKE measurements are used to study the evolution of the magnetic properties. We have provided direct evidence of stress-induced uniaxial magnetic anisotropy in such nanostructures by combining in-situ MOKE and real-time MOSS measurements. The proposed work has provided a basic understanding in this area, which may help in tailoring the properties according to the desired functionality. Furthermore, other measurements like atomic force microscopy (AFM), grazing-incidence small-angle x-ray scattering (GISAXS), x-ray diffraction (XRD), grazing incidence x-ray diffraction (GIXRD) etc. are performed for the complete study of the system under investigations.

The proposed study is covered in three parts:

### 1.4.1  Study of stress during growth:

Growth of e-beam evaporated polycrystalline magnetic films taking Co thin films as a model film on Si substrates is studied using MOSS, four-probe resistivity (FPR) and in-situ MOKE measurements. While MOSS and FPR provide information about internal stress and film morphology, MOKE provides information about magnetic properties, thus making it possible to correlate the evolution of film stress with that of morphology and magnetic properties of the film. The deposited films in the present study grow via Volmer–Weber mechanism, where islands grow larger to impinge with other islands, and percolation transition occurs around some particular film thickness, which eventually coalesces into a continuous film. It is important to study long-range tensile stress generated in the film during island coalescence, which may show some trends with increasing thickness. Simultaneous MOKE measurements as a function of azimuthal angle have provided information on uniaxial magnetic anisotropy in the thin film and have been understood in terms of long-range tensile strain-induced as a result of the coalescence of islands. The evolution of magnetic anisotropy with film thickness is understood in detail by taking magneto-elastic coupling into account, where the stress tries to align the magnetic moments to minimise the magneto-elastic as well as anisotropy energy.

### 1.4.2  Controlling magnetic properties by external stress:

Magnetic materials having inverse magnetostrictive effects have been studied extensively. It has been found that stress and stress anisotropy (magneto-elastic anisotropy) can significantly affect the magnetic properties of such magnetic materials. The internal stress caused by the lattice mismatch in epitaxial thin films is an essential factor affecting magnetic anisotropy. While in the case of polycrystalline and amorphous films, the origin of such magnetic anisotropy is often considered to be related to the residual stresses in the thin films, which might have evolved during the film growth process. It is to be noted that such magnetic anisotropy in polycrystalline and amorphous films is not only unpredictable but also found to be weak in strength and depends on the various deposition parameters such as vacuum during deposition, deposition rate, method of deposition, stray magnetic field, etc. In extreme cases, such stresses are found to be responsible for surface roughening, peeling off of the film from the substrates, film cracking or mass transport. Despite these disadvantages, there are still many opportunities for polycrystalline and amorphous magnetic thin films for fundamental studies of stress-dependent properties if the stress in these films can be controlled artificially.

In the present thesis work, we have induced tunable magnetic anisotropy in polycrystalline films using intentionally applied controlled stress to achieve desired magnetic anisotropy in such films. Stress in the film has been induced in-situ during growth by applying curvature to the substrates.

In the present work, polycrystalline Co (taken as a model thin film) films with varying compressive and tensile stress strengths have been prepared on the silicon substrate. The strength of applied stress has been correlated with the induced magnetic anisotropy and domain magnetisation during the magnetisation reversal process. Present work has provided a clear understanding and an effective way to produce desired magnetic anisotropies even in polycrystalline thin films, where crystalline anisotropy is absent.

### 1.4.3   Role of stress in the obliquely deposited thin film:

The magnetic properties of thin films also depend on their morphological behaviour and microstructures. For example, a thin film prepared with oblique-angle (or glancing angle) incidence forms a tilted nanocolumn structure, which is very much different from a continuous thin film grown under normal incidence. Generally, oblique incidence deposition results in the formation of grains in the plane of the film that these grains are elongated perpendicular or parallel to the incident flux direction depending upon the deposition angle and with an increasing aspect ratio at a larger deposition angle with respect to the surface normal. As a result, an in-plane UMA with an easy axis perpendicular or parallel (depending upon the deposition angle) to the plane of the incident flux (perpendicular to the film surface) induces the growth of the magnetic thin films. Thus, shape anisotropy is important in determining the magnetisation reversal in oblique depositions.   In contrast to the earlier work in this area, where the origin of magnetic anisotropy in the obliquely deposited magnetic film is understood and attributed to the shape and magnetocrystalline anisotropy, the present work has explored the role of stress in the origin of magnetic anisotropy in obliquely deposited polycrystalline thin films. Understanding the development of stress and the influence of stress on film growth during oblique incidence is very important because of its ability to control and manipulate magnetic anisotropy and surface morphology. In the proposed work, "GIXRD measurements are carried out to get information about microstructure as well as the growth-induced long-range stress in the FeCo and Co thin films and multilayers. By analysing the relative peak intensities in the diffraction pattern at different q's, the magnitude and the preferred structure orientation have been determined with high accuracy. The same can also be correlated with magnetic properties.

## 1.5    Aim and scope of present thesis:

Although the proposed work is complex in terms of preparation and characterisation, the versatile in-situ MOKE and MOSS characterisation facilities aim to obtain genuine properties of the thin films and relate the stress generation to the film properties. The combined in-situ characterisation techniques have been used in detail to study surface, interface and morphological properties. Such combined in-situ measurements play a promising role in the present study and provide a basic understanding of the area of stress-induced magnetic anisotropy, which also helps in tailoring the properties according to the desired functionality.

In the case of oblique angle deposition, one of the main aims is to find if the stress is present or not. If present, then how does it affect the film properties? It is expected that shape anisotropy, magnetocrystalline anisotropy (due to possible texturing), and stress-induced anisotropy can be tuned by varying oblique angles and thickness of the magnetic thin films. Initial preliminary measurements done by us using the MOKE showed strong in-plane uniaxial magnetic anisotropy, where the direction of the easy axis of magnetisation and strength had a definite relationship with the sample dimension and oblique angle kept during deposition, which motivated us to study this field in detail. The study of magnetic properties using MOKE on the same samples allow us to understand the role of stress and microstructure in the evolution of the magnetic anisotropy in the obliquely deposited film and multilayers. GISAXS, AFM and X-ray reflectivity (XRR) measurements are done to find the preferential orientation of morphology, thickness, roughness, and density. The same has been correlated with the observed magnetic anisotropy in the films. Variation of in-plane UMA with the structure and stress modification makes it possible to get insight into the tailoring of magnetic anisotropy in these systems.

# CHAPTER 2 — Experimental Techniques and Methodology

This chapter discusses various experimental techniques involved in the deposition and characterization of thin films and multilayers during this thesis work. In addition, a detailed overview of the ultra-high vacuum chamber used for in-situ characterizations has also been discussed.

## 2.1    Introduction

Various polycrystalline thin films and multilayers have been prepared to examine the possible cause of magnetic anisotropy in the present thesis. The present chapter brings a detailed overview of the working principles and theory of various preparation and experimental characterization techniques used for different samples studied in the thesis work. Section 2.2 provides details of the various methods used for thin film and multilayer deposition, whereas various structural, morphological and magnetic characterization techniques are discussed in section 2.3. In the present thesis work, many samples are studied in-situ under high and ultra-high vacuum (UHV) conditions. Therefore, details of the UHV chamber together with all the deposition methods and characterization techniques, are also presented in section 2.4. The same section also provides a discussion on the advantages of in-situ studies over ex-situ studies.

## 2.2    Thin-film deposition techniques

Thin-film technology is known to be one of the oldest arts and the newest sciences [1] in the field of condensed matter physics. Growth of a thin film persists in three major steps: (i) production of the depositing species (atomic, molecular or ionic) by respective suitable techniques, (ii) transport of these species to the substrate and (iii) condensation on the substrate. There are various thin film deposition processes that can be categorized into major sections: physical vapor deposition (PVD) and chemical vapor deposition (CVD) processes. A PVD process is fundamentally a vaporization sputtering coating technique involving the transfer of material on an atomic level. Usually, depositions are carried out at lower temperatures and without corrosive products. These techniques are precise, and the depositions are done at lower rates. However, residual stresses are always present in the films irrespective of deposition techniques.

In CVD systems, gases are heated by thermal energy in the coating chamber and carry out the deposition reaction [2], but the film grows at high temperatures, which leads to the formation of corrosive gaseous products, which may produce impurities in the film. Thus for growing metallic multilayers with low rms roughness and periodicity of few nm, the PVD techniques are well-suitable and thus commonly used. Further, PVD processes can be sub categorized and classification of these processes is shown in fig. 2.1.

In the present work, magnetic thin films and multilayers are prepared using **Electron beam evaporation** (EB) and **Ion-beam sputtering** (IBS) methods. These techniques are discussed in detail in the following sub-sections.

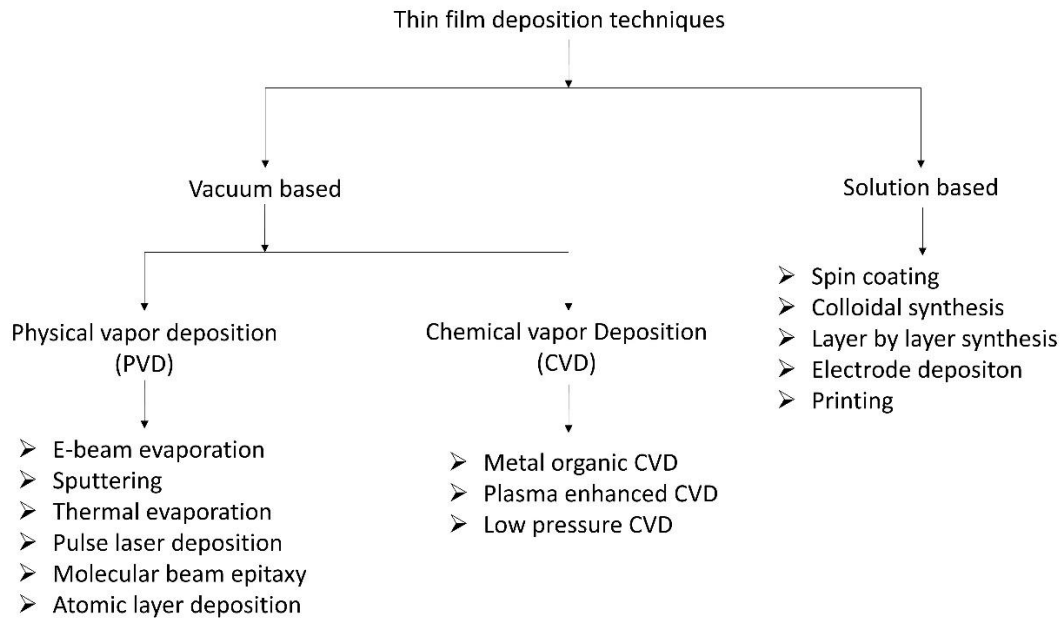

**Figure 2.1:** classifications of thin vapor deposition.

## 2.2.1    E-beam evaporation

Electron beam (E-beam) evaporation is one of the widely used PVD techniques [3,4]. This technique can yield a high deposition rate ranging from 1 Å/min to thousands of Å/min at relatively low substrate temperatures than other deposition techniques. The E-beam process offers enormous possibilities for controlling thin film morphology and structure with desired physical properties such as low contamination, high thermal efficiency, dense coating, high reliability and high productivity. The electrons are produced through thermionic emission using tungsten filament by passing electric current in this evaporation process. The electrons are then accelerated with the help of high voltage (3-40 kV) to achieve the required high kinetic energy. The magnetic field generated from a permanent magnet or electromagnet focuses these accelerated electrons on the target (or source) material. This E-beam transfers its kinetic energy to the target material, which heats it. Suppose the kinetic energy is large enough (by providing enough voltage to the filament which produces E-beam) to reach the sublimation temperature.

Then, it evaporates the material and deposits it on the substrate above it. Crucibles need to be water-cooled to provide local heating of the target material. Otherwise, excessive heating can contaminate the target material due to intermixing between evaporant target material and the crucible. The minimum vacuum required in the chamber to operate an electron gun is about $10^{-5}$ Torr/mbar [5]. Vacuum below this value during e-beam gun operation can damage e-gun filament through oxidation at filament surface. Figure 2.2 gives a schematic of the e-beam evaporation set-up.

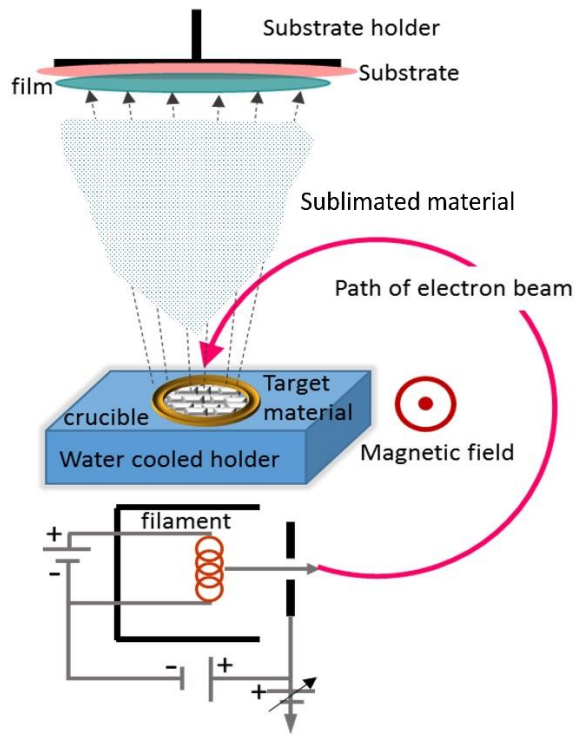

**Figure 2.2:** Schematic of e- beam evaporation set-up [6].

## 2.2.2    Thermal Evaporation

In the thermal evaporation methods, the target material is placed in a crucible made of another refractory material. Figure 2.3 shows the crucible heated to vaporize the target material.

In the resistive heating method, the target materials are placed on a material "boat" and heated by the resistive heating electrodes. We can manage the rate of evaporation by just adjusting the current supply. Though the set-up is simple, this heating method has a few disadvantages. First, the contamination from the degassing or evaporation of the boat and circuitry during the evaporation process at high temperatures. Second, it is sometimes difficult to find a suitable boat material, especially when evaporating refractory materials. Thus, it is clear that the boat must be made of materials with a much higher melting temperature than the source materials they carry. In various cases where a nonsuitable boat material is chosen, excessive heating will cause the boat to melt or break apart. Third, the alloying and intermixing problem may happen between the target and the boat material. This can result in the poor stoichiometry of the evaporated film.

So, to conquer some of the problems involved with the resistive heating method, inductive heating can be done. In this method, the source materials are placed in a crucible (fig. 2.2), having water-cooled radiofrequency (RF) inductive coils winding around it for power supply. The thermal evaporation process is usually performed in a vacuum of $\sim 10^{-6}$ mbar/Torr. We use a different thermal evaporator made of Pyrolytic boron nitride material to evaporate organic material Alq$_3$. A detailed design description is given in chapter no. 3.

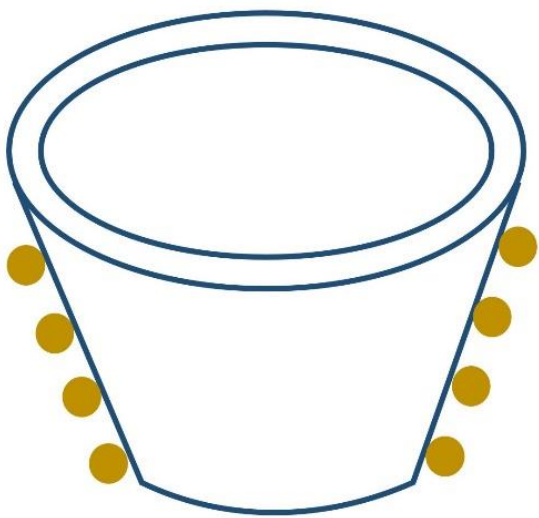

**Figure 2.3:** Schematic of the crucible with heating wire for thermal evaporation set-up.

## 2.3    Characterization techniques

### 2.3.1    Magneto-optical stress sensor (MOSS): laser-based reflection technique

Deposition of films is a non-equilibrium process, which may result in the material deposited may not acquire its most stable state. Such situations may be helpful; several metallic films would ball up if these films could achieve their equilibrium structure. Such a situation can arise if they are heated to high temperatures. However, the non-equilibrium state also indicates that the atoms in the films are not in their respective relaxed positions, which is responsible for the stress in the layer. During the fabrication of thin-film systems, both intrinsic (appearing during film growth) and extrinsic (i.e. appearing after film growth) residual stresses may occur. Such induced residual stresses can strongly affect the functional properties and the mechanical reliability of these thin-film systems. Also, excessive deformation caused by the residual stresses can drop the mechanical stability of thin-film systems during operation, leading to spallation and fracture. Furthermore, electronic and optical performances of devices (e.g. lasers and solar cells) reduce due to electrons and phonons scattering at defects caused by stresses. On the contrary, the electronic, magnetic, mechanical, and optical properties of various thin-film systems can also be benefitted by applying controlled film stresses.

To date, the potential to optimize the properties of thin films by tailoring their microstructure together with the stress state is still confined. It results in the lack of fundamental understanding of the stress development mechanisms and their proper correlation with the microstructure developed during thin film growth [7].

Multi-beam optical stress sensor (MOSS) technique is a reflection-based technique of the laser spots array from the surface of the specimen and works on the Stoney equation [8]. It is well known that the stress in a thin film is responsible for the curvature induced in the substrate. Therefore, according to the Stoney equation, curvature change (k - $k_0$) can be related to the film stress ($\sigma$) using as given below-

$$k - k_0 = \frac{6\sigma h_f}{M_s h_s^2} \qquad (2.1)$$

Here, $k_o$ and k are the curvatures of the unstressed and after stressed substrate, respectively. The other parameters are the film thickness ($h_f$), the substrate thickness ($h_s$) and the biaxial modulus of the substrate ($M_s$). $1/k$ is the radius of curvature R. The biaxial modulus of the film can be expressed in term of bulk modulus (E) by taking the Poisson ratio ($v$) in the equation as follow:

$$M_s = \frac{E}{1-v} \tag{2.2}$$

Thus, precise curvature measurement is the main point of this type of stress measurement technique. A schematic of the MOSS set-up for measuring sample curvature is shown in fig. 2.4. In the MOSS technique, the sample is targeted by an array of parallel laser beams. The main part of the measurement is to determine the spacing of the adjacent laser beams reflected from the sample surface when they reach the CCD camera. The beams will have unchanged spacing before and after being reflected on the flat surface. But for the curved surfaces, the laser beams will be deflected from the original path, and the spacing of the adjacent beams on the CCD will be changed, i.e. either it will increase or decrease.

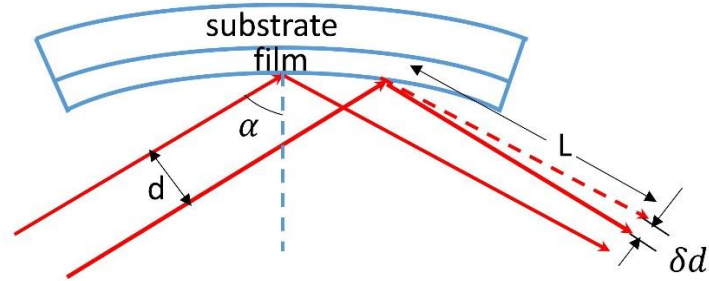

**Figure 2.4:** Laser beam reflection from a curved surface.

A CCD camera monitors the corresponding change in the spacing of the adjacent reflected beams arising from stress-induced curvature. In this measurement, the spacing between the adjacent beams directly measures the produced substrate curvature during film deposition. As shown in fig. 2.4, the stress-generated curvature change of substrate is related to the deflection of the beams according to the relation below:

$$\kappa - \kappa_0 = \left( \frac{\delta d}{d_0} \right) \frac{\cos \alpha}{2L} \tag{2.3}$$

$\delta d$ is the change in beam spacing due to the stress-induced curvature while the initial spacing

was $d_0$. The ratio $\delta d/d_0$ is the differential spacing. The substrate to camera distance is L, and the angle between the substrate normal and incident beam array is $\alpha$. Combination of this eq. 2.3 with the Stoney formula (eq. 2.1) shows that the stress is directly proportional to the deflections of the beams, as follows:

$$\sigma\, h_f = \left(\frac{\delta d}{d_0}\right) \frac{M_z h_s^2 \cos\alpha}{12L}$$

(2.4)

We obtain the respective thin film's stress-thickness vs thickness plot using this relation. A set of two etalons is used in MOSS to get a laser spot matrix. Below fig. 2.5 shows the schematic of the etalon functioning. Using this set-up, the maximum radii of curvature greater than 50 kM can be obtained.

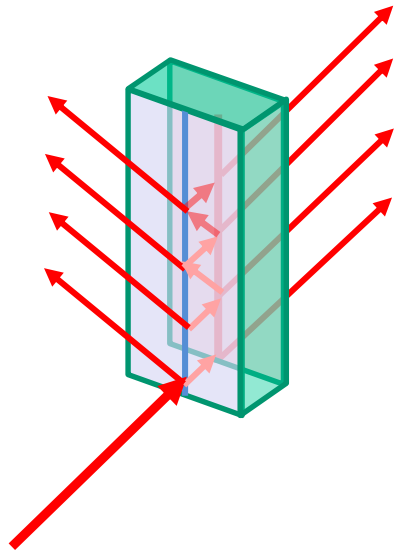

**Figure 2.5:** Etalon system provides multiple parallel laser beams to form a laser matrix.

## Advantages:

➢ This technique is based on the measuring spacing between the reflected laser beams; therefore, any tiny vibration in the system doesn't affect our results.

➢ Compared to the XRD technique, MOSS is a real-time stress measurement technique performed during deposition. In addition, MOSS provides stress variation in the thin film systems with increasing thickness, while XRD is a time-taking technique and is inadequate for in-situ measurements.

## Sensitivity

This MOSS set-up is sensitive to both film and substrate thickness. The determination of the smallest value of stress that can be measured should be specified relative to the film thickness, as the obtained curvature measures the product of stress and thickness. So, a film with a stress value of 1 GPa will generate the curvature1000 times more for a film thickness of 1 µm than it will for a film thickness of 1 nm. Also, the curvature to stress ratio is inversely proportional to the square of the substrate thickness. It means a film that generates a 10 m radius of curvature on a 100-micron thick substrate will induce a 1 km radius of curvature on a 1 mm thick substrate.

## MOSS assembly

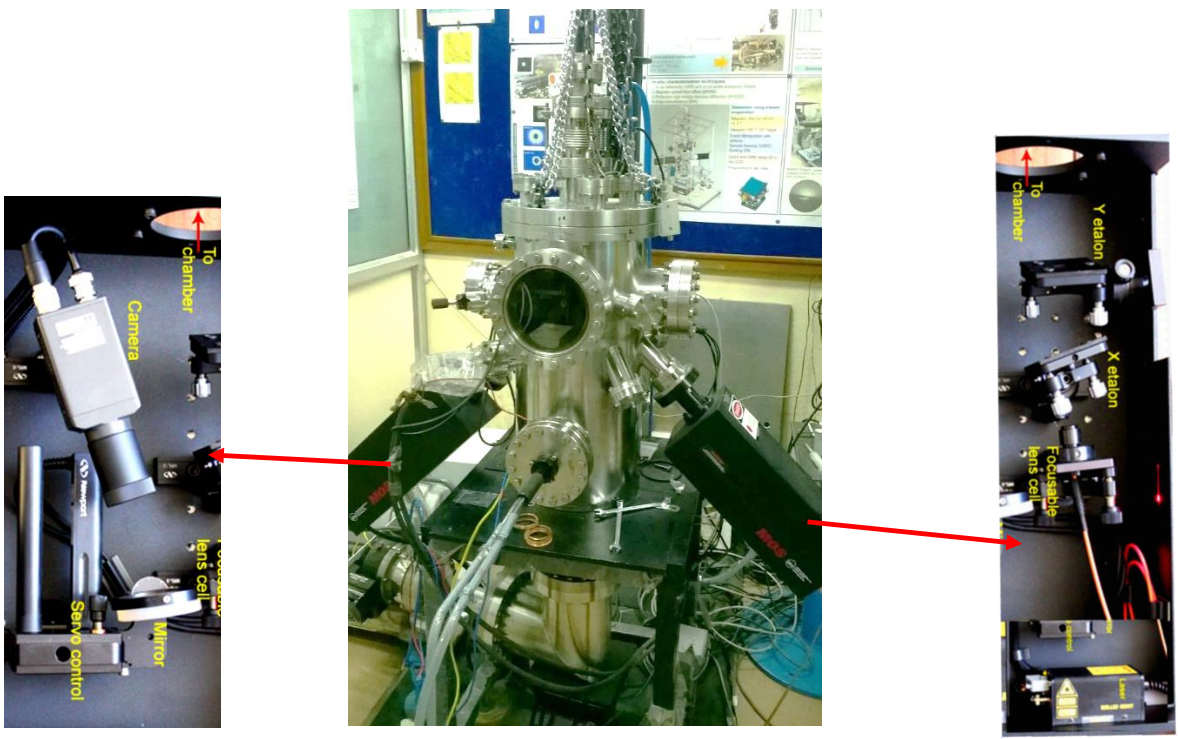

**Figure 2.6:** Actual view of the MOSS set-up.

The photograph of the MOSS chamber is shown in fig. 2.6. The schematic view shows all the assembly inside and outside the chamber. The sample holder attaches the substrate to a freely floating state. Contact pads, e.g. gold contact pads, can be applied for simultaneous resistance versus thickness and stress measurements. MOSS set-up (laser side and detector side) is attached to the e-beam chamber at the window, which provides the 45º angle to the laser and detector from the substrate normal. The crucible for material deposition is at the bottom of the MOSS chamber. The substrate is attached to rotating feedthrough for MOKE measurement. Figure 2.6 also shows the inside assembly of the laser and detector sides of the MOSS set-up. The laser side shows the laser and etalon assembly, while the detector side shows the servo motor-controlled mirror and the CCD camera. Servo-controlled mirrors compensate for the drift of the reflected beam induced in the substrate holder if heated.

### 2.3.2    Magneto-optical Kerr effect (MOKE)

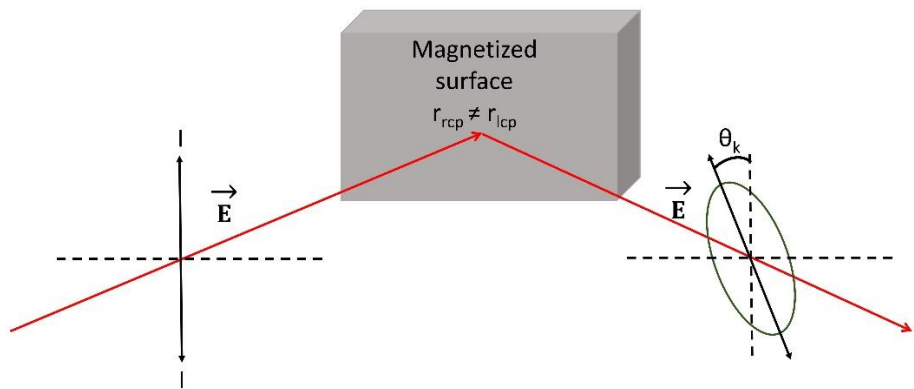

**Figure 2.7**: Schematic illustration of Kerr effect.

In 1877 John Kerr modified the polarization state of light with a magnetized metallic iron mirror. This effect in the reflection of light is presently known as the magneto-optical Kerr effect (MOKE), which is proportional to the magnetization M of the sample. Magneto-optic Kerr effect is considered one of the most prominent techniques to visualize the magnetic property of magnetic thin films and multilayers. This technique/phenomenon results from the polarization and intensity changes of the laser light after reflecting from the surface of a chosen magnetic material [9,10] means it is the consequence of the optical anisotropy of the materials. Change in the polarization of the light can take place by (a) rotation of the plane of polarization, i.e. Kerr rotation ($\theta_k$), and (b) formation of ellipticity ($\varepsilon_k$) in the linear polarized light. These two effects provide information on the magnetic behaviour of the sample (fig. 2.7). Various theories provide the origin of the Kerr effect as discussed below:

## Physical origin

The classical motion of an electron in a medium provides the physical origin of MOKE. Let us suppose a linearly polarized light having polarization and propagation along the x and z-axis, respectively, incident on a medium (fig. 2.8). The electric field of the light leads to the motion of electrons in the medium. The linearly polarized light is made of the superposition of equal amplitudes of left circularly polarized (LCP) and right circularly polarized (RCP) light. In the absence of an externally applied magnetic field, the expression of linear polarized light can be written as [11]:

$$E^{inc} = \frac{E_+^{inc}}{2}\left(\hat{e}_x + j\hat{e}_y\right)e^{j(wt-kz)} + \frac{E_-^{inc}}{2}\left(\hat{e}_x - j\hat{e}_y\right)e^{j(wt-kz)} \qquad (2.5)$$

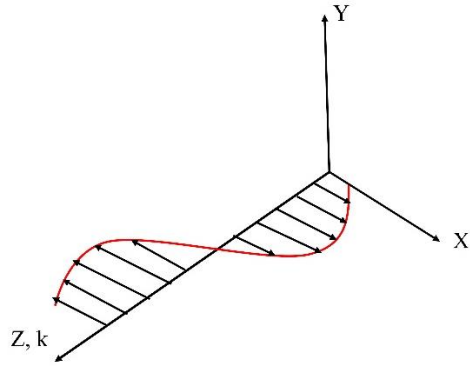

**Figure 2.8:** Linearly polarized light with polarization and propagation along +X and +Z directions respectively [11].

Electric fields corresponding to RCP and LCP drive electrons to the right and left circular motions, respectively. When no magnetic field is present in the system, the radii of both circular motions will be equal. As electric dipole moment is proportional to radii of circular motion, the dielectric constants of LCP and RCP will be the same. However, in the presence of an external magnetic field, additional Lorentz force is witnessed by electrons, which in turn, affect the radii of both left and right circular motions. The direction of force points in opposite directions, i.e. towards and away from the centre for LCP and RCP parts, respectively. Hence, the radius of left and right circular motions decreases and increases, respectively. This results in the finite difference in respective dielectric constants and refractive indices [12]. Respective refractive indices of LCP and RCP in the presence of a magnetic field are given as

$$n_{\pm}(\omega) = n(\omega \pm \omega_L) \tag{2.6}$$

here, '+' and '-' refers to RCP and LCP parts, respectively. Respective reflected amplitudes are given by [13]

$$E_+^{ref} = \frac{n_+ - 1}{n_+ + 1} \; E_+^{inc} \; \text{ and } \; E_-^{ref} = \frac{n_- - 1}{n_- + 1} \; E_-^{inc} \tag{2.7}$$

These different amplitudes result in the elliptically polarized light, which in turn results in the appearance of the Kerr effect. The angle between the linear polarization axis and the major axis of ellipse polarization gives Kerr angle $\theta_k$.

**Macroscopic electromagnetic/dielectric theory**

The macroscopic theory of the Kerr effect is expressed as dielectric tensor theory. 3×3 dielectric tensor $\varepsilon_{ij}$ with i,j=1,2,3 provides the dielectric property of a medium. This tensor can be further decomposed into antisymmetric and symmetric terms. The symmetric part of the tensor can be diagonalized on the appropriate rotation of coordinates. For the isotropic medium, the dielectric tensor simply reduces to the dielectric constant as all three Eigenvalues are the same. However, for the anisotropic system, normal modes of the symmetric part are actually linearly polarized light. It does not give rise to Faraday Effect and is assumed as isotropic [14]. To visualize the effect of the antisymmetric part, the dielectric tensor is written as Euler's formula [15,16]

$$\varepsilon = \varepsilon_{xx} \begin{pmatrix} 1 & -iQm_z & iQm_y \\ iQm_z & 1 & -iQm_x \\ -iQm_y & iQm_x & 1 \end{pmatrix} \tag{2.8}$$

where $Q = i\varepsilon_{xy}/\varepsilon_{xx}$ is Voigt Constant which is a magneto-optical constant, and $m_i$ represents the components of magnetization vector M. The normal modes of the antisymmetric term are left ($\varepsilon_L$) and right ($\varepsilon_R$) circularly polarized components of lights which are given as:

$$\varepsilon_L = 1 - Qm.\,k \tag{2.9}$$

$$\varepsilon_R = 1 + Qm.\,k \tag{2.10}$$

The difference between these components is responsible for the rise of the magneto-optic effect., Fresnel reflection matrix R, on solving the above matrix using Maxwell's equation given as [12,14,16]

$$R = \begin{pmatrix} r_{pp} & r_{ps} \\ r_{sp} & r_{ss} \end{pmatrix} \tag{2.11}$$

where s and p represents the s and p polarised components of to wave. The diagonal elements, $r_{pp}$ and $r_{ss}$, signify the extent of the reflection of the original polarization-state, whereas the off-diagonal elements, $r_{sp}$ and $r_{ps}$, represent the net elliptical polarization and rotation, i.e. the magneto-optical Kerr effect. The Kerr angles corresponding to s, and p polarized light is represented as [12,14,16]:

$$\Theta_K^p = \theta_K^p + i\,\varepsilon_K^p = \frac{r_{sp}}{r_{pp}} \tag{2.12}$$

$$\Theta_K^s = \theta_K^s + i\,\varepsilon_K^s = \frac{r_{ps}}{r_{ss}} \tag{2.13}$$

In general, for the arbitrary direction of magnetization and oblique incidence, the expression for Kerr rotation is given by [16]

$$\theta_k^{s,p} = \frac{\cos\theta_0(m_y\,tan\theta_1 + m_z)}{\cos\,(\theta_0 - \theta_1)} \cdot \frac{in_0 n_1 Q}{(n_1^2 - n_0^2)} \tag{2.14}$$

where $\theta_0$ and $n_0$ are the angles of incidence, and the refractive index of the nonmagnetic medium 0, respectively. Similarly, $\theta_1$, $n_1$ are that of the magnetic medium 1.

**Various geometries of MOKE**

Classification of MOKE is based on the direction of the magnetization vector with respect to the plane of incidence of light and the reflecting surface. Thus, MOKE can be classified into three geometries [11,15,17,18] as follow (Table 2.1):

**Polar geometry**: In this geometry, the direction of the magnetization vector is aligned perpendicular to the reflecting surface and parallel to the plane of incidence of light. Irrespective of the angle of incidence, magnetization present in the system is always perpendicular to the polarization of both s and p-polarized light, thus Kerr rotation is observed. Using this geometry explores the perpendicular magnetic anisotropy to the sample surface.

**Longitudinal geometry**: In longitudinal geometry, the magnetization vector is aligned parallel to the plane of light incidence and the reflecting surface. Here, polarization change is dependent on the angle of incidence of light on the sample surface. For normal incidence, either Lorentz force (p-polarized light) is absent, or directions of induced polarization and incident polarization (s-polarized light) are the same, hence Kerr effect vanishes. In view of the facts, the angle of incident of light is kept oblique to the reflecting surface for the experiments.

**Transverse geometry**: In this geometry, the magnetization vector is aligned perpendicular to the plane of incidence of light but parallel to the reflecting surface. The normal incidence of light doesn't provide the Kerr effect. For oblique incidence, the plane of polarization doesn't provide the Kerr effect, but the Kerr effect is observed in terms of change in reflected light intensity with change in the direction of the magnetization vector. This is because the reflected light intensity is proportional to the component of the magnetization vector aligned parallel to the reflecting surface and perpendicular to the plane of incidence of light.

| Name | (a)      Polar | (b)      Longitudinal | (c)      Transverse |
|---|---|---|---|
| Geometry |  |  |  |
| Detection | Out of plane | In-plane | In-plane |
| Polarisation variation | Rotation ellipticity  | | None  |
| Measurement | Polarization analysis | | Intensity measurement |

**Table 2.1**: Various geometries of MOKE measurements for p-polarized light. Incident plane of light is represented by black rectangle. Polarization variation row shows polarization planes of the incident and reflected lights in the plane perpendicular to the light path [19].

## MOKE experimental set-up

A schematic view of the MOKE set-up is shown in fig. 2.9. The set-up consists of various parts: (1) HeNe Laser (2) photoelastic modulator (PEM) (3) polarizer (4) photodiode detector (5) analyzer (6) sample holder (7) electromagnet (8) computer interface control and (9) lock-in amplifier.

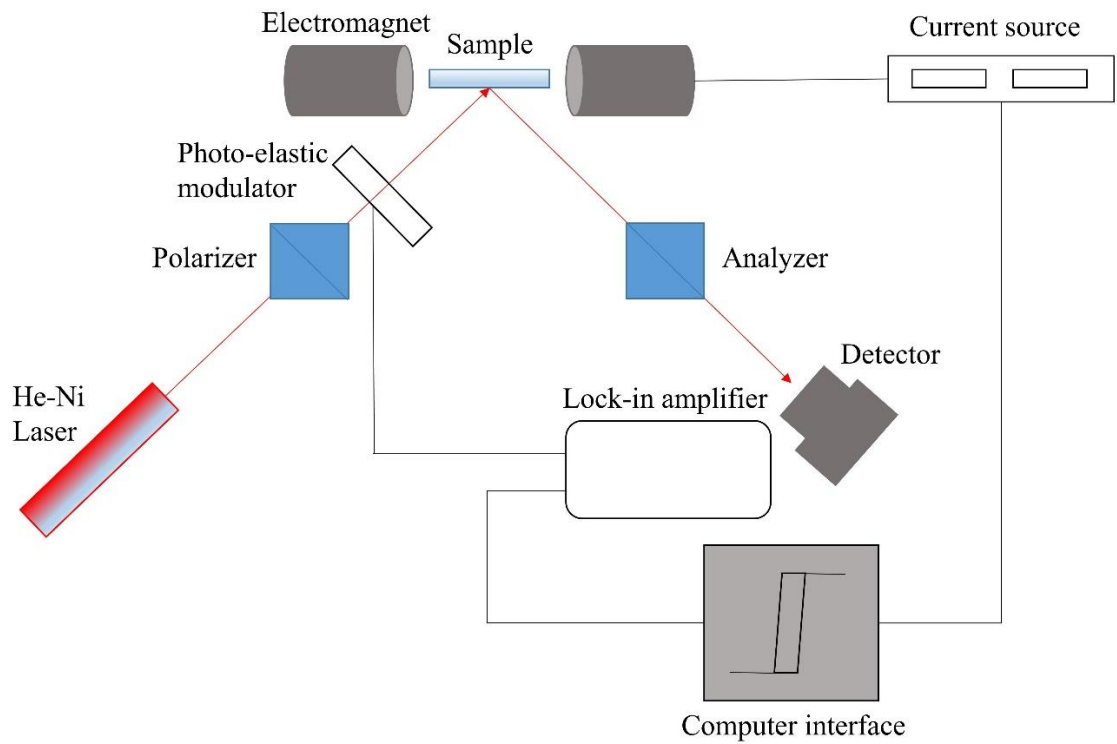

**Figure 2.9**: Schematic view of MOKE set up.

After passing through the polarizer, the polarized laser beam passes through a PEM and directs to the sample surface. PEM modulates the incident laser beam, so that phase shift (50 kHz for the present system) is produced on the component of the beam parallel to the PEM axis. The magnetic sample placed between electromagnet poles reflects the laser beam, which then passes through the analyzer. The polarization axis of the analyzer is kept at an angle $\delta$ from extinction with that of the polarization axis of the polariser. Finally, the photodiode detects the intensity of the reflected laser. The intensity measured using detector is given by equation [14]

$$I = |E_p sin\delta + E_s cos\delta|^2 \approx |E_p\delta + E_s|^2 = |E_p|^2|\delta + \varphi' + i\varphi''|^2 \qquad (2.15)$$

Since $\varphi'$ and $\varphi''$ are Kerr rotation and ellipticity respectively and are directly proportional to magnetization, the intensity with continuously varying magnetic fields determines the hysteresis loop. The detector provides the signal to the Lock-in amplifier, which then filters the noise with reference to the modulation frequency of PEM to maximize the signal-to-noise ratio. Finally, the signal is plotted as a hysteresis loop.

### 2.3.3    X-ray reflectivity (XRR)

XRR is a non-destructive and surface-sensitive technique used to analyze the structural profile of any kind of thin film, single or multilayer [20]. It is a grazing incident scattering technique which uses an x-ray beam, incident at very low angles (0.2- 3°) on top of the sample and measures the specularly reflected intensity as a function of angle. XRR technique is very sensitive to electron density gradients and can characterize the electron density profiles of thickness few tens of angstroms [21]. Also, XRR allows the determination of the surface and interface roughness and the thickness of the thin films and multilayers.

**Basic principle**

For hard x-ray radiation, the For hard x-ray radiation, n,  which comes into picture due to the interaction of x-rays with the material, can be described through the following relation [20,22,23]:

$$n = 1 - \delta - i\beta \qquad (2.16)$$

Where

$$\delta = \frac{r_e\lambda^2\rho_e}{2\pi} \qquad (2.17)$$

$$\beta = \frac{\lambda}{4\pi}\mu \qquad (2.18)$$

correspond to the scattering and absorption coefficient of the material under exposure, and these values depend on its electron density ($\rho_e$) and mass density ($\mu$).

The material's refractive index, using an x-ray as an incident beam, is less than unity. Thus, the total internal reflection will take place for the critical angle defined as $cos\theta_c = n = 1 - \delta$. Using Taylor approximation, this equation transforms as [20,22]

$$\theta_c^2 = 2\delta = \frac{r_e\lambda^2\rho_e}{\pi} \qquad (2.19)$$

For specular reflection, intensity is given by [20,22]

$$R(\theta) = \frac{I(\theta)}{I_0} \qquad (2.20)$$

here, $I_0$ and $I(\theta)$ are incident and reflected intensities at an angle $\theta$. As XRR measurements are performed at incident low angles, using continuity of electric and magnetic field of the electromagnetic wave at the interface together with Fresnel coefficients, intensity reflected from the smooth surface is given as [20,22]

$$R(\theta) = \left|\frac{\theta - \sqrt{\theta^2 - \theta_c^2 - 2i\beta}}{\theta + \sqrt{\theta^2 - \theta_c^2 - 2i\beta}}\right|^2 \qquad (2.21)$$

Since XRR is measured at specular reflection, wave vector $\vec{q} = (0, 0, q_z = 4\pi\sin(\theta)/\lambda)$, transforms the equation into

$$R(q_z) = \left|\frac{q_z - \sqrt{q_z^2 - q_{z_c}^2 - \frac{32i\pi^2\beta}{\lambda^2}}}{q_z + \sqrt{q_z^2 - q_{z_c}^2 - \frac{32i\pi^2\beta}{\lambda^2}}}\right|^2 \qquad (2.22)$$

When an incident x-ray falls beyond $\theta_C$ on an ideal surface, specular intensity decreases as a function of $\theta^4$. However, when an x-ray falls on the surface having roughness $\sigma$, reflected intensity is given by [20,22]

$$R = R_{flat}e^{-q_{z,0}q_{z,1}\sigma^2} \qquad (2.23)$$

here, $q_{z,0}$ and $q_{z,1}$ are the wave vectors in air and material, respectively.

**Reflectivity from single layer and multilayer**

According to the dynamical theory, the reflectivity of a single layer film deposited on a semi-infinite substrate is given by [20,22]:

$$R = \left| \frac{r_1 + r_2 e^{-2ik_{0z}t}}{1 + r_1 r_2 e^{-2ik_{0z}t}} \right| \tag{2.24}$$

Where the wave vector of the transmitted beam has $k_{0Z}$ as the vertical component (z-direction), t is the thickness of the layer and $r_{1,2}$ denote the Fresnel reflectivity coefficients of the free surface and the substrate interface, respectively.

If the multilayer has two elements, X and Y, with their respective thicknesses as $t_X$ and $t_Y$ and the reflection amplitude from X/Y interface and Y/X interface is opposite, then the Fresnel reflection coefficient can be given as $r_{XY} = -r_{YX}$. Thus, according to the Born approximation, the reflectivity of such multilayers is expressed as [20,22]

$$R = \left| r_{0X} + r_{XY}[(\phi_X^2 - \phi_X^2 \phi_Y^2 + \phi_X^2 \phi_Y^2 \phi_X^2 - \ldots \ldots (\phi_X^2 \phi_Y^2)^{N-1} \phi_{AX}^2] + r_{YS}(\phi_X^2 \phi_Y^2)^N \right|^2 \tag{2.25}$$

where $r_{0X}$ and $r_{YS}$ Represent the Fresnel coefficients of the free and the substrate surfaces, respectively. While the remaining intermediate terms correspond to the reflectivity from the in-between interfaces in the multilayer system and $\phi_X$, $\phi_Y$ correspond to the phase factors of the X and Y layers, respectively.

Treating periodic multilayer as one-dimensional crystal, the condition for maxima is given by modified Bragg law [20,22] as

$$2d\sqrt{sin^2\theta_1 - sin^2\theta_C} = m\lambda \tag{2.26}$$

with Period of multilayer $d = t_X + t_Y$.

In q space, d is given by

$$d \approx m \frac{2\pi}{(q_z)_{bragg}} \tag{2.27}$$

where m denotes the order of reflection. Bragg peak is impacted by thickness as well as the roughness of both layers X and Y. $m^{th}$ Bragg peak disappears when the multilayer follows the following relation

$$m = p \left( \frac{t_X}{t_Y} + 1 \right) \tag{2.28}$$

where p is an integer number. Total N-2 Kiessig fringes occur between two Bragg peaks [20,22].

**Information gathered through reflectivity pattern**

XRR provides information about the thickness, roughness and electron density of the film [20,22,23]. Interference occurs in the reflected x-ray, which gives rise to oscillations. These oscillations are known as Kiessig fringes [20,22,23]. Distance between the two consecutive Kiessig fringes in q space, i.e. oscillation period, is the thickness of the film expressed as

$$t = \frac{2\pi}{\Delta q_z} \qquad (2.29)$$

This means that increased thickness will provide a shorter period of oscillation. Figure 2.10(a) shows a simulated XRR pattern of Fe film of 500 Å thickness deposited on Si substrate.

The amplitude of XRR oscillations and rate of decay of XRR intensity gives information on the roughness of the film structure. In general, beyond $\theta_C$, specular intensity decreases as a function of $\theta^4$. Figure 2.10 (b) presents a simulated XRR pattern for Fe thin film with changing surface roughnesses. It is clear from the XRR pattern that intensity decays faster for the film with higher roughness, and the amplitude of oscillation is reduced.

Critical angle, $\theta_C$ or $q_c$ in q-space is the angle at which intensity drops sharply. This angle provides information about the electron density of the film. The amplitude of oscillation information provides the density difference between film and substrate. Figure 2.10(c) shows simulated XRR patterns where we have chosen Mg and Fe films on the Si substrate for compassion. It is observed that density is proportional to critical angle means with decreasing material density, $q_c$ decreases. Also, for Mg film, the amplitude of oscillations is higher as compared to Fe film because the density of Mg is close to the Si substrate. Figure 2.10 (d) represents simulated XRR pattern for Fe(30 Å)/Ag(30 Å)]$_{10}$ deposited on Si substrate. The Bragg peak position corresponds to the bilayer thickness.

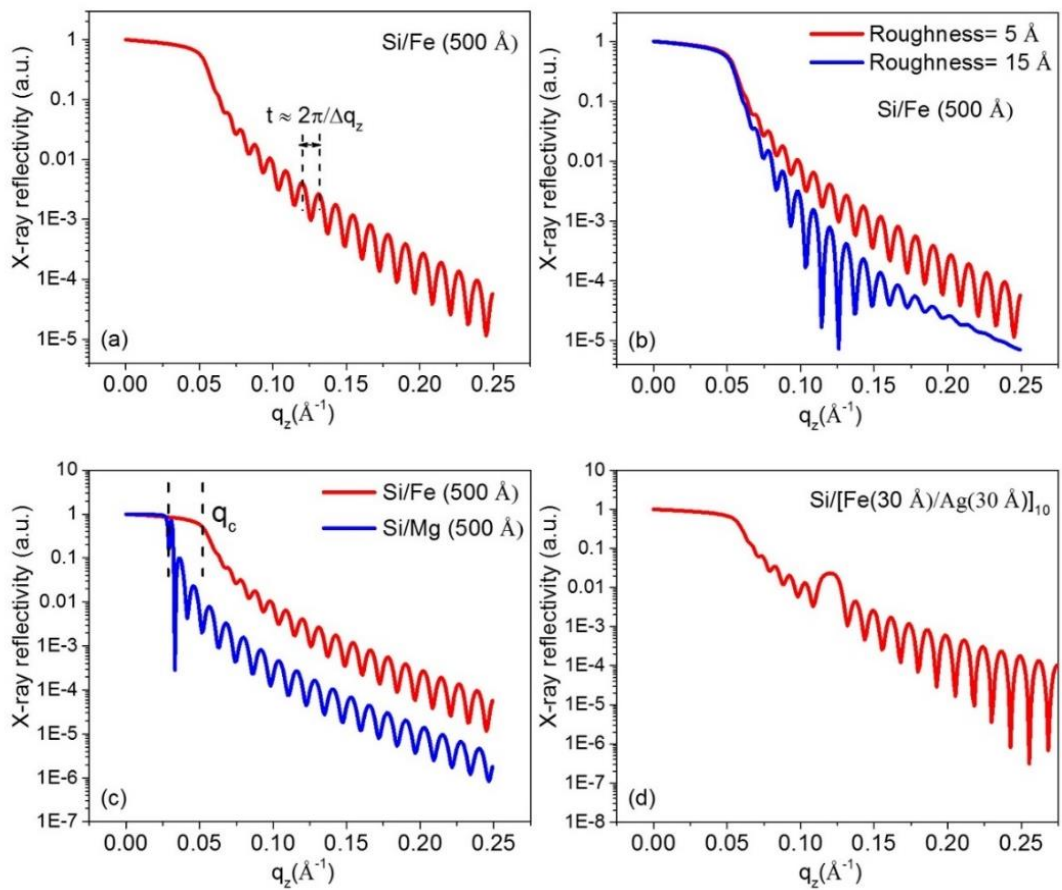

**Figure 2.10:** Simulated XRR patterns of (a) Fe film of 500 Å thickness, (b) 500 Å Fe film with varying roughness, (c) Fe, Mg films of thickness 500 Å, and (d) Fe(30 Å)/Ag(30 Å)]$_{10}$ deposited on Si substrate.

### 2.3.4    Reflection high energy electron diffraction (RHEED)

RHEED technique is used for the structural characterization of thin films. It has limited penetration and escapes the depth of electrons [24,25]. Hence, it is an effective surface-sensitive diffraction technique. Owing to its measurement time and unique geometry, this technique can be used for in-situ real-time monitoring of thin-film structural growth.

In the RHEED technique, a high-energy collimated monoenergetic electron beam falls on the thin film surface at a grazing angle (0.5- 3.0º) and scatters by surface atoms. These scattered electrons diffract and form a diffraction pattern in a reciprocal space on a fluorescent screen. This diffraction pattern depends on the surface's structure and atomic plane morphology. The pattern in the fluorescent screen is captured on charged coupled device (CCD) camera and interfaced with the computer system for further analysis. Typically, electrons with energy in the range of 10-30 keV are used in the RHEED technique. Figure 2.11 shows the schematic of the RHEED measurement.

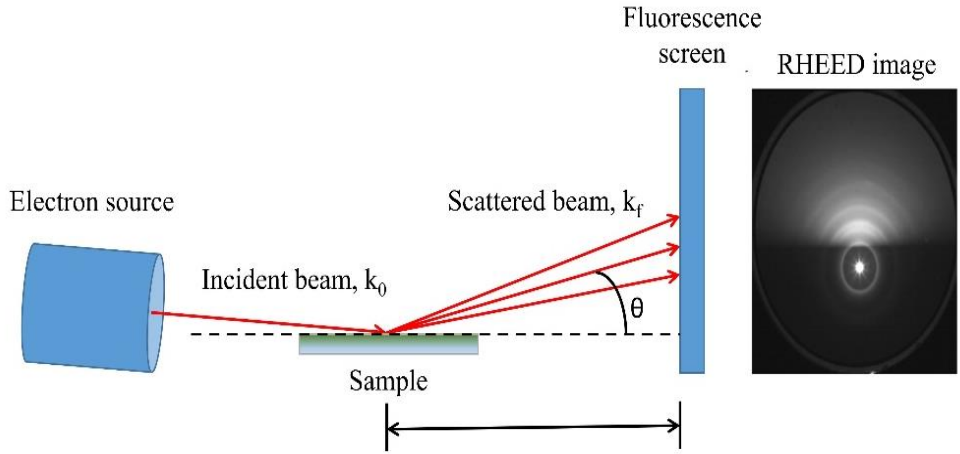

**Figure 2.11:** Schematic representation of RHEED set-up. θ is the diffraction angle while $k_0$ and $k_f$ are incident and diffracted wave vectors, respectively.

### Kinematical theory of diffraction and Ewald construction

The kinematical theory explains that diffraction occurs when the Laue equation is satisfied. According to the Laue equation, reciprocal lattice vector G is the difference between incident ($k_0$) and diffracted wave vector ($k_f$) [25,26]:

$$k_f - k_0 = G \qquad (2.30)$$

However, for elastic scattering, $|k_f| = |k_0|$ as energy is conserved. Hence geometrically, this scattering vector will lie on the surface of a sphere (known as the Ewald sphere) having a radius $k = 2\pi/\lambda$. In reciprocal space, the condition for diffraction to occur is that if the tip of vector $k_0$ is placed at the reciprocal lattice point, then the reciprocal lattice points lie on the surface of the Ewald sphere, satisfying the Laue condition gives diffraction.

The magnitude of wave vector k, for high energy electron, is expressed by [25]

$$k_0 = \frac{1}{\hbar}\sqrt{2m_0 E + \frac{E^2}{c^2}} \qquad (2.31)$$

As RHEED is performed at grazing incidence, the interaction of electrons with thin films occurs only for a few Å of layers from the surface. Hence 2D reciprocal lattice rods will be observed, not 3D. In addition, as the energy of the electron is very high (say 20 keV), the radius of the Ewald sphere will become 785 nm$^{-1}$, which is much larger than the reciprocal lattice unit vector ~ 1 Å$^{-1}$. It means the Ewald sphere will resemble almost flat in the scattering region. Thus, the intersection of reciprocal lattice rods and the Ewald sphere will form a line in the diffraction pattern. Figure 2.12 shows a schematic representation of the formation of the RHEED image.

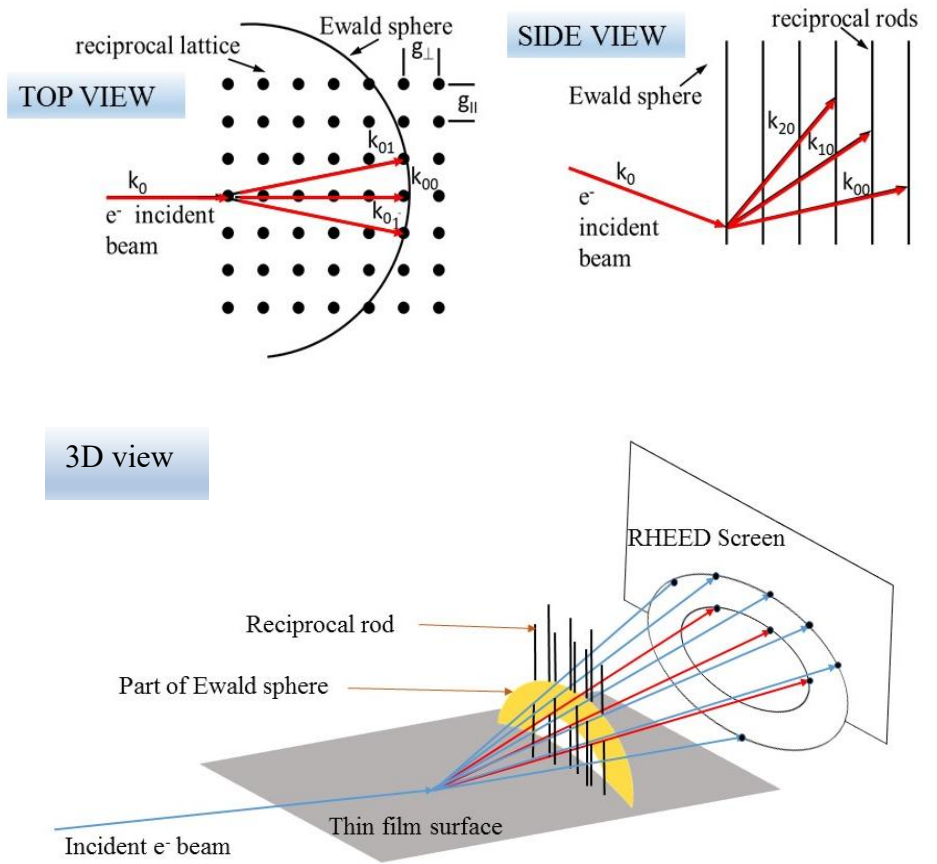

**Figure 2.12:** Schematic representation of the formation of RHEED image [25,27].

## Various diffraction geometries:

RHEED diffraction pattern contains two categories, i.e. reflection pattern and transmission pattern. These patterns depend on the scattering geometry (fig. 2.13) [28]. For smooth single crystals, diffraction occurs through reflection and generates a set of one-dimensional reciprocal lattice rods. On the other hand, if diffraction occurs for single crystal islands or polycrystalline films, then the transmission process is involved for diffraction due to roughness. Pattern involving spots is observed through diffraction for single crystal islands, and these spots correspond to a particular crystal orientation. On the other hand, polycrystalline films contain a random orientation of crystallites. Hence, concentric rings known as Debye rings are obtained in the diffraction pattern. These rings develop by the intersection of each reciprocal sphere with the Ewald sphere. Each ring corresponds to a different (hkl) plane of the crystal under investigation.

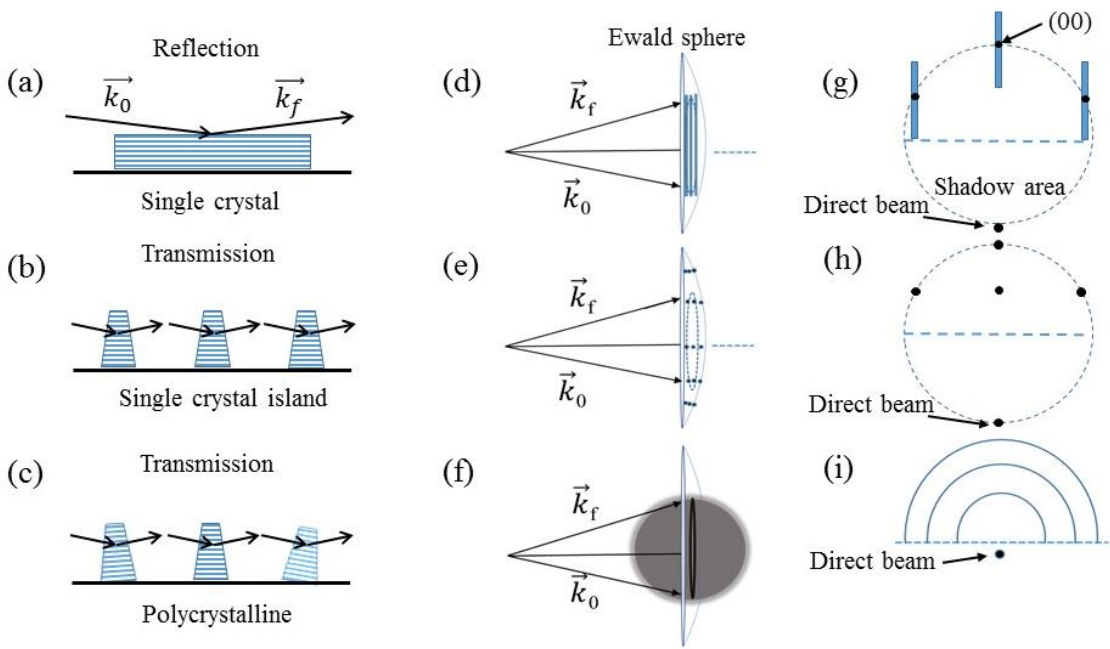

**Figure 2.13:** Schematic view of electron scattering, Ewald sphere formation and corresponding diffraction pattern, film morphology and crystalline structure. Scattering takes place in (a) reflection geometry for single crystal film having a smooth surface, ) and transmission geometry for (b) single crystal with islands at the surface and (c) polycrystalline films. $k_0$ and $k_f$ are the incident and scattered wave vectors [28].

### 2.3.5 X-ray diffraction (XRD)

XRD is a non-destructive technique in nature, and it is used to study the crystal structure, crystallite size and phase of the material. Constructive interference of monochromatic x-rays provides XRD when diffracted from a crystalline sample [29]. Diffraction occurs when the condition for Bragg's law is satisfied (fig. 2.14) i.e.

$$n\lambda = 2dsin\theta \qquad (2.32)$$

where λ, d and θ are the wavelengths of x-ray, spacing between planes and angle of incidence of x-ray, while n is an integer representing the order of diffraction. The pattern generated through XRD is a unique fingerprint of the crystal structure of the sample. X-ray diffractometers include an x-ray tube, a sample holder, and an x-ray detector.

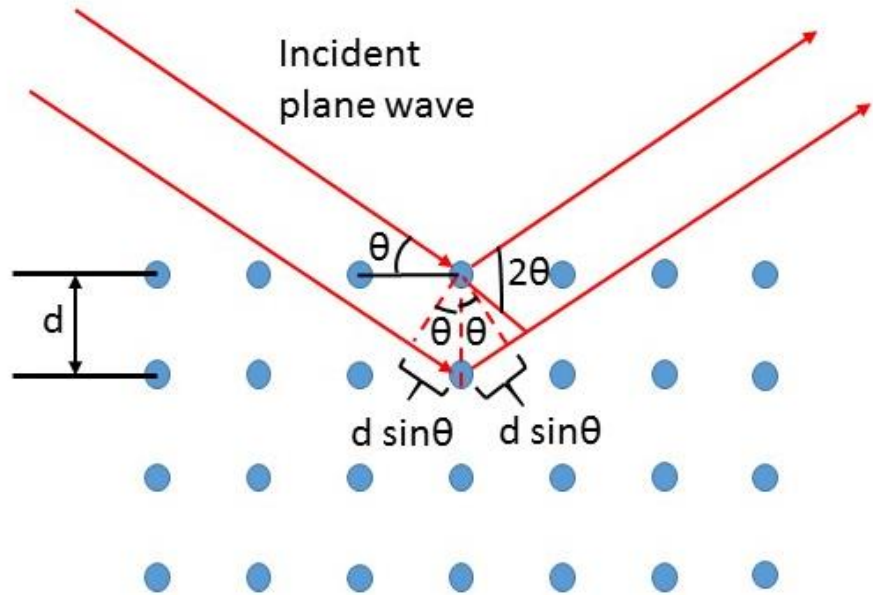

**Figure 2.14:** Schematic representation of x-ray diffraction following Bragg's equation.

In general, XRD measurement follows out of plane geometry, having two modes of measurement [30]:

(i)     Symmetrical (or specular) reflection

(ii)    Asymmetrical (or off-specular) reflection

In symmetrical reflection, geometry measures the momentum transfer vector q, which is always normal to the surface of the sample because the measurement is done for $\theta_{IN} = \theta_{OUT}$. This provides information on atomic planes parallel to the sample surface. This geometry is well known as θ-2θ geometry. But in the case of thin films, this diffraction mode can provide weak signals from the film compared to signals from the substrate. Hence, for such thin films, measurement through asymmetrical reflection is done, which uses a fixed small incidence angle which is chosen to be above the critical angle of total reflection, and only a detector scan (2θ scan) is performed. Fixing of incidence angle itself fixes the penetration of x-ray to film; hence substrate contribution in the XRD pattern is suppressed and enhances the film contribution. But in such mode of measurements, with the variation of 2θ angle, the net q vector changes the direction from normal to off-normal. Hence, this diffraction measurement mode is known as grazing incidence XRD (GIXRD). Figure 2.15 represents these two diffraction modes.

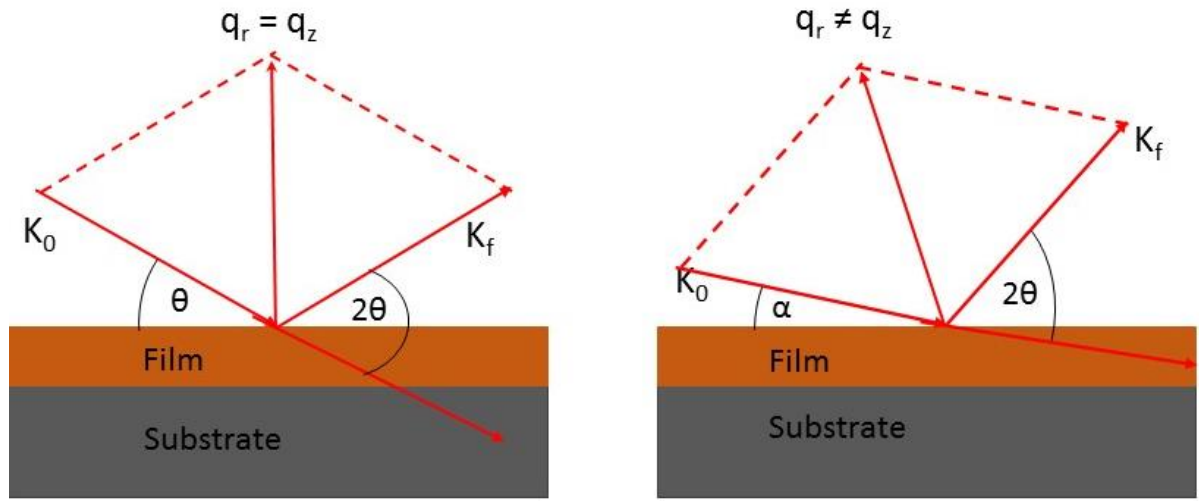

**Figure 2.15:** Schematic representation of two different modes of out-of-plane diffraction geometries [30].

### 2.3.6    Atomic force microscopy (AFM)

In the field of scanning probe microscopy (SPM), the landmark publication by Binnig, Quate, and Gerber in 1985 about the invention of atomic force microscopy (AFM) has made it possible to use interatomic forces to image the topography of the sample under investigation on the nanometer scale.

Whatever the answers that researchers need, Topography alone can not always arrange, and thus, the topology of the sample surface often does not correlate to the properties of the same sample. For such reasons, further advanced imaging modes have been invented to provide quantitative data on various surfaces. Presently, different material properties can be obtained using AFM techniques, including capacitance, electrical forces, conductivity, friction, magnetic forces, surface potential, viscoelasticity, and resistance.

As STM uses tunnelling current as the imaging probe, it can only image conducting and semiconducting surfaces. While AFM uses van-der Walls interaction as a probe, hence it has the advantage over STM to image almost any kind of surface, including ceramics, polymers, glass, composites and biological samples. Figure 2.16 shows the SPM machine used by us.

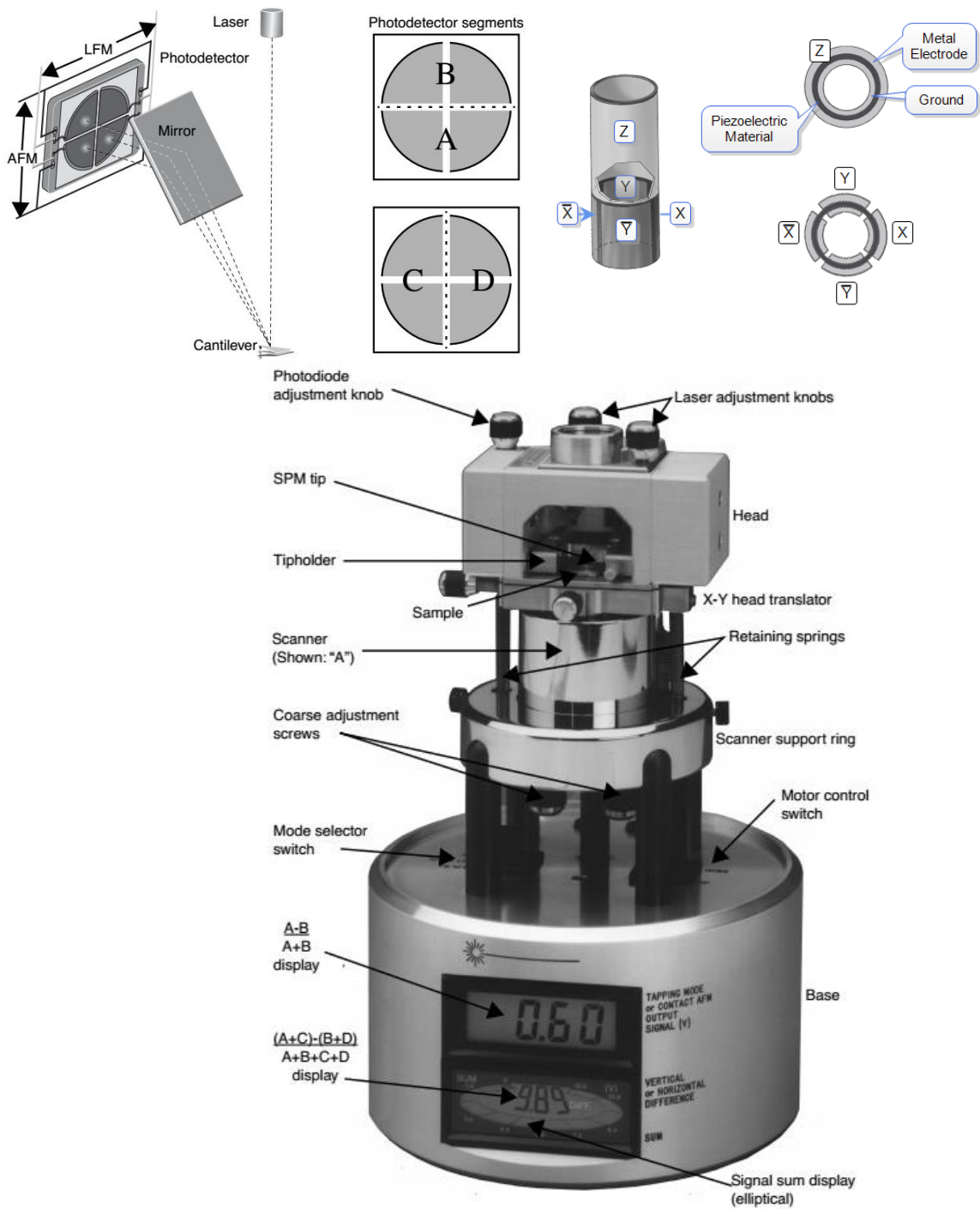

**Figure 2.16:** Schematic of SPM with its components and working. Image is taken from manual of DI-Veeco Instruments, USA.

**Working principle**

AFM assigns a fine triangular cantilever with a length of 120 microns and a tip in its apex. This tip radius can be 10 to 20 nm depending on the nanoprobes, and the base of the apex can be around 4 microns. Laser reflection is done through the gold-coated flat top. The red laser light has a wavelength of 680 nm. The photodetector has a four-quadrant array, as shown in fig. 2.16. Top-bottom arrays of detector viz. A and B form the differential mode for AFM, while left and right viz. C and D involve in Lateral Force Microscopy (LFM). Z scanning is performed through a cylindrical piezo, and XY motion is performed through a quadrant piezo. Imaging of high-resolution surfaces for atomic resolution, e.g. to reveal atomic terraces,  is done through contact mode.

A schematic diagram of set-up to perform force spectroscopy of AFM is shown in fig. 2.17. It uses a fourth piezo to oscillate the cantilever above the measuring sample at a particular lift height already pre-determined.

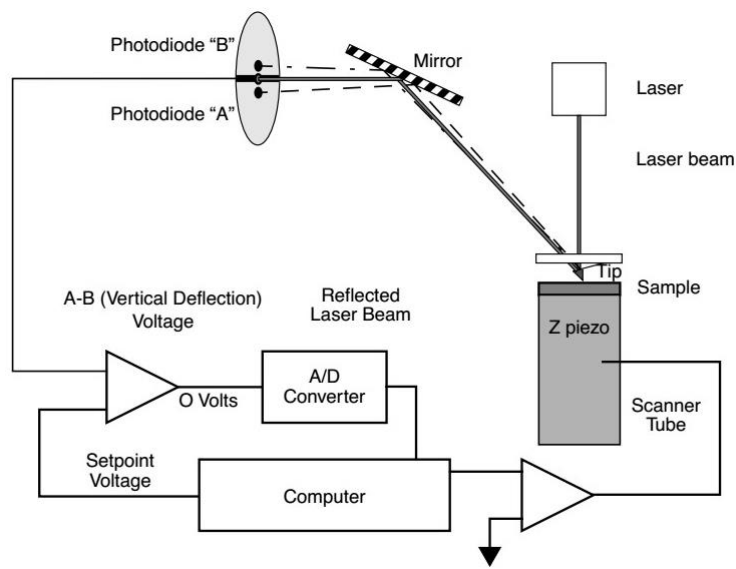

**Figure 2.17:** Schematic of the working of AFM in Tapping mode. Image is taken from Manuals of DI-Veeco Instruments, USA.

### 2.3.7 Grazing incidence small-angle x-ray scattering (GISAXS)

For analyzing the morphology and distribution of buried particles or islands on a substrate, the Grazing Incidence Small Angle X-ray Scattering (GISAXS) technique has emerged as a potent tool in the last decade. The capability of GISAXS to characterize implanted systems [31], granular multilayered systems [32], as well as semiconductor quantum dots, is now well established [33,34]. Latest technical developments [35,36] allowed this technique to be applied in-situ, during particle growth to the topics relevant to surface science. Using this technique, quantitative studies of surface nanostructuring or numerous details of thin-film growth became possible, from sub-monolayer to beyond percolation. Furthermore, high-quality statically averaged information relevant to the growth process can be obtained using the Distorted Wave Born Approximation with suitable programs [37,38] in a single experiment which, otherwise, would have been obtained through lengthy *ex-situ* measurements.

**The scattering geometry:**

In X-ray scattering, the Probing of nanostructures is done by exploring the so-called reciprocal space (Fourier space). The concept of X-ray scattering from nanostructures samples is similar to the traditional X-ray diffraction method, except that (i) as the volume of matter under investigation is small, synchrotron X-ray radiation is usually needed for higher intensity, and (ii) grazing angle incidence with respect to the sample surface is used for the incident X-ray wave vector $\mathbf{k}_i$ to minimize the unnecessary background scattering (both elastic and inelastic) originating from the bulk part, and this enhances the near-surface scattering. The schematic geometry of such grazing incident scattering measurements is shown in fig. 2.18, where the incident angle $\alpha_i$ is usually small and kept constant during the measurement, generally most of the time close to the angle of total external reflection $\alpha_c$. The scattered wave vector $\mathbf{k}_f$, makes an angle $2\theta_f$ (in the horizontal plane) with respect to the transmitted beam and an angle $\alpha_f$ with respect to the surface of the sample. The wavevector transfer is expressed as $\mathbf{q} = \mathbf{k}_f - \mathbf{k}_i$, and has two components, $\mathbf{q}_\parallel$ and $\mathbf{q}_\perp$, parallel with and perpendicular to the surface, respectively. The absolute value of $\mathbf{q}_\perp$ is a function of $\alpha_i$ and $\alpha_f$: $|\mathbf{q}_\perp| = q_z$. $\lambda$ is the X-ray wavelength used. In GISAXS and GIXS both, the wavevector transfer coordinates are connected to the angular coordinates according to the following relations:

$$q_x = k_0[\cos(2\theta_f)\cos(\alpha_f) - \cos(\alpha_i)] \tag{2.33}$$

$$q_y = k_0[\sin(2\theta_f)\cos(\alpha_f)] \tag{2.34}$$

$$q_z = k_0[\sin(\alpha_f) + \sin(\alpha_i)] \tag{2.35}$$

$$k_0 = |k_f| = |k_i| = 2\pi/\lambda \tag{2.36}$$

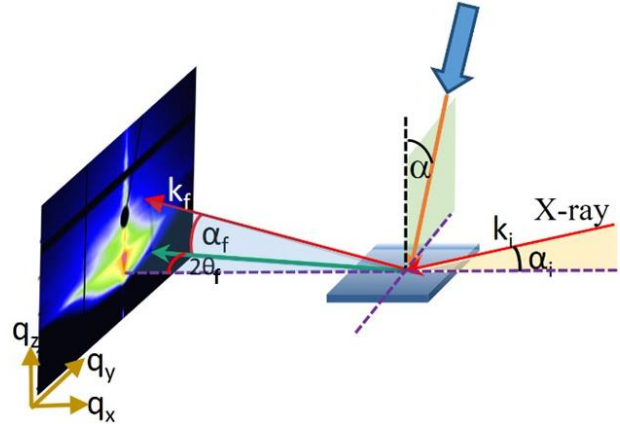

**Figure 2.18:** Grazing incidence X-ray scattering geometry.

The wavevector transfer will be small when all angles are small; hence, large dimensions are probed in real space.

## 2.4    A versatile in-situ UHV chamber for in-situ study

### Technical details

A versatile in-situ vacuum chamber available in the lab is used to perform the study done in the present thesis in order to characterize the thin film under study in real-time during the growth or during heating/cooling. A load lock chamber is attached to the UHV chamber, which is used for loading the sample and transferring it to the manipulator inside the UHV chamber. The UHV chamber is equipped with an e-beam evaporation technique at the bottom for deposition, while XRR, MOKE, RHEED, and four-probe resistivity set-ups are present for characterizations. In addition, an ion gun is present at a glancing angle for surface sputtering/cleaning. These techniques have already been discussed in the earlier sections of the chapter. Base pressure of ~$2\times10^{-10}$ mbar in the main chamber and $1\times10^{-8}$ mbar in load lock is achieved via a combination of rotary, turbo molecular and ion pumps. The actual photograph of the UHV chamber is shown in fig. 2.19.

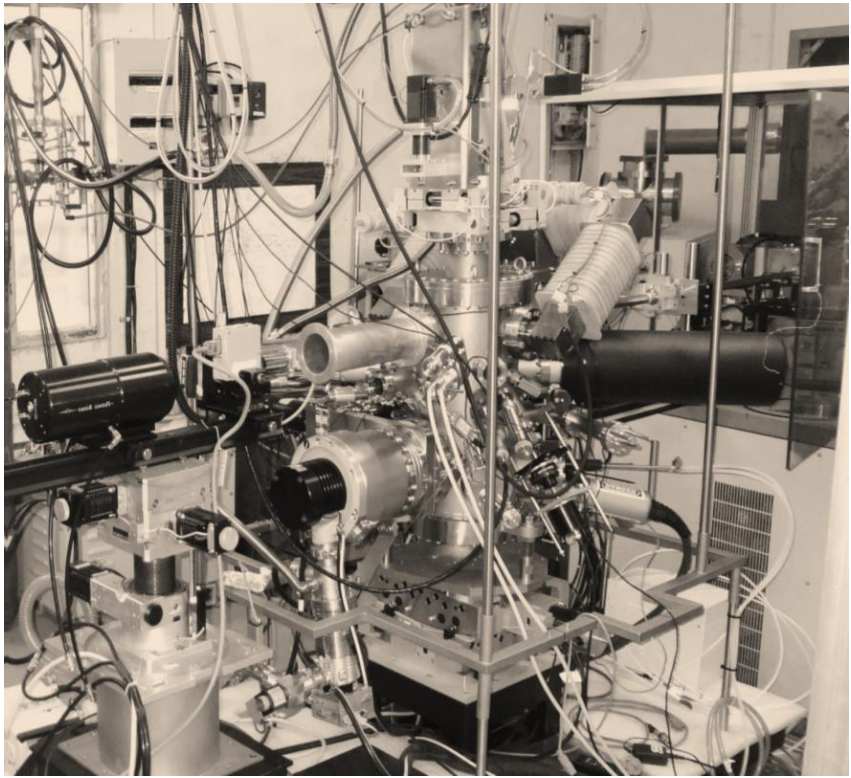

**Figure 2.19:** Actual image of in-situ UHV chamber.

      Five-axis (x, y, z, rotation and tilt) manipulator and sample holder are situated on top of the 100 CF flange. The Manipulator axis has azimuthal rotation (up to 355°) and tilt (±12.5°) together with Z (vertical) motions for sample alignment for various experiments. Cooling collection in the manipulator and heater in the sample holder provides sample cooling up to ~50 K and heating to ~1200 K, respectively, along with four connections on top of the manipulator for resistivity and magneto-resistance (MR) measurements. An electromagnet with 2500 Oe field strength is used for MOKE measurement. MOKE set-up is arranged for the longitudinal geometry. Cu $K_\alpha$ x-ray lab source with collimated beam passes through the Beryllium window on the chamber for the XRR experiments. The whole chamber is mounted on a HUBER stage with a tilt motion along the x-ray direction. This tilt motion is used for θ scan and the scintillation detector is mounted on a detector stage with combined linear and tilt motion for 2θ scan. For a particular measurement technique among these techniques, the positions of the motors of these various axes are fixed, and all the techniques can be carried out during the sample growth.

**Advantages of in-situ studies using UHV chamber:**

1. Real-time study of structural, magnetic and transport properties with film growth without exposure to ambient.

2. Study of these properties with film thickness and/or annealing temperature etc, can be done during and after the deposition of the film.

3. The heating/cooling option and magnetic field make it suitable for exchange bias studies, and deposition can be done in the presence of the magnetic field.

4. Azimuthal angle-dependent magnetic properties (i.e. magnetic anisotropy) study can be done with the help of a sample manipulator, which has an option for angular rotation.

5. Compared to the earlier studies in the literature, where different measurements are performed separately, and the conclusion is drawn by combining all data, in the present UHV set-up, the study of magnetic thin films/multilayers is studied simultaneously with deposition in the same chamber.

**3.1    Introduction**

**3.2    Portable Mini-Chamber for Temperature-Dependent Studies using Small Angle and Wide Angle X-ray Scattering**

   **3.2.1  Design and description**

   **3.2.2  Experimental details**

   **3.2.3  Geometry of measurements**

   **3.2.4  Capability demonstration**

         **3.2.4.1  GISAXS measurements**

         **3.2.4.2  GIWAXS results**

   **3.2.5  Option for X-ray fluorescence along with GISAXS and GIWAXS measurements**

**3.3    Thermal evaporator for organic film deposition**

   **3.3.1  Requirements**

   **3.3.2  Design and fabrication**

   **3.3.3  Capability demonstration**

**3.4    Other developments**

   **3.4.1  GI-SAXS set-up at BL-07 beamline RRCAT**

         **3.4.1.1  Schematic of the set-up and its functioning**

         **3.4.1.2  Capability demonstration**

   **3.4.2  Copper heaters for in-situ thin-film heating**

         **3.4.2.1  Copper heater (nichrome wiring)**

         **3.4.2.2  Heater with tungsten wiring**

**3.5    Conclusion**

**References**

The present work describes the design and performance of a vacuum-compatible portable mini chamber for temperature-dependent GISAXS and GIWAXS studies of thin films and multilayer structures. The water-cooled body of the chamber allows sample annealing up to 900 K using ultra-high vacuum compatible (UHV) pyrolytic boron nitride heater, thus making it possible to study the temperature-dependent evolution of structure and morphology of two-dimensional nanostructured materials. Due to its lightweight and small size, the chamber is portable and can be accommodated at synchrotron facilities worldwide. A systematic illustration of the versatility of the chamber has been demonstrated at beamline P03, PETRA-III, DESY, Hamburg, Germany. Temperature-dependent grazing-incidence small-angle x-ray scattering (GISAXS) and grazing incidence wide-angle x-ray scattering (GIWAXS) measurements were performed on oblique angle deposited Co/Ag multilayer structure, which jointly revealed that the surface diffusion in Co columns in Co/Ag multilayer enhances by increasing temperature from R.T. to ~ 573 K. This results in a morphology change from columnar tilted structure to densely packed morphological isotropic multilayer.

## 3.1    Introduction

Thin films, in particular nanometer-scale structures, have been the subject of numerous studies because of their potential applications in thin-film technology [1]. The surface/interface roughness, stresses, strain, morphology and structure of thin films control many important physical and chemical properties of such types of structures [2,3,4,5]. Particularly, in the case of ferromagnetic thin films, surface and interface morphologies significantly influence magnetic properties, such as magnetic hysteresis, magnetic anisotropy, coercive force, magnetoresistance, and structure of the magnetic domains. Besides the preparation conditions [2], temperature is an important parameter that strongly affects the structure and morphology of the thin films [6]. For example, depending on substrate temperature during deposition, good quality films with smooth surfaces can be deposited [7]. In the case of FePt thin film, annealing leads to the transformation of the face-centred cubic (fcc) disorder phase to a highly L10 ordered face-centred tetragonal (fct) phase, which is responsible for the appearance of hard magnetic properties in the FePt layer [8]. Furthermore, annealing in the presence of a magnetic field is used for inducing strong magnetic anisotropy in thin films, where the strength of the magnetic anisotropy sensitively depends on structure and film morphology. Several experiments in the literature aiming to correlate magnetic properties with interface structure and morphology have been performed [9]. In such studies, surface morphology and structure evolution have been studied by depositing a series of samples and post-annealing at selected temperatures. However, the possible unwanted variation in deposition conditions or post-deposition treatments in preparing a series of samples may induce changes in morphological features like grain size and texture. Therefore, in such experiments, it is difficult to unambiguously separate the effect of temperature on morphological and structural properties.

A dedicated mini chamber has been designed and developed to avoid such problems and study temperature-dependent properties on a single sample. Here, temperature-dependent morphological and structural changes can be studied simultaneously during annealing under high vacuum (HV) conditions. A systematic illustration of the potential of the set-up has been demonstrated on a multilayer sample: $Co_{oblique}(44\text{Å})/Ag_{normal}(15\text{Å})$ multilayer (10 bilayers) deposited on Si(100) substrate. The evolution of the structural and morphological features, temperature-dependent grazing-incidence wide-angle x-ray scattering (GIWAXS) and small-angle x-ray scattering (GISAXS) measurements were carried out at P03 beamline, Petra III, DESY, Germany [10]. The results were later correlated with the magnetic properties of the multilayer.

In the next part, to create spin valve structures in our lab, deposition of organic materials like $Alq_3$, C60 etc., is required. For this purpose, we have also fabricated a special type of thermal evaporator to deposit smooth films of such materials. Also, to heat the samples, we have also discussed the design of copper heaters in the last session.

## 3.2    Portable Mini-Chamber for Temperature-Dependent Studies using Small Angle and Wide Angle X-ray Scattering

### 3.2.1    Design and description

The overall design of the vacuum-compatible mini chamber is decided according to the demand for simultaneous structural characterization using grazing incidence GISAXS and GIWAXS of thin films with an option of in-situ heating. The design of the top part of the chamber, containing two Kapton windows, is shown in fig. 3.1(a). Two Kapton windows of size $50\times25mm^2$ and $80\times35mm^2$ have been made on the top dome of the chamber in opposite directions for the incident and scattered x-rays. Exit window size is kept bigger to cover suitably large scattering angles along the exit direction of the x-ray, which is essentially required for GIWAXS measurements. The top view of the chamber is shown in fig. 3.1(b), where we can see the horizontal angles covered by incoming and outgoing x-ray through the windows are 60° and 120°, respectively. It covers ~20° angle in the vertical direction. A photograph of the chamber along with the vacuum port description is shown in fig. 3.2(a). SS304L material is used for the fabrication of the chamber. It has a diameter of 80 mm and a height of 170 mm. The chamber has several ports to accommodate various components, namely, a heater, vacuum pump, a UHV gauge, electrical feedthrough and Kapton windows for x-rays. The top dome is prepared by fixing it on a 63mm C.F. flange. It houses two Kapton windows for x-ray entrance and exit. The designed chamber has a volume of about 2 liters, and the surface area exposed to vacuum is about 66100 $mm^2$. Considering the gas load due to the surface area, a rough estimate of the time required evacuating the complete chamber to a pressure ~$10^{-8}$ mbar with a 100 l/s turbo molecular pump (TMP) is less than an hour, whereas the time required to pump out the gas load due to the volume of the chamber is almost negligible.

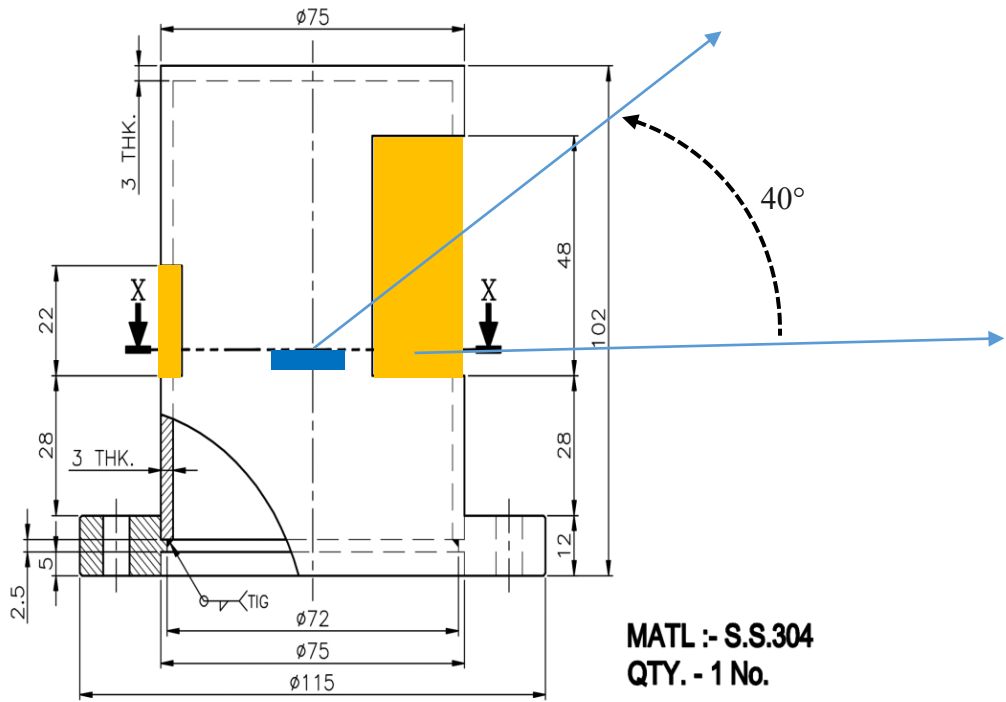

(a)

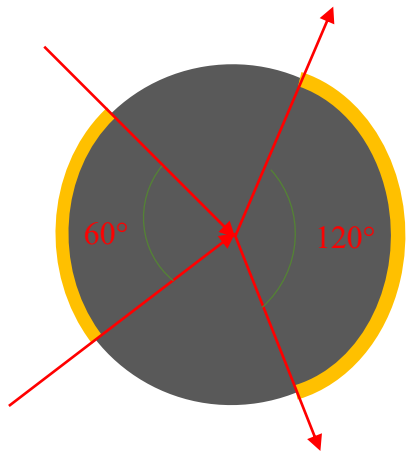

(b)

**Fig. 3.1:** (a) 2D AUTOCAD design of the XRD- mini chamber (b) top view of the mini chamber showing the angles covered by Kapton windows.

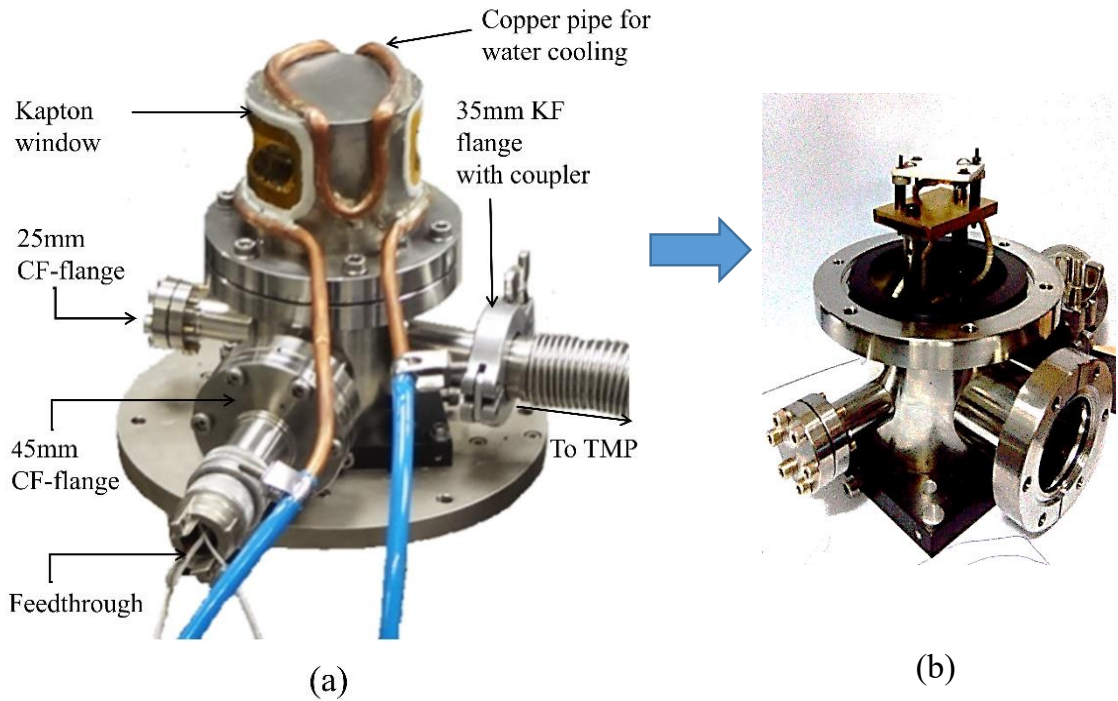

(a)

(b)

**Fig. 3.2:** (a) Actual photograph of the chamber and its various components (b) inside heater assembly of the chamber.

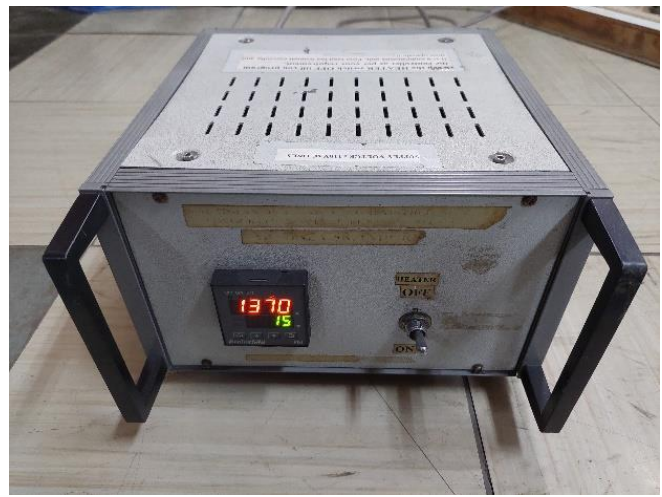

**Fig. 3.3:** Homemade PID controller for the precise temperature control of the heater.

A compact cold cathode gauge (Model: FRG-700) has been used for vacuum measurement in a pressure range up to $1 \times 10^{-8}$ mbar. The ultimate vacuum $\sim 5 \times 10^{-8}$ mbar is achieved in the chamber within 30 minutes using 100 l/s TMP. Figures 3.2(b) and 3.3 show the mini chamber's inside view and the PID controller, respectively. A UHV-compatible pyrolytic boron nitride heater has been used for sample heating, which offers quick heating of the sample [11] with a rate of about 100 K/s. It has a very low thermal mass and a chemically inert dielectric surface. Two stainless steel (S.S.) rods are used to provide the heater's base and keep the heater surface at the level of Kapton windows for x-ray scattering experiments. The heater surface dimensions are about $35 \times 25 \times 2$ mm$^3$. The K-type thermocouple is attached to the surface of the heater to record the sample temperature. A copper pipe is welded with the chamber body for the water flow to maintain its temperature near room temperature (RT) during the annealing of the sample. Water flow also prevents damage to the Kapton window due to a possible increase in the surrounding temperature.

### 3.2.2 Experimental details

In order to demonstrate the strength of the chamber, a multilayer structure with obliquely deposited Co layer and normally deposited Ag layer [Co$_{oblique}$(44Å)/Ag$_{normal}$(15Å)]$_{10}$ was deposited on Si(100) substrate in a UHV chamber using e-beam evaporation technique. The deposition angle (θ) was kept alternatively at θ=75° (oblique angle) and θ=0° during evaporation of Co and Ag materials, respectively. Here θ is the angle of deposition flux direction with respect to substrate normal. Since obliquely deposited Co layers are expected to be in a columnar structure, the final multilayer structure consists of alternating layers of columns (Co) and isotopic Ag layers. It may be noted that this multilayer is designated as Co$_{obl}$/Ag$_{nor}$ in the rest of the chapter.

### 3.2.3 Geometry of measurements

For temperature-dependent GISAXS and GIWAXS measurements, a multilayer was attached to the heater plate inside the chamber. The mini chamber was placed on a sample stage at P03 beamline, PETRA-III, DESY, Hamburg [10], with the capability of aligning the sample using various motorized motions (6-axis motions in HEXA pod). Geometry for the GISAXS measurements is shown in fig. 3.4, where $\alpha_i$ and $\alpha_f$ are the incident and reflected angles of x-ray with respect to sample surface and $\vec{k_i}$ and $\vec{k_f}$ are the incident and outgoing wave vectors. The centre of rotation of the chamber was kept at the sample surface height in such a way that two–dimensional GISAXS and GIWAXS patterns could be measured through the Kapton windows in a vertical and horizontal scattering plane without variation in the beam position on the sample surface (in case of change in $\alpha_i$). For the present experiments, incident photon energy of 13 keV was used at the sample position. PILATUS 1-M (Dectris Ltd., Switzerland) and PILATUS 300k detectors with a pixel size of $(172 \times 172)$ $\mu m^2$ were used for GISAXS and GIWAXS data recording, respectively. The sample-to-detector distance (SDD) was set as 2245 mm for GISAXS and 128 mm for GIWAXS measurements. Incident angle $\alpha_i$ was fixed to 0.45°, which is well above the critical angle of the Co film at 13keV (corresponding $\lambda$=0.95373Å) in the present study. Measurements were performed on $Co_{obl}/Ag_{nor}$ multilayer with increasing temperature under a base vacuum of $2 \times 10^{-5}$ mbar.

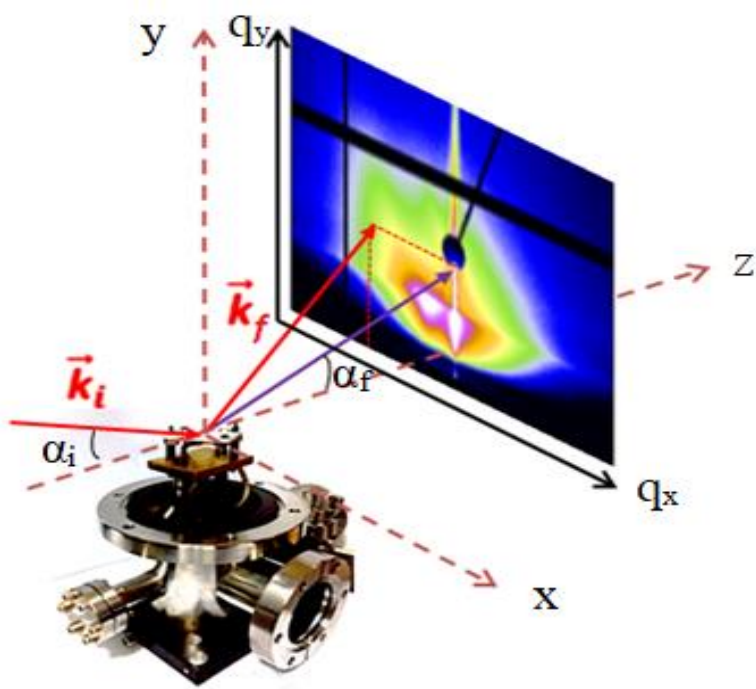

**Fig. 3.4:** Geometry for GISAXS measurements.

### 3.2.4    Capability demonstration

#### 3.2.4.1    GISAXS measurements

Figure 3.5 (a) shows the representative GISAXS patterns of $Co_{obl}/Ag_{nor}$ multilayer at RT. In the GISAXS pattern, asymmetric distribution of the scattered intensity on both sides of the specular rod has been observed. In order to understand the temperature dependence more clearly, 1D profiles extracted from the temperature-dependent 2D GISAXS image are shown in fig. 3.5 (b). In fig. 3.5(b), broken 1D profiles (zero photon counts) near q= -0.74Å$^{-1}$ correspond to the intensity drop underneath the module of the modular detector and do not influence the data analysis. 1D GISAXS profile presents two intensity maxima visible along the q direction, separated by the specular rod. It may be noted that the maxima (satellite peaks) at the left side of the specular peak are relatively intense. It may be noted that this maximum shifts toward lower q values (towards specular) and become weaker as temperature increases. Interestingly, at a temperature above 648 K, both side of the specular rod becomes symmetric. In GISAXS measurements, decreasing asymmetric distribution of the scattered intensity with increasing temperature confirms the termination of tilted columns (formed due to oblique angle Co deposition) [12] due to diffusion at a temperature above 573K.

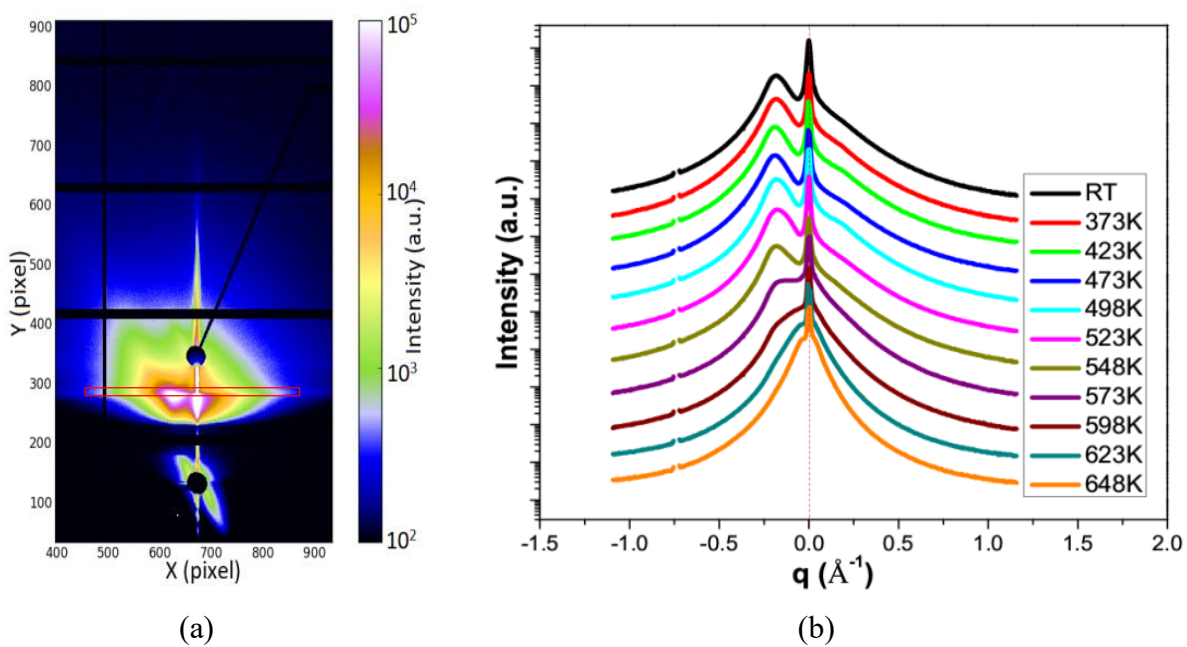

(a)                                    (b)

**Fig. 3.5:** (a) GISAXS patterns of $Co_{obl}/Ag_{nor}$ multilayer at RT. and (b) GISAXS 1-D profiles with increasing temperatures. GISAXS 1-D profiles are taken at the red marked area in fig. 3 (a) corresponds to the Yoneda wings region.

### 3.2.4.2    GIWAXS Results

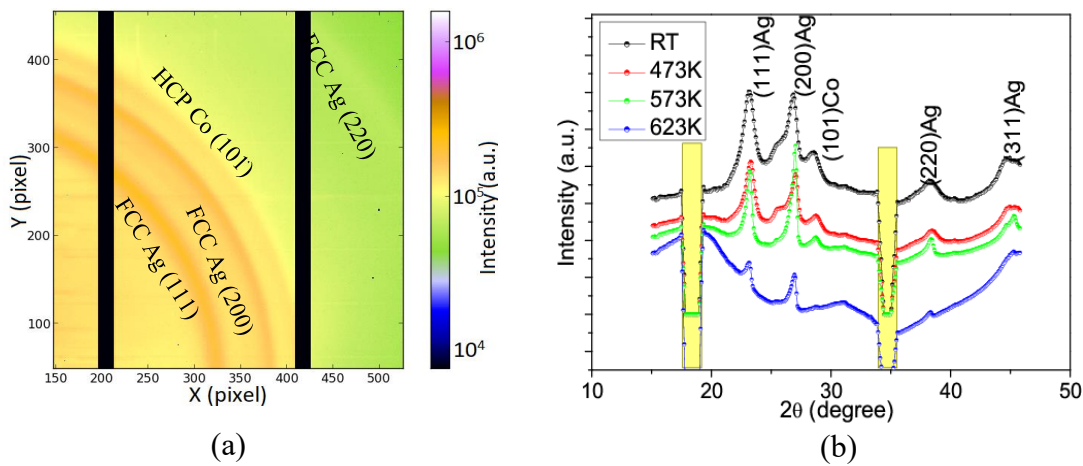

(a)                                                        (b)

**Fig. 3.6:** (a) GIWAXS patterns of $Co_{obl}/Ag_{nor}$ multilayer at RT and (b) GIWAXS 1-D profiles with increasing temperatures. The yellow areas in GIWAXS 1-D profiles correspond to the detector's intermodular gap (IDG), where no photons can be detected.

Representative GIWAXS patterns of $Co_{obl}/Ag_{nor}$ multilayer at RT are shown in fig. 3.6(a). Continuous arcs in the GIWAXS image confirm the polycrystalline nature of the $Co_{obl}/Ag_{nor}$ multilayer. It may be noted that the full diffraction rings in GIWAXS measurements are not visible due to the fixed opening of the Kapton window. Arcs with corresponding planes are marked in the image. 1D profiles extracted from the temperature-dependent 2D GIWAXS images are shown in fig. 3.6(b). In fig. 3.6(b), extracted temperature-dependent 1D data from simultaneous GIWAXS measurements using DPDAK software confirms the growth of Co layers in the hcp phase and Ag layers in the fcc phase in $Co_{obl}/Ag_{nor}$ multilayer. Furthermore, sharpening the diffraction peaks with temperature confirms an increase in particle size.

### 3.2.5    Option for X-ray fluorescence along with GISAXS and GIWAXS measurements

We have also developed a circular Kapton window at the top of the dome for fluorescence measurements, having a diameter of ~30 mm. The respective image is shown in fig. 3.7. Our lab-mates performed the fluorescence experiments using this chamber and published the data [13].

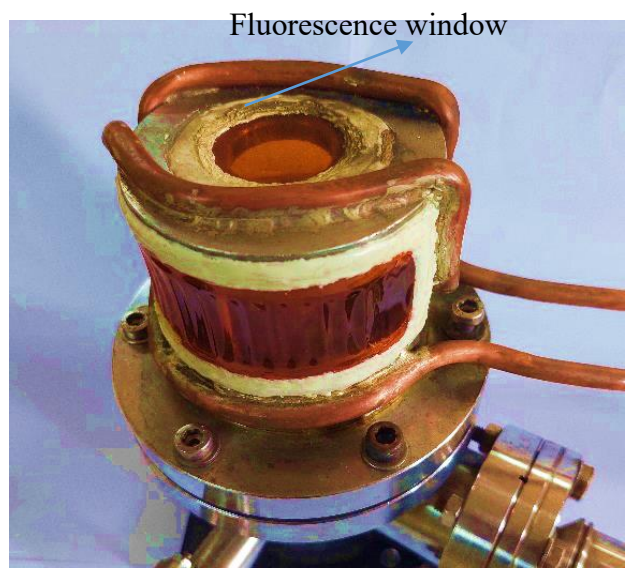

Fluorescence window

**Fig. 3.7:** Top Kapton window for the x-ray fluorescence measurements.

## 3.3 Thermal evaporator for organic film deposition

One of our motives is to create spin valve structures using organic materials deposition and oblique angle deposition. An evaporator was required for such organic materials depositions. For serving such a purpose, we have fabricated a crucible containing a pyrolytic boron Nitride heater as follows:

### 3.3.1 Requirements

To deposit the organic materials like $Alq_3$ and C60, the temperature required is above 350°C for smooth deposition. To fulfil such a purpose, an organic heater is required to reach this temperature. For this purpose, the material selected is the pyrolytic boron nitride, a good conductor of heat and electrically insulator. Such material works well in UHV without any degassing effect. The rate of heating in such material is very high, as 100k/s. It contains low thermal mass with a chemically inert surface. We can directly evaporate any organic material above it. The wiring in this material is done using platinum wire. A thermally and electrically insulator ceramic material is used for the heater base. A k-type thermocouple is attached at the corner to read the temperature. Nut- bolts are used to attach the wires to a ceramic plate and to connect the heater assembly to the main chamber.

### 3.3.2    Design and fabrication

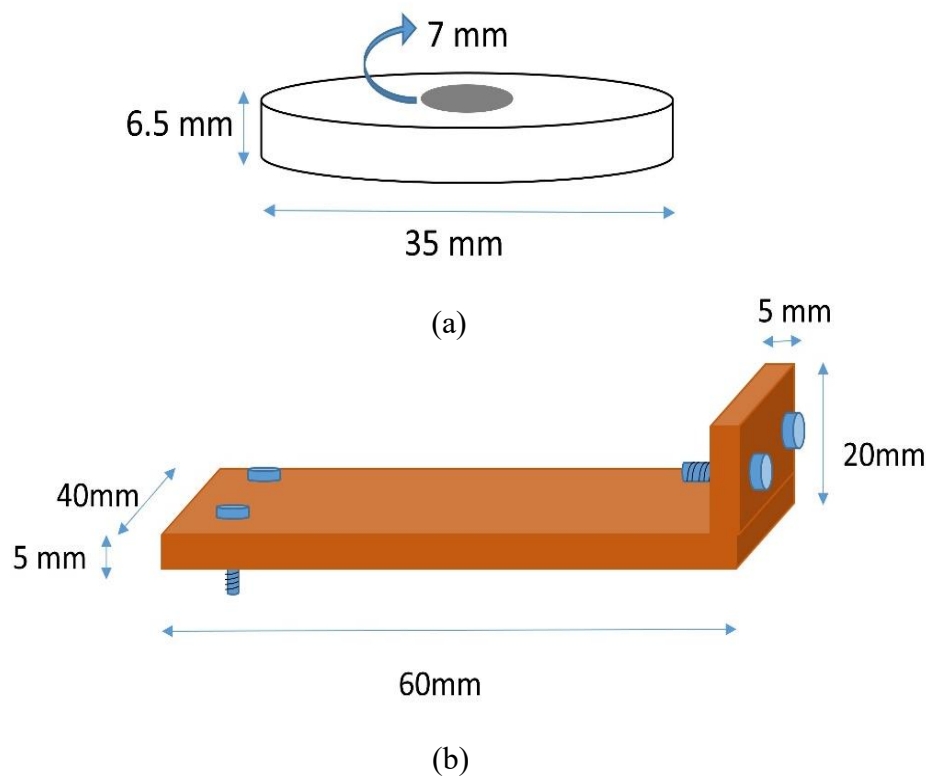

(a)

(b)

**Fig. 3.8:** Schematic design of (a) pyrolytic boron nitride thermal evaporator (b) base of the heater.

This evaporator assembly is an in-house design and developed in the lab. It contains two parts: (i) the heater and (ii) the base part. A heater is an evaporator to deposit organic material like $Alq_3$ or $C_{60}$ on the substrates attached at the top of the chamber. For that, we have cut a boron nitride cylindrical rod, which is easy to modify in any shape and, most importantly, suits well for our purpose. It is machined in the above disc form (fig. 3.8a), where dimensions are mentioned. At the centre of the evaporator disc, a reversed frustum-shaped crucible is carved, having the top and bottom diameters of 7 mm and 4 mm with 5 mm in height. Such frustum-shaped crucible is found to be more suitable than simple cylindrical-shaped crucibles because, during fast heating, organic material bounces off the cylindrical crucible, while frustum-shaped crucibles perform well in such depositions. We have used platinum wire with a resistance of 11 Ω to heat the pyrolytic boron nitride evaporator.

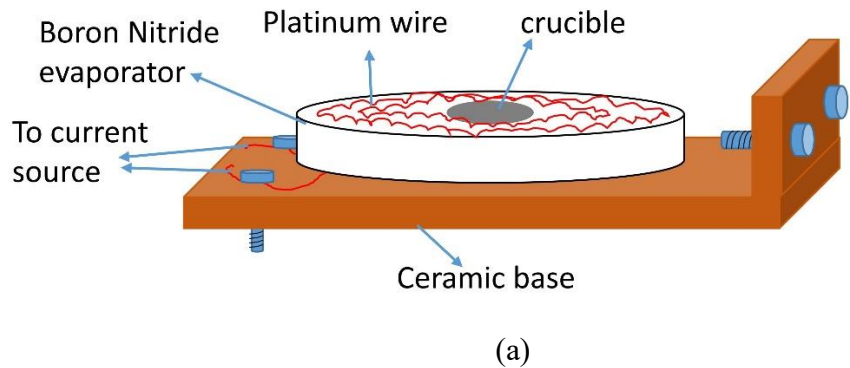

(a)

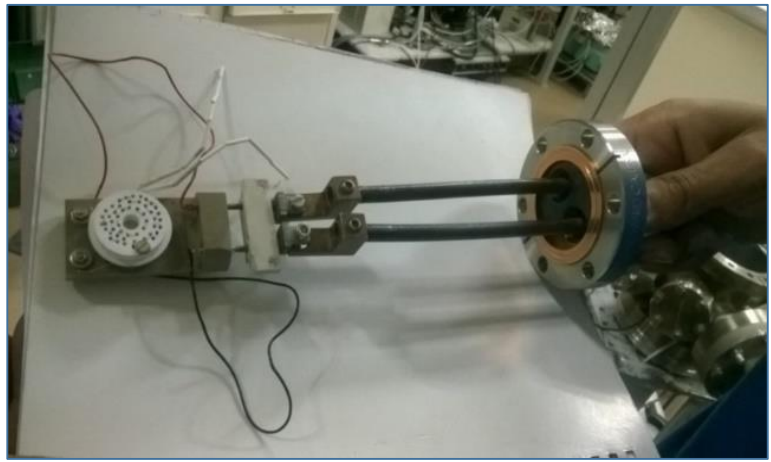

(b)

**Fig. 3.9:** (a) Schematic diagram of the thermal evaporator (b) actual image of the evaporator.

To attach the platinum wire to the boron nitride material, we made various holes from top to bottom inside the material, with a diameter of 0.4 mm. To provide the base for the evaporator, we have chosen a ceramic material that is electrically and thermally insulated. After carving it into a suitable bed-shaped design is shown in fig. 3.8(b) above, where the dimensions are mentioned, the ceramic material is heated at 1200°C temperature inside a furnace for 12 hours. This process makes it ready for the required purpose to be served. Finally, the Pyrolytic boron nitride evaporator is attached on top of the ceramic material with the help of M2 size bolts and nuts. The final schematic of the evaporator is shown in fig. 3.9(a) while an actual image is shown in fig. 3.9(b).

In fig. 3.10(a), we can see the main MOSS chamber, inside which the evaporator is attached (fig. 3.10b). The heater is attached with the help of the S.S. plate to linear feedthrough, and we can move it from the corner of the chamber to the centre, i.e. adjacent to the e-gun for the deposition as the substrate position is directly above it. Heater and thermocouple wires are separated with the help of ceramic hollow rods and attached to electrical feedthrough, which is attached to a 35CF flange where we can provide the current through a controller (fig. 3.10c)

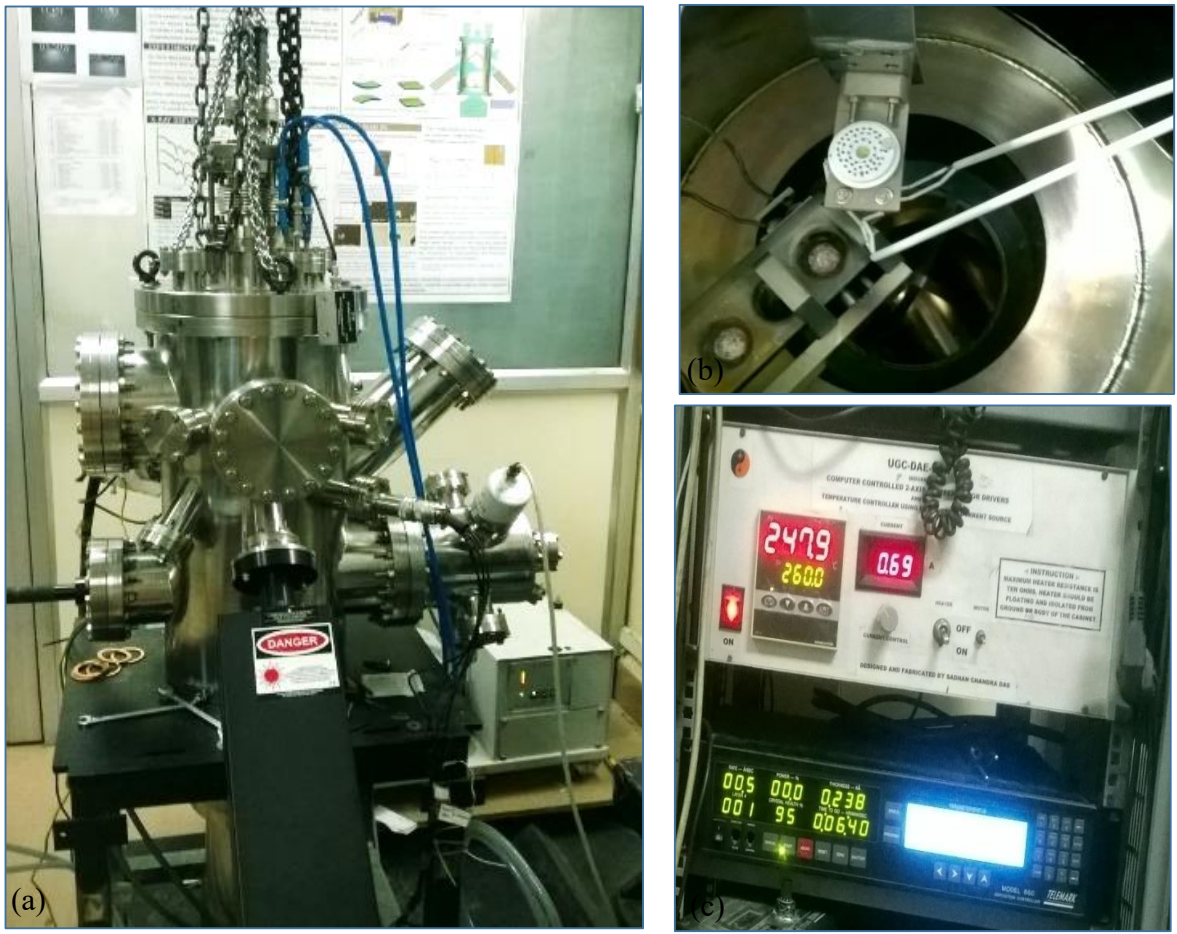

**Fig. 3.10:** (a) Main deposition chamber, (b) attached thermal evaporator inside the chamber, (c) controller assembly.

### 3.3.3    Capability demonstration:

### Deposition of 150 nm Thick Alq₃ film on Si substrate

We have deposited various samples of Alq3 and Alq3/Co combinations. The XRR data of Alq3 deposited is shown in fig. 3.11 with fitting. It is found that for the thickness deposited 180 nm for Alq3, we can see the presence of oscillations up to very high range of q. the roughness found is 0.8 nm which means the organic material deposited using the designed evaporator is of good quality film.

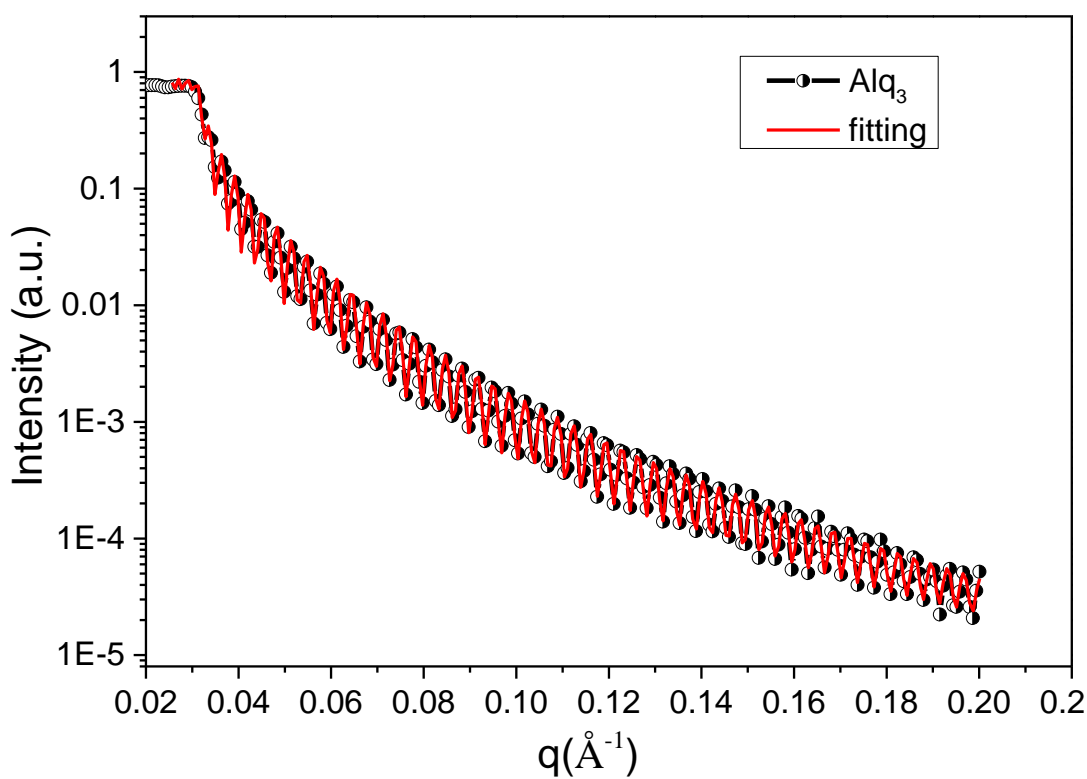

**Fig. 3.11:** XRR data of 180 nm Alq₃ deposited using a thermal evaporator.

### Some other studies utilizing the above set-up:

Some other studies carried out other by groups using homemade evaporators

1. **Growth behavior, physical structure, and magnetic properties of iron deposited on Tris(8-hydroxy quinoline)-aluminium:** K. P. Mondal, S. Bera, A. Gupta, D.

Kumar, V. R. Reddy, G. Das, A. Singh and Y. Yamada-Tamakura, Appl. Sur. Sci. 562 (2021)150169; DOI: https://doi.org/10.1016/j.apsusc.2021.150169

2. **Effect of growth rate on quality of Alq₃ films and Co diffusion:** K. P. Mondal, S. Bera, A. Gupta, D. Kumar, A. Gome, V. R. Reddy, N. Ito, Y. Yamada-Takamura, P. Pandit and S. V. Roth; J. Phys. D: Appl. Phys. 54 (2021) 155304; DOI: https://doi.org/10.1088/1361-6463/abd9eb

3. **TiO₂ as diffusion barrier at Co/Alq₃ interface studied by x-ray standing wave technique:** V. P. Londhe, A. Gupta, N. Ponpandian, D. Kumar and V. R. Reddy; . Phys. D: Appl. Phys. 51 (2018) 225303; https://doi.org/10.1088/1361-6463/aabf7a

## 3.4    Other developments

### 3.4.1    GISAXS set-up at BL-07 beamline RRCAT

#### 3.4.1.1    Schematic of the set-up and its functioning

A schematic of the set-up is shown in fig. 3.12 below. The whole measurement assembly contains the x-ray beam, slit system, sample alignment tower, flight tubes, masking system for the direct beam stop and 2d-detector. Each slit system includes four slits controlled independently using two horizontal and two vertical motors. Horizontal motors cut the beam vertically, while vertical motors cut the beam horizontally. Using these motors, we shaped an x-ray beam with the required size of $0.5 \times 0.5$ mm$^2$. Another cross slit next to the first cross slit is used to minimize parasitic scattering by the first slit. These two slits are attached through a high vacuum tube to reduce any chance of losing the x-ray counts due to diffusion in the air. The sample is attached to the five motion alignment assembly tower, including linear motion, rotation and tilt. Scattered x-ray passes through the 1.8-meter long vacuum tube in its path. Direct and specular beams are cut through the direct beam stopper as they have very high counts and can damage the 2d-detector. We have used a photonic science 2d detector to record x-ray scattering in x-z space. The actual photo of the set-up at the lithography beamline can be seen in fig. 3.13.

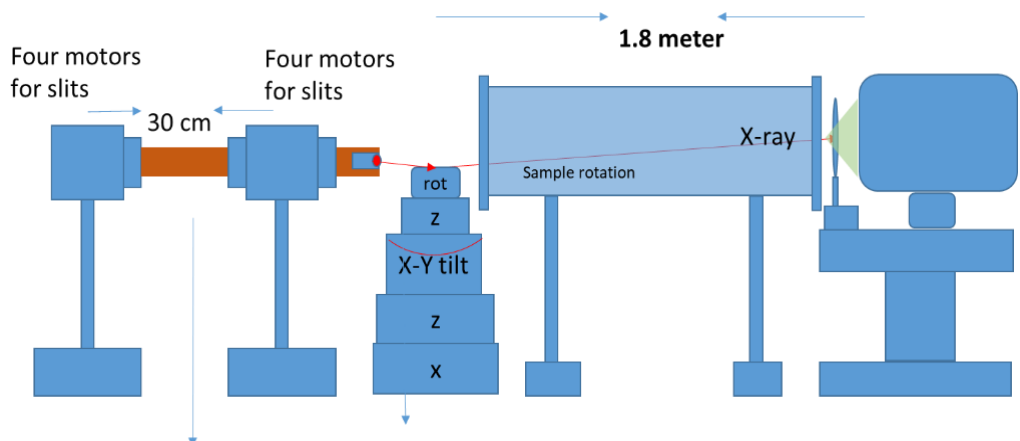

**Fig. 3.12:** Schematic diagram of the GISAXS set-up developed at BL-07, Indus-II, RRCAT.

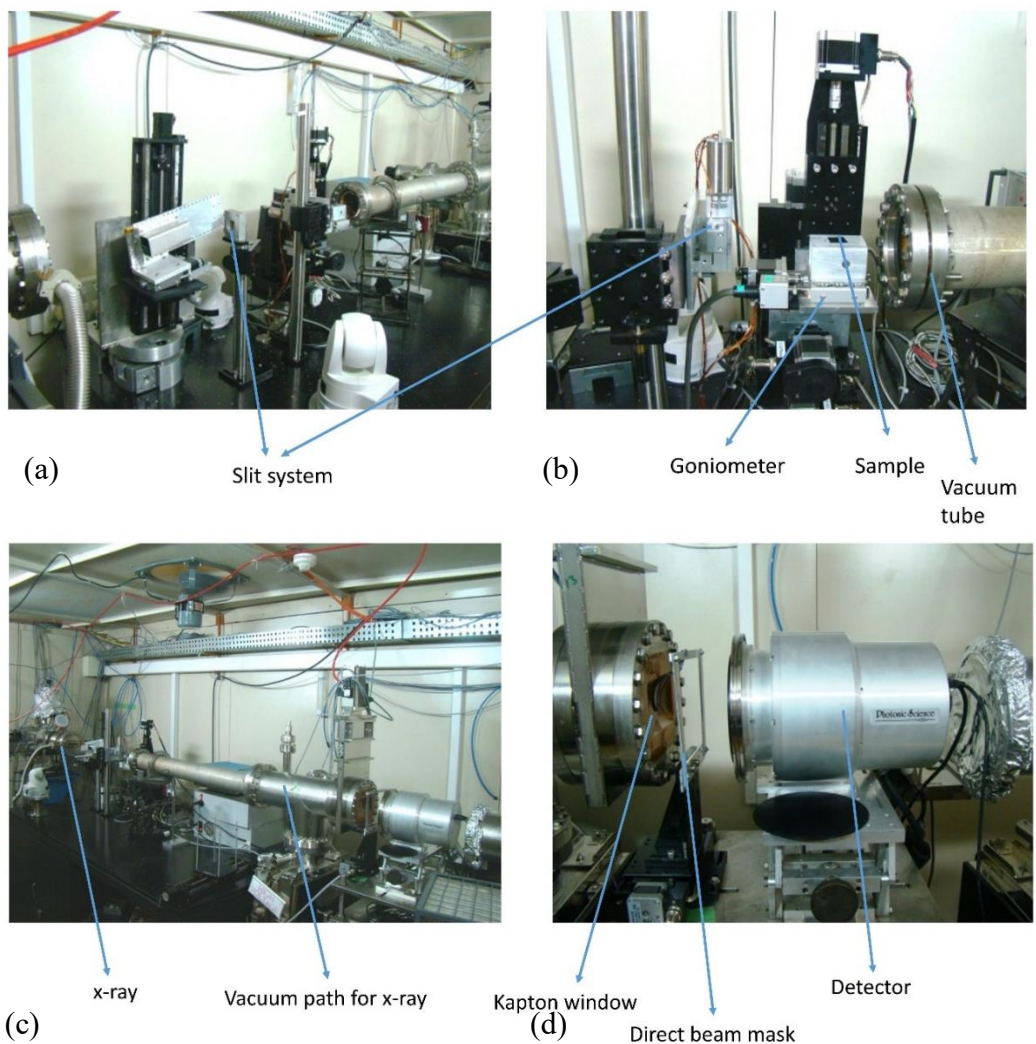

**Figure 3.13:** Actual photograph of GISAXS set-up at lithography beamline, Indus-2, RRCAT.

### 3.4.1.2    Capability demonstration

A multilayer structure [Co$_{oblique}$(44Å)/Ag$_{normal}$(15Å)]$_{10}$ was deposited on Si(100) substrate. To prepare this multilayer, the angle of incidence (θ) was kept alternatively at θ=75° (oblique angle) and θ=0° during evaporation of Co and Ag materials, respectively. Since obliquely deposited Co layers are expected to be in a columnar structure, the final multilayer structure consists of alternating layers of columns (Co) and isotopic Ag layers. We have obtained the GISAXS data for this sample using the developed beamline shown in fig. 3.14a while the extracted 1D data at various horizontal positions are shown in fig. 3.14b.

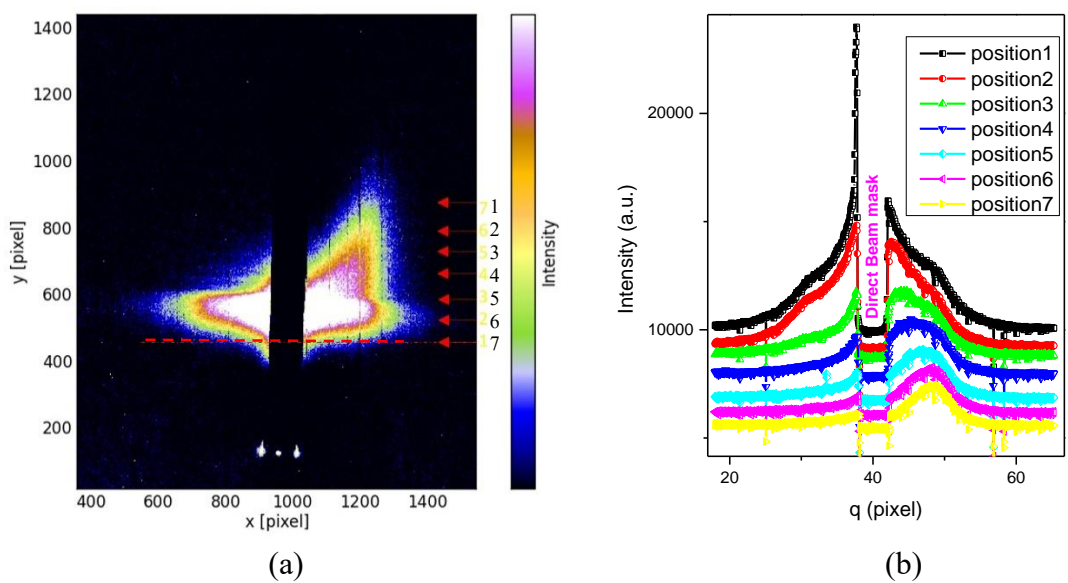

(a)                                                            (b)

**Figure 3.14:** (a) GISAXS data obtained at beamline at RRCAT and (b) the corresponding extracted 1-D data at various positions of (a).

### 3.4.2    Copper heaters for thin-film heating:

### 3.4.2.1    Copper heater (nichrome wiring):

To anneal the samples in the vacuum for the temperature-dependent study, we have designed the copper heater to serve this purpose. A schematic image of the first heater is shown in fig. 3.15a. Here, we chose oxygen-free high thermal conductivity (OFHC) copper for the heater design. We have carved copper with dimensions 40×30×6 mm$^3$. From the length edge, we have designed a slot of 4 mm in height with the help of a lathe machine. Inside this slot, we have attached several ceramic rods (fig. 3.15b), which are thermally conductor but electrically insulator. This ceramic rod contains four holes in which we can pass the heater wire. We have used a nichrome wire with a resistance of 18.4 $\Omega$ for this purpose. The actual image of the heater can be seen in fig. 3.15c. At the top of the copper heater, we have attached clips for mounting the sample to be heated. Also, we have attached a k-type thermocouple at the corner of the heater. Finally, we designed a ceramic base for the copper heater and heated it for 12 hours at 1200°C inside a furnace for final use. The maximum temperature attained by the heater is 520°C because above this temperature, the heat transfer from the ceramic rod to the copper plate is not so fast; hence wire inside the rod reaches above melting temperature when the thermocouple is reading 520°C on the copper plate. Hence safe temperature for operating the heater is found to be below 500°C.

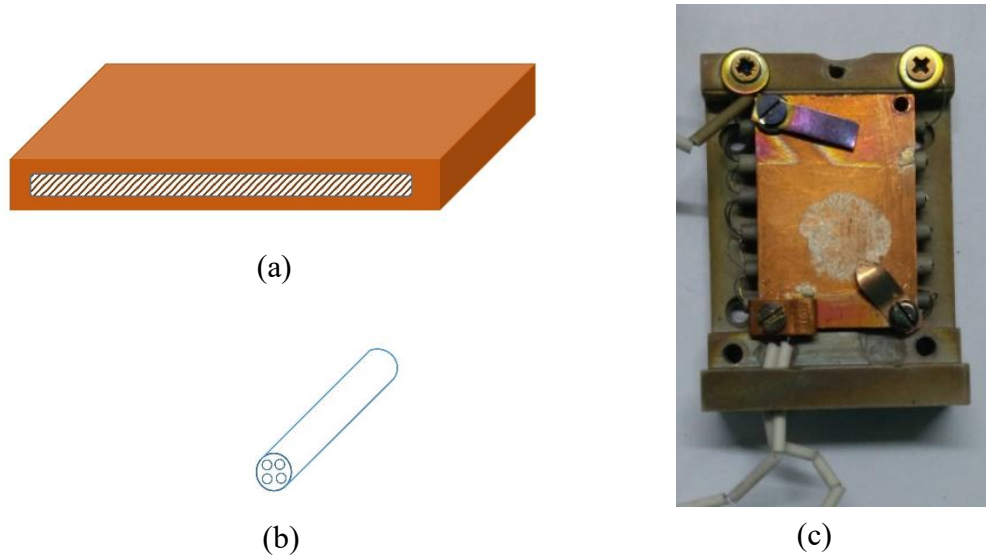

(a)

(b)                                      (c)

**Figure 3.15:** (a) Designed OFHC copper plate for the heater, (b) ceramic tube used for separating the nichrome wire, (c) actual image of the heater.

### 3.4.2.2    Heater with tungsten wiring:

Similar to the previous, we have also developed another copper heater for another vacuum chamber, as shown in fig. 3.16. The dimensions of the copper heater are 60×50×0.6 mm$^3$. Accordingly, we have designed the ceramic base for the heater, which is a thermally and electrically insulator. Proper holes are provided for the thermocouple and sample attachment on the heater. A tungsten wire with a resistance of 1.1 $\Omega$ is used for heating thin film samples. The maximum temperature of  350°C can be achieved using this heater assembly.

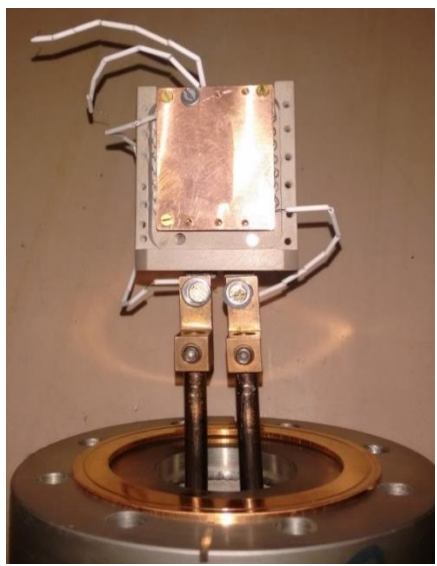

**Figure 3.16:** Designed OFHC copper heater.

## 3.5 Conclusion

The present chapter describes a mini chamber's design and performance for temperature-dependent studies of thin films and multilayer structures using simultaneous grazing incidence small and wide-angle x-ray scattering measurements. The water-cooled body of the chamber allows sample annealing up to 900 K using a UHV-compatible pyrolytic boron nitride heater, thus making it possible to study the temperature-dependent evolution of structural and morphological properties of 2D nanostructured materials. A systematic illustration of the versatility of the chamber has been demonstrated on oblique angle deposited $Co_{obl}/Ag_{nor}$ multilayer structure on Si(100) substrate at P03 beamline, PETRA-III, DESY, Germany. In-situ temperature-dependent GISAXS and GIWAXS measurements were done to study morphology and structure, respectively. A combined analysis revealed that the temperature increase enhances Co columns' diffusion. Temperature about 648K, resulting in a morphology change from columnar tilted structure to densely packed morphological isotropic multilayer.

The present chapter describes the development of the GISAXS beamline at BL-07, Indus-2, RRCAT, Indore. Further, we have designed an organic evaporator for organic deposition materials. We have described the process of development in detail and shown the data obtained from material deposition. Finally, this chapter describes the design and development of two copper heaters for in-situ heating. One of which uses nichrome wire, having a temperature range from RT to 500 ℃ (773 K), while another copper heater which uses a tungsten heater for wiring, has a temperature range from RT to 350 ℃ (663 K).

# CHAPTER 4

## Residual stress and uniaxial magnetic anisotropy during in-situ growth

*Origin of the magnetic anisotropy in the polycrystalline thin film*

In this Chapter, the growth of e-beam evaporated Co films on Si (001)/SiO2 substrate has been studied using a real-time multi-beam optical stress sensor (MOSS), four-probe resistivity (FPR) and in-situ magneto-optic Kerr effect (MOKE) measurements. While MOSS and FPR provide information about internal stress and film morphology, MOKE provides information about magnetic properties, thus making it possible to correlate the evolution of film stress and morphology with uniaxial magnetic anisotropy (UMA) in the film. The film grows via Volmer–Weber mechanism, where islands grow larger to impinge with other islands and eventually merge into a continuous film at around 10 nm thickness. The development of tensile stress in the film is found to be associated with the island coalescence process. The film exhibits a well-defined UMA, which varies synchronously with internal tensile stress developed during growth, suggesting that the observed UMA originates from the minimization of magnetoelastic and morphological anisotropic energies in the presence of internal stress. The absence of UMA in Co film having Ag underlayer in the form of islands is understood in terms of random short-range stress in the film caused by random pinning by the Ag islands.

## 4.1    Introduction

Magnetic anisotropy is a fundamental property of magnetic materials, which determines the alignment of the spins along a preferential direction, called the easy axis of magnetization. From the technological viewpoint, it is an important property of 2D magnetic materials, which often do not have their counterpart in the bulk material. For example, thin films, which possess spin alignment perpendicular to the film plane, i.e. perpendicular magnetic anisotropy (PMA), have been used in magnetic recording media [1] such as hard disk drives (HDDs), whereas in-plane uniaxial magnetic anisotropy (UMA)   is becoming increasingly important for applications in the rapidly developing field of spintronics [2].

The key sources of magnetic anisotropy (MA) in a magnetic thin film are its (i) crystal structure [3], (ii) shape [3] and (iii) strain [4,5,6]. Anisotropy that arises from periodic lattice arrangement (crystal structure) is often realized in the epitaxial magnetic thin films such as Fe [7], Co [8], FeCo [9] and SrRuO3 [10] thin films deposited on substrates such as GaAs (001), MgO (100) and SrTiO3 (001). Here, spin-orbit interaction (SOI) is the resulting preferred magnetization direction with respect to the crystallographic structure of materials and gives rise to magneto-crystalline anisotropy (MCA)[11]. In epitaxial thin films [4,5,10], large PMA [12] and in-plane UMA [13,14] are reported in several studies [4,5,10], which arises due to stress developed via lattice mismatch between substrate and thin film.

In the case of polycrystalline and amorphous thin films, MCA is not expected due to the random orientation of grains. Therefore, understanding the origin of UMA is generally difficult and can't be understood in terms of crystal orientation. The origin of UMA in polycrystalline films is often related to the preparation conditions [15] and substrate properties [16]. In some of the studies, the direction of deposition is found to have a definite relationship with the direction of the easy and hard axis of UMA in such a magnetic thin film [17]. In a recent study by T. Kuschel et al. [16], Co films of different thicknesses are deposited on glass substrates, where ex-situ investigation on a series of samples revealed that the preparation conditions like temperature, deposition rate, obliqueness of deposition, strain in the substrate and possible texturing are not responsible for the origin of UMA. Although the substrate shape is reported to be responsible for the rotation of the anisotropy direction, the origin of UMA is not clearly understood. On the other hand, in an in-situ thickness-dependent study of polycrystalline Co film on Si substrate, the appearance of UMA for thickness greater than 10 nm is attributed to the long-range internal stress developed during the film deposition [18]. In the case of ion beam sputtered nanocrystalline Co thin films, UMA exhibits a non-monotonic behavior with thermal annealing, which is understood in terms of variations in the internal stress, surface roughness, and grain structure [19].

Similarly, in the case of ion beam sputtered amorphous FINEMET film, stress is known to play an important role in the origin of UMA [20,21,22]. Here, the internal stress in films is expected to couple with the magnetostriction, thereby giving rise to UMA. The origin of such stress in ion beam sputtered film is related to the anisotropic ejection of the sputtered atoms from the target material [17]. On the other hand, A. T. Hindmarch et al. [2] have shown that UMA in amorphous CoFeB thin films is consistent with a bond-orientation anisotropy [23,24], which propagates the interface-induced UMA through the thickness of the interface.

A considerable amount of work has been done with the aim of understanding the origin of UMA in polycrystalline and amorphous thin films [9,11,15,16,17,18,19,20,21,22], and most of the studies propose the stress (either growth induced or extrinsic) as a possible cause of the magnetic anisotropy present in the films. Still, the direct relationship between intrinsic stress and the magnetic anisotropy developed during the growth of the film is absent. In fact, several studies [25,26,27,28,29] on stress development in thin-film have been carried out, where stress is found to vary significantly with film thickness. Thus, because of the above facts, it would be important to simultaneously determine intrinsic stress as a function of film thickness and correlate it with the UMA unambiguously. For this purpose, films must be deposited in identical preparation conditions, and thickness dependence studies must be done on the same substrate to avoid unwanted samples to sample morphological variation in different depositions.

In the present work, polycrystalline Co thin films deposited on Si (001)/ SiO2 substrate are studied in-situ to understand the origin of magnetic anisotropy and corroborate with the internal stress evolved during the film growth. Stress evolution during film growth is examined using the real-time curvature-based MOSS technique, whereas in-situ MOKE measurements have been used to study the evolution of the magnetic properties. Combined in-situ MOKE and real-time MOSS measurements successfully demonstrate the direct relationship of stress with the developed uniaxial magnetic anisotropy in polycrystalline Co thin films. In contrast to the ex-situ measurements, where thickness dependence involves growing a series of samples of varying film thicknesses, in the present study, by continuously monitoring real-time stress and magnetic hysteresis of a growing Co film in-situ, a complete thickness dependence curve is obtained in a single sample. The present study also investigates fundamental mechanisms of stress generation and evaluates magnetic anisotropy with film thickness.

## 4.2    Experimental details

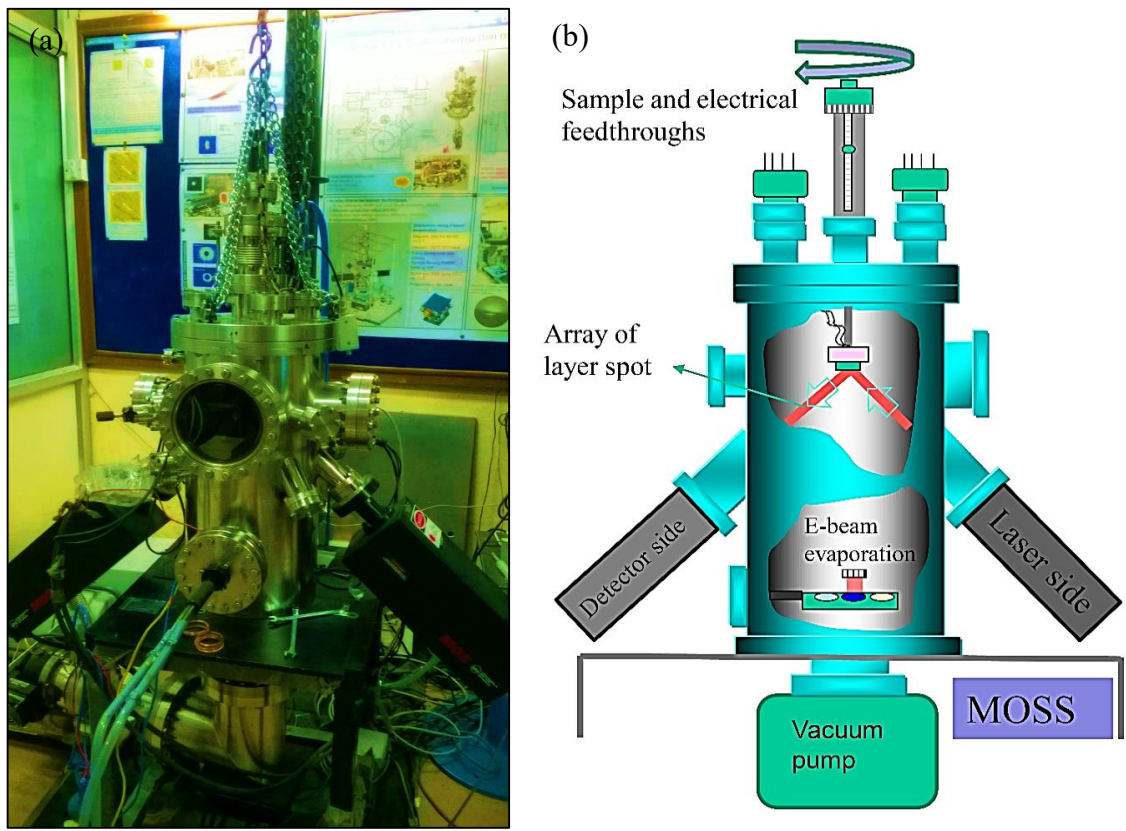

**Figure 4.1.** (a) Original and (b) Schematic view of the MOSS chamber with MOSS components (laser and detector).

Co thin films ranging from a few nanometers (nm) to 45 nm were prepared on Si (001) substrate and studied in-situ in a UHV chamber. The Si substrate generally contains an amorphous oxide ($SiO_2$) layer on its surface [18], which prevents Co from coming in contact with Si to form a silicide layer at the Si/Co interface. Therefore, in the present study, Co film was grown on a bare Si substrate without removing the oxide layer from the surface. UHV chamber is equipped with a facility for thin film deposition using electron beam evaporation and characterization using MOKE, MOSS and FPR measurements (fig. 4.1). In order to study the evolution of stress in thin film and to correlate the same with the origin of UMA, all measurements were done one after the other in identical conditions. The base pressure in the chamber was $\sim 1 \times 10^{-8}$ mbar. The deposition rate was monitored using a quartz crystal monitor and kept fixed at $\sim 0.1$ Å/s during deposition.

**Figure 4.2.** (a) Original and (b) Schematic view of the sample holder with connections for Resistivity measurement. Laser is a part of MOSS components. The details about sample placement in the holder with the connections for resistivity and laser array on the sample are also included in the figure. (c) top view of the sample with connections. Two perpendicular stresses present in the thin film, are also shown.

A specially designed sample holder is shown in fig. 4.2(a), where the sample remains freely floating without any externally applied stress, and the connection for the resistance measurements are shown in fig. 4.2 (c). MOSS (KSA MOS Thermal Scan, k-Space Associates, Inc., Dexter, MI, USA) technique was used to study stress developed in the film during growth. It is a laser-based technique where the array of parallel laser beams, as shown in fig. 4.2(b), reflects from the substrate. After reflection, the change in the radius of curvature (R) is measured on a charge-coupled device (CCD) camera by measuring the changes in the spacing between adjacent laser beams. As shown in fig. 4.2 (b), for a perfectly flat surface, the reflected beams have the same spacing ($d_0$), whereas laser spots are deflected by an amount $\delta d$ from a curved surface. The radius of curvature is given by $1/R = (\cos \alpha/2L) (\delta d/d_0)$; [11,30] where L is the distance from the sample to the CCD camera and $\alpha$ is the angle of incidence of the laser with respect to the surface normal. More detail about MOSS can be found in the paper by Eric Chason et al. [31]. The stress in the film was calculated using Stoney's equation [32], where $1/R$ is directly proportional to the product of the film stress ($\sigma$) and the film thickness ($d_f$) as,

$$\sigma\, d_f = (M_s d_s^2/6)\, (1/R) \tag{1}$$

Where $d_s$ and $d_f$ are substrate and film thicknesses, respectively, $M_s$ is the biaxial modulus of the substrate. MOSS can provide the change in R along (x-direction) and normal (y-direction) to the plane containing the incident and reflected laser beam, and thus $\sigma_x d_f$ and $\sigma_y d_f$ can be obtained simultaneously in both directions during film growth [31]. To perform FPR measurement simultaneously with MOSS measurement, as shown in fig. 4.2(c), Gold (Au) contact pads were deposited on the substrate before Co deposition. Fixed dc current (I) of 10 mA from a constant current source was applied to the current leads of the four-point probe, and the film resistance was measured by measuring voltage from the sample. It is important to note that the current and voltage leads are indirectly in contact with each other through Au contact pads (fig. 4.2c). Therefore, when the film is discontinuous, or the current is not flowing through the sample, the voltmeter reads compliance voltage ~21 V for KEITHLEY 2425 current source. The FPR Interface program divides the voltmeter value by the set current value (V/I) to obtain the film resistance during growth. The actual and schematic views of the sample holders used for such measurements are shown in fig. 4.2(a) and 4.2(b) respectively.

Hysteresis loops of the film were obtained using MOKE in longitudinal geometry. Experiments were repeated in identical conditions for the evolution of UMA with film thickness after replacing MOSS components with the MOKE components, as explained in reference [33]. Here magnetic field for MOKE measurements was produced using Helmholtz coils (HCs) [33]. No current to the HCs was given during the growth process to avoid any influence of magnetic field on the film growth. To get information about UMA, a series of hysteresis loops were collected for different thicknesses at different in-plane azimuthal directions of the sample. After in-situ experiments, AFM measurements were performed ex-situ to get information about the surface morphology of the films. Further, in order to understand the origin of UMA more distinctly and to correlate the same with the developed stress and morphology during growth, experiments were also performed after depositing Ag islands (pinning centre) on Si(001)/SiO$_2$ substrate prior to the Co deposition. Additionally, to study the Co film structure during film growth, reflection high energy electron diffraction (RHEED) measurements were also performed separately by keeping identical growth conditions.

## 4.3    Results and discussion

### 4.3.1    In-situ growth of polycrystalline Co thin film on Si/Sio$_2$ substrate

#### 4.3.1.1    Stress and resistivity measurement during growth

Figure 4.3 gives stress thickness product ($\sigma \times d_f$) as a function of the nominal Co film thickness $d_f$, along two directions of the substrate viz; i) $\sigma_x \times d_f$ along 'x'- and ii) $\sigma_y \times d_f$ along 'y' direction. The figure also includes thickness-dependent normalized resistance ($R_{Co}$), which has been obtained through FPR measurements carried out simultaneously with MOSS measurement. It is clear from FPR measurements that $R_{Co}$ is not significantly affected up to a thickness of about 2 nm. A further increase in the thickness exhibits a rapid decrease in the thickness range of about 2 nm to 6 nm. $R_{Co}$ decreases monotonically after coverage of 10 nm thick film. The observed thickness dependence of the stress-thickness plot shows the systematic change in both x and y directions of the sample; it reaches its maximum value of 6.5 GPa-nm in 'x' direction and about 4.5 GPa-nm in 'y' direction. With further deposition, it decreases until it reaches ~4.5 GPa-nm in 'x' and ~3.0 GPa in 'y' direction at about 30 nm film thickness. Further deposition up to a thickness of 40 nm causes a slight increase in the stress-thickness values in both directions.

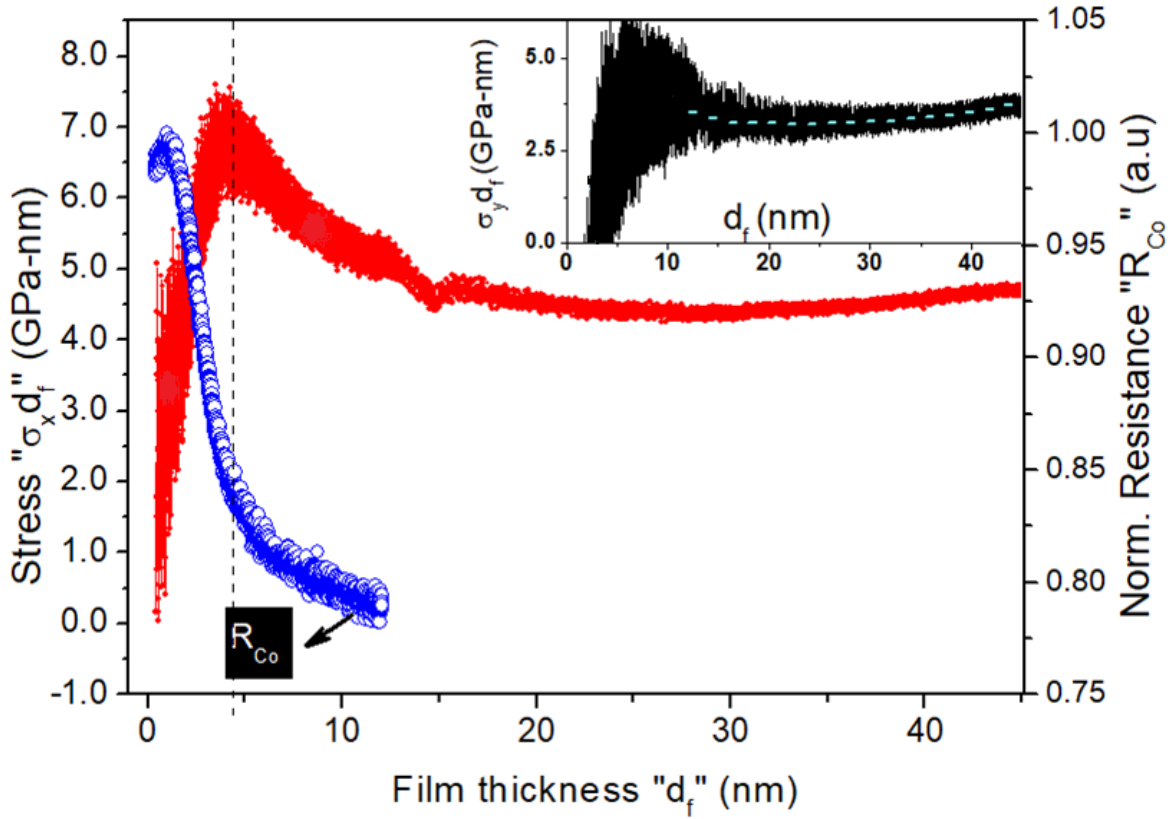

**Figure 4.3.** In-situ real-time evolution of resistance $R_{Co}$ (right axis, blue plot) and stress-thickness product- $\sigma_x \times d_f$ (left axis, red plot) as a function of the nominal film thickness $d_f$. Inset gives $\sigma_y \times d_f$ vs $d_f$ plot.

### 4.3.1.2 In-situ MOKE measurement

To study the evaluation of magnetic anisotropy during Co film growth and to correlate the same with average stress developed during film growth, MOKE measurements were carried out for identically grown film at various thicknesses in the same set-up. To get information about UMA using MOKE at a particular thickness, it is necessary to obtain several hysteresis loops (at least 10-15 in numbers) in different azimuthal directions with respect to the applied magnetic field. Therefore, about 10 to 15 minutes of interruption time in the growth process was needed at each thickness for such measurements. As the islands are more dynamic/active just before and after the percolation stage, therefore it causes self-change in island morphology through the self-diffusion process in this region [34,35]. Thus, UMA measurements near the thickness of island percolation are intentionally avoided so as to get genuine information about

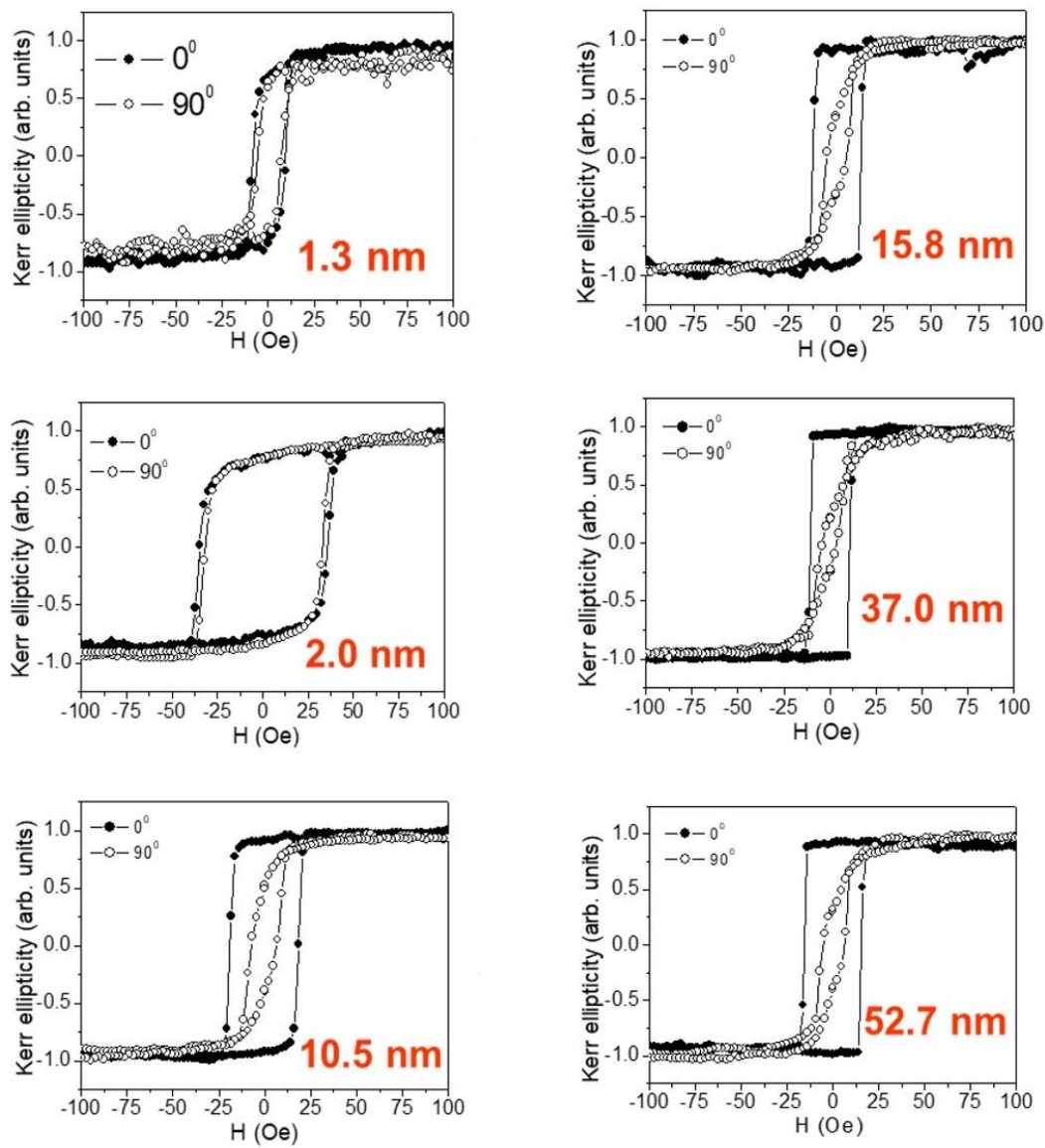

**Figure 4.4.** Hysteresis loops along x (hard axis of magnetization; θ=90⁰) and y (easy axis of magnetization; θ=0⁰) directions for Co film at various thicknesses marked.

UMA. Figure 4.4 gives representative hysteresis loops along x (designated as θ=90˚) and y (designated as θ=0˚) directions at different thicknesses, and corresponding azimuthal angle dependence of $H_c$ are plotted in the polar plots in fig. 4.5. It is clear from the figure that for ~1.3 nm thick Co film, there is no observable difference in the shape of the loops and coercivity ($H_c$). However, the film with a thickness of ~10 nm or more exhibits noticeable variation in the shape of the hysteresis loops. It suggests that low thickness (before the percolation stage) film is isotropic in nature, and films with thickness ~10 nm or more possess

UMA in the film plane. It may also be noted that the hysteresis loops collected along direction 'y' are square in shape, whereas the rounding-off of the hysteresis loop along the x-direction indicates the presence of UMA in the film with an easy axis of magnetization along the y-direction. $H_c$ polar plots (fig. 4.5) also confirm that the two-fold symmetry in the $H_c$ versus azimuthal angle (θ) develops in the film having a thickness greater than 10 nm and retains up to higher thickness.

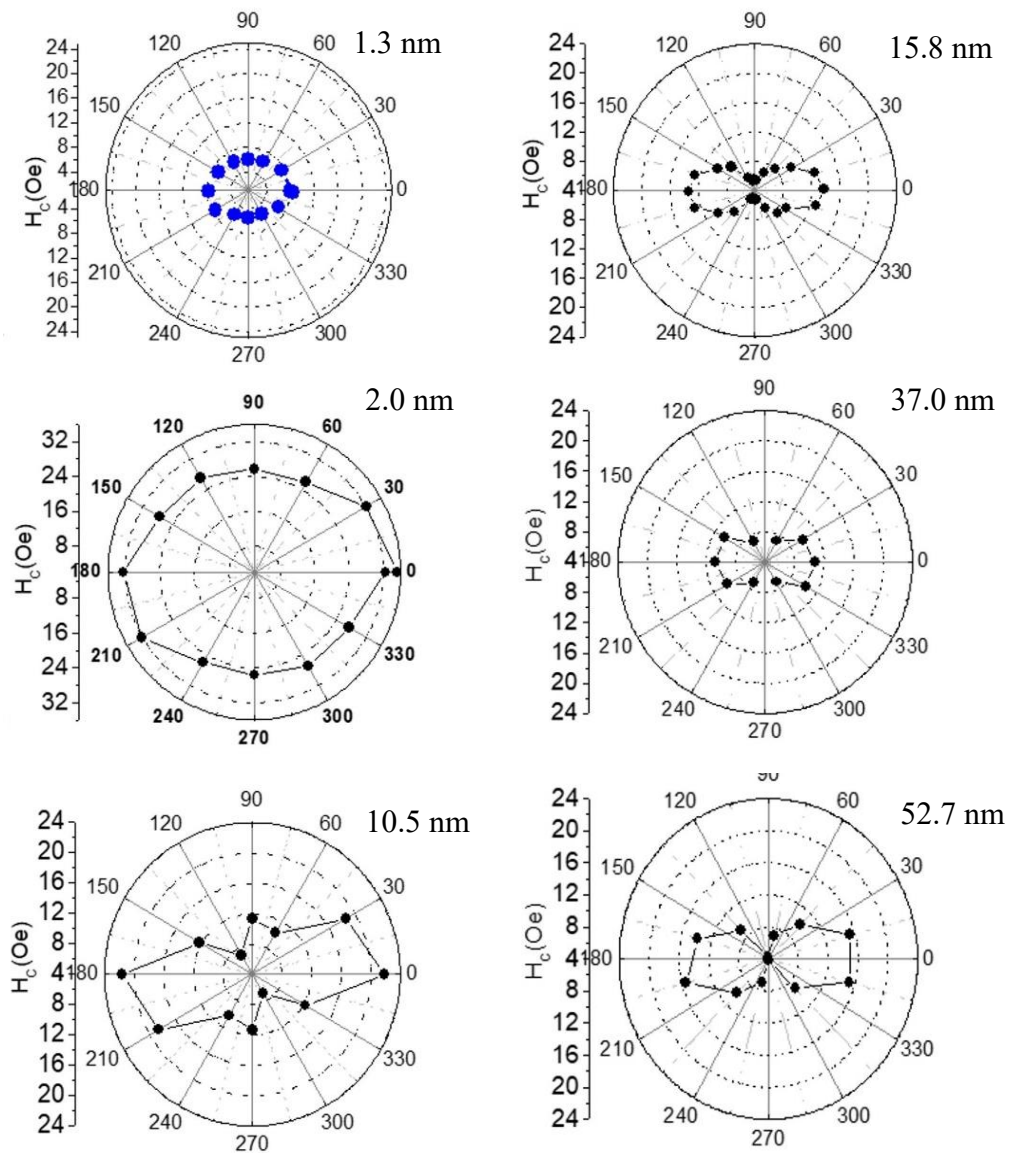

**Figure 4.5.** Azimuthal angle dependence of coercivity ($H_c$) is plotted in the polar plots for the corresponding thickness marked.

The quantitative information about the UMA with increasing film thickness is obtained using $H_c$ variation as a function of azimuthal angle ($\theta$), and it can be written as [19],

$$H_c = H_a \cos^2\theta + (H_{cw}^2 - H_a^2\cos^2\theta \sin^2\theta)^{1/2} \qquad (2)$$

Where $H_a$ is the magnetic anisotropy field, $H_{cw}$ is the domain-wall-pinning coercivity, and $\theta$ is the in-plane angle between the applied field (H) and the easy axis of magnetization. The azimuthal angle dependence of $H_c$ at various thicknesses is fitted with equation (2) by taking $H_a$ and $H_{cw}$ as fitting parameters. One of the representative $H_c$ data fitting for the film having a thickness ~15.2 nm is shown in fig. 4.6. The angular dependence of $H_c$ for all thicknesses is fitted, and parameters $H_a$ and $H_{cw}$ are plotted in fig. 4.7a and 4.7b, respectively. Both $H_{cw}$ and $H_a$ show similar behaviour with increasing film thickness up to 20 nm. $H_a$ decreases from 12 Oe to 5 Oe as the film thickness increases from 10 nm to 26 nm. Further, an increase in Co film thickness to 45 nm results in a slow increase in $H_a$ up to 8 Oe. Similarly, $H_{cw}$ decreases with increasing thickness and reaches its minimum value at about 21 nm film thickness. Initial higher values of $H_{cw}$ may be attributed to the domain wall pinning at the surface/interface, which decreases with increasing film thickness.

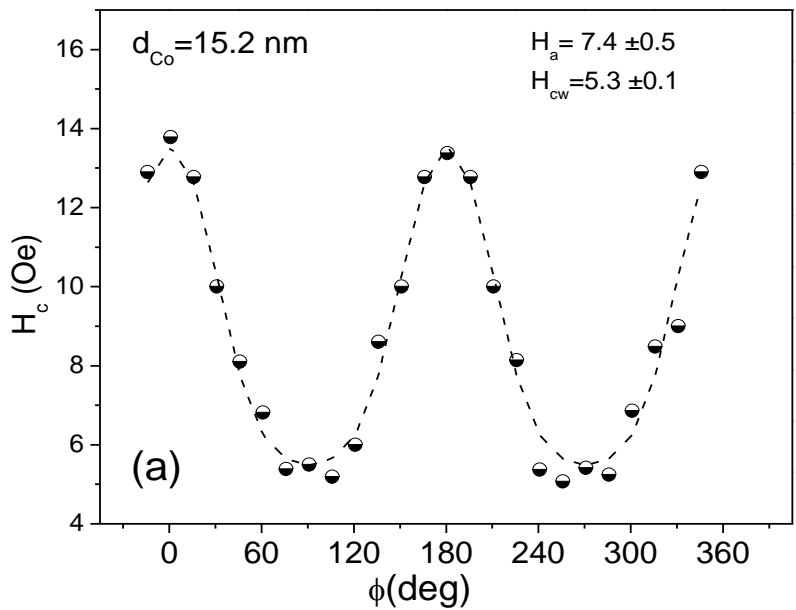

**Figure 4.6.** Azimuthal angle dependence of the $H_c$ fitted with $H_a$ and $H_{cw}$ as a fitting parameter for film thickness $d_{Co}$=15.2 nm.

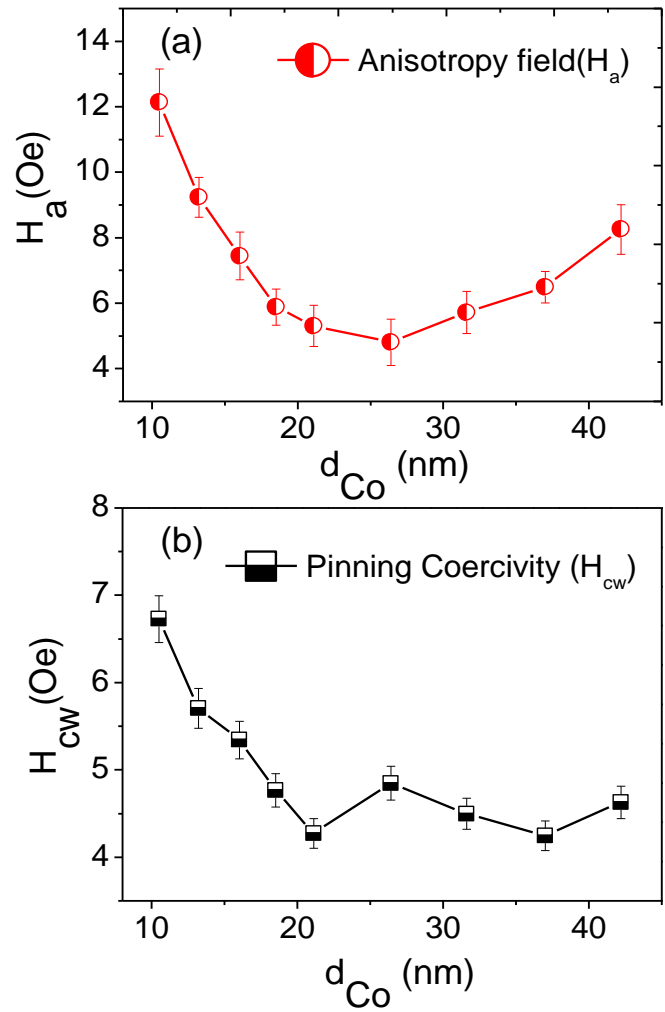

**Figure 4.7.** Thickness-dependent variation of (a) $H_a$ and (b) $H_{cw}$.

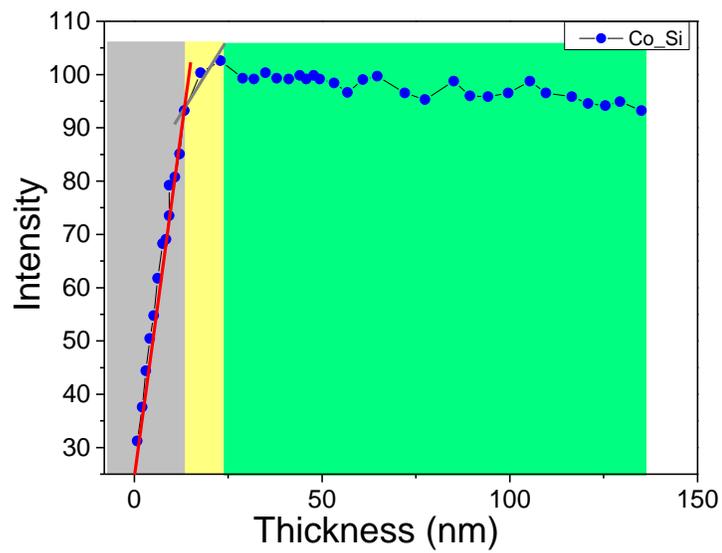

**Figure 4.8.** Intensity of MOKE loops vs thickness of Co film plot.

To get the information about any magnetic dead layer present in the interface, the intensity of the reflected laser (used in MOKE) from the sample vs thickness of the film is measured, and the corresponding intensity vs thickness plot is shown in fig. 4.8. Up to the Co film thickness of 13.3 nm, reflected laser intensity increases linearly (region covered by a grey area). Cut at the thickness axis by the linear fitting of this region provides zero intensity at zero thickness of Cobalt film, which means there is no magnetic dead layer (cobalt silicide) present at the interface. It may be possible because of the significant thickness (2-3 nm) of the native $SiO_2$ layer present on top of the Si substrate. The yellow area represents the gradual region where MOKE intensity increases slowly and reaches its maximum at a Co thickness of 23 nm, which is the saturation point. As marked in the figure, above 23 nm thickness, MOKE intensity remains constant (saturation region: green area).

### 4.3.1.3    In-situ RHEED measurement

In the case of hcp Co, which possesses an inherent UMA with the c-axis being the easy direction, any crystallographic texture, if present, can also contribute to the UMA. Therefore in the present case, such structural information about Co film is obtained through RHEED measurements under UHV conditions. As these measurements could not be performed along with MOKE and MOSS measurements due to geometrical and technical constraints, a separate set of experiments were performed by keeping all deposition conditions identical to the previous experiments. Representative RHEED images were collected for different Co film thicknesses: 1 nm, 10 nm and 40 nm and presented in fig. 4.9. Observation of different concentric rings corresponding to different diffraction planes (110), (103) and (112) of the hcp Co film confirms the polycrystalline nature of the film [36]. Since the polycrystalline film is composed of many diverse crystallites with different orientations, therefore such films are usually rough, and background electron scattering makes diffraction rings a bit hazy. Nevertheless, the rings are found continuous in the lower half of the screen at all thicknesses, which confirms the absence of any texturing in Co thin film [37]. In the present film, the absence of crystallographic texture nullifies any role of structure in the origin of UMA.

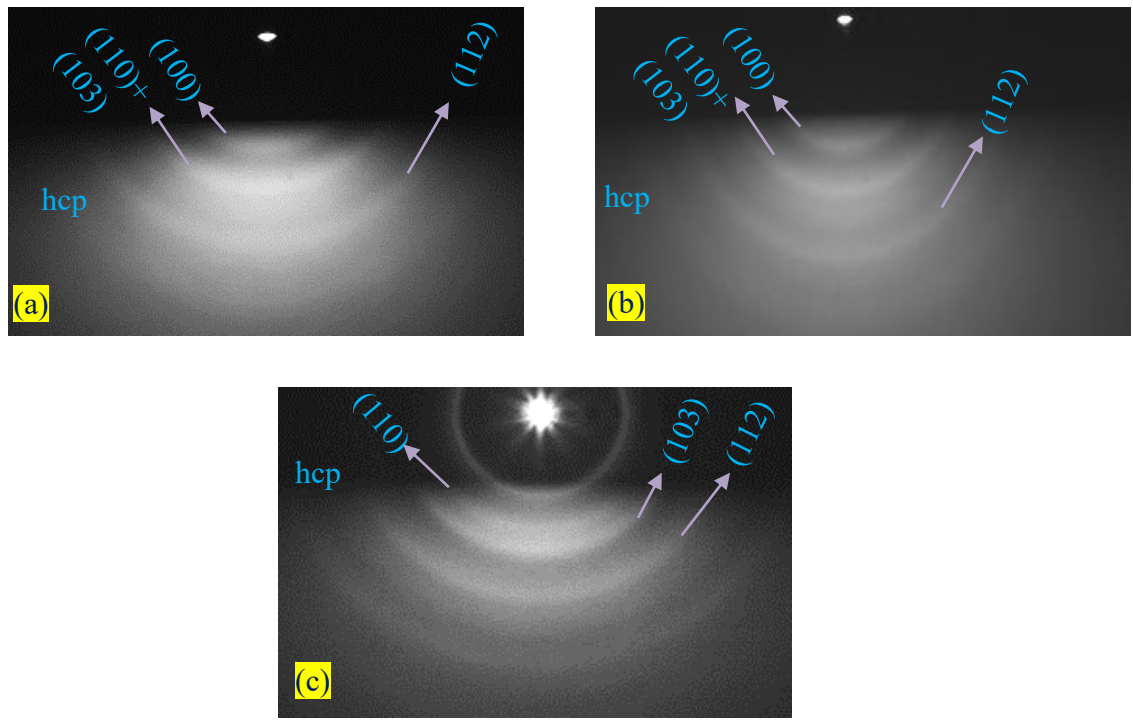

**Figure 4.9.** RHEED patterns of (a) 1nm, (b) 10 nm, and (c) 45nm thick Co film.

Attempts were made to correlate the origin of magnetic anisotropy present in such polycrystalline thin films with varying preparation conditions, where most of the studies were performed ex-situ by growing a series of samples of different thicknesses [15,16,17,18,19]. In this way, it is difficult to grow films in identical growth conditions, making it difficult to obtain unambiguous information about the origin of UMA and its appropriate correlation with other growth parameters. In contrast to the literature, in-situ MOKE, FPR and MOSS studies with Co growth make it possible to correlate the evolution of film stress and morphology with the UMA unambiguously, as explained below;

- The full-thickness range of Co film is grown on Si (001)/ SiO$_2$ substrate to avoid the effect of substrate crystalline orientation on the cause of UMA in the grown film.
- The same substrate size during the experiments nullifies any significant role of substrate size.
- Growth parameters such as temperature, deposition rate, chamber pressure etc., were constant throughout the film growth.

- The sample holder was kept about 40 cm far from the electron beam gun [33] to avoid the effect of a stray magnetic field (H <0.3 Oe) on the film growth.

- During growth, Helmholtz coil current (used for MOKE measurements) was kept at zero to avoid the coil magnetic field's influence on Co film.

Thus, present experiments clearly establish that the origin of the UMA and its variation during growth is mainly due to stress developed through morphology evolution during growth.

### 4.3.1.4 AFM measurement

AFM measurements provide surface morphology of the sample after deposition of 45 nm Co thin film. AFM images are taken in various scales at various portions of the Co film, as shown in fig. 4.10. It is clear from AFM images of lower scales that grains are preferentially elongated along the vertical "x" direction (θ = 90°) (blue arrow in fig. 4.10c) as compared to the horizontal "y" direction (θ = 0°). To understand more clearly, horizontal and vertical line profiles (fig. 4.11) are taken from the 400nm scale AFM image. This line profile clearly shows the wider size of grains (wider oscillations) along the "x" direction as compared to the "y" direction (narrow oscillations), which is also in agreement with the stress plot, which shows higher tensile stress in "x" direction as compared to "y" direction. Such surface grain size asymmetry may be due to anisotropy in nucleation and growth or preferential percolation in the x-direction similar to the reference [38].

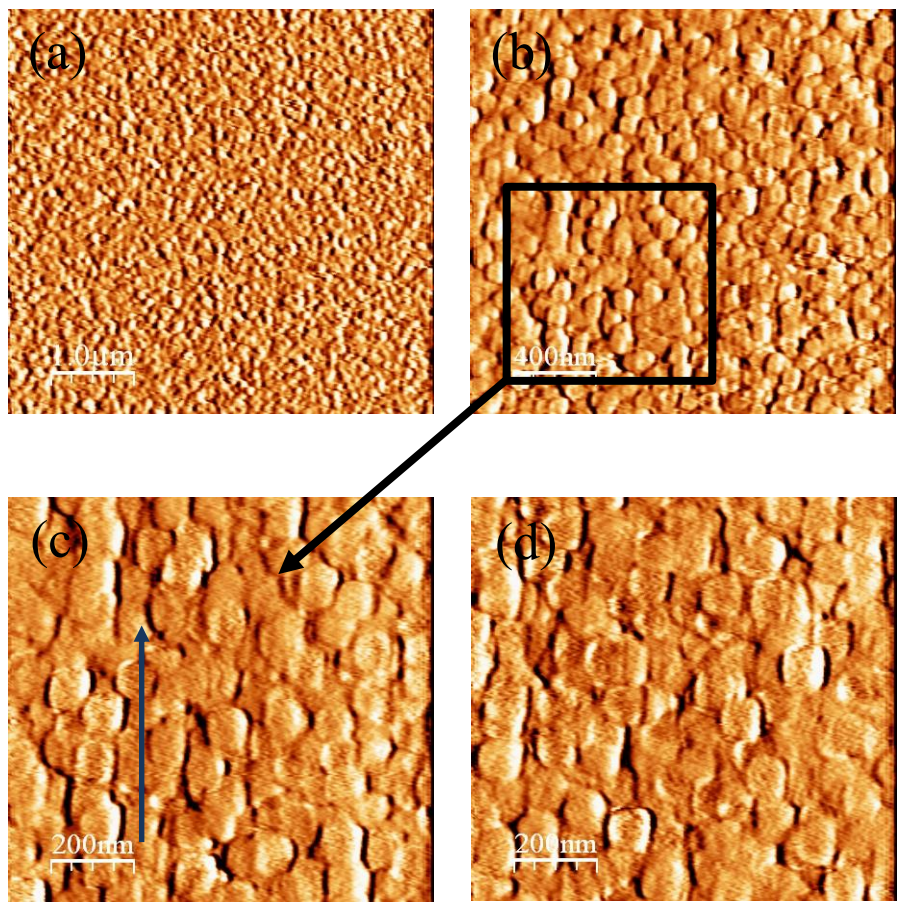

**Figure 4.10.** AFM image of 45 nm thick Co thin film is various scale (a) 1.0 μm scale (b) 400nm scale (c) 200 nm scale (scan taken at portion marked by the black box in (b)) (d) 200 nm (taken at some other portion of the film).

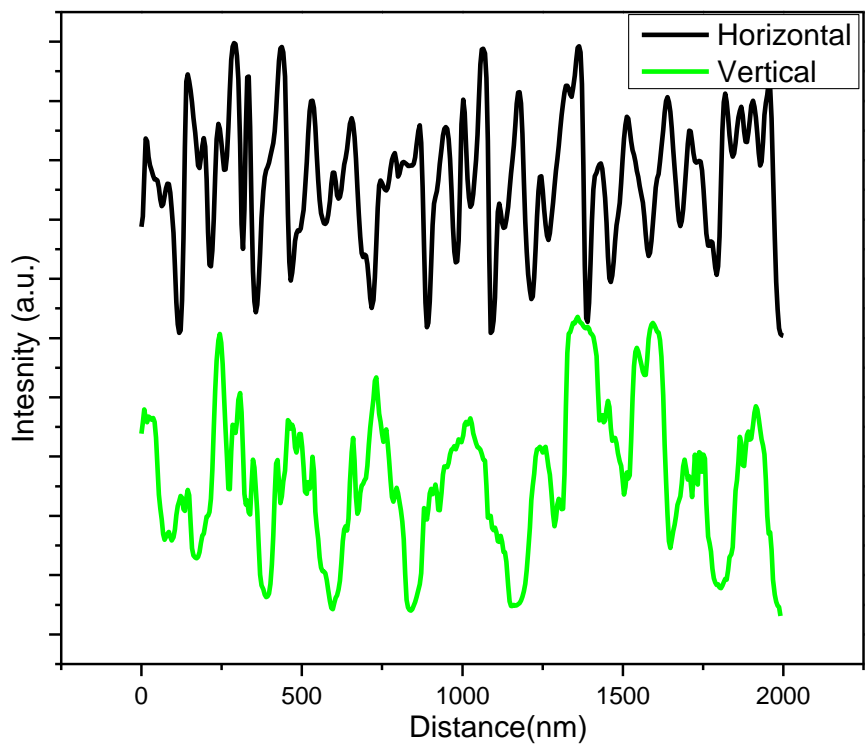

**Figure 4.11.** Horizontal and vertical line profiles taken from 400 nm scale AFM image of Co Film.

### 4.3.2    Co thin film on Ag islands: reduced long-range stress

With an aim to grow stress-free Co film, long-range stress in the film is reduced by providing short-range random pinning using an appropriate buffer layer. For this purpose, the Ag buffer layer is grown on the surface of the Si/SiO2 prior to the deposition of the Co layer. It is known that the surface energy of the Ag is much larger as compared to the SiO2 surface; therefore, Ag material is expected to grow in the island state, which is expected to reduce long-range stress in the film. The related findings are shown in the following section.

### 4.3.2.1    In-situ real-time transport measurement during Ag growth

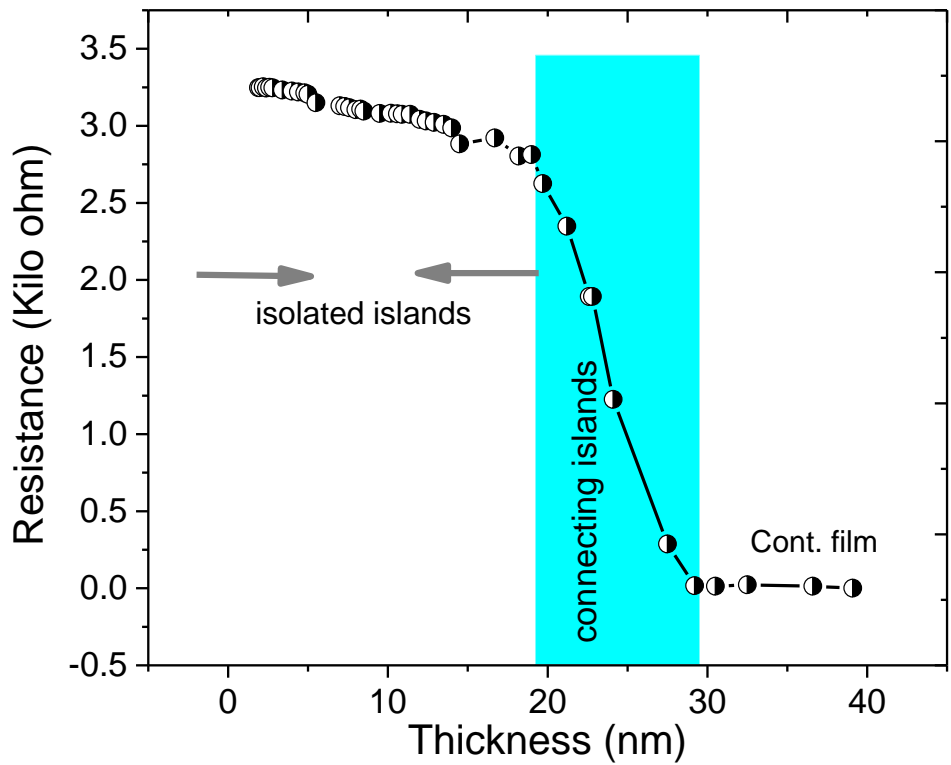

**Figure 4.12.** Resistivity vs thickness plot of Ag film deposited on Si substrate.

To deposit Co on Ag islands, we should have the limiting thickness below which Ag remains in island form. We made four Ag contacts on Si(100) substrate for current supply and voltage measurements for that purpose. We supplied 5.0 mA current to the substrate and started the deposition of Ag. Initially, Ag Forms islands on Si substrate, which are isolated from each other hence resistance is very high, as shown in fig. 4.12. The resistance is very high for Ag thickness up to 19 nm but decreases slowly from 3.26 KΩ to 2.80 KΩ. It shows that current is flowing along the path which passes through islands and islands area is increasing with thickness and current has found more conducting path to flow. As thickness increases beyond 190Å, resistance decreases very fast (percolation region), which indicates that islands have started connecting and thickness up to 19 nm, all the islands have connected, and Ag film becomes continuous.

#### 4.3.2.2    Stress evolution with increasing thickness using MOSS

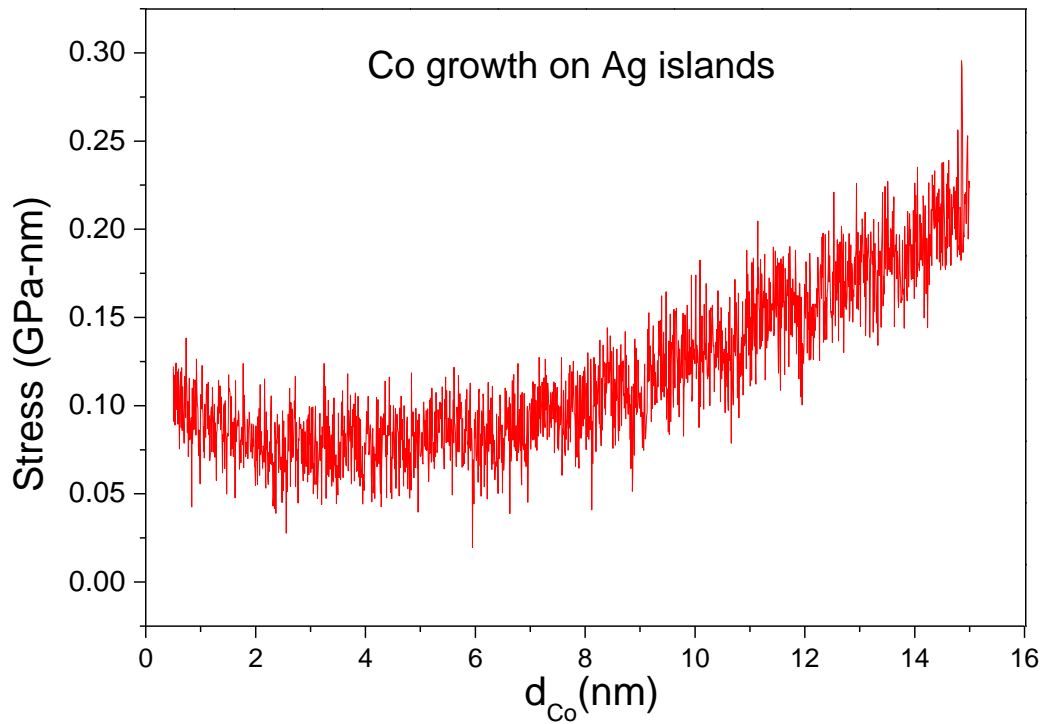

**Figure 4.13.** Variation of stress in Co film grown on Ag islands film.

In order to further understand and confirm the role of the stress on the origin of UMA, Co film was grown on Si/SiO₂/Ag (10 nm) structure, where the Ag buffer layer was in the form of isolated islands. FPR measurements are used to confirm the island nature of the film, where percolation (path for electron flow) between the islands does not takes place up to 10 nm coverage due to the separation between Ag islands. When we deposited Co film on Ag islands, we found no significant tensile stress evolution on the film surface as found in the previous case where there were no Ag islands used as a base layer. Stress-thickness vs thickness plot obtained during Co growth on Ag islands, as shown in fig. 4.13, clearly shows no significant change in the stress up to 15 nm thickness.

#### 4.3.2.3    Magnetic properties of Co film on Ag islands

In contrast to Co on Si/SiO₂ sample, when Co film is grown on Ag islands, it is found that magnetic hysteresis loops are the same in all the directions in the film plane, which means

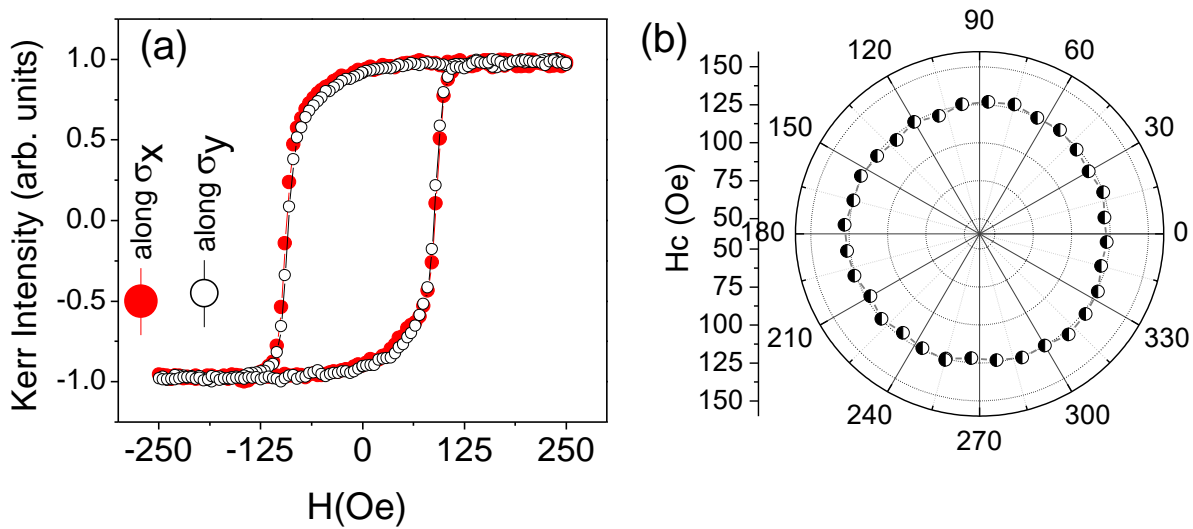

**Figure 4.14.** (a) Hysteresis loops along $\sigma_x$ ($\theta=90^0$) and along $\sigma_y$ ($\theta=0^0$) directions and (b) isotropic polar plot of coercivity.

Co film is magnetically isotropic in nature. Figure 4.14(a) shows the hysteresis loops along two perpendicular directions $\sigma_x$ and $\sigma_y$ in the film plane. The increased coercivity $H_c$ confirms large domain wall pinning in the Co film due to Ag islands. Coercivities of the hysteresis loops taken in the film plane are shown as a polar plot in fig. 4.14(b) confirms the magnetic isotropy of the film.

### 4.3.2.4 Surface morphology: AFM measurement

AFM measurements are performed on the bilayer film to understand the Co film morphology on Ag islands. AFM measurement in various scales, including 1.0 µm, 400 nm, and 200 nm, are presented in fig. 4.15. All these AFM images are taken at various portions of the film plane and clearly show two important points: (i) grains are random in shapes having boundaries in random directions, unlike Co film on Silicon substrate where grain boundaries were preferably aligned along with two perpendicular directions and (ii) bigger grains marked blue shape in fig. 4.15(c) are grown around small grains marked by white shape in fig. 4.15(c). These two types of grains are clear in fig. 4.16. These small grains are understood as the Ag islands having sharp heights due to island nature, while bigger grains belong to Co film growing around Ag islands. Thus the surface morphology obtained through AFM measurement is isotropic in nature.

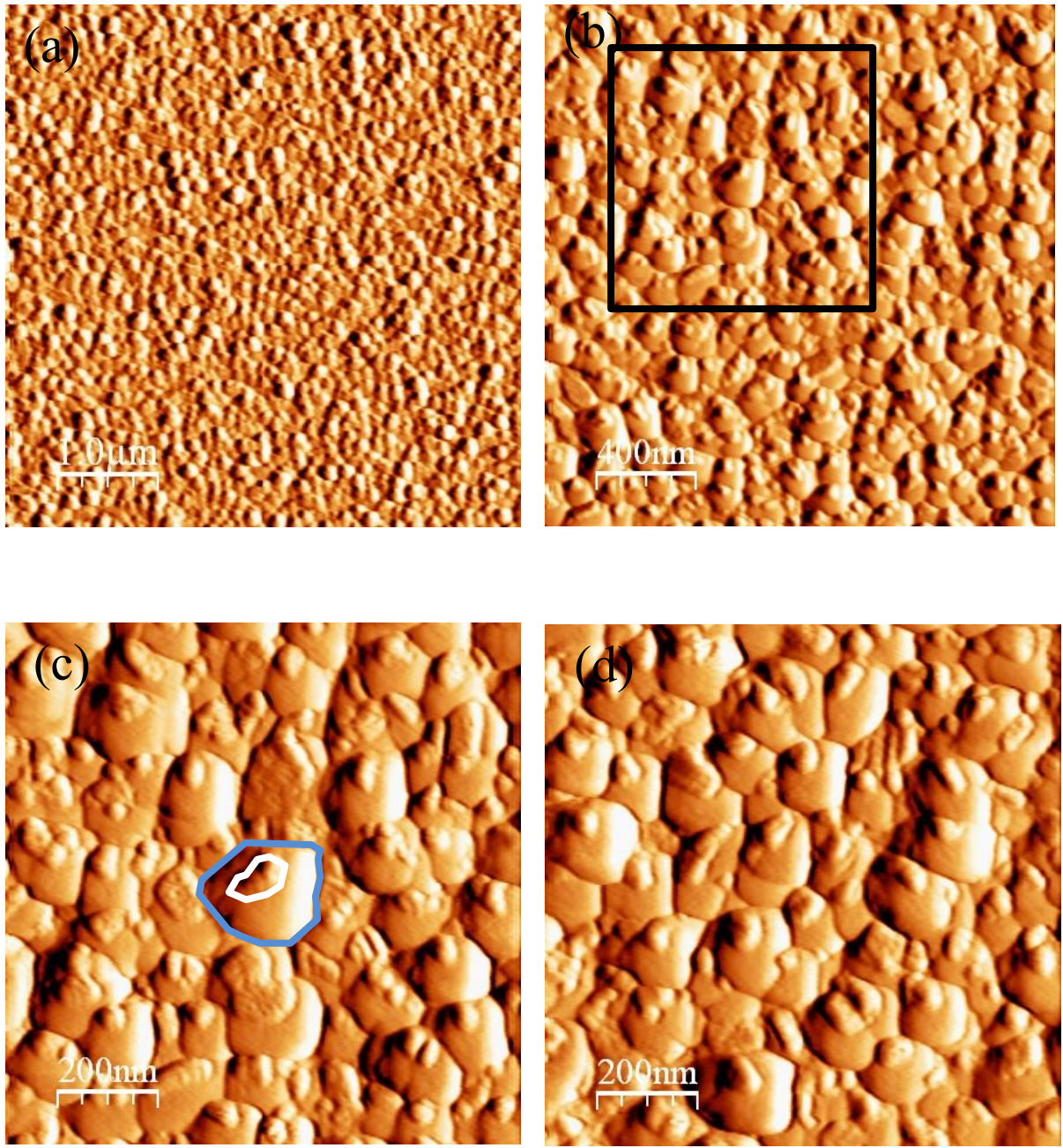

**Figure 4.15.** AFM image of 15 nm thick Co film on 10 nm Ag islands in various scales (a) 1.0 μm scale (b) 400nm scale (c) 200 nm scale (scan taken at portion marked by the black box in (b)) (d) 200 nm (scan taken at some other portion of the film).

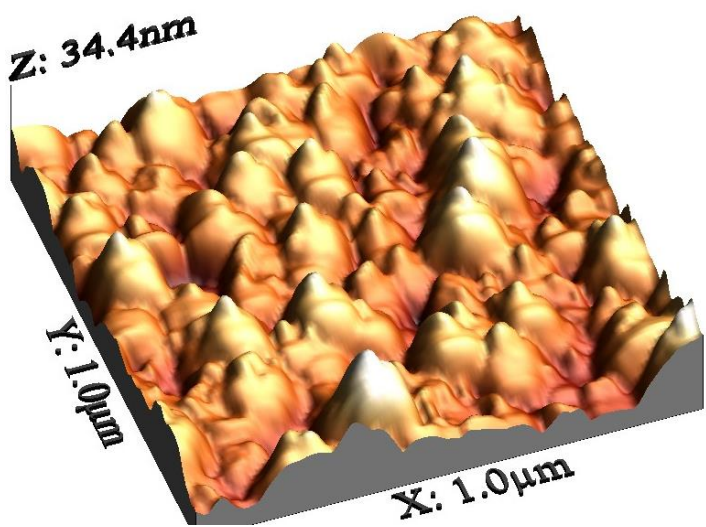

**Figure 4.16.** 3-D AFM image of 15 nm thick Co film on 10 nm Ag islands: 200 nm scale.

### 4.3.2.5    XRR measurements of Co thin film on Ag islands

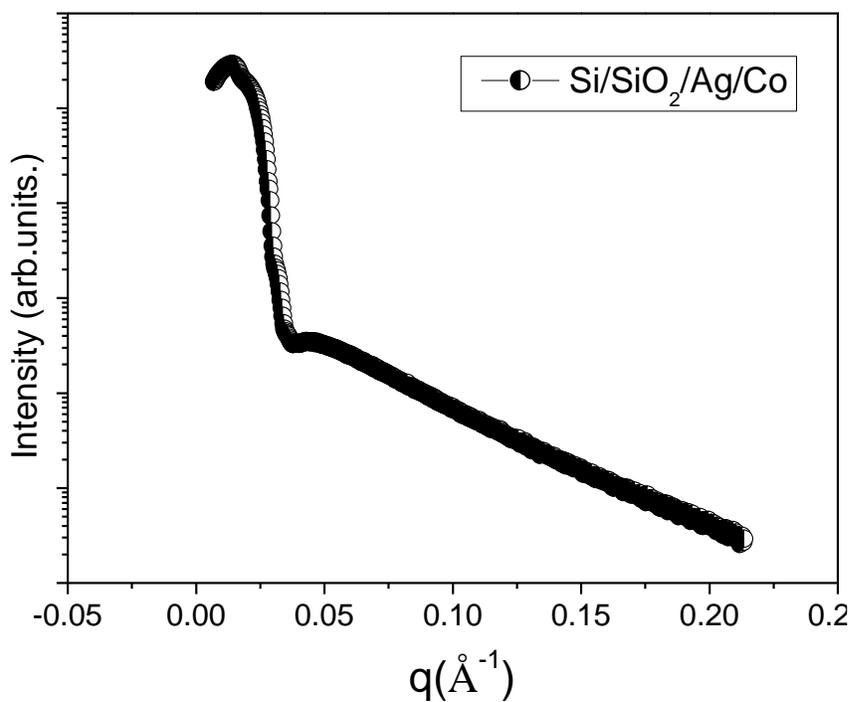

**Figure 4.17.** XRR plot of Ag/Co thin films.

We performed XRR for Ag/Co bilayer thin film but found no clear oscillations, which confirms the bilayer is highly rough in nature (fig. 4.17). The presence of sharp Ag islands is the main cause of this roughness. AFM results also confirm the results obtained through XRR. The RMS roughness found from the AFM analysis is 4.7 nm for this bilayer structure which is sufficiently higher than the RMS roughness of 1.1 nm found in the case of Co thin film.

## 4.4    Discussion- the origin of magnetic anisotropy & role of stress

As the surface free energy of Co is substantially higher than that of $SiO_2$ [34,39], the Volmer–Weber type of growth is expected in the present case, where islands of Co nucleate on the Si substrate. The same has also been observed from resistivity measurements, where conduction of electrons from island to island does not takes place up to 2 nm coverage of the film, which is attributed to the separation between Co islands. It may be noted that up to 2 nm thickness (fig. 4.3), where the current does not flow through the film due to the discontinuous nature of the film, the voltmeter reads the compliance voltage of the current source in FPR measurement and therefore does not have any connection with Co film in this region. A rapid decrease in $R_{Co}$ around the thickness of about 2 to 6 nm (fig. 4.3) confirms the formation of a percolation path between the isolated islands (i.e. island coalescence). With further deposition, the film becomes continuous and shows a monotonic decrease in $R_{Co}$. During polycrystalline film growth, crystallites nucleate with different crystallographic orientations and the grain boundaries are formed when the islands grow to impinge and coalesce [40]. These boundaries have excess free energy and are responsible for tensile stress generation during coalescence [41]. In literature, it is also demonstrated that at various growth stages, tensile stress development is associated with the island coalescence process, with peak tensile stress corresponding to the completion of the coalescence process [41]. In the present case, large tensile stress is developed as islands coalesce and is retained even up to 45 nm thickness of the film. It may be noted that even the hysteresis loop with finite coercivity starts appearing in the films near the percolation stage (before the coalescence), but UMA is not observed. It may be due to the expected high roughness of the film in the island state, even in the presence of stress, the UMA is absent and starts appearing near the continuous film region, where the film is relatively smooth. The fact that stress $\sigma_x$ in one of the in-plane directions) is higher than $\sigma_y$-direction over the full thickness range suggests that the coalescence of islands preferentially takes place along the x-direction. This observation confirms that the anisotropic stress distribution is responsible for the origin of UMA in Co film [34,42]. Slightly off normal deposition [43] of Co film might be the possible reasons for the anisotropic films growth, which may cause island coalescence along x-direction preferentially and lead to anisotropic stress distribution in the film plane. However, the precise reason for the presence of anisotropic stress distribution is not clear.

The relationship of stress-induced UMA with net tensile long-range stress $\Delta\sigma = (\sigma_x - \sigma_y)$ along x direction can be understood in terms of the magneto-elastic interaction energy $E_M$ [3,44] as:

$$E_M = 3/2 \, \lambda(\sigma_x - \sigma_y) \, Sin^2\phi \qquad (3)$$

Where $\sigma_x$, and $\sigma_y$ is the stress contribution along the x and y direction of the sample, $\phi$ is the angle between the direction of saturation magnetization ($M_s$) and $\Delta\sigma$ ($\sigma$ is positive for tensile and negative for compressive stress), $\lambda$ is magnetostriction constant. This relation is often used to determine the effect of stress on magnetic behavior [11], where the material response to stress depends on the sign of the product of $\lambda$ and $\sigma$. For polycrystalline Co films, the average magnetostriction constant ($\lambda$) is negative in magnitude [3], therefore, the product of $\lambda$ and $\Delta\sigma$ is negative. In this case, the minimum of $E_{me}$ occurs when $M_s$ and $\Delta\sigma$ are at the right angle ($\phi=90^0$), which suggests that the net tensile stress $\Delta\sigma$ along x-direction would try to align the magnetic moments preferably in the direction normal to it. This result is in accordance with our recent study [11], where UMA was induced in polycrystalline Co films by inducing uniaxial stress in the film by removing the bending force from the substrates after film deposition. It may be noted that in the case of Co film on Ag islands, the disappearance of the UMA may be attributed to the absence of long-range residual stress due to the presence of random short-range stress in the film caused by random pinning by the Ag islands [11,19]. In this case, the applied field needed to saturate the magnetization would exceed $H_a$. The drastic increase in the coercivity of Ag(islands)/Co film further confirms huge domain wall pinning in the Co layer due to Ag islands.

## 4.5    Conclusion

In conclusion, the growth of e-beam evaporated polycrystalline Co films has been studied in-situ with the aim of investigating the origin of uniaxial magnetic anisotropy. During growth, anisotropic morphology and anisotropic tensile stress in the film are found to be associated with the island coalescence process. The position of the peak tensile stress results from the competition between stress generation due to island coalescence and stress relaxation due to an increase in thickness with deposition. As a result, the film exhibits a well-defined UMA, which varies concurrently with internal tensile stress developed during growth. Stress couples with non-zero magnetostriction in film and results in the appearance of UMA through the minimization of magnetoelastic and anisotropic energies. The present in-situ real-time investigation led to a better and unambiguous understanding of the origin of the UMA in polycrystalline thin films. Consequently, applying controlled stress in the film increased the possibility of tuning magnetic anisotropy according to the desired functionality.

# CHAPTER 5

## Tailoring of magnetic anisotropy by external stress

*__External stress, a factor to induce the magnetic anisotropy in thin films__*;
*"Demonstration through polycrystalline Co thin film"*

**5.1**      **Introduction**

**5.2**      **Experimental details**

       **5.2.1 Preparation of polycrystalline Co thin film**

       **5.2.2 Applying the controlled stress in thin films**

       **5.2.3 Calculation of stress through Stony equation**

**5.3**      **Results**

       **5.3.1 Film thickness and Surface morphology**

       **5.3.2 Structural Analysis: X-ray diffraction**

       **5.3.3 X-ray photoelectron spectroscopy for elemental analysis**

       **5.3.4 Surface magneto-optical Kerr effect for thin-film magnetism**

**5.4**      **Discussion**

**5.5**      **Conclusion**

**References**

Artificial tailoring of magnetic anisotropy by manipulation of interfacial morphology and film structure is the fundamental interest in spintronic and magnetic memory devices. This chapter explores the possibilities of inducing and enhancing magnetic anisotropy in polycrystalline thin films using external stress. Co films on Si substrates have been taken as a model system for the present study. Tensile and compressive stresses were induced in the Co films by relieving the curvature. It has been found that the presence of stress leads to an in-plane magnetic anisotropy in the film, and its strength increases with increasing stress. The easy axis of magnetization in the films is found to be parallel/ transverse to the compressive /tensile stresses, respectively. The origin of magnetic anisotropy in stressed films is understood in terms of magneto-elastic coupling, where the stress tries to align the magnetic moments to minimize the magneto-elastic and anisotropy energy. Tensile stress is also responsible for the surface smoothening of the films, which is attributed to the movement of the atoms associated with the applied stress. The present work provides a possible way to tailor magnetic anisotropy and its direction in all types of polycrystalline and amorphous films using external stress.

## 5.1    Introduction

Uniaxial magnetic anisotropy is one of the most important properties of ferromagnetic thin films, where preferential alignment of the magnetic spins plays an important role in the rapidly developing fields of thin-film technology. Initially, the considerable interest in magnetic anisotropy was motivated by the technological demand to increase magnetic recording density, where large anisotropy was responsible for pushing down the superparamagnetic limit [1,2] and providing stable magnetization in nano-sized magnetic structures [2,3]. Due to the renewed interest in the emerging field of spintronics, magneto-optic recording and sensor technology [4], it has become important to study thin films that exhibit strong, and tunable in-plane magnetic anisotropy gets desired functionality of the devices [5].

The main source of uniaxial magnetic anisotropy (UMA) in ferromagnetic materials is the spin-orbit interaction (SOI) in single crystals. Therefore, it could be often realized in the epitaxial ferromagnetic films and is known as magneto-crystalline UMA [6]. In view of this fact, UMA has been studied in epitaxial thin films in detail [7], whereas less attention has been paid to the polycrystalline and amorphous thin films, where grains are randomly oriented, and therefore, the origin of the observed anisotropy cannot be the magneto-crystalline in nature. Several studies in the literature revealed that a weak UMA is often present in most polycrystalline and amorphous films, irrespective of whether they are produced by evaporation or sputter deposition techniques [8,9]. The origin of such magnetic anisotropy is often related to the residual stresses in the thin films, which might have been generated during the film growth. It is important to note that such UMA in polycrystalline and amorphous films is unpredictable in terms of direction and weak in strength and depends on the deposition parameters such as deposition rate, vacuum during deposition, stray magnetic field, method of deposition etc. [10]. In extreme cases, such stresses are even found to be responsible for peeling off the film from the substrates, film cracking, surface roughening, or mass transport [11]. Recently, there has been considerable interest in inducing tuneable UMA in polycrystalline and amorphous films using intentionally applied controlled stress [5,12,13], so that desired UMA in such films can be achieved.

In the present work, polycrystalline Co films with varying strength of compressive and tensile stress were prepared on the silicon substrates. The strength of applied stress is correlated with the induced magnetic anisotropy and domain magnetization during the magnetization reversal process. It has been realized that controlled external mechanical stress could effectively produce desired magnetic anisotropies even in amorphous and polycrystalline thin films, where crystalline anisotropy is absent.

## 5.2    Experimental details

### 5.2.1    Preparation of polycrystalline Co thin film

Co films of the nominal thickness of 180 Å were deposited on a flat as well as a set of intentionally curved (concave or convex) Si (111) substrates with different radii of curvature (R). As shown in fig. 5.1(a), a specially designed sample holder was used to induce controlled curvatures in the silicon substrates so as to achieve the required stress in the film upon relieving the curvature (fig. 5.1b to 5.1d). Films were deposited using electron beam evaporation with a base pressure of $\sim 10^{-8}$ mbar, while pressure during deposition was $\sim 10^{-6}$ mbar. Deposition parameters, including a deposition rate of 0.1 Å/s, were kept identical in all depositions to maintain similar structural and morphological features in all the films. The layer thickness and deposition rates were maintained using a quartz crystal thickness monitor. It is essential to mention that all substrates were rotated azimuthally during deposition to avoid any stray field present at the sample position.

### 5.2.2    Applying the controlled stress in Co thin films

Stress in the films was induced corresponding to the different radii (R) of convex (compressive stress) and concave (tensile stress) surfaces after retracting force from the substrates.

### 5.2.3    Calculation of stress through Stony equation

Stress ($\sigma$) on the Co films is calculated using relation $\sigma = (E_f\ t)/2R(1-\vartheta^2\ )$ [13], where $E_f$ and t are Young's modulus of Co film and the thickness of the substrate, including the film thickness, $\vartheta$ is the Poisson ratio of Co and R is the radius of curvature of the Si substrate. The standard value of $E_f$ =209GPa and $\vartheta$= 0.31 for Co was taken from the references [13,14] to calculate stress in the film. R for all the substrates was measured using a multi-beam optical stress sensor (MOSS) system [15, 16]. In the MOSS approach, an array of parallel laser beams is made to reflect from the surface. For a perfectly flat surface, the reflected beams have the same spacing ($d_0$). For a curved surface, as shown in fig. 5.1, the beams are deflected by an amount $\delta d$ relative to their flat surface. The radius of curvature is given by [15,16]

$$\frac{1}{R} = [\frac{\cos\alpha}{2L}](\frac{\delta d}{d_0}) \qquad (1)$$

where L is the distance from the sample to the CCD camera and $\alpha$ is the angle of incidence.

Stresses ($\sigma$) = ±100MPa and ±150MPa were induced in a set of films after removing the bending force from the substrates (designated as UNS, T100, C100, T150 and C150; where UNS is the unstressed film, which corresponds to the flat substrate and T, C stand for tensile and compressive stresses respectively). It may be noted that $\sigma$ is considered positive and negative for the film subjected to tensile and compressive stress, respectively. All samples were characterized for their magnetic properties by longitudinal MOKE set-up [17]. In order to study the effect of stress on the magnetization reversal process, hysteresis loops were also measured by rotating the sample azimuthally with respect to the direction of the external magnetic field. Additionally, domains were imaged in C150 film by high-resolution magneto-optical Kerr microscopy (M/s Evico Magnetics, Germany). For this film, hysteresis loops were also obtained by deriving the magnetization signal from the average domain image intensity [18]. The thicknesses and deposition rate of the films were obtained using X-ray reflectivity (XRR) measurements. The polycrystalline nature of Co films was confirmed using grazing incident X-ray diffraction (GI-XRD) measurements. The surface morphology of selected Co films was imaged by atomic force microscopy (AFM) using nanoscope III, version 'E' from digital electronics. To check the contamination in the films, X-ray photoelectron spectroscopy (XPS) measurements were done using Mg K$\alpha$ (h$\nu$ = 1253.6 eV) radiation.

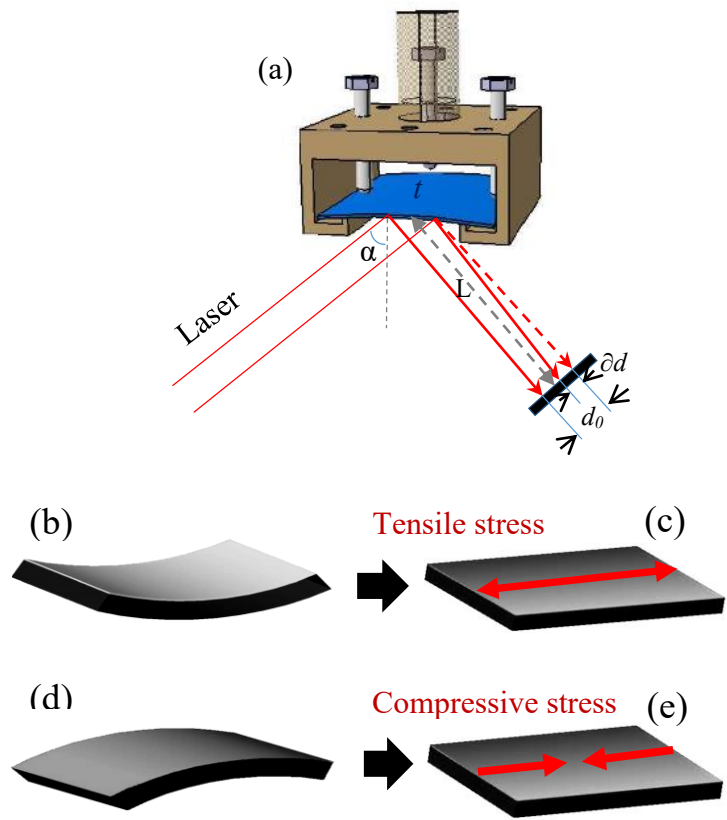

**Figure 5.1:** (a) Pictorial view of the MOSS sample holder to curve the substrates. Schematic of the curved substrates before the deposition: (b) concave and (d) convex shape. After deposition, (c) tensile and (e) compressive stresses are induced in the films after removing the force on the substrates.

## 5.3    Results

### 5.3.1    Film thickness and surface morphology

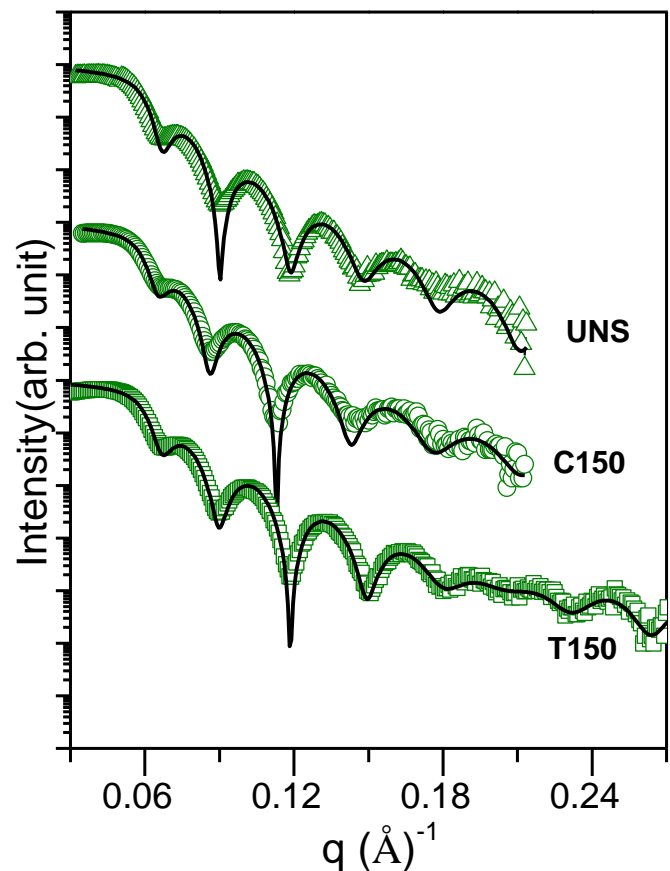

**Figure 5.2**: Fitted XRR patterns of unstressed and stressed films (sample name mentioned on the respective plot).

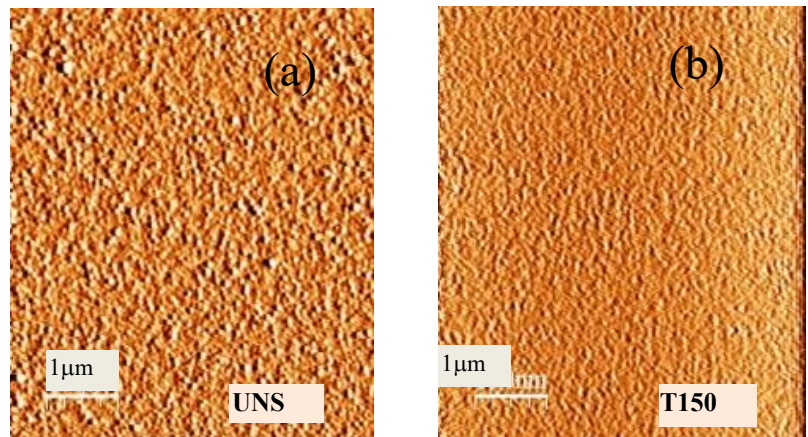

**Figure 5.3**: AFM images of (a) unstressed and (b) stressed films.

| Sample | $r_{XRR}$ (Air/Oxide) | $r_{XRR}$ (Oxide/Co) | $r_{AFM}$ (Air/Oxide) | $d_{CoO}$ | $d_{Co}$ |
|--------|------------------------|-----------------------|------------------------|-----------|----------|
| **UNS** | 11 Å | 9 Å | 10 Å | 17 Å | 168 Å |
| **C150** | 13 Å | 10 Å | 11 Å | 14 Å | 181 Å |
| **T150** | 5 Å | 6 Å | 4 Å | 16 Å | 176 Å |

**Table 5.1**. The fitting parameters of reflectivity data of the samples UNS, C150 and T150; $d_{Co}$ is the Co film thickness; $r_{XRR}$ and $r_{AFM}$ are surface roughnesses obtained from XRR and AFM, respectively. Errors in layer thickness, as well as roughnesses, are ±0.5 Å.

Figure 5.2 gives the XRR pattern of unstressed (UNS) and stressed films (C150 and T150). The XRR patterns were fitted using Parratt's formalism [19]. It may be noted that XRR measurements were done in the ambient. Therefore, to get a good theoretical fit of the data, it was found necessary to incorporate a thin surface layer with somewhat lower electron density. This low-density layer could most likely be attributed to the formation of an oxide layer on the Co surface. All XRR data presented in fig. 5.2 are fitted by taking Co thickness, oxide layer thickness, air/ oxide roughness, oxide/Co roughness and Co/SiO$_2$ roughness as fitting parameters and the results of fitting are presented in table 5.1. The thicknesses ($d_{Co}$) of T150 and C150 films are found to be 181 Å and 176 Å, respectively. It may be noted that the surface roughnesses of the films T150 and C150 are found notably different from that of the UNS film.

The Co film with tensile stress (T150) was seen as significantly smooth as compared to the unstressed (UNS) film. On the other hand, the surface roughness of C150 was found to be relatively more. The same has also been confirmed by AFM measurements (Fig. 5.3a and 5.3b), where the surface morphology of T150 film was found to be significantly smooth compared to UNS and C150 films.

### 5.3.2    Structural analysis: X-ray diffraction

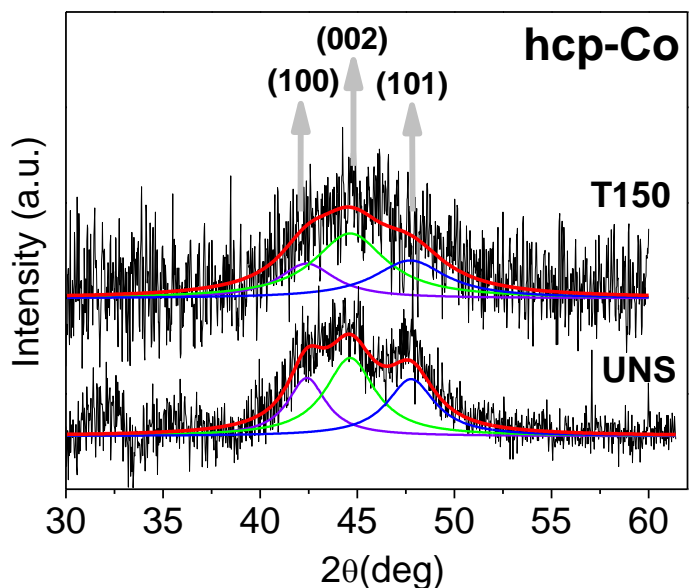

**Figure 5.4**: Fitted X-ray diffraction patterns of unstressed and stressed films.

Figure 5.4 gives X-ray diffraction (XRD) patterns of stressed (T150) and unstressed (UNS) Co films. It is clear that the XRD pattern of both the films exhibits a broad hump which could be fitted very well using an overlap of three peaks of (1 0 1), (0 0 2) and (1 0 0) planes of hcp-Co [8,20]. Present XRD measurements revealed that grown films are polycrystalline. Relatively broad peaks in stressed film can be attributed to small crystallite size, structural disorder, and internal stresses, which contribute to the line width.

### 5.3.3    X-ray photoelectron spectroscopy for elemental analysis

Figures 5.5 give the core level spectra of C 1s, O 1s and Co 3p, respectively, before and after sputtering surface for 14 min using low energy Ar ions. C 1s spectrum shows two peaks at 283eV and 284.5 eV. The peak at 284.5 eV corresponds to carbon contamination at the surface, while the peak near 283eV has been attributed to carbon bonded to Co [21].   The O 1s spectrum also consists of two peaks. The peak around 529.5 eV corresponds to oxygen bonded with Co, while the other peak at higher binding energy has been attributed to the adsorbed oxygen. After 14 min sputtering, decreased area of oxygen and carbon peaks show substantially reduced contamination in the Co film. Qualitative information of the same has also been obtained by relative peak Intensities of Co, C and $O_2$ after correcting for atomic sensitivity factors yield and the ratio of Co: C: $O_2$ ratio is found to be 1:0.08:0.03. The growth of Co films in polycrystalline nature may be attributed to the adsorption of these residual gases (Carbon- 8% and Oxygen-3%) during Co deposition [22].

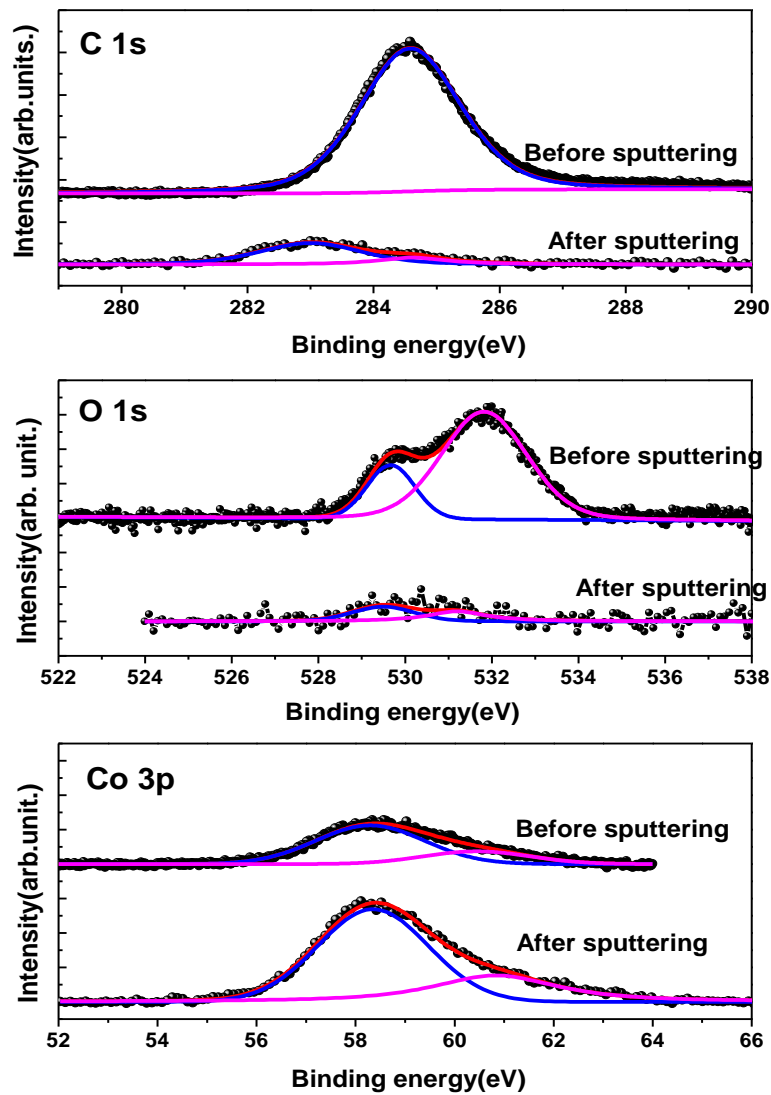

**Figure 5.5:** Core level C 1s, O 1s, and Co 3p XPS spectra of sample UNS (unstressed films) before and after surface etching using low-energy Ar ions. The solid lines fit the experimental data. The curves were shifted vertically for clarity.

### 5.3.4    Surface magneto-optical Kerr effect for thin-film magnetism

The hysteresis loops of UNS film for the field applied along the easy axis and hard axis is shown in fig. 5.6(a). The coercivity ($H_c$) variation as a function of θ is illustrated in fig. 5.6(b) as a polar plot. Slight variation in the squareness of the hysteresis loop and $H_c$ as a function of θ  indicate weak uniaxial magnetic anisotropy (UMA) in the UNS film, which could be due to weak internal stresses generated during the growth of Co film.

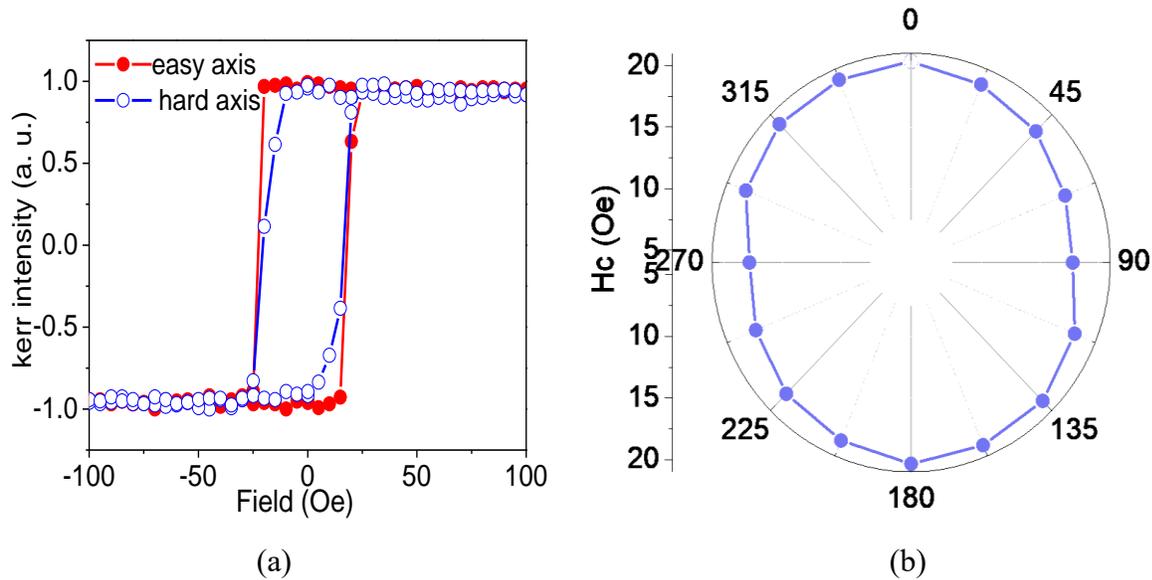

(a)                                                                (b)

**Figure 5.6**: (a) Hysteresis loops of UNS film with magnetic field parallel and perpendicular to the easy axis of magnetization. (b) Variation of $H_c$ as a function of azimuthal angle in a polar plot.

Figure 5.7 and 5.8 gives the hysteresis loops and coercivity variation for tensile and compressive stressed films, respectively. It may be noted that a significant variation in the hysteresis loops can be seen in all stressed films in the direction perpendicular ($\theta$=90°) and parallel ($\theta$=0°) to the stress. When the magnetic field is perpendicular to the tensile stress T100 (fig. 5.7a) and T150 (fig. 5.7b) in the films, the value of remanence is close to unity, but coercivity increases with increasing tensile stress, which is more clear by $M_R/M_S$ vs θ plot (fig. 5.7c) and $H_C$ polar plot (fig. 5.7d).  On the other hand, the values of remanence and coercivity decrease with an increase in the tensile stress when the tensile stress and field are in the same direction.

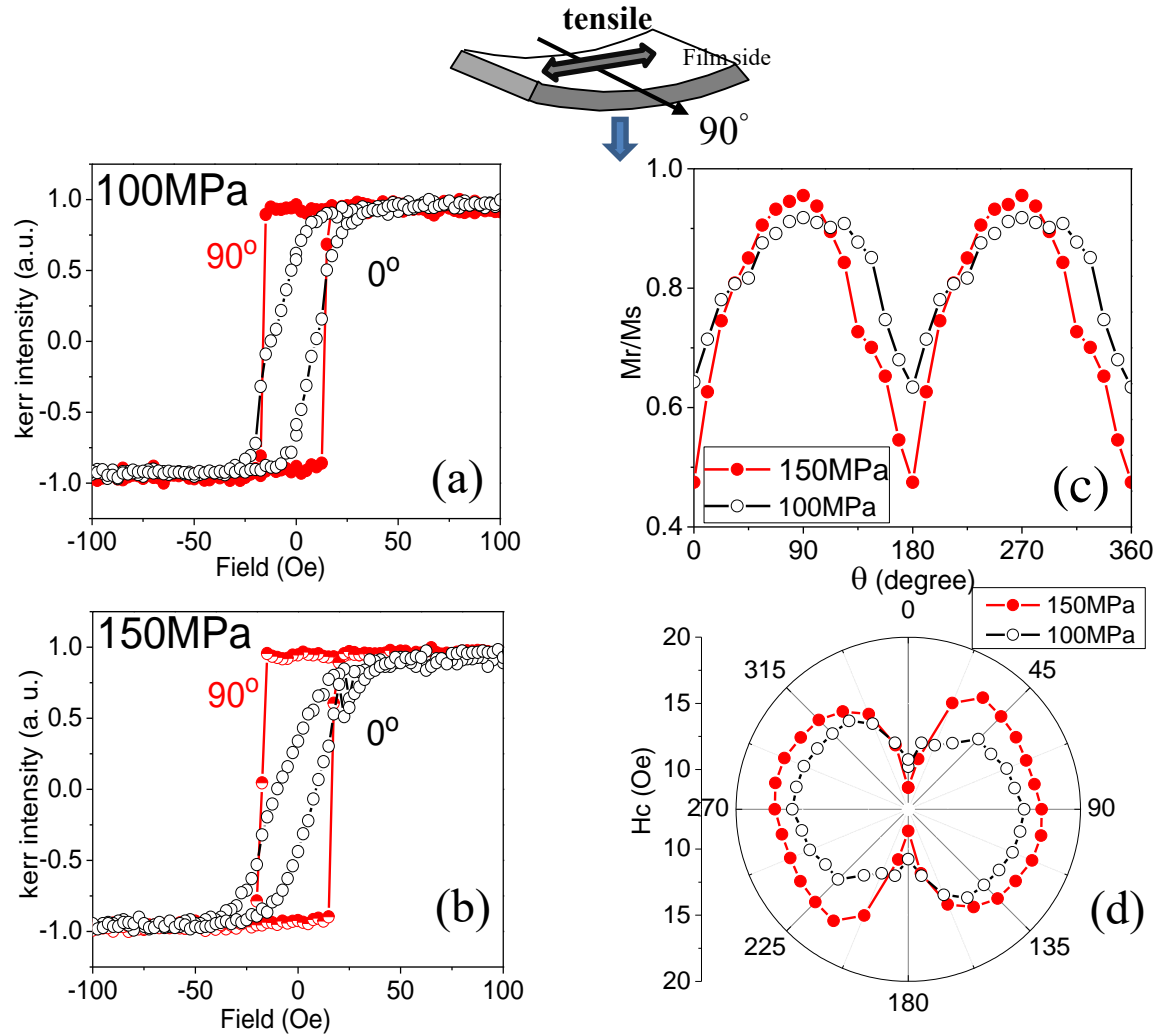

**Figure 5.7:** Selected hysteresis loops for stressed Co films with the magnetic field applied perpendicular ($\theta=90°$) and parallel ($\theta=0°$) to the tensile stresses of strength (a) $\sigma=100$MPa and (b) $\sigma=150$MPa. (c) $M_R/M_S$ vs $\theta$ plot (d) Polar plots of $H_c$ vs azimuthal angle ($\theta$) for both sets of samples.

In the case of films C100 (fig. 5.8a) and C150 (fig. 5.8b), the easy axis of magnetization is parallel to the stress direction. The remanence is high when the field is along the direction of stress and decreases when the field is perpendicular to stress in the film (fig. 5.8c), whereas coercivity increases with increasing compressive stress for both cases of either field along or perpendicular to the direction of stress (fig. 5.8c). The qualitative information about the magnetic anisotropy in stressed films is obtained by the difference in the area "A" between loops corresponding to the easy and hard directions of the magnetization. This area is proportional to the anisotropy energy of the film [23,24].The values of "A" for samples C100 and C150 are 13 and 52, whereas for the samples T100 and T150- these are 28 and 61, respectively. Present measurements revealed the presence of UMA in both films, which increases with increasing strength of tensile or compressive stress. It may be noted that the easy axis of magnetization is perpendicular to the direction of applied tensile stress in the film. In contrast, it is parallel to the direction of applied compressive stress. Polar plots of $H_c$ variation with azimuthal angle (fig. 5.8d) also clearly confirm the presence of UMA. Further, to understand magnetic microstructure, MOKE microscopy measurements were carried out on the film C150 by applying a magnetic field parallel and normal to the direction of the compressive stress axis.

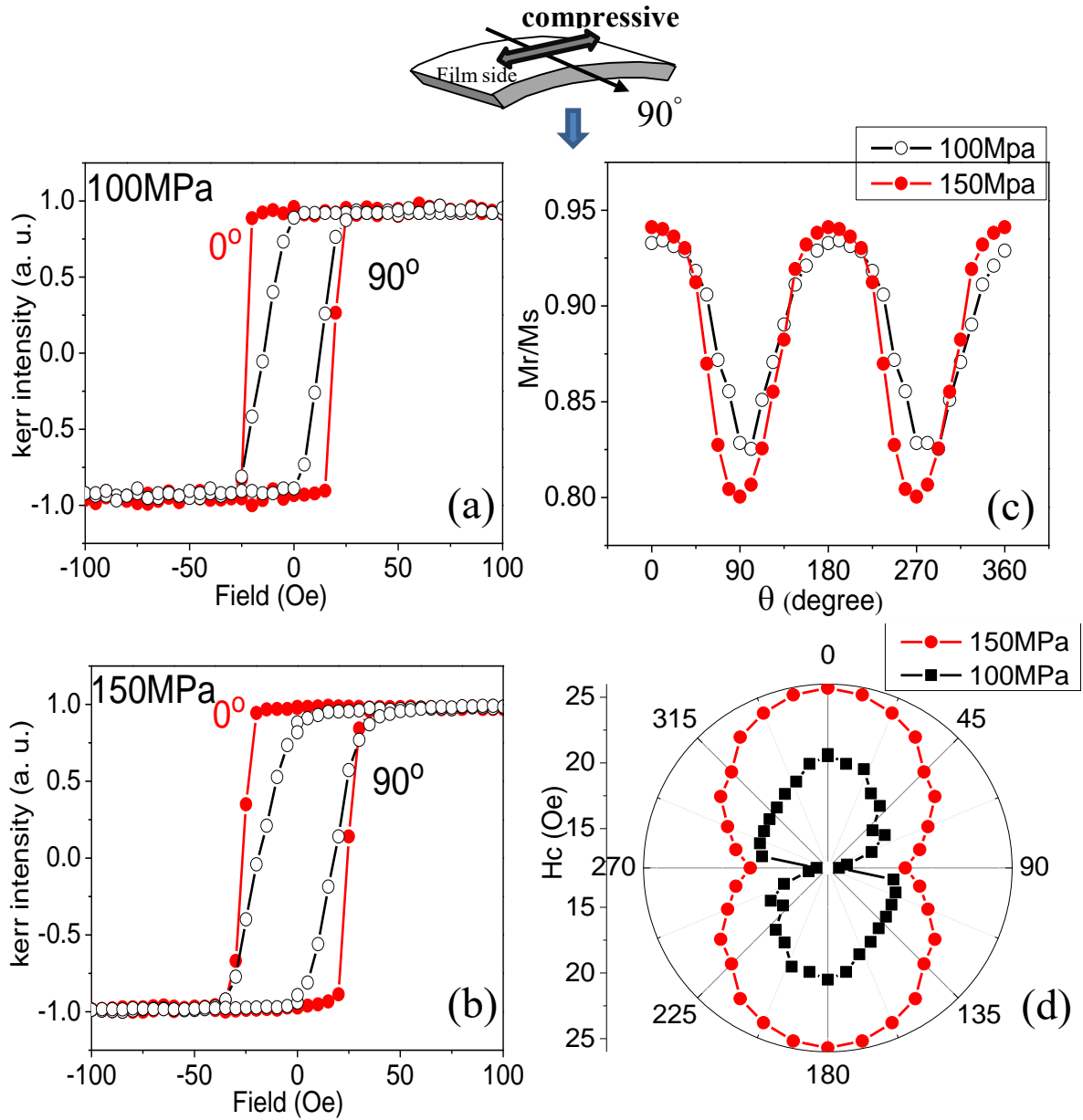

**Figure 5.8:** Selected hysteresis loops for stressed Co films with the magnetic field applied perpendicular ($\theta=90°$) and parallel ($\theta=0°$) to the compressive stresses of strength (a) $\sigma=100$MPa and (b) $\sigma=150$MPa. (c) $M_R/M_S$ vs $\theta$ plot (d) Polar plots of $H_c$ vs azimuthal angle ($\theta$) for both sets of samples.

Figure 5.9 shows the hysteresis loop and several domain images collected during the magnetization reversal process along the applied compressive stress direction. It can be understood from the hysteresis loop and from image (a) that sample is in a positive saturation state. The domain starts nucleating around the coercive field (b)-16.1 Oe and (c)-17.2 Oe. Finally, above the coercive field, the sample reaches its negative side of saturation at a field value of H≈ 18 Oe. The domain images show that the magnetization reversal process occurs by the mechanism of domain nucleation and growth via domain wall motion.

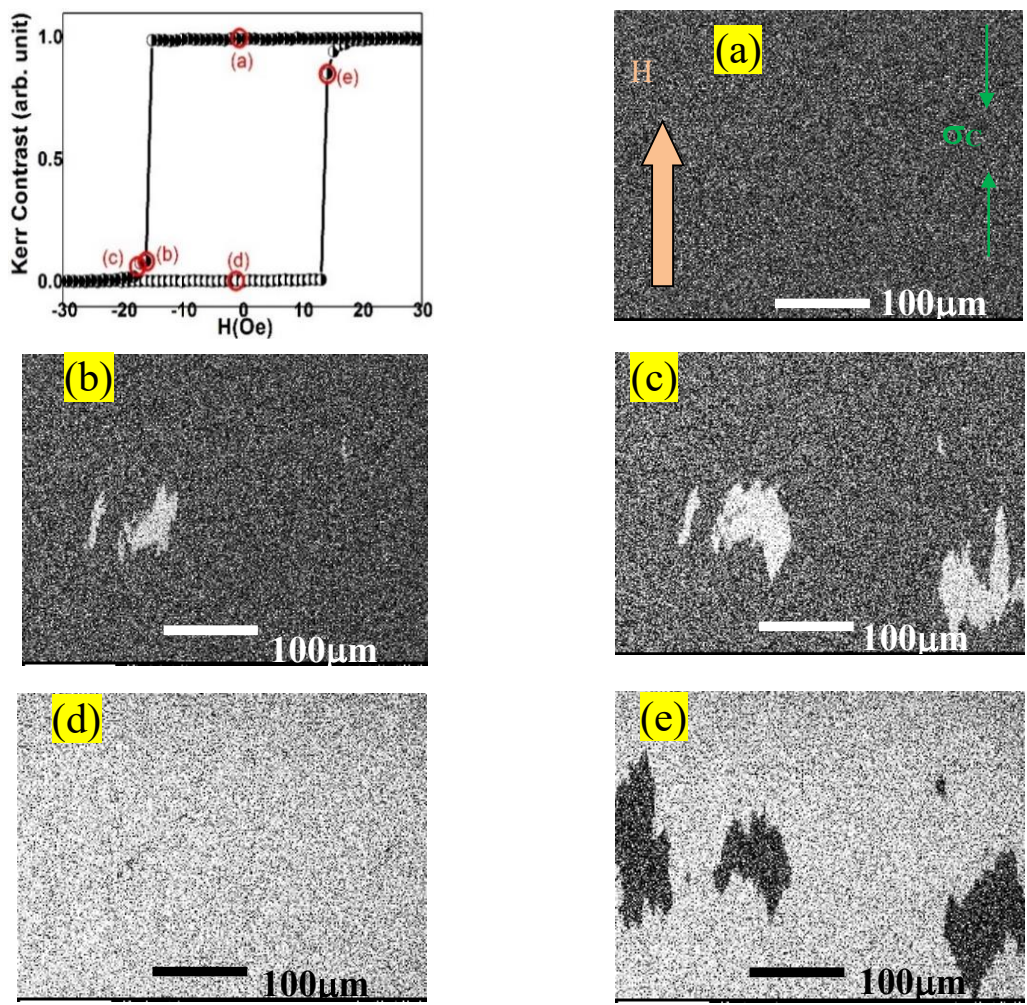

**Figure 5.9**: Hysteresis loop measured by longitudinal Kerr microscopy of C150 along the compressive stress direction (along the easy axis of magnetization). (a)-(e) are domain images taken at the field marked on the hysteresis loop.

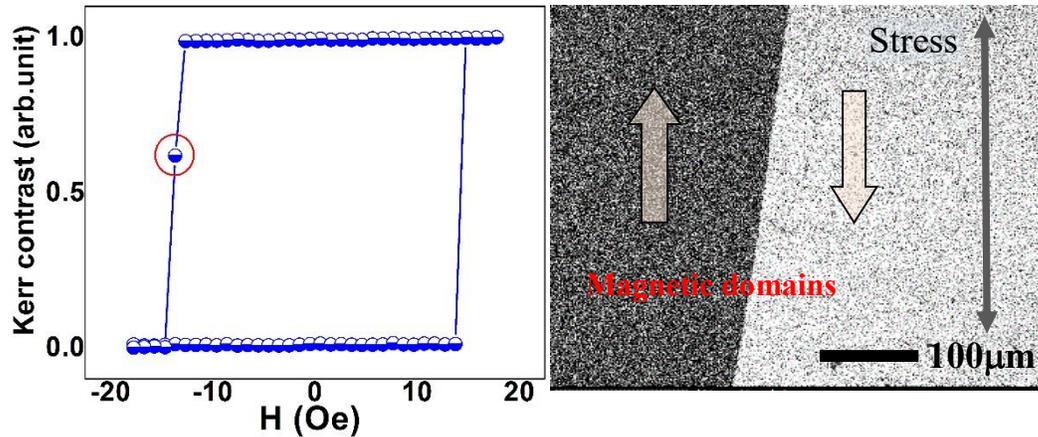

**Figure 5.10:** Kerr microscopy in an almost demagnetized state of the Co film. Domain image was taken at the field marked on the hysteresis loop.

Figure 5.10 gives Kerr image with black and white contrast in almost equal proportion, which indicates zero magnetization of the C150 film. From the sharp-edged domain image, it is confirmed that magnetization reversal is occurring via $180^0$ domain wall motion.

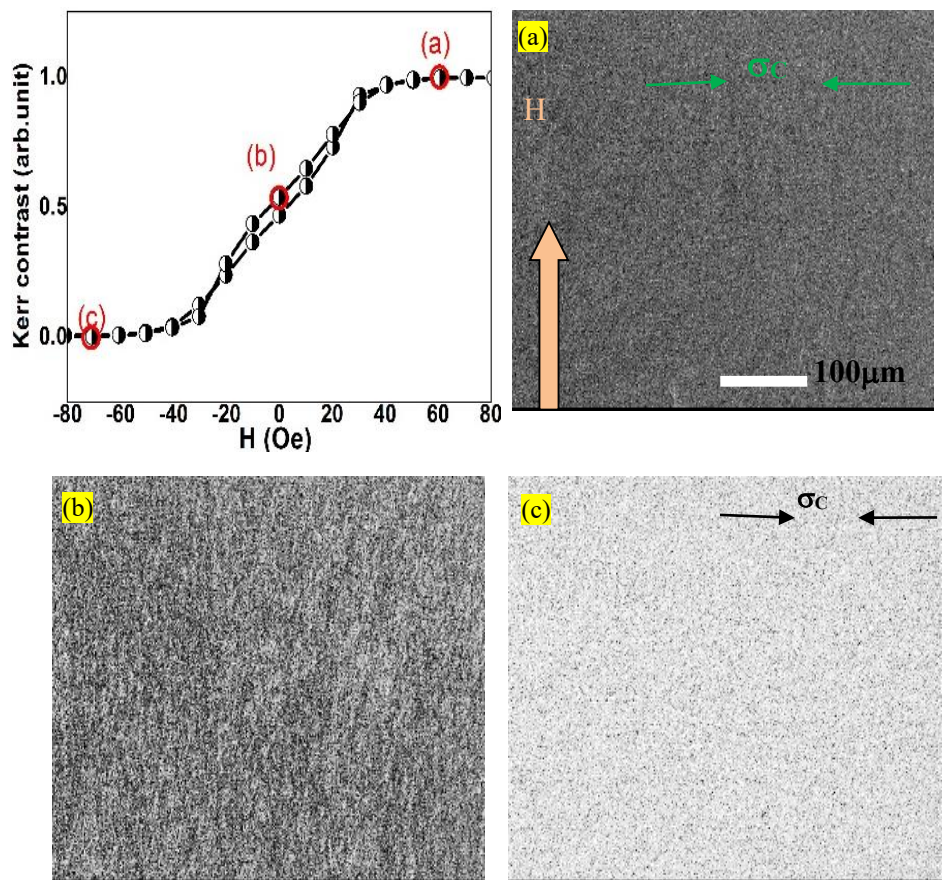

**Figure 5.11**: Hysteresis loop measured for the sample C150 perpendicular to the direction of compressive stress (along hard axis of magnetization). (a)-(c) are domain images taken at the field marked on the hysteresis loop.

Figure 5.11 gives Kerr microscopy measurements at a right angle to the applied compressive stress axis. The magnetization reversal takes place through domain wall rotation. This rotational part is reflected in the rather linear change of magnetization as marked in the hysteresis loop. It may be noted that the magnetic reversal process along easy and hard axis of magnetization in stressed Co film is in accordance with the Stoner–Wohlfarth model [25], where the magnetization along the easy axis of magnetization takes place purely through domain wall motion and the rounding off of the hysteresis curve along hard axis indicates magnetization reversal through rotation of domain.

## 5.4    Discussion

In magnetic thin films, magnetic dipolar and spin-orbit interactions are the two main sources responsible for the origin of such in-plane UMA. Spin-orbit interaction induces a small orbital momentum, which couples the total magnetic moment to the crystal axes and gives rise to magnetocrystalline anisotropy [10]. It is clear that magneto-crystalline anisotropy can only be present and can show uniaxial nature if films are epitaxial/single crystal in nature [10,26]. In the present case, XRD measurements revealed the polycrystalline nature of the Co films, where grains are randomly oriented in all directions, and long-range crystallinity is absent. Therefore, magneto-crystalline anisotropy is absent in the current set of Co films. There are several studies in the literature where the magnetic dipolar anisotropy in the films was induced using the stray dipolar fields generated due to the asymmetric morphology (rippled geometry) of the film surface [27,28,29], where the contribution of the stray dipolar fields has been used to induce in-plane magnetic anisotropy in the polycrystalline thin film. In the present study, where the surface morphology of the films is random, dipolar interaction cannot be responsible for the in-plane preferential alignment of the spins. It is clear that the origin of the observed magnetic anisotropy in the present films cannot be the magneto-crystalline and dipolar. In the present case, it may be attributed to the long-range stresses in the films, where the direction of the easy axis had a definite relationship with the direction of applied stress. Similar magnetic anisotropy was also observed in amorphous magnetic thin films [9], where long-range stress was expected to be related to anisotropic ejection of the sputtered atoms relative to the direction of the ion beam [30,31,32].

The origin of magnetic anisotropy in polycrystalline Co film and its definite relationship with long-range stress can be understood in terms of the magneto-elastic interaction energy. The magnetoelastic energy for isotropic magnetostrictive materials is expressed as [33]

$$E_{me} = \frac{3}{2}\lambda\sigma sin^2\theta \tag{1}$$

where θ is the angle between the direction of magnetization ($M_s$) and stress (σ), λ is magnetostriction constant. From the equation, it is clear that $E_{me}$ is zero when Ms and σ are in the same direction and increase to a maximum of $\frac{3}{2}\lambda\sigma$ when they are at the right angle, provided that $\lambda\sigma$ is positive (fig. 5.12b). If this quantity is negative, the minimum energy occurs when Ms and σ are at the right angle. (fig. 5.12a) Now it is easy to understand the alignment of the easy axis of magnetization in polycrystalline Co film with respect to the direction of compressive and tensile stress.

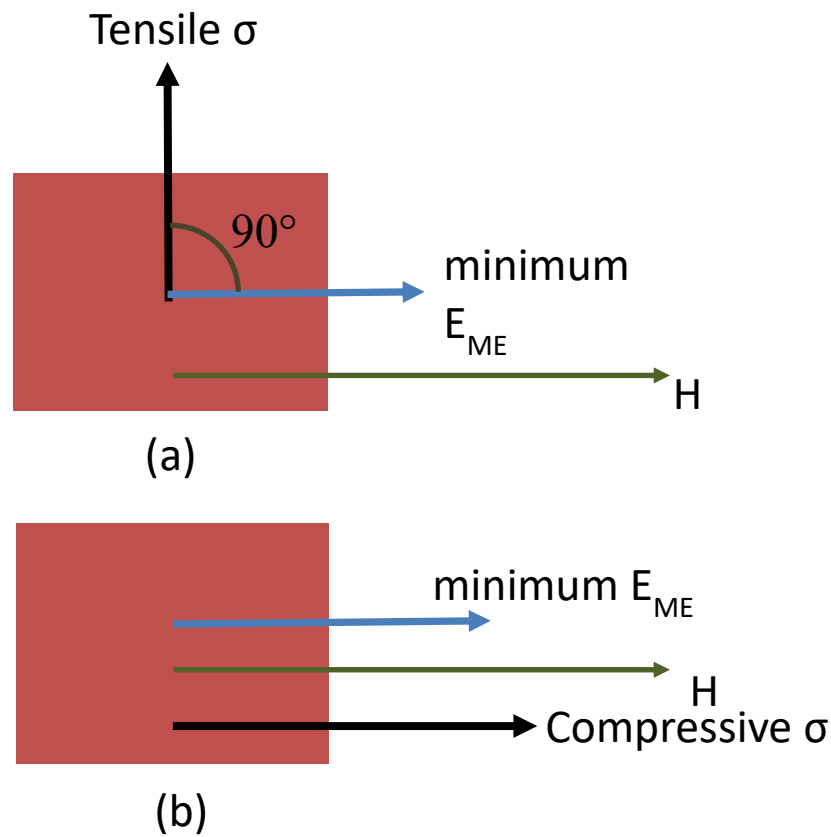

**Figure 5.12**: Minimization of magneto-elastic coupling energy $E_{ME}$ direction when field H and (a) tensile stress are perpendiculars and (b) compressive are parallel.

It is important to note that in epitaxial films, magnetostriction is anisotropic in nature; the magnitude is different along different planes. In the case of polycrystalline films, grains are orientated in all possible directions; therefore, in such cases, magnetostriction has an average value. In the case of polycrystalline and amorphous Co film, averaged magnetostriction constant ($\lambda$) is negative in magnitude [33]. Therefore alignment of the easy axis of magnetization is most likely resulted from negative $\lambda$ and in-plane compressive ($-\sigma$) and tensile ($+\sigma$) stress. The application of compressive stress in the Co film leads to the domain wall movement in such a way that the volume of the magnetized domain increases along the direction of the stress axis. Only a small field is now required to saturate the Co film because the transition of magnetic domains takes place through the relatively easy process of $180^0$ wall motion.

Recently stress-induced uniaxial magnetic anisotropy has been observed in polycrystalline $Fe_{81}Ga_{19}$ films deposited on flexible bowed-substrates [12,13]. In this study, the observed magnetic anisotropy is mainly attributed to the minimization of magnetoelastic and anisotropic energies in the presence of stress. The uniaxial magnetic anisotropy in polycrystalline Co films observed in the present case is in conformity with these earlier results.

The observation of Co surface smoothening due to tensile stress is very encouraging and interesting. Smoothening and roughening of the Co film surface occur may be due to the long-range movement of the atoms associated with applied stress. Tsung-Ceng Chan, et al. [34] have also found a significant effect of stress generated due to the passage of electrical current on surface roughness of Cu thin film. They have suggested that stress changes the surface morphology and film texture, which leads to the change in the surface roughness. Detailed experiments are required in the present case in order to understand stress-induced surface smoothening. The increased $H_c$ and decreased uniaxial anisotropy in the case of the compressively stressed film can be understood if one conjectures that the stressed film contains both long-range and short-range stresses. The long-range stresses give rise to the uniaxial anisotropy due to magnetoelastic coupling, while short-range stresses are random in direction [7]. With increased surface roughness in the case of C150 film, the short-range stresses and domain pinning start contributing against uniaxial magnetic anisotropy. As a result, the effective uniaxial anisotropy along the direction defined by the long-range stresses decreases, and the field needed to switch magnetization reversal increases.

## 5.5    Conclusion

In conclusion, polycrystalline Co films with the varying magnitude of long-range stresses were prepared by e-beam deposition on bent Si (111) substrates. Stress in the film was induced by removing the bending force from the substrates after film deposition. Depending upon the curvature of the film (concave or convex), the induced stress can be tensile or compressive. Tensile stress in the films was found to be responsible for the surface smoothening of the film, which is attributed to the movement of the atoms associated with the applied stress. Detailed MOKE and Kerr microscopy measurements revealed the presence of uniaxial magnetic anisotropy in both tensile and compressively stressed Co films. In the case of tensile stressed film, the easy axis of magnetization is found to be normal to the direction of the applied stress axis, whereas in the case of compressively stressed films easy axis of magnetization is along the direction of the applied stress axis. Importantly, the strength of magnetic anisotropy was found to increase with increasing stress. The origin of magnetic anisotropy in polycrystalline Co films is understood in terms of the minimization of magneto-elastic energy. The present method of inducing magnetic anisotropy in polycrystalline films, where crystalline anisotropy is absent, could be a possible way to tailor magnetic anisotropy to get desired functionality.

*A unique method of producing magnetic anisotropy and thermal stability in polycrystalline thin films*

Iron-Cobalt (FeCo) columnar, multilayered structure is prepared by depositing several thin FeCo layers by varying the angle between the surface normal and the evaporation direction as 0° (normal) and 60° (oblique), alternatively. In situ X-ray scattering and magneto-optical Kerr effect (MOKE) measurements established the evolution of magnetic properties with that of the morphology and structure of the multilayer. The strong shape anisotropy and compressive stress of nanocolumns in alternative FeCo layers resulted in a well-defined uniaxial magnetic anisotropy (UMA) with the easy axis of magnetization along the projection of the tilted nanocolumns in the film plane. The stress in the film provides minimization of magnetoelastic energy along the in-plane column direction, which couples with the columnar shape anisotropy energies and results in the preferential orientation of the magnetic easy axis along the oblique angle deposition direction in the film plane. Drastic reduction in the in-plane UMA after annealing at 450 °C is attributed to the merging of columns and removal of stresses after heat treatment. The present study opens a new pathway to produce magnetically anisotropic multilayer structures using single material and thus may have prominent implications for future technological devices.

## 6.1    Introduction

Magnetic thin films and multilayers have gained a lot of research attention due to their important role in the development of nanoscience and nanotechnology [1,2,3,4]. These 2D nanostructures can exhibit peculiar magnetic properties as compared to their bulk counterparts, including magnetic anisotropy [1-4], spin-dependent transport [5], intrinsic magnetic damping properties [6], interface Dzyaloshinskii–Moriya interaction [7], topological spin textures [8,9], spin Hall effect [10] and spin-orbit torque [11] etc. They have potential applications in microscale actuators [3], sensors [3], magnetic recording media [12,13,14], current-induced magnetization switching devices [15,16], high-density magnetoresistive random-access memory [17] and so on. Among all, magnetic anisotropies such as in-plane magnetic anisotropy (IMA) and perpendicular magnetic anisotropy (PMA), are the basic properties of ferromagnetic materials which have opened up a huge field in basic and applied research for the developments of advanced and modern devices mentioned above.

It is always a challenge to achieve the required in-plane UMA in a magnetic thin film that controls the performance and operation of these films. In the case of epitaxial magnetic thin films such as Fe [18,19], Co [20], FeCo [21], and SrRuO$_3$ [22], UMA is often realized when deposited on various single crystal substrates such as Ag (001), Cu (100), MgO (100) and SrTiO$_3$ (001) etc. It has the possibility to be tailored by appropriately choosing magnetic materials and substrate orientation [23]. Here, spin-orbit coupling (SOC) originated through ordered crystallographic structure is responsible for the preferred magnetization direction with respect to the crystal structure and is known as magnetocrystalline anisotropy (MCA) [24]. Unlike epitaxial thin films, polycrystalline films do not exhibit long-range structural order [24]. Therefore, it isn't easy to realize MCA or UMA from randomly oriented grains in polycrystalline thin films. The main sources of UMA in polycrystalline and amorphous thin films are shapes and magnetostriction, which can be obtained through modifying the morphology and the strain in the films [25,26,27,28,29,30]. In general, such anisotropies can be induced in most of the polycrystalline as well as amorphous thin films by, e.g. oblique angle deposition (OAD) [25], magnetic annealing [31], stress (magnetostatic) [29,30] and surface patterning [26-28] etc., irrespective of whether they are produced by evaporation or sputter deposition techniques [25-30]. The ripple pattern provides limited magnetic anisotropy as they cannot be formed throughout the depth of the film and hence limits its strength [32]. Controlled external stress in thin films is also used to induce UMA with the help of magnetostriction, but such bending of the substrate has its own limitations due to the formation of cracks and voids in the film, which in extreme cases leads to decreased performance, deformation and failure of the thin-film devices.

Among these methods, OAD is not only easy to handle but an extremely useful method to create tuneable anisotropy [33], in terms of its direction and strength, in different types of ferromagnetic thin films [34,35,36,37,38,39,40]. Here, the deposition geometry of the substrate, source and material flux control the morphological development of the film starting from the atomic level [41]. The interaction between the OAD atoms and the atoms on the surface modifies the trajectory of the deposited atoms and causes a self-shadowing effect (SSE) [3,42]. Due to the SSE, the OAD of ferromagnetic thin films on flat rigid substrates can result in elongated grains perpendicular [43,44] and parallel [45] to the direction of the incident atomic flux [46]. Depending on the angle of deposition [34-40] and thickness of the film [34,43], good separation between the columns can be achieved, and UMA can be induced due to the shape of the columns. In view of these facts, considerable research has been carried out, where oblique angle deposition of various polycrystalline and amorphous magnetic materials like $Co_2FeAl$ [47] Fe [48], Co [34,39,43], Ni [46], FeCoB [49] and Py/Ru/FeCo/IrMn [50] was done and resulted in reasonable UMA. It has been established that at OAD growth, by keeping the angle more than 60°, the axis of magnetization stays confined to the incident plane, whereas at the lower angles, the easy axis switches normal to the incident plane [34-40]. In addition, the direction of the easy axis also depends on the thickness of the film [34,43], at higher thicknesses, columns collapse with each other. This mechanism is responsible for the UMA decrease due to the increase in the exchange coupling between columns. Moreover, the slanted columns and magnetic domain create out-of-plane magnetization components along with the in-plane component of the magnetic anisotropy, even though the magnetic field is applied in the film plane. Both of these longitudinal and polar magnetization components participate in magnetization reversal [51,52] and are always present in UMA of such OAD deposited thin films. Therefore, it is almost difficult to avoid the above-mentioned drawback of the OAD to get pure and relatively high UMA in polycrystalline or amorphous thin films.

Due to the excellent soft magnetic properties and very high saturation magnetization [25,53], among all the intermetallic alloys, FeCo is one of the potential candidates for magnetic recording media and high-frequency applications etc. [25,54]. Yet, in general, FeCo films are magnetically isotropic in nature due to negligible crystalline magnetic anisotropy [55]. In the present case, several $Fe_{50}Co_{50}$ nanocolumnar structures are deposited at an oblique angle (OAD) in between the normally deposited (NAD) FeCo multilayer (FeCo MLT), aiming to induce pure and high in-plane magnetic anisotropy even at the higher thickness of $Fe_{50}Co_{50}$ multilayer. The individual low thickness of FeCo layers throughout the multilayer stacks provides a dominated in-plane UMA contribution [36], whereas the strength in the UMA is increased by combining the stress-induced magnetostrictive anisotropy and shape anisotropy along the sample direction (along in-plane OAD direction). Temperature-dependent UMA of the sample is correlated with film morphology as well as structure, obtained through *in situ* grazing-incidence small-angle X-ray scattering (GI-SAXS) together with *ex-situ* atomic force microscopy (AFM) and transmission electron microscopy (TEM) measurements. It has been demonstrated that obliquely deposited several thin FeCo underlayers in between normally deposited FeCo layers can be used to control and induce strong UMA in the FeCo multilayer or similar polycrystalline magnetic thin film structure, where magneto-crystalline anisotropy is absent.

## 6.2    Preparation of nanocolumnar FeCo multilayered structure

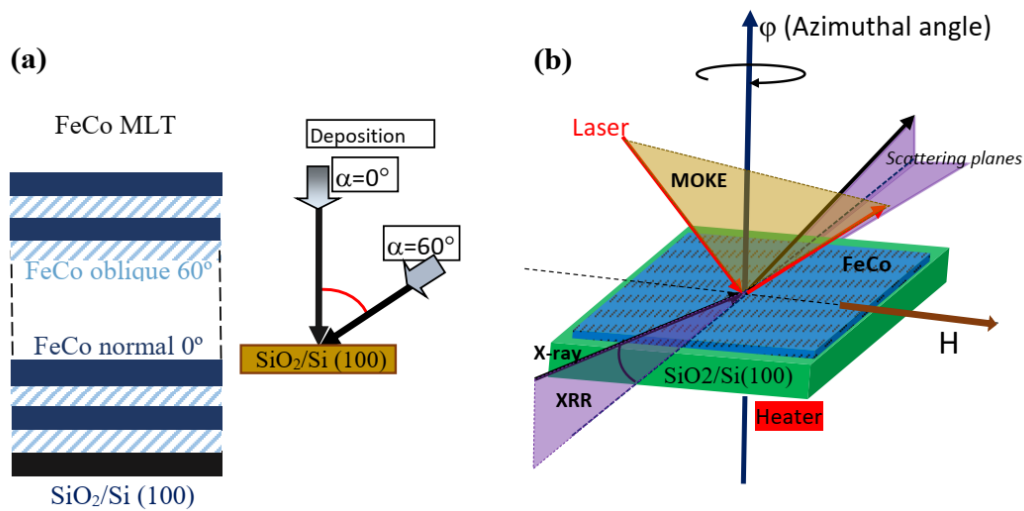

**Figure 6.1:** (a) Schematic diagram of FeCo nine bilayers stack on Si (100) substrate containing native oxide layer, where adatoms flux directions in two geometries; α= 0° (normal deposition) and 60° (OAD) from the substrate normal are shown. (b) Experimental geometry of XRR and MOKE measurements during *in situ* measurements.

The FeCo bilayers of structure [FeCo 39Å; OAD at α=60° /FeCo 51Å; at α=0°] are deposited in the form of multilayer (9 bilayer stack) on $SiO_2/Si$ (100) substrate using electron-beam co-evaporation under base pressure of $1 \times 10^{-7}$ mbar. The schematic representation of the FeCo multilayer is shown in fig. 6.1(a). During the deposition, the angle of the evaporated material is kept at α= 60° and 0° alternatively with respect to the substrate normal and denoted as $FeCo_{α=60°}$ and $FeCo_{α=0°}$, respectively, throughout the manuscript. The thickness of the multilayer is monitored using calibrated quartz crystal monitor in normal as well as oblique angle deposition conditions. The magnetic anisotropy of the multilayer is characterized by magneto-optic Kerr effect (MOKE) measurements in the longitudinal geometry with the magnetic field, H, applied along with different azimuthal directions (φ) in the plane of the film (see fig. 6.1b). In order to explore the thickness (d), roughness (σ), and electron density (ρ) profile of the individual layers in the multilayer, X-ray reflectivity (XRR) measurements are performed using Bruker D8 diffractometer using Cu Kα radiation along the in-plane OAD direction in the film plane. Temperature-dependent evolution of magnetic anisotropy is obtained using *in situ* MOKE under the base pressure of $8 \times 10^{-10}$ mbar.

Hysteresis loops and XRR patterns are also collected *in situ* during heating of the multilayer up to the temperature of 450 °C, which makes it possible to correlate the evolution of UMA with that of film morphology. Figure 6.1b shows the schematic of measurements, where geometries of all the measurements, such as MOKE and XRR, are shown. To get further insights into the magnetic anisotropy and its correlation with columnar structure and its interdiffusion, GISAXS measurements are performed at P03 (MiNaXS) beamline of the PETRA III storage ring at DESY, Hamburg (Germany) [56] at an incidence angle of 0.45°. A dedicated mini chamber [57] is used here to observe the real-time *in situ* evolution of the morphology. For GISAXS experiments, photon energy of E =13 keV is used, whereas data recording is done using PILATUS 1M (Dectris Ltd., Switzerland) 2D detector with a pixel size of $(172 \times 172)$ μm². Cross-sectional TEM and AFM studies are also carried out to observe film microstructure and surface morphology, respectively. Grazing incidence X-ray diffraction (GIXRD) measurements on the pristine as well as after annealing at 450 °C temperature are carried out using beam energy of 15.8 keV (corresponding λ=0.78Å) at ECXRD beamline (BL-09) Indus-2 synchrotron source, RRCAT, India. For convenience, diffracted intensity is mapped with a two-dimensional (2D) detector (at grazing incidence angle ~0.4º) to get comprehensive information about the in-plane microstructure with respect to OAD direction.

## 6.3    FeCo multilayer characterization

### 6.3.1    XRR characterization for structural analysis

Figure 6.2(a) gives XRR measurements of as-prepared nine bilayer stacks of FeCo films prepared at $\alpha = 0°$ and $60°$, alternatively. Here, X-ray intensity reflected from the sample is plotted with the magnitude of the momentum transfer vector $q_z = (4 \pi \sin\phi)/\lambda$, where $\phi$ and $\lambda$ are X-ray incidence angle and wavelength, respectively. Weak periodic oscillations (Kiessig fringes) along with a Bragg peak at around $q_z = 0.082$ Å$^{-1}$ in the XRR curves correspond to the total thickness of the film and bilayer thickness (FeCo$_{\alpha=60°}$+FeCo$_{\alpha=0°}$), respectively. Observation of the Bragg peak in a single material deposition is somewhat strange due to the formation of alternate low and high-dense FeCo layers at different angles of depositions (fig. 6.1a). In order to get quantitative information, XRR curves are fitted with well-known Parratt's formulism [58] to obtain thicknesses and densities of the FeCo$_{\alpha=60°}$ and FeCo$_{\alpha=0°}$ layers in the multilayer. It may be noted that compared to the conventional ex-situ XRR set-up, where knife-edges are used to nullify the sample footprint effect [59], geometrical constraints in the present in-situ UHV set-up do not allow the use of any knife-edge [60], which leads to the change in the slope and damping of oscillations in experimental XRR patterns. Hence, XRR data is not getting fit properly in the y-direction and is therefore restricted to extracting quantitative information of the surface and interface roughness of the film. On the other hand, the critical angle and Bragg peak position on the x-axis are well fitted and provide the correct electron density and bilayer thickness values. Detailed morphology/roughness of the multilayer has been extracted separately using GI-SAXS measurement, as discussed in the later sections.

The fitted XRR pattern in as prepared stage (pristine sample) along with obtained electron scattering length density (ESLD) is shown in fig. 6.2(a) and fig. 6.2(b), respectively. ESLD depth profile of the multilayer, as shown in fig. 6.2(b) confirms the formation of FeCo$_{\alpha=60°}$ layer with 17.1% less ESLD than FeCo$_{\alpha=0°}$ layers. The film thickness of both layers is obtained independently as FeCo$_{\alpha=60°}$ = 39 Å and FeCo$_{\alpha=0°}$ = 51 Å. In general, periodic arrangement of low and high dense layers can enhance the reflectivity at a particular angle due to Bragg reflection [61]. In the present case, the periodically repeating high and low dense bilayer structure of FeCo$_{\alpha=0°}$ and FeCo$_{\alpha=60°}$ layers is responsible for the appearance of the Bragg peak at q$_z$=0.082 Å$^{-1}$ in the XRR pattern.

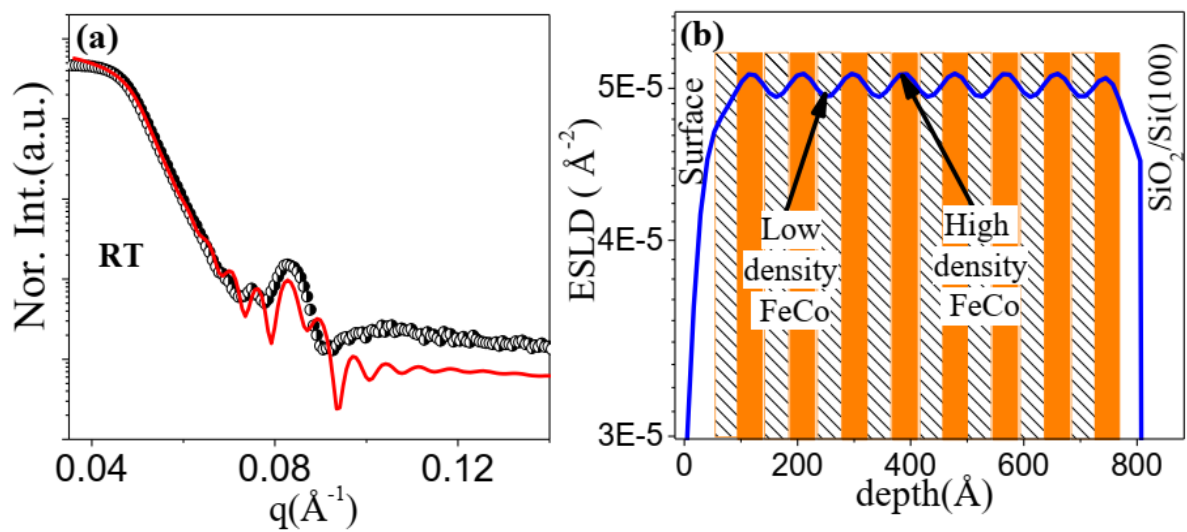

**Figure 6.2:** (a) Fitted XRR pattern for the FeCo multilayer at RT (pristine sample) (b) corresponding electron scattering length density (ESLD) profile obtained from fitting.

### 6.3.2 AFM and TEM for morphological/microstructure studies

Figure 6.3 (a) gives AFM image of the as-prepared sample, where tilted morphology is observed along the deposition direction. In general, it is difficult to see tilted columns using AFM measurements [43,62,63,64,65]. In the present case, the relatively low thickness of individual FeCo$_{\alpha=60°}$ layer is responsible for the tilted morphology close to the film surface. In cross-sectional TEM measurements, as shown in fig. 6.3(b), low and high dense layers can be seen clearly. As the TEM measurements are performed along the edge perpendicular to the OAD direction, column alignment cannot be seen in the TEM image, whereas alternate arrangements of low and high dense FeCo layers can be seen. In comparison with the oblique angle deposition method of preparing porous and dense multilayer stacks by A. Garcia-Valenzuela et al. [66], where smooth and homogeneous interfaces are achieved using migration mechanisms by promoting a relatively high energy ion impingement using magnetron sputtering, the present FeCo stakes are prepared using electron beam evaporation method. In the case of the sputtering atoms/ions accelerated towards the film with energies in the order of a few hundred eVs [67]. On the other hand, electron beam evaporated ions/atoms have energies only about a few eVs, which is insufficient to induce any atomic migration process to form a smooth layer on the surface of the porous layer. Because of the facts, as confirmed by XRR and TEM measurements, the present low dense and high dense layers are not well separated.

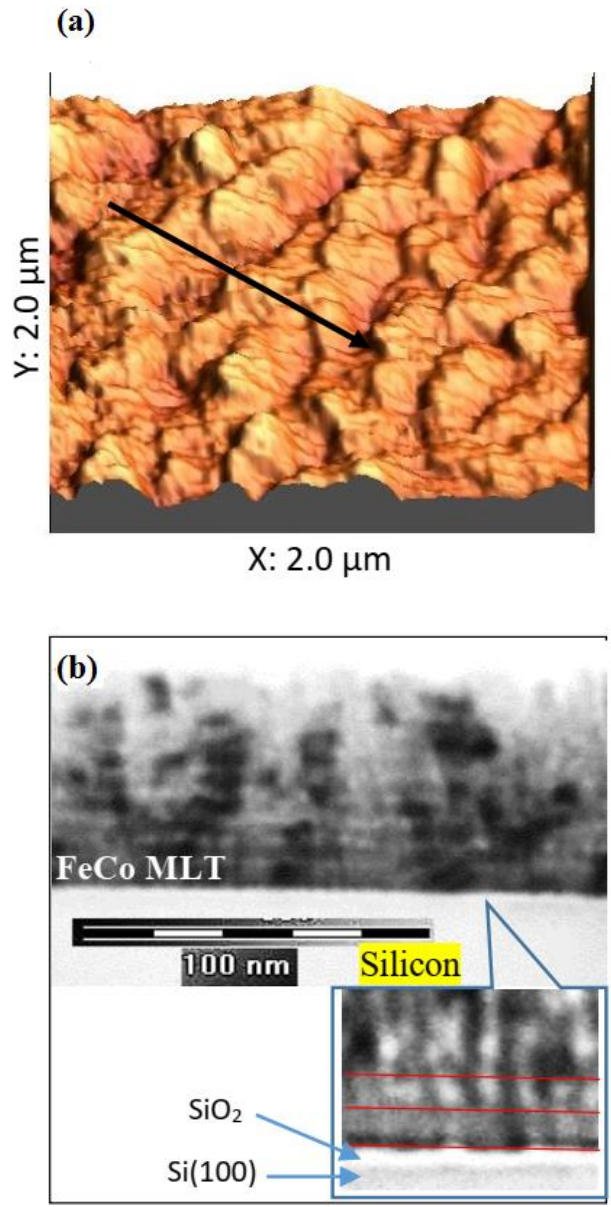

**Figure 6.3:** (a) AFM and (d) cross-sectional TEM image of the as-deposited FeCo multilayer. In the AFM image, the black arrow (60° from the surface normal in the ZX plane) shows the direction of oblique angle deposition, while the bottom right inset in (b) shows the zoomed image.

### 6.3.3    Azimuthal angle-dependent MOKE: Strong UMA

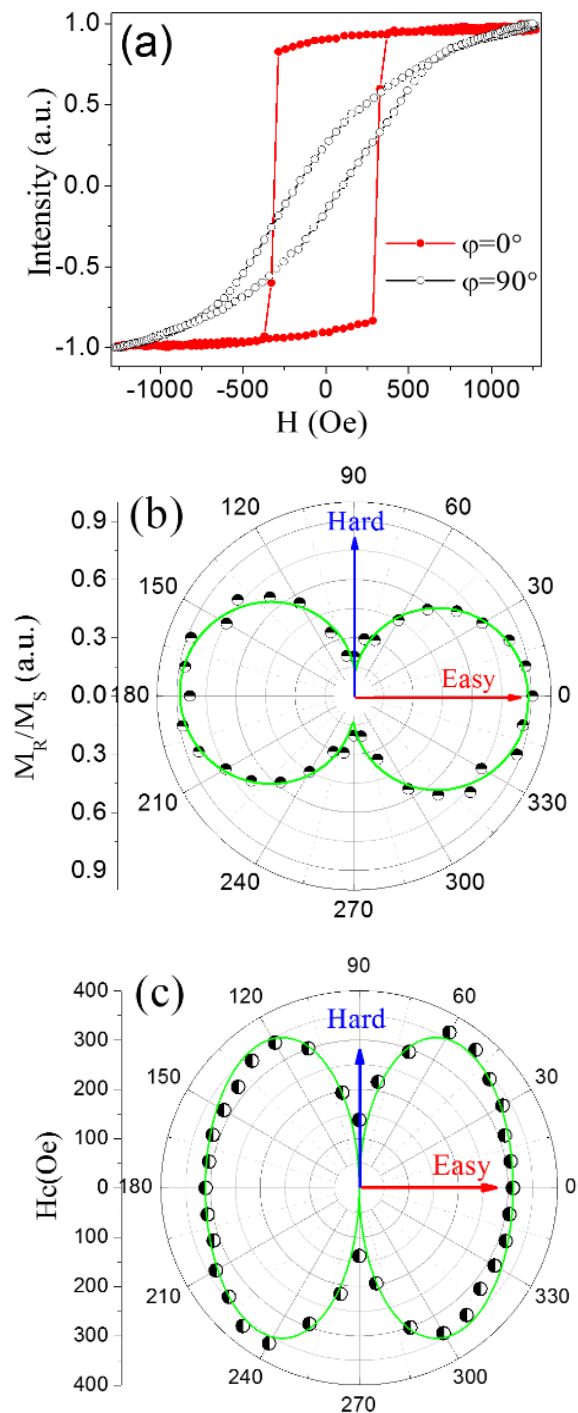

**Figure 6.4:** (a) MOKE loop along easy (0°) and hard (90°) directions for the pristine sample. (b) Polar plot of normalized remanence where solid green curve represents the fitting of normalized remanence using cosine function (c) angular dependence of coercivity in the as-prepared sample, where solid green curve represents the fitting using the two-phase model.

Figure 6.4(a) shows the hysteresis loops taken at room temperature (RT) along easy and hard directions, respectively. There is a strong variation in the shape of the hysteresis loop with φ. For φ = 0°, which is along the film plane projection of OAD, the hysteresis loop is almost square, whereas with increasing azimuthal angle, the hysteresis loop rounds off along φ = 90° direction, suggesting the anisotropic magnetic behaviour in this multilayer.[68] The observed angular dependence of the $M_R/M_S$ at RT is also shown in polar plot in fig. 6.4(b), which clearly shows that the film possesses UMA in the film plane. The direction of easy axis magnetization (EAM), as obtained by fitting ($M_R/M_S$) *versus* φ curve (fig. 6.4b) using adequate cosine function [69], is found to be along the direction of the in-plane projection of OAD. It is in accordance with the literature, where the direction of EAM lies along with the in-plane projection of OAD [34,36].

Figure 6.4(c) shows the angular dependence of coercivity at RT, which also shows a uniaxial symmetry about the directions parallel (φ=0°) and perpendicular (φ=90°) to the OAD, confirming the uniaxial magnetic anisotropy in the sample. For the magnetic field orientation φ rotated away from the easy axis (φ=0°), the coercivity first increases slowly but decreases sharply when approaching the hard axis (φ=90°). In general, the uniaxial nature of magnetic anisotropy in thin films is explained based on the Stoner-Wohlfarth (SW) model, considering the coherent rotation of magnetization. It predicts a monotonous decrease of coercivity with increasing φ from 0° to 90° [70,71]. In the present case, the angular dependence of coercivity can't be interpreted using SW model. The two-phase model [70,71,72], which includes both coherent rotation and domain wall nucleation, is employed to understand the magnetization reversal in the present case. The angular variation in coercivity in a two-phase system can be described as [70,71,72]

$$H_C(\varphi) \;=\; H_C(0°)\,\frac{(N_x + N_N)\cos\varphi}{N_z\,\sin^2\varphi + (N_x + N_N)\cos^2\varphi} \tag{1}$$

Where $N_z$ and $N_x$ are the demagnetizing factors in the directions parallel (φ=0°) and perpendicular (φ=90°) to OAD, respectively. $N_N = H_a/M_s$ is an effective demagnetizing factor; $H_a$ and $M_s$ are the anisotropy field and saturation magnetization, respectively. If the value of $(N_N + N_x)/N_z$ is close to zero, the magnetization reversal mechanism here is dominated by the coherent rotation. For an infinite value of this ratio, the magnetization reversal mechanism is mediated by the domain wall nucleation. The angular dependence of coercivity, as given in fig. 6.4(c), is well fitted by equation (1). An obvious deviation between the calculation and the experimental value of $H_C$ at around angle $\varphi=90°$ is mainly due to the assumption in the two-phase model, which considers a single-crystal system with an ideal uniaxial anisotropy and predicts zero coercivity at $\varphi=90°$. This model is used to explain the unusual behaviour of magnetic anisotropy in various magnetic thin film systems, where two phases at a low field regime, including the remnant state, have been employed to account for the magnetization reversal in polycrystalline magnetic thin films [70,71,72]. In the present case, the magnetic moments of $FeCo_{\alpha=60°}$ layers are not expected to be strictly parallel to $FeCo_{\alpha=0°}$ layer; therefore, the angular variation in coercivity is well fitted with a two-phase model rather than SW model as shown in fig. 6.4(c).

### 6.3.4    In-situ temperature-dependent study

#### 6.3.4.1    Variation in UMA with temperature

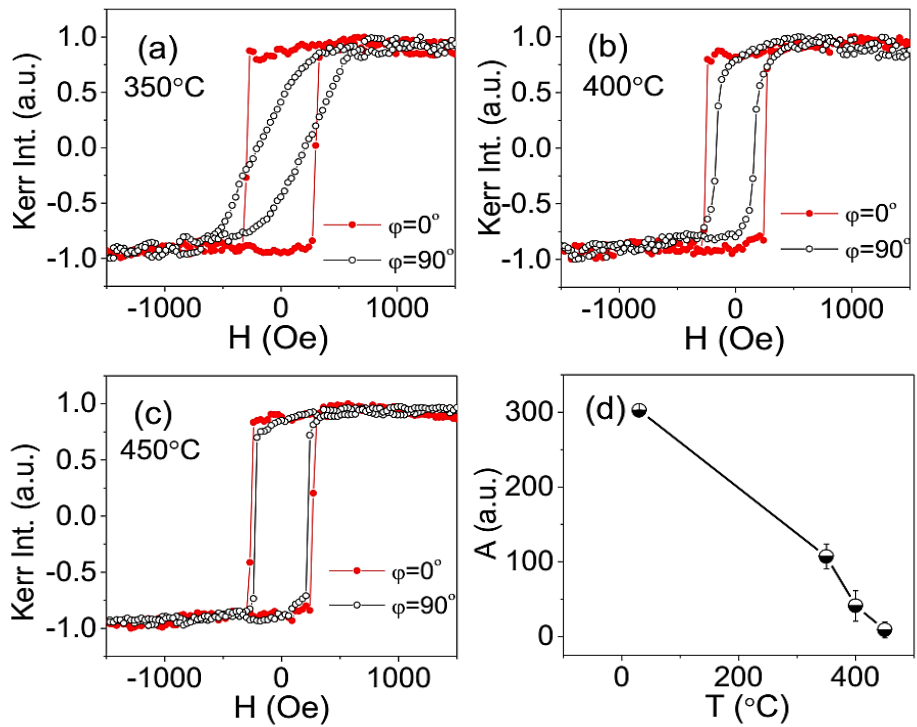

**Figure 6.5:** MOKE loop along ($\varphi=0°$, easy axis) and perpendicular ($\varphi=90°$, hard axis) directions of the OAD for the sample heated at (a) 350 °C (b) 400 °C and (c) 450 ° C, respectively. (d) Area difference A *versus* temperature plot.

Figure 6.5(a) to 6.6(c) give the temperature-dependent hysteresis loops obtained along ($\varphi=0°$) and perpendicular ($\varphi=90°$) directions after annealing at temperatures 350 °C, 400 °C and 450 °C, respectively. Decreasing differences (coercivity and remanence) between the loops along $\varphi=0°$ and 90° direction suggest the gradual decrease in UMA with thermal annealing. Qualitative information on UMA is obtained from the difference in areas A between loops corresponding to the easy and hard directions of the magnetization, which is proportional to the anisotropy constant ($K_u$) of the film [30]. The calculated temperature dependence of A is plotted in fig. 6.5(d). After annealing at 350 °C and 400 °C, UMA decreased to 35% and 13%, respectively with respect to its initial value (A~ 300) in the as-prepared sample. After annealing at 450 °C, almost similar loops along $\varphi=0°$ and 90° direction confirm the disappearance of magnetic anisotropy in the sample.

### 6.3.4.2    In-situ temperature-dependent XRR study

To elucidate the evolution of the observed magnetic anisotropy and to correlate it with the structure of the sample, temperature-dependent XRR measurements are done *in situ* under ultra-high vacuum (UHV) conditions. Representative temperature-dependent XRR patterns, collected after different stages of thermal annealing up to 450 °C, are shown in fig. 6.6(a). Two main features can be noticed: (i) the height of the Bragg peak decreases and disappear around 450 °C, and (ii) Kiessig fringes become more periodic and well defined with increasing temperature In order to get quantitative information, all XRR patterns are fitted using Parratt's formalism by taking thicknesses, electron densities and roughnesses of individual layers, i.e., $FeCo_{\alpha=60°}$ and $FeCo_{\alpha=0°}$. ESLD depth profiles of the multilayer with increasing annealing temperature are obtained and plotted in fig. 6.6(b). It may be noted that the difference in the electron densities between $FeCo_{\alpha=60°}$ and $FeCo_{\alpha=0°}$ gradually decreases with annealing temperature. Therefore, a film with almost uniform electron density throughout the sample depth is obtained after annealing the sample at about 450 °C. Variation in the electron densities ($\rho_{\alpha=60°}$ and $\rho_{\alpha=0°}$) of individual layers ($FeCo_{\alpha=60°}$ and $FeCo_{\alpha=0°}$) with increasing annealing temperature, is plotted in fig. 6.7.

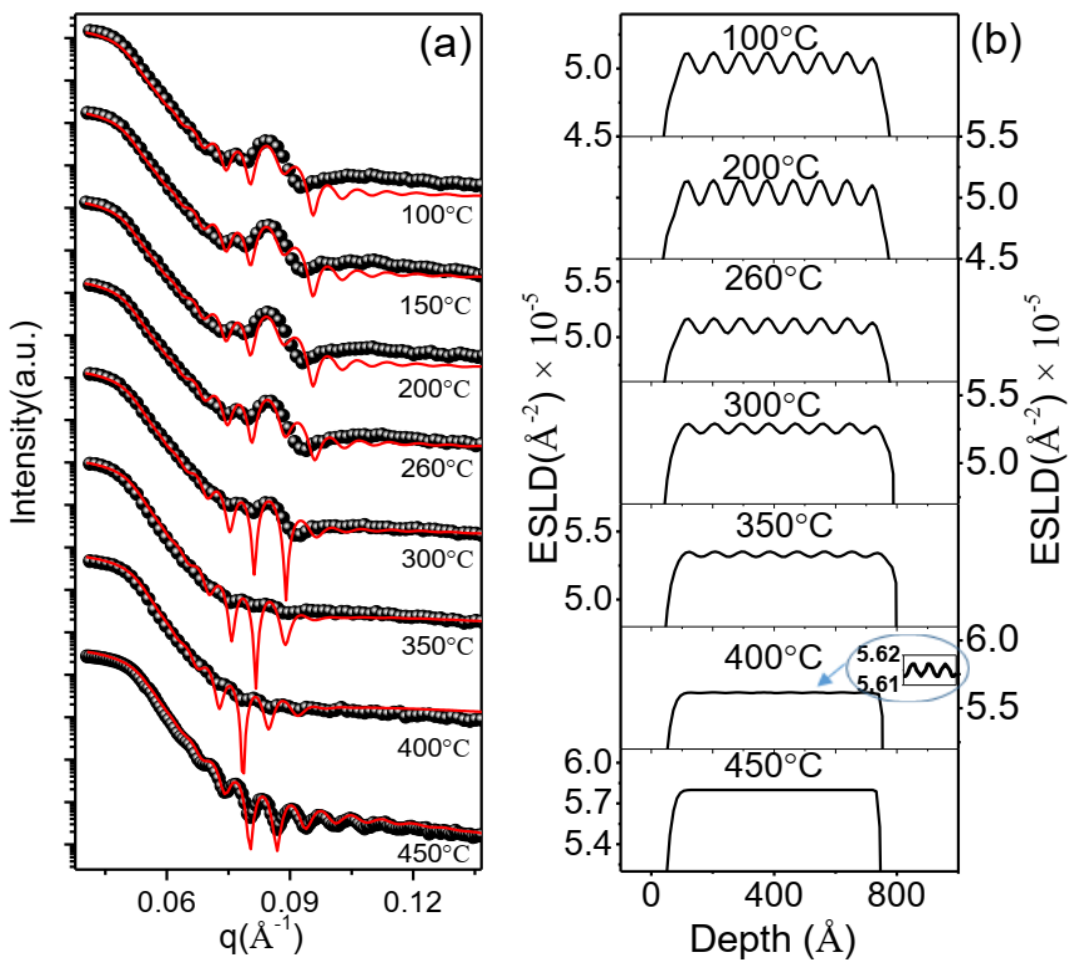

**Figure 6.6:** (a) Fitted *in situ* X-ray reflectivity data and (b) obtained corresponding ESLD profiles of the FeCo multilayer along the depth with increasing annealing temperature.

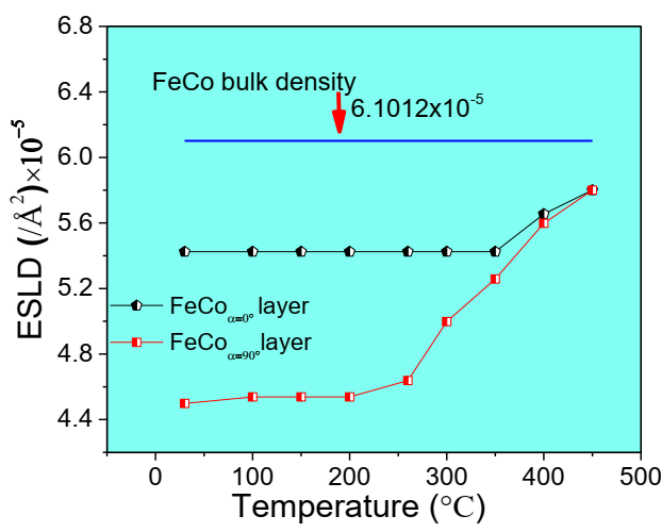

**Figure 6.7:** Variation of ESLD of FeCo$_{\alpha=0°}$ and FeCo$_{\alpha=60°}$ layers as compared to bulk FeCo density with increasing annealing temperature.

It is apparent that $\rho_{\alpha=60°}$ of the FeCo$_{\alpha=60°}$ layer remains almost unaffected up to the temperature of 250 °C; whereas an increase in the temperature beyond 250 °C leads to an increase of $\rho_{\alpha=60°}$ rapidly up to 450 °C. On the other hand, the electron density ($\rho_{\alpha=0°}$) of FeCo$_{\alpha=0°}$ layers increase slightly after 350 °C. Beyond 400 °C, $\rho_{\alpha=0°}$ and $\rho_{\alpha=60°}$ both reach almost the same values; this confirms the self-diffusion in FeCo layers, which promotes uniform FeCo densities throughout the depth. [73] Therefore, the decrease in the height of the Bragg peak with increasing temperature, as shown in fig. 6.6(a), corresponds to the decrease in the electron density contrast between FeCo$_{\alpha=60°}$ and FeCo$_{\alpha=0°}$ layers. After annealing at 450 °C, well-defined periodic Kiessig fringes and a uniform electron density along with the depth confirm a single layer of FeCo film.

### 6.3.4.3    Temperature-dependent Structural variation: GIXRD study

Figure 6.8 (a) to (d) gives the $q_z$ vs $q_r$ plots of GIXRD, transformed using grazing-incidence X-ray Scattering graphical user Interface software (GIXSGUI) [74]. These GIXRD 2D diffraction images are collected for as prepared sample as well as the annealed sample at 450°C by keeping the X-ray beam at grazing incidence angle ~0.4° along the in-plane direction of OAD ($\varphi=0°$ ) and perpendicular $\varphi=90°$ to it. Continuous isotropic half-ring patterns, indexed as (110) and (200) diffraction planes of bcc FeCo-phase, confirms randomly oriented FeCo grains, i.e. polycrystalline nature of the film. The data integrated using GIXSGUI software is plotted in fig. 6.8(e). Highly asymmetric diffraction peaks (different slops on both side of the peak), which remains asymmetric even after annealing at 450 °C, can be noticed. Groudeva-Zotova *et al.* [75], have also observed a similar asymmetry in diffraction peaks in magnetron-sputtered Fe$_{50}$Co$_{50}$ films after high dose Sm ion implantation. They have correlated this asymmetry to the ion beam induced strain in the film. In the present case, asymmetry in the peaks may be related to the anisotropic or elongated microstructure, which is created by the combination of normal and oblique angle deposition.

To obtain qualitative information about the possible anisotropic growth of grains and the in-plane structure with respect to the OAD direction, the bcc (110) peaks are extracted from the GIXRD pattern in non-coplanar geometry ($\delta=0°$) as shown in fig. 6.9(a). It is important to note that for $\delta=0°$, the incident beam makes an angle of ~0.4° in the vertical direction.

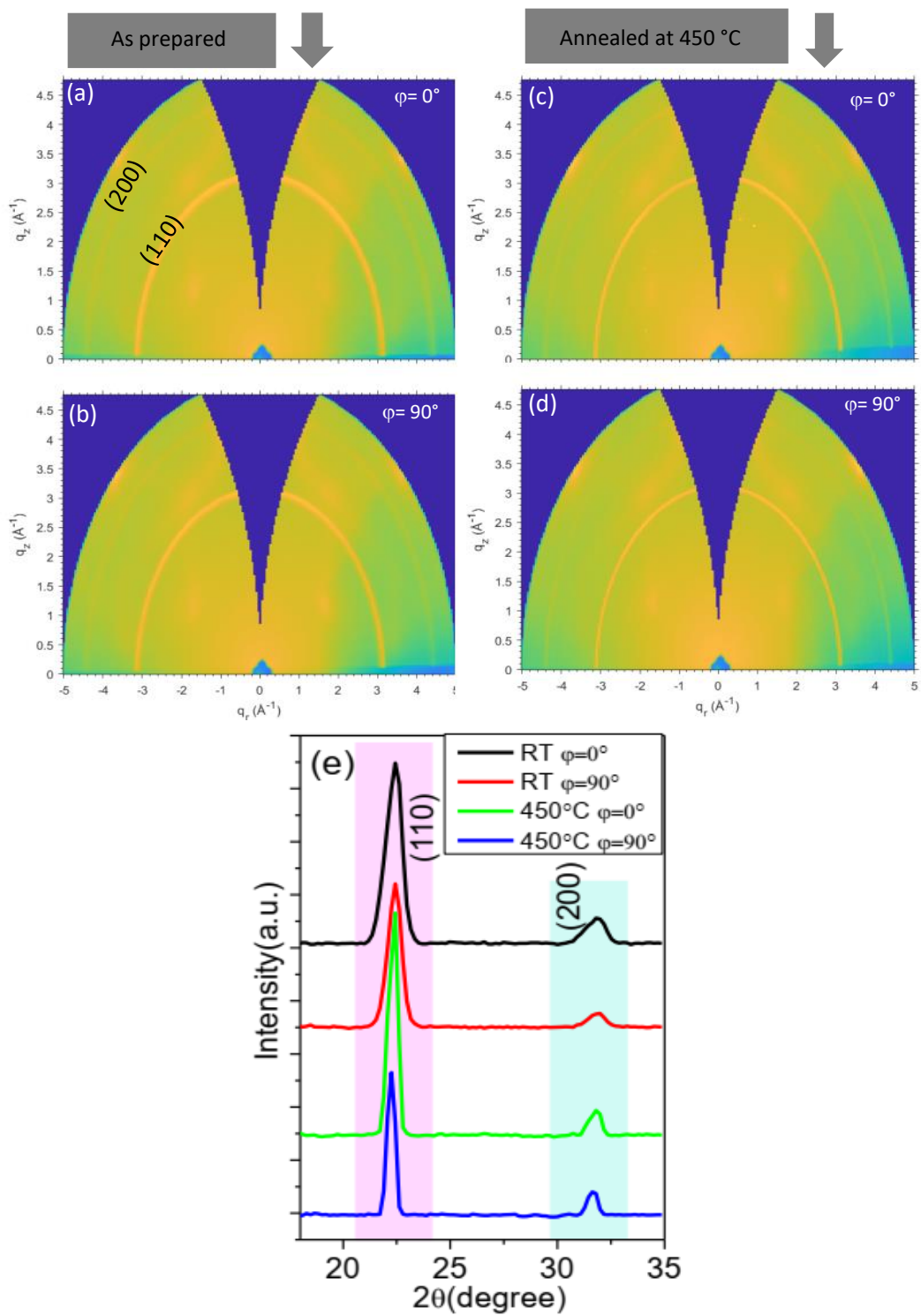

**Figure 6.8:** GIXRD patterns for pristine (a-b) as well as after annealing at 450°C sample (c-d) along ($\varphi=0^0$) and perpendicular ($\varphi=90^0$) to the columns' direction (e) intensity *versus* two theta plots obtained through integrating X-ray intensity using FIT2D software.

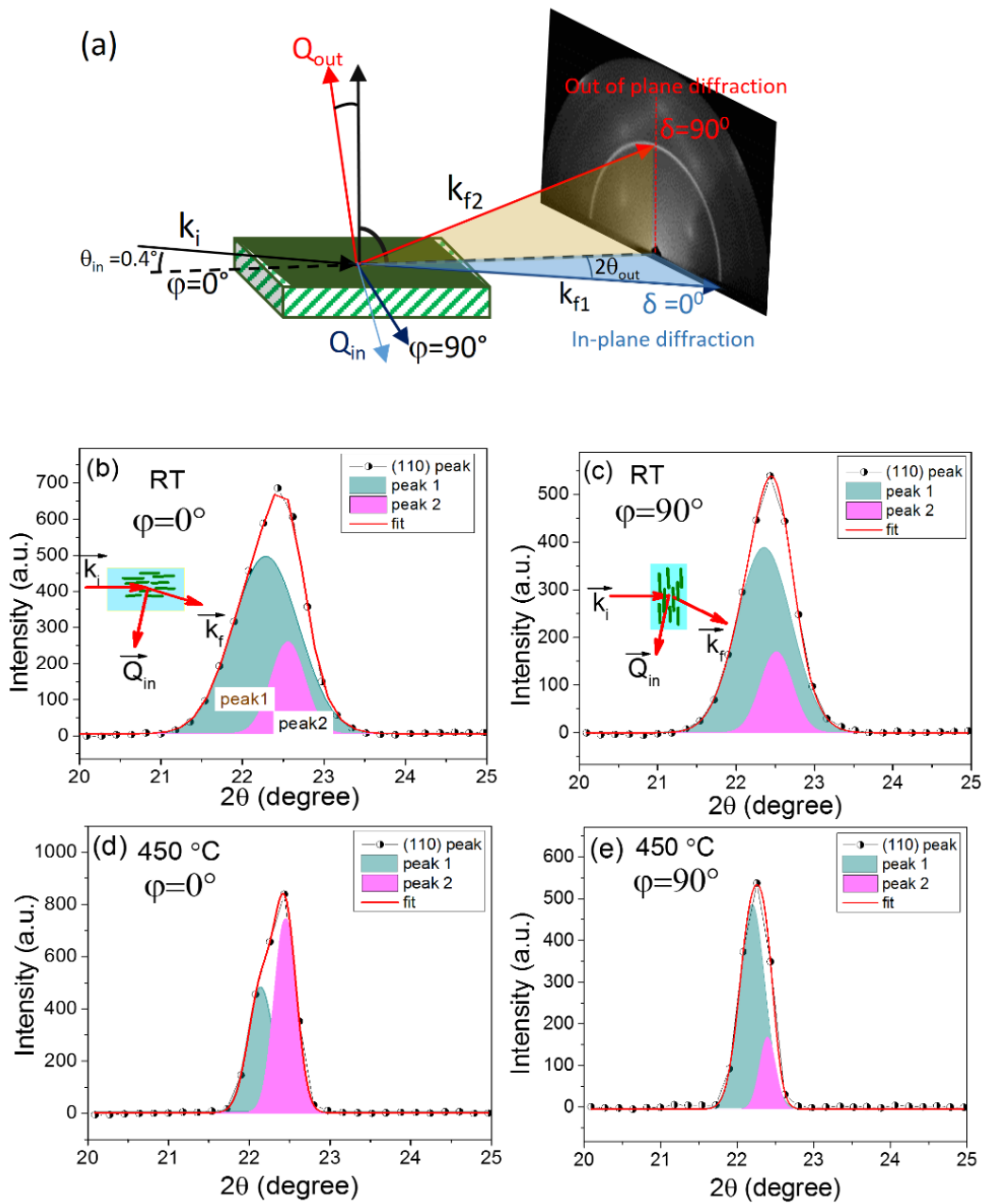

**Figure 6.9:** (a) GIXRD measurements' geometry, including indication about the out-of-plane and in-plane diffraction. Fitted GIXRD (110) peak for pristine sample (b-c) as well as after annealing at 450 °C (d-e), along ($\varphi$=0°) (b),(d) and perpendicular ($\varphi$=90°) (c),(e) to the columns directions.

Hence, the scattering vector ($Q_{in}$) lies almost in the film plane, and therefore, information is obtained about the scattering planes perpendicular to the film surface (in-plane structural information). In the present case of X-ray incident angle $\theta_{in} \sim 0.4°$ along $\varphi$=0°direction and diffracted peak at $2\theta_{out} \sim 22.4°$ for bcc FeCo (110), $Q_{in}$ lies almost normal to the OAD direction.

On the other hand, $Q_{in}$ lies almost along the in-plane OAD direction for $\varphi=90°$. Therefore, depending on the alignment of scattering vector $Q_{in}$, microstructural information along and normal to the OAD direction can be obtained in the film plane.

| | Peak (110) | | a (Å) | Min. crystallite size (nm) |
|---|---|---|---|---|
| **As prepared FeCo MLT** | ($\varphi=0°$) Along OAD | Peak 1 | 2.836(3) | 7.5 |
| | | Peak 2 | 2.870(5) | 4.3 |
| | ($\varphi=90°$) normal to OAD | Peak 1 | 2.841(1) | 8.4 |
| | | Peak 2 | 2.860(1) | 5.2 |
| **After annealing at 450 °C** | ($\varphi=0°$) Along OAD | Peak 1 | 2.850(3) | 11.1 |
| | | Peak 2 | 2.889(7) | 10.1 |
| | ($\varphi=90°$) normal to OAD | Peak 1 | 2.861(12) | 12.5 |
| | | Peak 2 | 2.889(21) | 8.8 |

**Table 6.1**: Lattice parameters and minimum crystallite size for different peaks for δ=0° (non-co-planar geometry) extracted from GIXRD measurements along ($\varphi=0°$) or normal ($\varphi=90°$) to OAD direction.

Figure 6.9 (b-e) shows bcc (110) peaks of FeCo multilayer for as prepared and 450°C annealed sample, obtained by integration of 2D diffraction image (fig. 6.8 a to d) in non-coplanar geometry (δ=0°). The asymmetry of all the peaks, corresponding to the bcc (110) planes of FeCo, are well fitted with the help of the two peaks model, and the lattice constant, and crystallite size corresponding to both peaks are obtained using the Scherrer formula [76]. The in-plane lattice constants corresponding to the two peaks for the as-deposited film were deduced as 2.836(3) Å and 2.870(5) Å, respectively, for $\varphi=0°$ geometry. The first peak value is reduced by 0.67(10) % that of $Fe_{50}Co_{50}$ alloy bulk material, 2.855(1) Å [77,78], whereas the second peak lattice parameter is increased by 0.52(18) %. It suggests the co-existence of compressive and tensile stresses in alternative FeCo layers in the multilayer. The lattice parameters and the minimum crystallite size corresponding to both peaks are obtained after fitting and are listed in table 6.1.

A perusal of table 6.1 suggests the following implications:

(i)     The crystallite sizes of the annealed samples are higher as compared to the as-prepared film. This indicates that the grain size increased after annealing.

(ii)    The grain size is different for tensile stressed (peak 2) and compressively stressed (peak 1) FeCo layers of the FeCo multilayer. The compressively stressed layer has about 40% larger grain size as compared to the tensile stressed layer.

(iii)   In comparison with bulk FeCo, the reduced lattice parameter of FeCo (peak 1) confirm compressive stress. It is found to be more compressive in the direction perpendicular to OAD. On the other hand, the increased lattice parameter of FeCo (peak 2) confirms tensile stress. It is found to be more tensile normal to OAD direction. From lattice parameters, it is clear that the strength of compressive stress is higher than tensile stress. It means the multilayer possesses resultant compressive strength in the direction perpendicular to the in-plane OAD direction.

(iv)   A significant reduction in the separation between peak 1 and peak 2 after annealing suggests a significant reduction of the overall stress.

### 6.3.4.4    Temperature-dependent GISAXS measurements

Further details about the buried nanostructures and their variation with temperature are obtained using GISAXS measurements in a mini chamber under vacuum conditions [57]. A schematic of the GISAXS geometry is shown in fig. 6.10(a), where $k_i$ and $k_f$ denote the incident and reflected wave vectors of the X-ray beam of wavelength $\lambda$=0.9537Å. The X-ray incidence angle is given by $\alpha_i$, while in-plane and out-of-plane scattering angle is given by $2\theta$ and $\alpha_f$, respectively. The deposition direction falls in the y-z plane and makes an angle $\alpha$ with respect to substrate normal. Scattering perpendicular and parallel to the film surface is given by $q_z$ and $q_y$ axis. Here, non-specular intensity away from the specular rod contains information on the film structure. Horizontal cuts taken at the Yoneda peak region along $q_y$ direction provide information on the lateral structures of the material parallel to the film surface, while the off-specular vertical cut along $q_z$ provides information on the structures perpendicular to the substrate [79].

The representative GISAXS 2D images of FeCo multilayer in as prepared stage and at various temperatures are shown in fig. 6.10(b) to (e). Increased X-ray reflected intensity along $q_z$ direction at $q_y$=0 nm$^{-1}$ is due to the specular reflection, which is blocked by the beam stop to avoid the saturation of the detector. Asymmetric distribution of the scattered intensity on both sides of the specular rod ($q_y$= 0 nm$^{-1}$) can be noticed in the as-prepared state. This asymmetry is almost disappeared after heating at 350ºC (fig. 6.10e). The angle (β) of tilted morphology (columnar structure) with substrate normal, which has been obtained based on the angle of intensity elongation I (fig. 6.10b) with respect to the horizontal plane, is found to be β=35º with respect to the deposition angle α=60° from the surface normal.

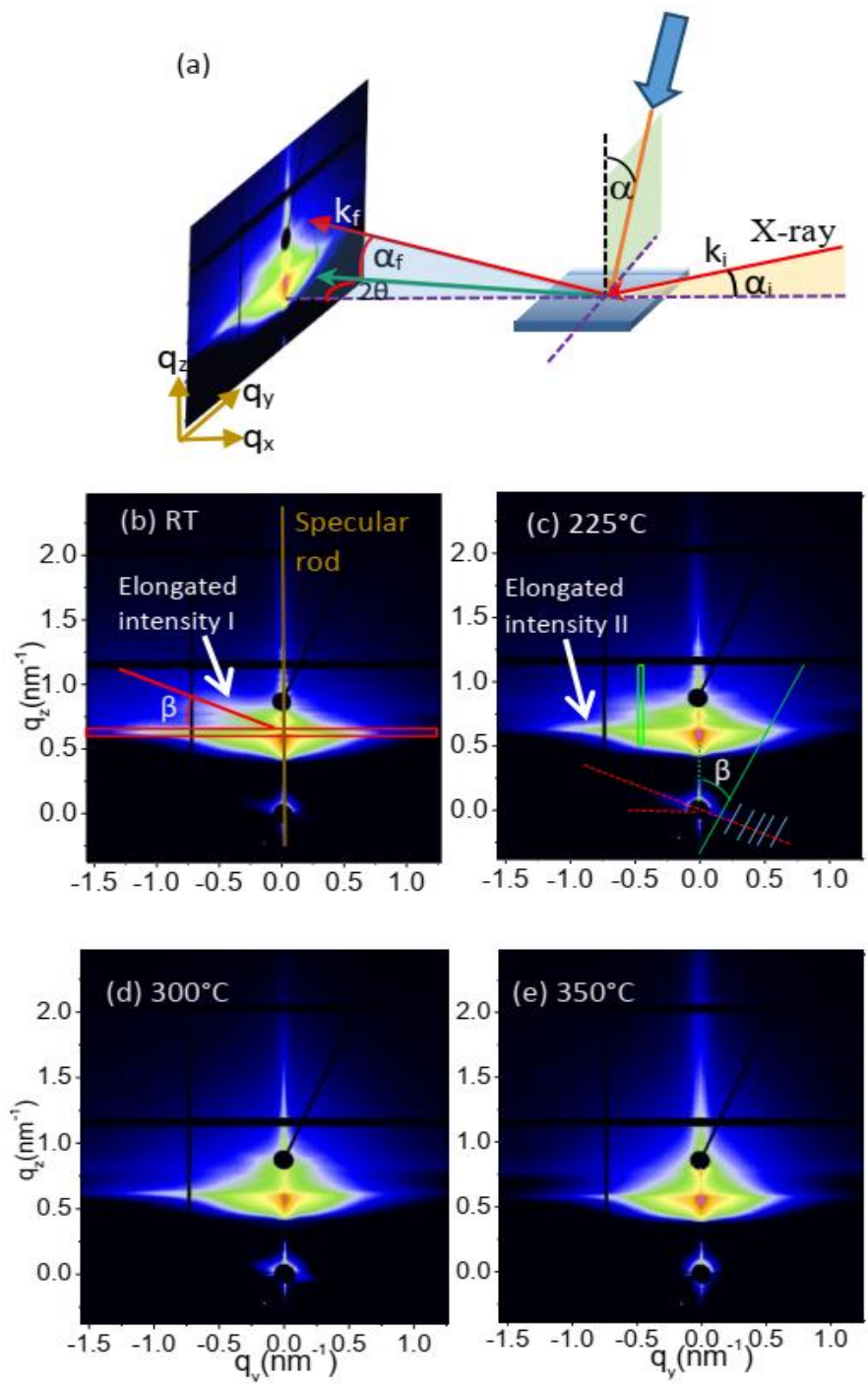

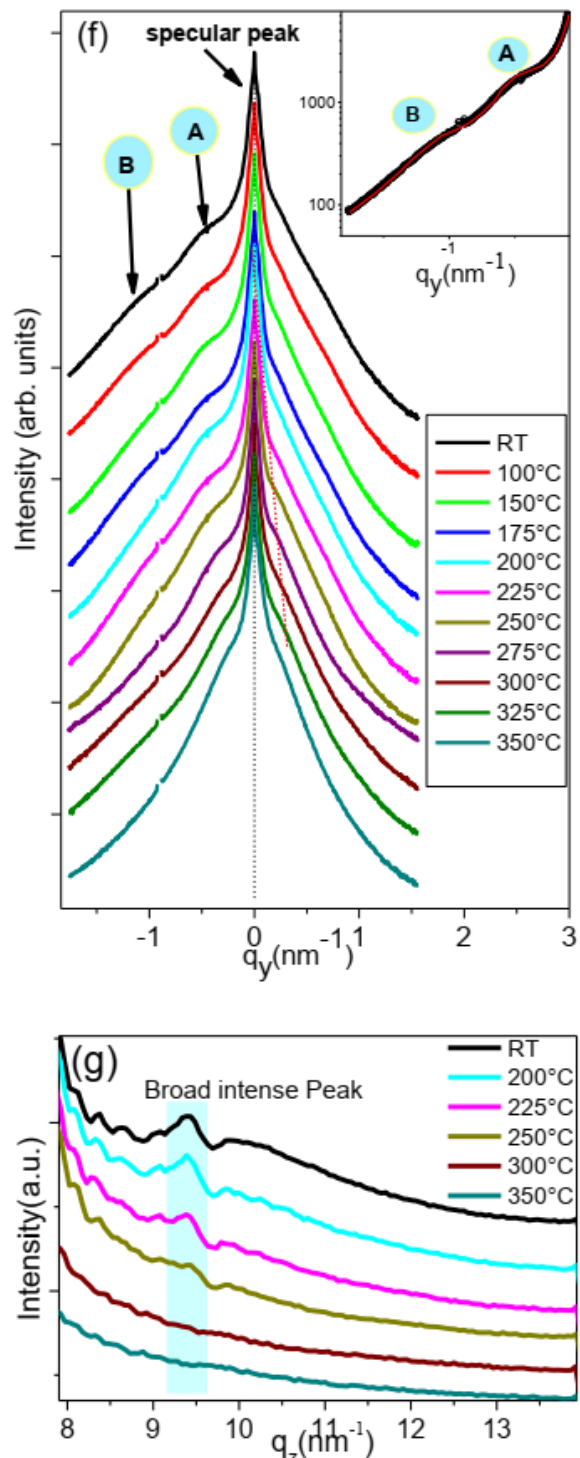

**Figure 6.10:** (a) GISAXS geometry, *in situ* GISAXS images at (b) RT, (c) 225 °C (d) 300 °C and (e) 350 °C. (f) shows 1D intensity vs $q_y$ profiles at the Yoneda wing region (red rectangle in fig. 9b) with increasing temperature (normalized with time) (g) 1D intensity vs $q_z$ plots are derived from the vertical green rectangle marked in fig. 9(c). (Normalized with exposure time). β in fig. (b) and (c) is the angle between the columns and the substrate normal.

In order to understand the temperature dependence more clearly, I *versus* $q_y$ profiles are extracted from the 2D GISAXS patterns for the horizontal rectangle planes as marked in fig. 6.10(b) and are plotted in fig. 6.10(f). Scattered intensity is found relatively high along the I *versus* -$q_y$ direction as compared to the positive $q_y$ direction. Two broad and weak Satellite peaks "A" at around $q_y$= -0.48 $nm^{-1}$ and "B" at around $q_y$= -1.16 $nm^{-1}$ indicate the presence of laterally ordered nanostructural arrangements that correspond to lateral spacings of about 13nm and 5.4 nm, respectively. As the temperature increases, peak "A" becomes relatively narrower and shifts toward smaller $q_y$ values, whereas it almost disappears at 350 °C. All these observations confirm the formation of columnar nanostructures, which grow laterally until they diffuse to form bigger columns with increased intermediate separation during annealing. Eventually, above 350°C, columnar nanostructures diffuse almost completely to form a uniform layer. Moreover, the I *versus* $q_z$ plot, as shown in fig. 6.10(g), is derived from vertical cuts near the satellite in the GISAXS images (fig. 6.10 c). The intensity distribution along $q_z$ shows weak oscillations along with a broad, intense peak at around $q_z$= 9.4 $nm^{-1}$. The intensity oscillations are related to the thickness modulation similar to Kiessig fringes or interference fringes in specular X-ray reflectivity, whereas the broad, intense peak is due to the presence of bilayer periodicity [73] created by the alternative arrangement of dense and porous FeCo layers. Systematic damping of the features with increasing temperature confirms the increase in roughness and intermixing between porous (low dense) and dense layers.

## 6.4    Discussion: High UMA and its dependence on stress and morphology

In oblique angle deposition, Volmer–Weber type of growth is expected in the initial stage of film growth due to the self-shadowing effect. [33] In the present case, islands of FeCo initially nucleate on the Si substrate at various nucleation centres. With further deposition, these islands become elongated columns with different crystallographic orientations due to OAD. Here, high surface-to-volume ratios of these columnar structures are found responsible for compressive stress in the film [80]. In general, FeCo (bcc) do not possess inherent uniaxial magneto-crystalline anisotropy [55]; therefore, the observed UMA in the present case of FeCo multilayer is mainly due to the shape anisotropy generated due to the growth of columnar structure and anisotropic compressive stress in the film. The absence of any preferential texturing in the present case also clearly confirms the absence of magneto-crystalline anisotropy, which is in accordance with the literature [55].

If we approximate the column as a prolate spheroid [39], the shape anisotropy is expressed as:

$$E_s = \frac{(N_a - N_c)M_s^2 \sin^2 \varphi}{2} \tag{2}$$

where $N_a$ and $N_c$ are the demagnetization factor along the plane perpendicular to the column and along the column, respectively. $\varphi$ is the angle between the direction of saturation magnetization ($M_s$) and columnar structure. The strength of $N_a$ and $N_c$ is related to the aspect ratio of common length over the column width. Due to self-shadowing, columnar structures of FeCo grains grow with a tilt elongated ($N_a/N_c > 1$) towards the deposition angle. Hence, shape anisotropy is the main factor for magnetic anisotropy in the in-plane direction of OAD projection.

In the present case, compressive stress is developed in the multilayer. The resultant compressive stress $\sigma_{0°}$ (along in-plane columns' direction) is lower as compared to $\sigma_{90°}$ (perpendicular to columns). Anisotropic compressive stress in the film plane is due to the morphological anisotropy in the film. It results in stress-induced (magnetostrictive) magnetic anisotropy [24], otherwise, homogeneously distributed in-plane stress hardly affects in-plane magnetization distribution. In order to understand the stress-induced UMA and its direction with respect to the shape anisotropy in the sample, net compressive long-range stress $\Delta\sigma = (\sigma_{90°} - \sigma_{0°})$ normal to OAD direction can be presented in terms of the magneto-elastic interaction energy $E_M$ [24] as:

$$E_M = \frac{3}{2}\lambda(\sigma_{90°} - \sigma_{0°})\sin^2 \psi \tag{3}$$

where $\psi$ is the angle between the direction of saturation magnetization ($M_s$) and $\Delta\sigma$ (positive for tensile and negative for compressive stress), $\lambda$ is the magnetostriction constant. With the help of eq. (3), it is easy to understand the alignment of the easy axis of magnetization due to the compressive stress in FeCo film with respect to the direction of in-plane OAD. Since the $Fe_{50}Co_{50}$ alloys have positive magnetostriction $+\lambda$ [81], stress ($\Delta\sigma$) is negative in the present case. Therefore alignment of the easy axis of magnetization is most likely resulting from $+\lambda$ and $\Delta\sigma$ in terms of magnetoelastic anisotropy. From eq. (3) it is clear that for a negative value of $\lambda \times \Delta\sigma$, the minimization of $E_{me}$ occurs when $M_s$ and $\Delta\sigma$ are perpendicular [24], hence alignment of the easy axis of magnetoelastic anisotropy will be preferred along morphology induced shape anisotropy (along OAD direction). As the minimization of the magnetoelastic, as well as shape anisotropy energy, would result in the alignment of the easy axis in the same direction (parallel to the in-plane projection of OAD), hence the in-plane direction of OAD is found to be the easy magnetization axis. Since anisotropy is reduced drastically due to the merging of columns after annealing the multilayer sample at 450ºC, the magnetostrictive contribution to the magnetic anisotropy is also expected to be negligible.

Umlor *et al.* [36] have demonstrated a similar method to introduce the uniaxial magnetic anisotropy on a cobalt film deposited at normal incidence with the use of a cobalt underlayer deposited at oblique incidence. In other studies, UMA is also seen in sputter-deposited thin cobalt film at normal incidence on obliquely deposited nonmagnetic Ta [82] and Pt [83] layers. In both cases, UMA in the normally deposited magnetic layer was found to be connected with elongated corrugations on the underlayer. The present work demonstrates the use of several obliquely deposited thin FeCo underlayers in between normally deposited FeCo layers to control and induce strong UMA in the FeCo multilayer structure. To further demonstrate the same, a series of four single FeCo thin films of the same thickness with varying oblique angle angles ($\alpha$) at 0º, 45º, 60º and 80º are prepared in identical deposition conditions to FeCo multilayer. The strength of magnetic anisotropy in FeCo multilayer is compared with a single OAD FeCo layer and with some potential studies on obliquely deposited single layers in the literature. To the best of our knowledge, FeCo thin film prepared in the form of multilayers is found to have much larger UMA.

We know that UMA in oblique angle deposited layers have out-of-plane anisotropy contribution due to slanted columns with respect to the surface normal. The low individual thickness of FeCo layers (4nm) throughout the multilayer stacks provides pure in-plane UMA contribution [36]. Importantly, the presence of compressive stress in thin FeCo layers due to small columns/islands provides the possibility to combine the effect of stress-induced magnetostrictive anisotropy and shape anisotropy along the in-plane column direction (along in-plane OAD direction). Otherwise, in higher thicknesses of FeCo, tensile stress may be present [80] and will try to align the stress-induced easy axis of magnetization normal to the direction of shape anisotropy, which may drastically reduce UMA in the multilayer. Moreover, separated columns in the thin films produce more UMA as compared to collapsed columns at a higher thickness due to increasing dipolar interaction between columns. Hence both the shape and strain anisotropy contributions can be utilized to increase the effective magnetic anisotropy in FeCo nanostructures.

## 6.5    Comparison of UMA in FeCo$_{MLT}$ with the single-layer FeCo thin film

A series of four single FeCo thin films with varying oblique angle angles ($\alpha$) at 0°, 45°, 60° and 80° are prepared in identical deposition conditions to FeCo$_{OAD}$($\alpha$=60°)/FeCo ($\alpha$=0°)]$_{\times 9}$ multilayer (denoted as FeCo multilayer) to compare in-plane UMA. The thicknesses of the single OAD FeCo layers are kept almost equal to the thickness of all OAD layers in the FeCo multilayer. In this report, we have calculated the magnetic anisotropy constant ($K_u$) using the hysteresis loops obtained along easy and hard magnetization directions and compared it with the FeCo multilayer. $K_u$ is also compared qualitatively with some potential studies on single layers in the literature. Several materials at different oblique angles are prepared to get strong UMA in thin polycrystalline films. To the best of our knowledge, FeCo multilayer prepared in the present work is found to have relatively larger in-plane UMA.

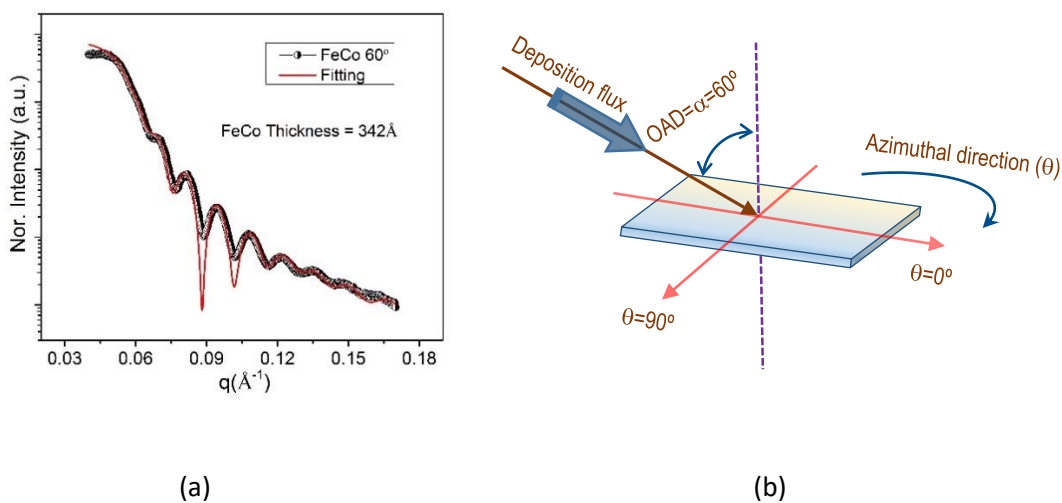

(a)                                                                (b)

**Figure 6.11:** (a) XRR pattern of oblique angle ($\alpha$=60°) deposited FeCo single layer film (b) Schematic of the OAD deposition and MOKE measurement geometry of obliquely deposited single FeCo film. $\theta$ is the azimuthal angle in the film plane, where $\theta$=0° is the direction along the in-plane direction of OAD.

Figure 6.11(a) gives fitted XRR pattern for the FeCo film deposited at oblique angles $\alpha = 60°$. Parratt fitting [58] of the data provided film thickness of about 342 Å and surface roughness of 14 Å.

MOKE measurements of all the single-layer FeCo films were performed in different azimuthal directions (fig. 6.11 b) to get the strength of UMA. Figure 6.12 gives some representative MOKE hysteresis loops collected along the $\theta = 0°$ (OAD direction) and normal to it ($\theta = 90°$). The OAD angles significantly modify the direction dependence of the hysteresis loops; hence, affecting the strength of in-plane UMA. Oblique angle dependence of hysteresis loops are as per related studies in the literature [45,84,85,86].

For the qualitative information of the UMA, the in-plane uniaxial anisotropy constant ($k_U$) is obtained using MOKE hysteresis loops along $\theta = 0°$ and $90°$ directions by utilizing the following relation [87],

$$k_U = \frac{M_S H a}{2} \tag{4}$$

Where $M_S$ is the saturation magnetization, and $H_a$ is the in-plane anisotropy field of magnetic layers. $H_a$ is quantitatively defined as the variation in the magnetic fields, which needs to reach saturation magnetization along the hard ($\theta = 0°$) and easy ($\theta = 90°$) direction of magnetization [88]. It can be obtained with the help of easy and hard hysteresis loop using the following relation-

$$H_a = H_{sat}(\text{hard loop}) - H_{sat}(\text{easy loop}) \tag{5}$$

It may be noted that FeCo multilayers are prepared by depositing nine FeCo layers at OAD 60° (each layer of 3.9 nm thick) alternatively, in between normally (OAD 0°) deposited nine FeCo layers (each layer of 5.1 nm thickness). The total thickness of all OAD nine layers is about 35.1 nm (351 Å), almost close to the FeCo (OAD 60°) single layer (thickness=342 Å) with < 2.6% deviation in thickness. Figure 6.13 (taken from fig. 3a, now with $H_{sat}$ marked on it) gives the MOKE loops of OAD 60° FeCo multilayer along easy ($\theta = 0°$) and hard ($\theta = 90°$) directions. The strength of in-plane magnetic anisotropy ($K_u$) is calculated for OAD 60° FeCo single layer (fig. 6.12c) and OAD 60° FeCo multilayer (fig. 6.13) for the comparison.

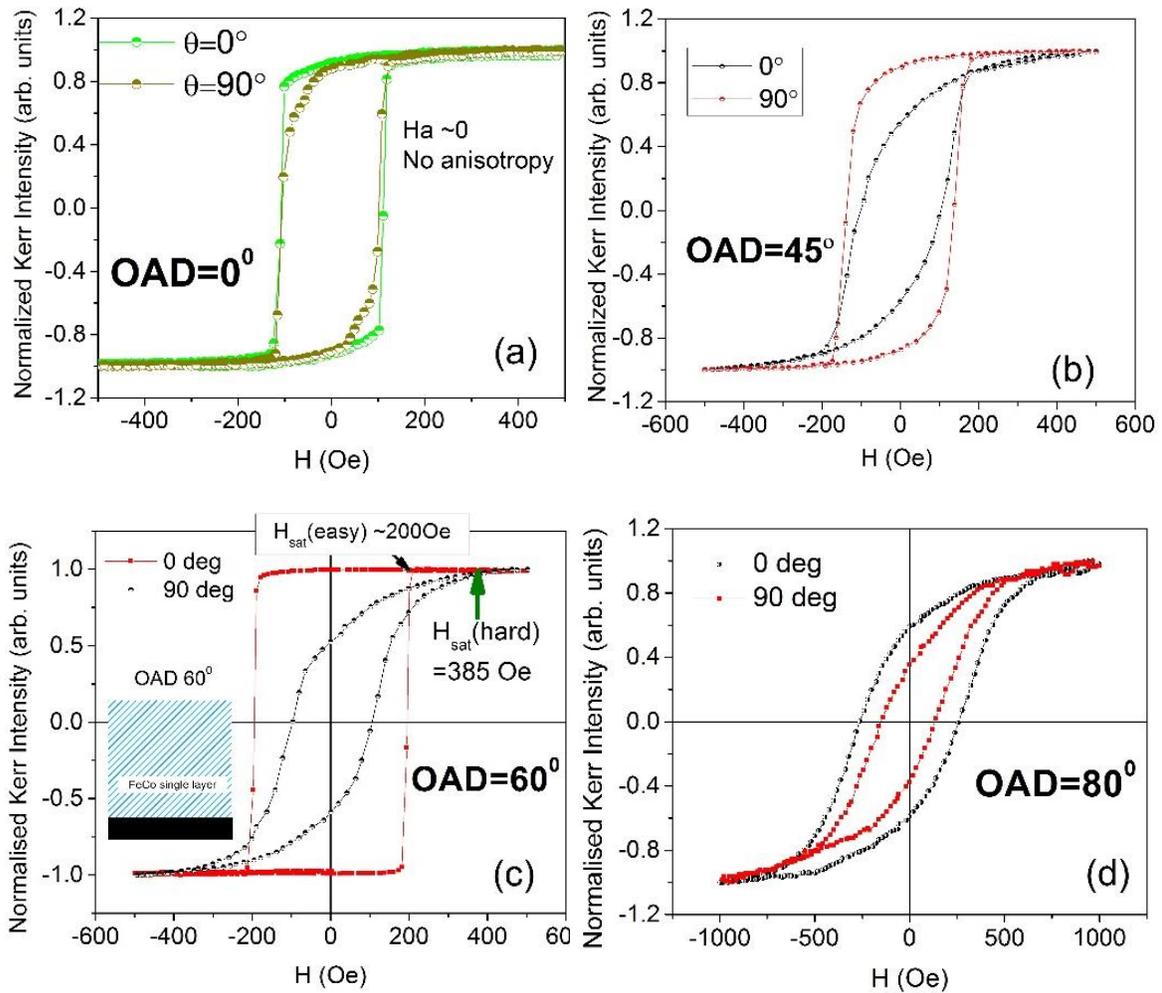

**Figure 6.12:** MOKE loops for Single layer FeCo thin films of ~36 nm thickness deposited at (a) OAD=0° - normal to the substrate (b) OAD=45° (c) OAD=60°, and (d) OAD= 80°.

$K_u$ is calculated using eq. (4) by taking the standard value of $M_s$ for FeCo = 1450 emu/cm³ [84]. $H_a$ = 185 Oe (385Oe-200 Oe) and 900 Oe (1250Oe-350 Oe) for FeCo 60° single layer (fig. 6.12c) and FeCo multilayer (fig 6.13), respectively. Therefore, $K_u$ for the FeCo multilayer in the as-prepared state and single FeCo OAD 60° layer are found as –

$$k_u = 6 \cdot 52 \times 10^5 \; erg/cm^3 \; \text{(FeCo multilayer)}$$

$$k_u = 1 \cdot 34 \times 10^5 \; erg/cm^3 \; \text{(FeCo single layer)}$$

The in-plane magnetic anisotropy constant of the OAD FeCo multilayer is about 4.8 times more than the FeCo 60° single layer of almost the same thickness. Moreover, compared to similar studies in the literature [45,84,85,86], present FeCo multilayers possess relatively much larger UMA.

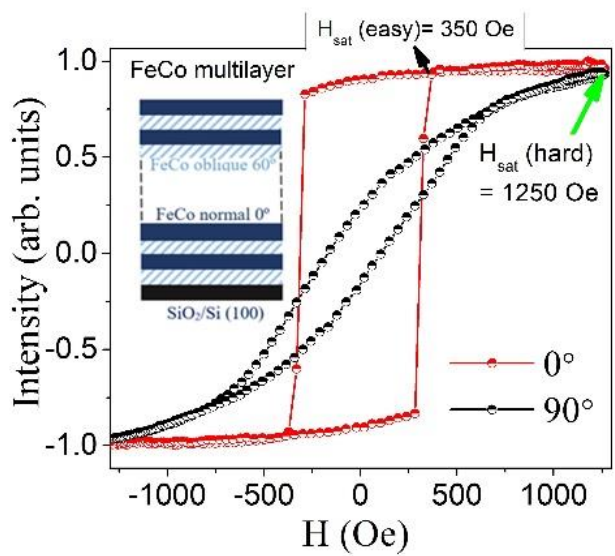

**Figure 6.13:** MOKE loop along easy (0°) and hard (90°) directions for as prepared FeCo multilayer sample.

**<u>Comparison with the studies in the literature</u>**- The strength of in-plane magnetic anisotropy in FeCo multilayer is also compared with potential studies in the literature.

Y. Fukuma et al. [84] have prepared single-layer FeCo films with an in-plane UMA by growing them on an obliquely sputtered thin Ru underlayer. They obtained $H_a$ as high as 25 Oe and magnetic anisotropy constant, $K_u \sim 1.8 \times 10^4$ ergs/cm$^3$ for a relatively much thinner FeCo layer of about 200 Å as compared to our FeCo multilayer. They have claimed the maximum strength of magnetic anisotropy about $K_u \sim 1.2 \times 10^5$ ergs/cm$^3$ only in 30 Å thin FeCo thin film. Also, according to this study, as the thickness increases, magnetic anisotropy decreases.

In another recent work, Z. Ali et al. [85], have investigated the magnetic properties of 34 nm thick permalloy (NiFe) thin films with the increasing oblique angle (<70º) deposition. The effect of OAD has been studied by performing Vibrating Sample Magnetometer (VSM), XRD etc., measurements. Maximum anisotropic field $H_a$ ~300Oe (0.03T) was obtained in 30 nm thick FeNi film which was deposited at OAD 60º. It is much less than the value of $H_a$ obtained in the case of our OAD 60º FeCo multilayer. Similarly, Ni films of 50 nm thickness (single layer) were deposited by dcMS and HiPIMS methods. At the same time, the OAD is varied from 0° to 70° [45]. Magnetic anisotropic field, $H_a$= 70 Oe for OAD 60º Ni thin film of 50 nm thickness is much less than the $H_a \sim 900$ Oe as obtained in case of OAD 60º FeCo multilayer. Co thin films of 3 nm thickness were also deposited on MgO (0 0 1) substrates using a DC magnetron sputtering system [86]. Uniaxial anisotropy is induced in the samples due to the deposition under the oblique angle of incidence. In-plane anisotropy constant, $K_u$ at OAD 60º is found to be only about $1.53 \times 10^5$ erg/cm$^3$.

## 6.6    Conclusion

A soft magnetic columnar multilayered structure of FeCo material is prepared by depositing FeCo layers by keeping the evaporation direction along the surface normal and at an angle of 60° from the surface normal, alternatively. In contrast to the literature, where FeCo films are usually magnetically isotropic in nature, strong shape anisotropy resulted in a well-defined UMA with the easy axis of magnetization along with the projection of the tilted nanocolumns in the film plane. The present work demonstrates the unique method of preparing FeCo multilayer with controlled and very strong UMA (4.8 times higher as compared to the OAD single layer FeCo film). Furthermore, compared to the literature, this study directly correlates the stress dependence of magnetic anisotropy in OAD films/multilayers. Resultant compressive stress in the obliquely deposited layers normal to the oblique angle deposition provides minimization of magnetoelastic energy along the in-plane column direction. This energy couples with the columnar shape anisotropy energies and thus results in preferential orientation of magnetic easy axis along the oblique angle deposition direction. Thus, the origin of strong UMA is understood in terms of the shape and magnetostrictive anisotropy corresponding to the ensemble of nanocolumns in the multilayer structure. Drastic reduction in the in-plane UMA after annealing at 450 ºC is attributed to merging columns and significant removal of anisotropic stresses after the heat treatment. It is established that the present method controls and induces UMA in the FeCo layered structures and provides an option of producing an in-plane uniaxial magnetic anisotropy in thin films, which are usually magnetically isotropic in nature. Finally, the study presented here opens a new gateway to creating a multilayer structure using single material with density variation, which may have epochal implications for future technological devices in various fields.

# CHAPTER 7

## Oblique angle deposited (OAD) Nano-columnar layered structure for magnetic anisotropy

### Understanding of the combined role of magnetocrystalline and shape anisotropy in OAD Co multilayered structure

A systematic study of Co-based multilayer structure [Co$_{oblique}$(4.4nm)/Co$_{normal}$ (4.2 nm)] - 10 bilayers], where each alternative Co layer is deposited at an oblique angle of 75°, has been studied with an aim to induced high in-plane uniaxial magnetic anisotropy (UMA) and to enhance the thermal stability. Multilayer is grown in an HV chamber using e-beam evaporation and characterized in-situ in another UHV using magneto-optical Kerr effect (MOKE) and reflection high energy electron diffraction (RHEED) for their temperature-dependent magnetic and structural properties. Well-defined strong UMA with the easy axis of magnetization along the projection of the tilted nanocolumns in the film plane is observed. It is present in the multilayer even after annealing at 450°C. To correlate the evolution of UMA with that of morphology and structure of the film, temperature-dependent grazing incidence small angle X-ray Scattering (GISAXS) measurements were performed in-situ at synchrotron radiation source, PETRA III, DESY, Germany. Observation of in-plane UMA in this multilayer is attributed to the combination of shape and magneto-crystalline anisotropy (MCA) due to orientated/textured columnar growth at oblique angle deposition. Crystalline texture in the film minimizes spin-orbital coupling energy along the column direction, which couples with the columnar shape anisotropy energies and results in preferential orientation of easy magnetic axis along the in-plane projection of the columns in the film plane. Reduction in UMA after annealing is attributed to merging columns and texturing removal after heat treatment.

## 7.1    Introduction

Oblique angle deposition (OAD), also known as "glancing angle deposition" (GLAD) controls the morphological development of thin films through "surface shadowing" effects which start from the atomic level [1]. Shadowing controls the nucleation of atoms and the nanostructure evolution of the films by preventing the deposition of film's atoms or molecules in regions situated just behind initially developed nuclei structures [2,3]. Such process results in tilted columnar film structures with porosity depending on the geometrical condition such as angle of deposition, thickness deposited etc.

Although extensive work has been carried out on various magnetic polycrystalline films, the Co-based system remains a model system for studying UMA because Co has excellent growth properties as a thin film, and very flat interfaces can be prepared by sputtering and molecular beam epitaxial techniques[4]. It may be noted that most of the previous work done on Co is mainly on normal depositions, a very few Co-based polycrystalline systems have been reported, where detailed investigation of oblique angle deposited films. Considering all the above issues, it has been planned to study UMA in a Co-based polycrystalline thin film structure to understand how the growth, stress and texturing affect the magnetic behavior of the film.

There are various papers on OAD for various materials, but in such systems role of stress is not properly understood. In our study of FeCo multilayer [5], stress plays an important role in inducing magnetic anisotropy in OAD thin films. But that study is done for the oblique incidence of 60°. At lower oblique deposition angles, anisotropy is not that much higher as for higher deposition angles [6,7,8]. So it is also our motive to study how stress affects the UMA for higher oblique deposition angles. Also, in the previous study of FeCo multilayer [5], no magneto-crystalline anisotropy (MCA) was present. No such anisotropy is reported for the OAD of FeCo films because FeCo film is magnetically isotropic. But in the case of Co, obliquely deposited films show some texturing behavior. According to the literature, the (002) peak (c-axis) of the Co hcp structure falls along the columns direction, which is the easy axis of magnetization hence creating MCA in the columns direction [9,10]. So in the growth method we have used here in this study, we also have to find out whether any texturing is present and, if present, how this texturing in the OAD part of this complex structure affects the magnetic anisotropy.

In the present study, we have applied the same approach to deposited Co multilayer, which we previously used for FeCo multilayer deposition [5]. It is the oblique-normal alternate depositions to make multilayer structure using single Co magnetic material. The Co multilayer [Co$_{(\alpha=75°,oblique)}$(4.4 nm)/ Co$_{(\alpha=0,normal)}$(4.2 nm)] (10 bilayers) on Si(100) substrate having native oxide induces strong in-plane UMA. We have demonstrated using x-ray reflectivity (XRR) study that the presence of Bragg peak confirms that up to 10 bilayers, the system was able to follow the same morphology. However, the system has a complex structure (oblique-normal multilayer structure). A strong UMA in the plane of vapour deposition direction is observed as compared to the other Co-related magnetic multilayers, i.e. Co/Cu deposited obliquely at 84° [11,12] and Co/Au deposited obliquely at variable OAD angles up to 60° [13, 14]. Effect of annealing on in-plane UMA, structure thickness, roughness and density of the multilayer has been studied in a UHV chamber using in-situ magneto-optical Kerr effect (MOKE), Reflection high energy electron diffraction (RHEED) and ex-situ XRR studies respectively. Grazing incidence x-ray diffraction (GIXRD), atomic force microscopy (AFM) and grazing incidence small angle x-ray scattering (GISAXS) measurements at various temperatures are performed to show the temperature-dependent structural and morphological changes in the multilayer and correlated with UMA.

## 7.2    Preparation method and experimental details

The Co multilayer structure [Co 4.4 nm; OAD=75° /Co 4.2 nm; NAD]$_{10}$ are deposited on SiO2/Si(100) substrates under base pressure of $2\times10^{-7}$ mbar using e-beam evaporation as deposition source. NAD represents the normal angle deposition (normal deposition). The deposition geometry and the grown bilayer stack are shown in fig. 7.1. 75° and 0° angles of incident flux with respect to the substrate normal are used alternatively as the deposition angles for the incident evaporated material to get a multilayer structure. Figure 7.1(a) shows the schematic of deposition, while the final multilayer is shown in fig. 7.1(b). The thickness of each layer is monitored using calibrated quartz crystal monitor in both configurations: normal and oblique angle deposition. XRR measurements are performed using Bruker D8 diffractometer having Cu K$\alpha$ radiation to obtain the structural information: thickness (t), electron density (ρ) and roughness (σ) of the individual layers. A separate measurement chamber with a base pressure of 4.6×10⁻⁹ mbar is used for in-situ temperature-dependent MOKE measurements to observe how the magnetic anisotropy varies with temperature. Hysteresis loops using MOKE and structural information using RHEED patterns are collected in-situ during heating the Co multilayer up to 500 °C. All these measurements make it possible to correlate the UMA with the film structure and morphology.

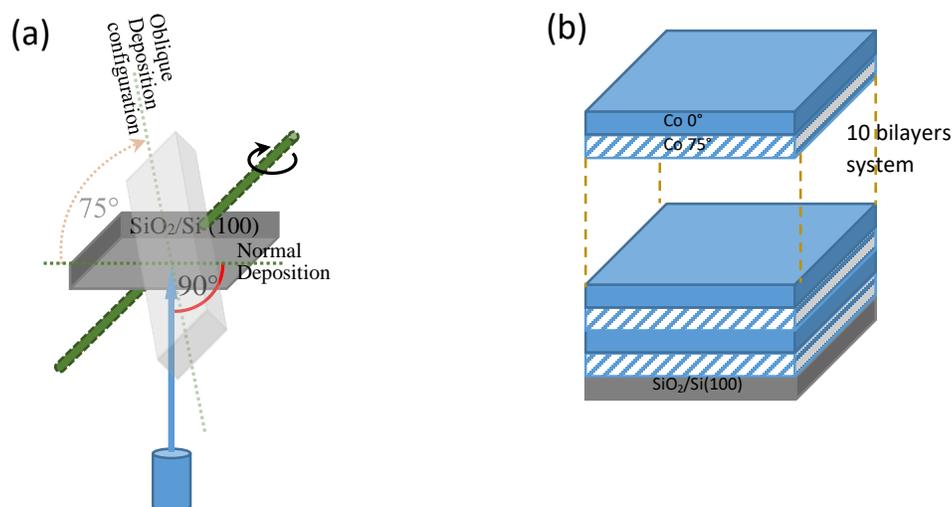

**Figure 7.1.** (a) Schematic diagram of the deposition geometry for both normal and oblique deposition of Co layers, green rod shows the axis about which the substrate is rotated (b) final multilayer having 10 bilayers stack (Co multilayer) on Si (100) substrate containing native oxide layer.

GISAXS measurements are performed at the P03 (MiNaXS) beamline, Petra-III storage ring at DASY, Hamburg (Germany). An incident angle of 0.45° and photon energy of 13 keV are used for such measurements. The platform of portable mini x-ray diffraction (XRD) chamber is used for GISAXS measurements to observe the morphology of the various samples heated at various temperatures up to 500 °C. The complete detail of the chamber and its working geometry is already published [15]. PILATUS 1-M (Dectris Ltd., Switzerland) detector, which has a pixel size of (172 × 172) μm², is used for GISAXS data recording. AFM studies are also carried out separately to observe film morphology. GIXRD measurements of the pristine and annealed samples are done with x-ray beam of wavelengths λ=0.505917 Å and λ=0.78445 Å at x-ray synchrotron BL9, Indus2, RRCAT (India). Scattered X-ray intensity is mapped for grazing incident angle of ~0.4° using a two-dimensional (2D) image plate detector to receive information about the microstructure along the OAD column direction.

## 7.3   In-situ characterizations

### 7.3.1   MOKE study: Temperature-dependent anisotropy variation

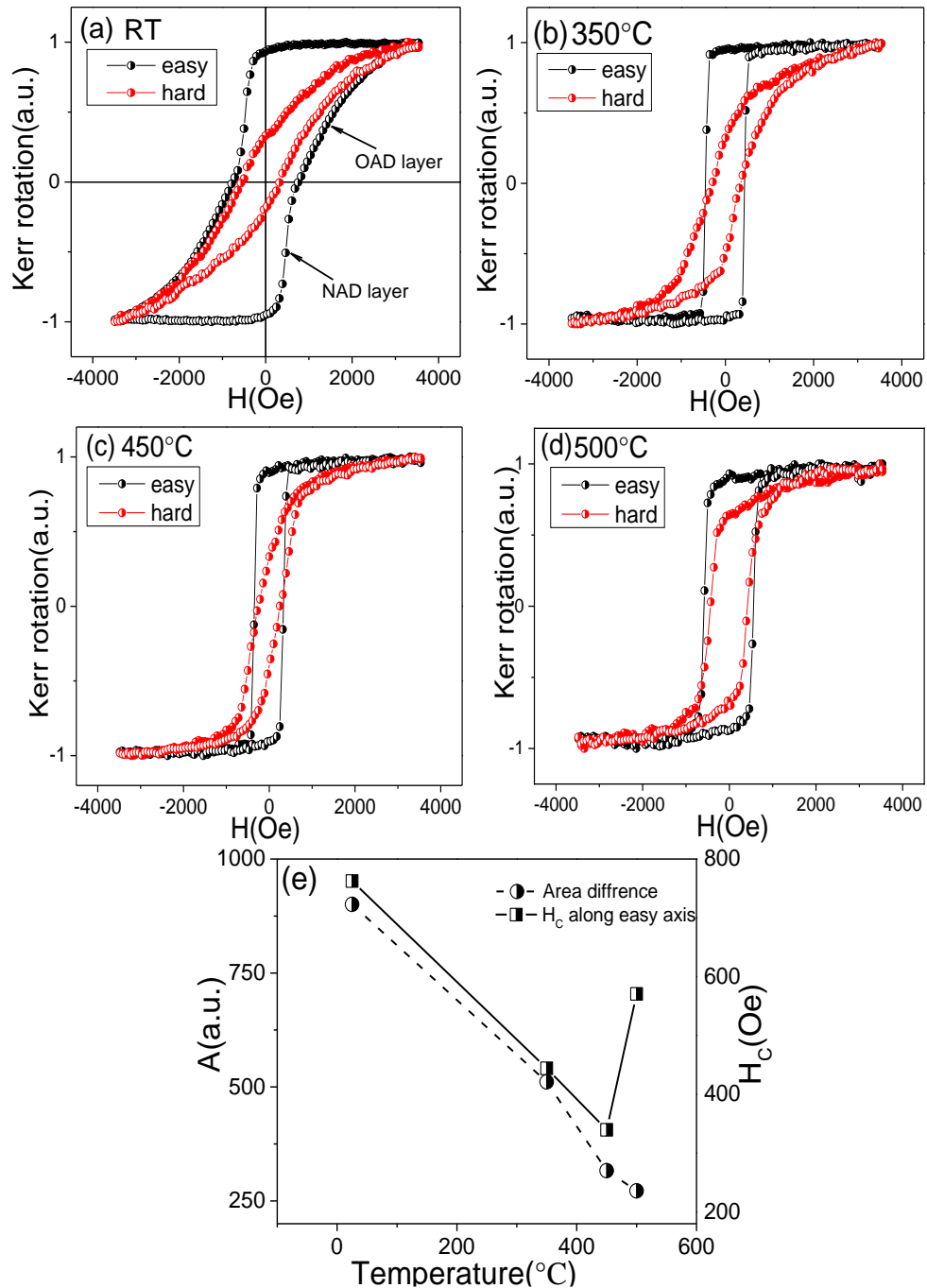

**Figure 7.2**: Hysteresis loops along the easy and hard axis at (a) room temperature (pristine), (b) 350 °C, (c) 450 °C and (d) 500 °C, while (e) shows the area difference between the easy and hard axis and $H_C$ along the easy axis at various temperatures.

| Temperature | Area difference A (a.u.) |
|:-----------:|:-----------------------:|
| RT | 900.60 |
| 350 °C | 511.03 |
| 450 °C | 316.29 |
| 500 °C | 272.09 |

**Table 7.1**: Area difference A vs temperature.

In-situ temperature-dependent MOKE measurements are performed to have the magnetic information of the multilayer, as shown in fig. 7.2(a) to 7.2(d). Hysteresis loops are taken in the longitudinal geometry. It is found that at room temperature (fig. 7.2a), when the field is applied along the direction of the in-plane projection of the columns, the hysteresis loop comes in near square form. At the same time, it becomes rounded when the field is applied in the perpendicular direction. This near square loop is also found in literature in Co/Cu multilayer deposited obliquely [11,12] but does not explain the cause of such shape. In our case easy loop contains two components (i) square part and (ii) rotated part. It is because the NAD layer having less porosity shows less coercivity. OAD layers contain pinning centres due to their porous nature, and the coupling of the OAD layer with the NAD layer shows such behavior.

At 350 °C (fig. 7.2b), the hysteresis loop along the in-plane projection of the columns becomes a perfect square. The steepness of hard loops gradually increases with increasing temperature, and at 500 °C (fig. 7.2d), it transforms to near square loops, which means UMA is diminishing with heating. Coercivity continuously decreases up to 450 °C (fig. 7.2a to c) and then increases at 500 °C (fig. 7.2d), as plotted in fig. 7.2(e). Magnetic anisotropies are calculated using the area difference method (difference in areas of easy vs hard loops) [16] and plotted vs temperature in fig. 7.2(e) and shown in table 7.1, which confirms the significant reduction in anisotropy at 500 °C.

### 7.3.2   RHEED: Temperature-dependent surface structural study

To obtain crystallographic information, RHEED measurements are performed along the column plane (0°) and perpendicular to the column plane direction (90° direction) in-situ simultaneously with MOKE measurements, as shown in fig. 7.3. Main features of the RHEED measurements are as follows: (i) Concentric rings confirm that structure grows in the polycrystalline hcp phase, (ii) Texturing is clearly visible when columns direction is placed perpendicular to the e- beam direction, (iii) Texturing in the (002) peak (c-axis of the hcp structure) is higher along the direction marked by the blue arrow. This arrow makes the average angle 40º with respect to the film's normal, (iv) at a temperature of 500 °C (fig. 7.3f), texturing is found missing. In the case of Co OAD, the direction of the c-axis is found to be along the column direction, which generally contributes to the easy axis of the MCA in the column's direction [10,17,18].

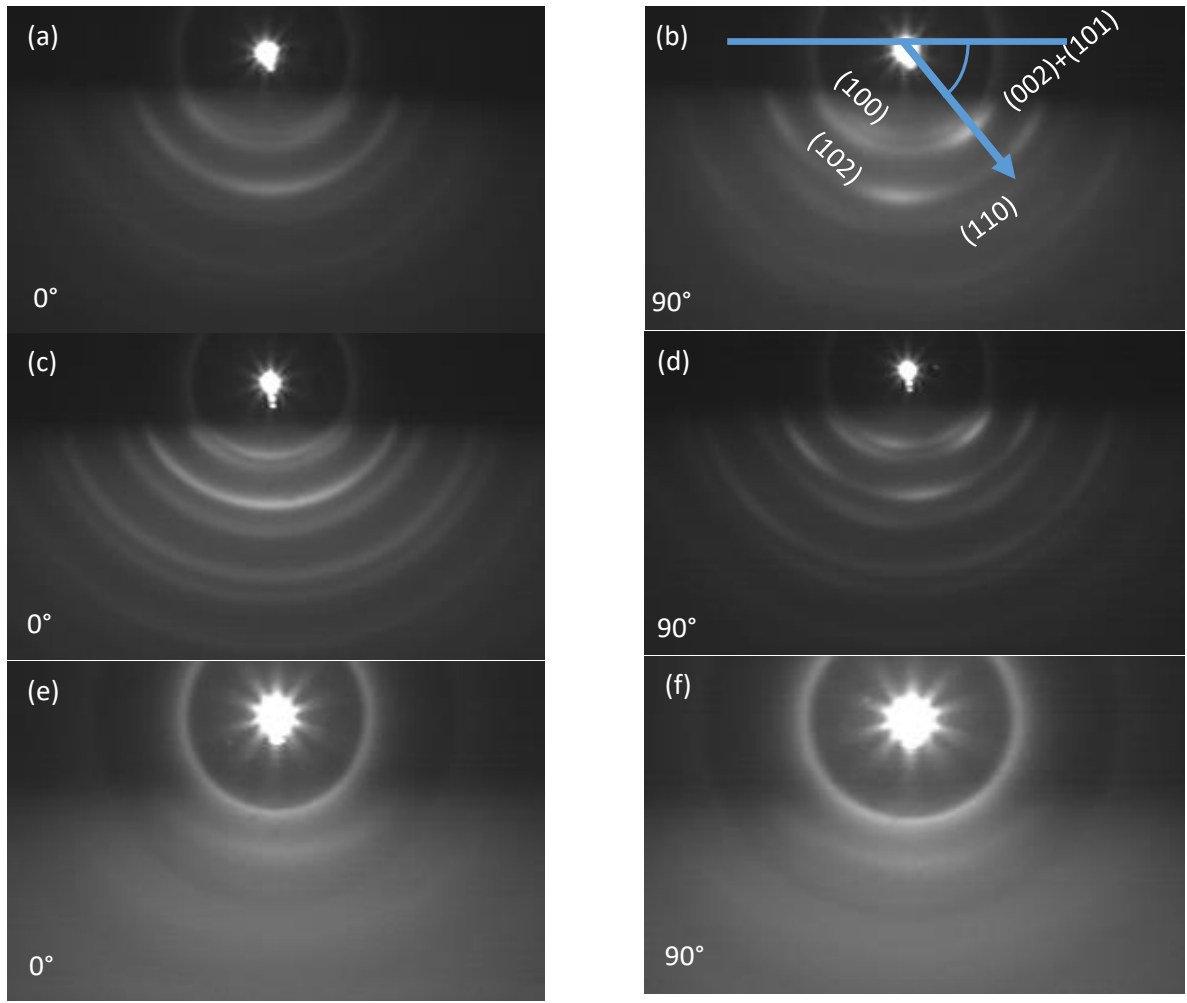

**Figure 7.3:** RHEED images along 0° and 90° at (a), (b) room temperature, (c),(d) 400 °C and (e),(f) 500 °C.

## 7.4    Ex-situ measurements

### 7.4.1    GIXRD: Depth-resolved structural variations with temperature

As RHEED is a surface-sensitive technique, so a depth-resolved structure analysis is required. For this, ex-situ GIXRD measurement for the pristine sample is done along two perpendicular directions similar to RHEED, as shown in fig. 7.4(a) and 7.4(b). X-ray wavelength λ=0.505917 Å is used for the measurements. Concentric rings correspond to hcp Co, while bright spots correspond to the Si substrate. Similar weak texturing (as the ring is continuous) is found for the (002) peak in the case of GIXRD (fig. 7.4b), confirming the texturing is present throughout the depth in the OAD layer. The average texturing angle is similar to RHEED measurements (~40º). Also, the 1-D plot is shown in fig. 7.4(c), and we draw two conclusions from the 1-D plots: i) along 90º, the intensity of (002) peak is higher with respect to the GIXRD along 0º. It means the (002) peak is textured for GIXRD measurements taken along a 90º direction. ii) No peak shift found along both directions confirms the absence of stress in the multilayer.

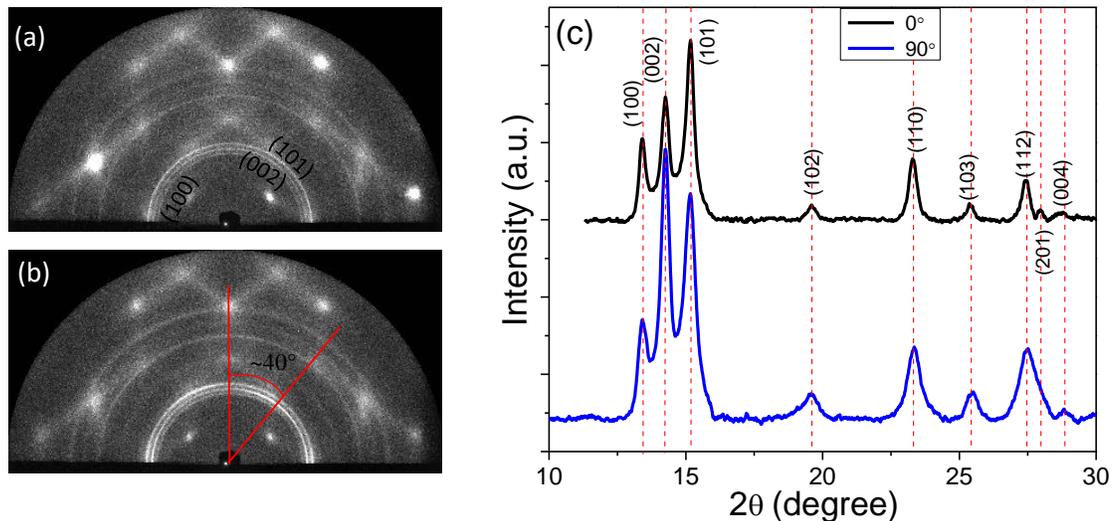

**Figure 7.4:** GIXRD image (for x-ray λ =0.505917 Å) of Co multilayer along (a) 0º and (b) 90º directions (c) corresponding 1-D plots.

Now, we just rotated the 90º position of the sample by 180º. In this position, columns again remain perpendicular (90°) to the x-ray beam direction, but the direction of texturing should change in the GIXRD image. We performed this measurement at different x-ray wavelength λ=0.78445 Å. The measurements are carried out for pristine and heated samples (400 ℃, 500 ℃) along the columns (0º) and perpendicular to the column's direction (90º) as shown in fig. 7.5(a-f). As expected, texturing is observed in the opposite direction for data of 90º positions (fig. 7.5b and 7.5d). The extra peaks are observed at a temperature above 400 ℃. When the 1D data (fig. 7.5g) is extracted from the measurements along the 0º direction, it is found that the FCC phase starts to grow at 400 ℃ and becomes dominant at 500 ℃ as the peak heights of HCP are significantly decreased at 500 ℃. Both HCP+FCC phases coexist up to 500 ℃. Recrystallization considerably reduces texturing in the (002) peak at 500 ℃ (fig. 7.5f). Particle size is also calculated from GIXRD data at various temperatures. For the hcp (002) peak, particle size at room temperature, 400 °C and 500 °C is found to be 5.4 nm, 8.3 nm and 15.2 nm, respectively.

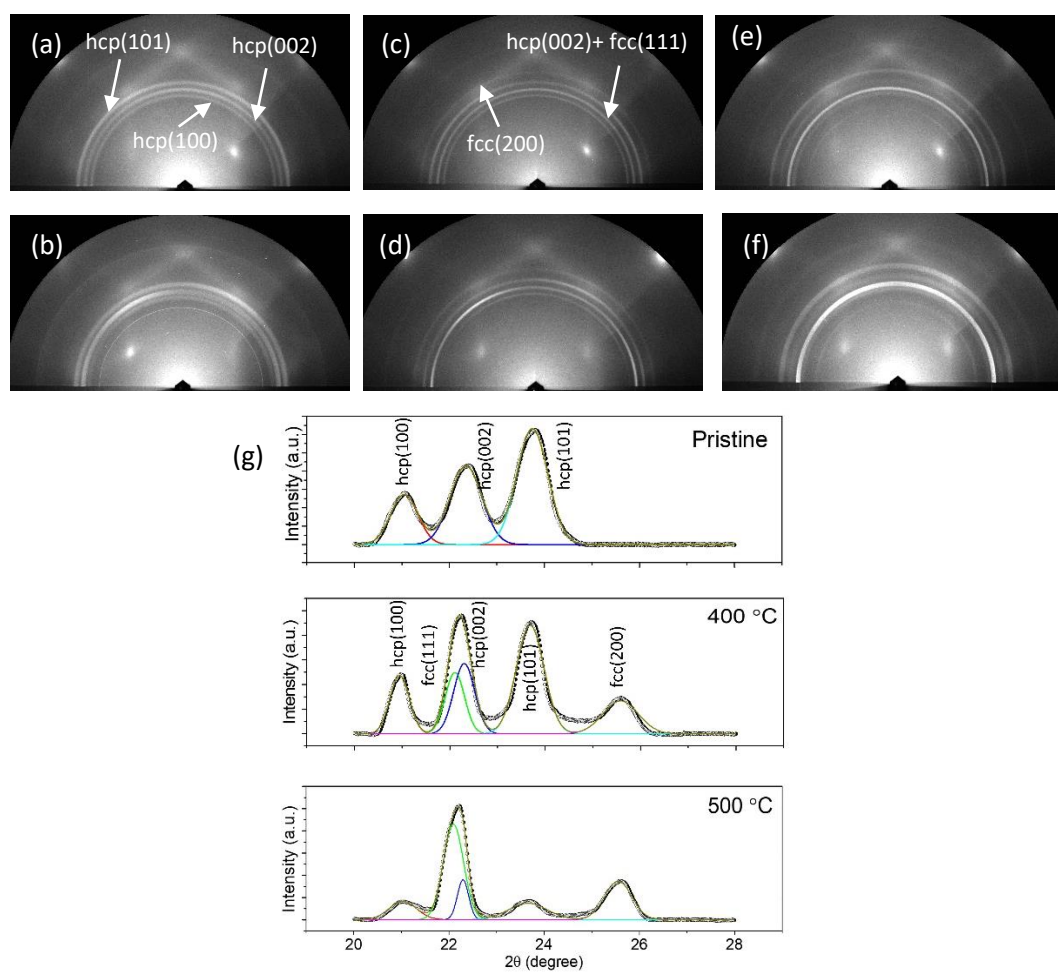

**Figure 7.5:** Temperature dependent GIXRD (for x-ray λ =0.78445 Å) images at (a-b) room temperature (pristine sample), (c-d) 400 °C and (e-f) 500 °C for x-ray along (a-c-e) 0º and (b-d-f) 90º to the columns (in-plane projection), (g) fitting of the 1D data shows phase change.

## 7.4.2 XRR and AFM: Density and morphology study

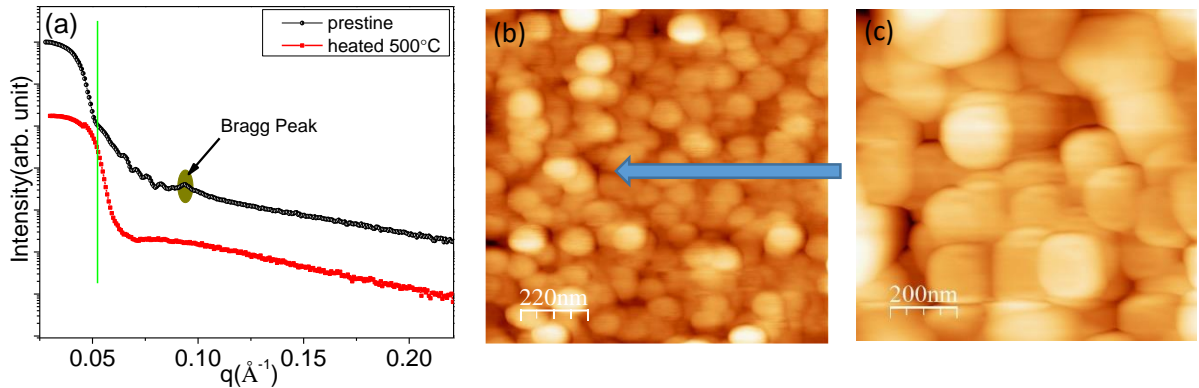

**Figure 7.6:** (a) XRR plots of pristine and annealed samples of Co multilayer. The vertical green line shows the shift in the critical angle of the multilayer system. AFM images of pristine (b) and annealed samples (c) at 500 °C of Co multilayer. The arrow shows the in-plane direction of the oblique angle deposition.

XRR measurements are done for both the pristine sample and for the sample heated in UHV at 500 °C obtained from temperature-dependent MOKE and RHEED measurements. The in-situ UHV chamber's full detail can be found elsewhere [19]. The outcome of the performed XRR measurement is shown in fig. 7.6(a). A clear Bragg peak can be seen for the pristine sample, which confirms the multilayer structure of the sample having layers of different electron densities though the material used for the deposition is only Co. The difference in the densities hence multilayer formation is the outcome of the deposition method we used: alternate OAD and NAD. We know that the OAD forms the layer having oblique or tilted nanocolumns, which is the result of the shadowing effect [1] and hence possesses the porous layer with lower electron densities, while the NAD layer possesses the higher electron density.

The heated multilayer doesn't show any clear thickness oscillations. The possible reason for such oscillation of less XRR amplitude may be the increased roughness after heating. Also, it can be clearly seen from the XRR plots that the overall electron density of the multilayer is increased (increased critical angle: green line in fig. 7.6a) and reaches near the bulk Co density, which suggests the diffusion of the OAD layer. This process causes the diminishing of the Bragg peak and will turn the multilayer into a single-layer film. Another reason for the absence of the Bragg peak can be the increased roughness as XRR of the heated sample shows no oscillation, but increased electron density is a clear sign of Bragg peak disappearance at 500 °C.

The AFM images of the pristine and 500ºC heated sample are presented in fig. 7.6(b) and 7.6(c). The Blue arrow shows the in-plane projection of the OAD direction. Also, the grains look elongated along the deposition direction, but the tilted columnar structure can't be seen. It may be due to the reason that the top layer is normally deposited. The heated sample shows a larger grain size without elongation in any direction. The rms roughness for the pristine and heated samples is found to be 2.4 nm and 7.6 nm, respectively. This increased roughness well matches the XRR plot for a heated sample as it doesn't show any clear thickness oscillations.

### 7.4.3 GISAXS: Study of temperature-dependent morphological variations

For more insights into the buried nanostructures and their temperature-dependent structural variation, a mini vacuum chamber [15] is used to perform separate GISAXS measurements. Figure 7.8(a) represents the GISAXS geometry used where x-ray beam of wavelength $\lambda$=0.95373 Å is used for such an experiment with incidence angle $\alpha_i$=0.45° from the sample surface. $k_i$ and $k_f$ are the incident and reflected wave vectors. Tilted columns fall in the vertical pink plane perpendicular to the plane containing vectors $k_i$ and $k_f$ (fig. 7.7a). In GISAXS measurement, the information of the film structure along the vertical direction and lateral structure of the material parallel to the film surface are obtained by the vertical cut along $q_z$ at non-specular intensity away from specular rode and horizontal cut taken at Yoneda peak region along $q_y$ direction respectively.

Figure 7.7(b) to (f) shows the 2D GISAXS images for samples annealed at various temperatures. Direct beam stopper and stopper for specular reflection at $q_y$=0 nm$^{-1}$ and along $q_z$ direction are used to avoid saturation of the GISAXS detector. As we explore the obtained GISAXS results, The GISAXS image for the pristine sample shows clear asymmetry in scattered intensity distribution about the specular rod ($q_y$=0.0 Å$^{-1}$). The asymmetric distribution of scattered intensity remains present for the multilayer annealed up to 450 ºC and almost vanished for the sample annealed at 500 ºC.

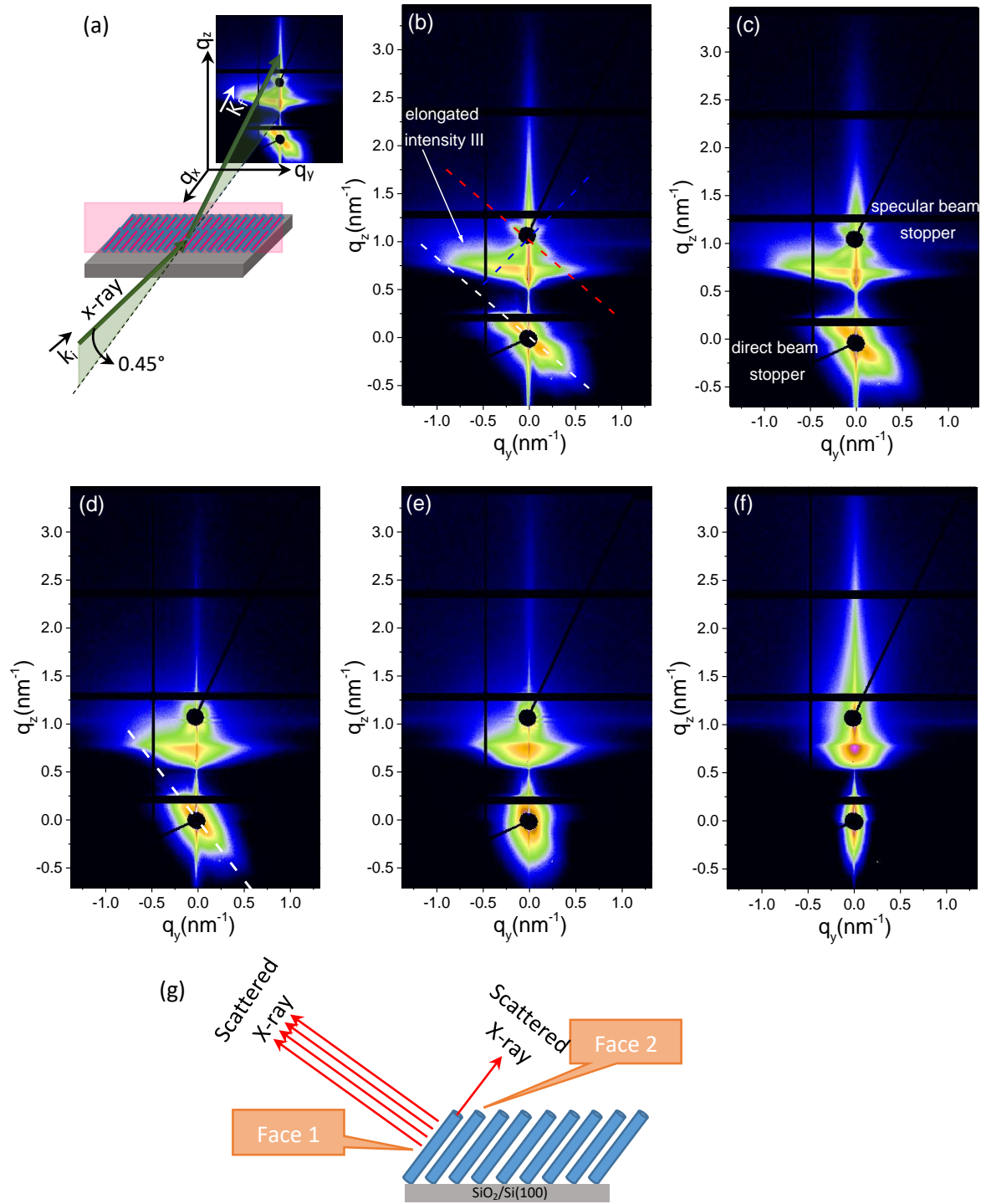

**Figure 7.7:** (a) GISAXS geometry, 2-D GISAXS images at (b) RT, (c) 250 °C, (d) 350 °C, (e) 450 °C and (f) 500 °C (not in-situ heating during GISAXS measurements, these are GISAXS measurements for already heated samples). (g) Both the faces of the oblique columns scatter the x-ray in respective directions (the direction of the incident x-ray is perpendicular to the plane containing the columns).

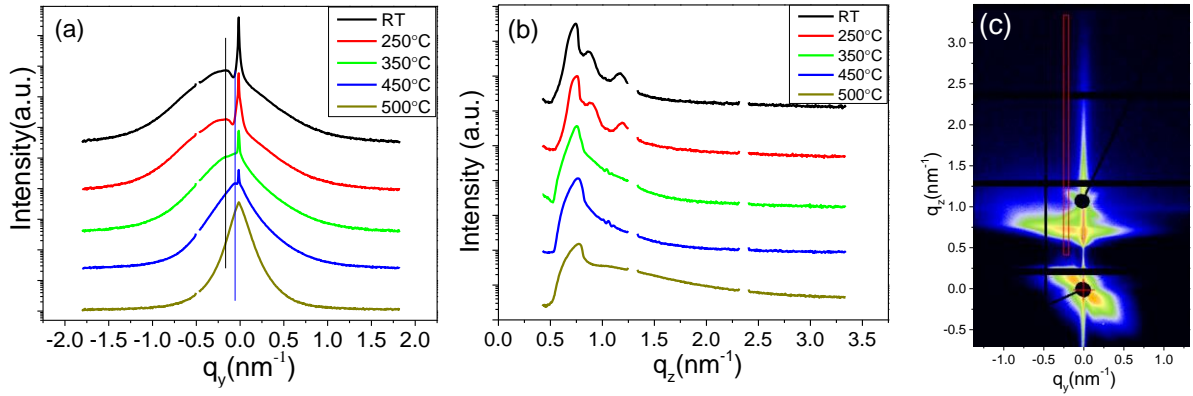

**Figure 7.8:** (a) 1-D GISAXS line plot for respective temperatures at Yoneda wing region (b) vertical 1-D GISAXS line plot for respective temperatures for respective vertical red line in (c).

Two crossed elongated intensities are observed for the pristine sample: (i) elongated intensity I, marked by a dashed red line (fig. 7.7b), is the reflection of x-ray beam from face 1 of the columns (fig. 7.7g) where the angle of the dashed red line made with horizontal direction provides the angular tilt of the columns with respect to the substrate normal and, this type of elongated intensity is also visible on direction beam position (marked by the dashed white line in fig. 7.7b), (ii) elongated intensity II (dashed blue line in fig. 7.7b), which are the reflection from the face 2 of the columnar structures (fig. 7.7g). Additional elongated intensity observed is the elongated intensity III, which comes from the scattering through both NAD and OAD layers. Such kind of elongated intensities present in the GISAXS images is the measure of tilted structures present in the multilayer. The images also clearly show that the present asymmetry in intensity almost disappeared for the sample heated up to 500 ℃ (fig. 7.7f). For the pristine sample, the tilt angle between columns and substrate normal is β=40º, while the deposition angle is α=75°. This inequality (α≠β) arises due to adatom mobility, which is a highly material-dependent property [20,21].

1D profiles (I vs q plots) are obtained from the 2D GISAXS patterns to find the in-plain and out-of-plain correlation with increasing temperature, as shown in fig. 7.8(a) and 7.8(b). Figure 7.8(a) clearly shows that the I variation for $-q_y$ range significantly differs from the I variation for $+q_y$. Intensity maxima along the $-q_y$ direction is visible. This intensity maxima (satellite peak) is relatively intense as compared to the maxima peak at the right, which is hardly visible. Intensity maximum (shoulder-like peak) represents the internal columnar or lateral distance between the columnar structures. Intensity maxima at $q_y \sim$ -0.18 nm$^{-1}$ correspond to the lateral spacing of ~34.8 nm. As the temperature increases, intensity maxima shift toward lower $q_y$ values (i.e. towards Yoneda peak) for temperature above 250 °C (peak shifts from vertical black line to the blue line in fig. 7.8a), and become weaker when temperature increases to 500 °C confirming the increasing separation (centre to centre) of the columns confirming that columns are diffusing to into the bigger columns and eventually diffuse to form a column-less layer. It is to be noted that it is a general trend to take a horizontal cut at the Yoneda peak region (in our case at $q_z = 0.75$ nm$^{-1}$) along $q_y$ direction to analyze the lateral separations of formed structures in the films [5]. But it is clear; see the corresponding fig. 7.7(f) for 500 °C at the region near $q_z = 1.0$ nm$^{-1}$ and along $q_y$ direction that the intensity distribution is still not fully symmetric, but we can treat this asymmetry as almost symmetric hence the columnar structure is almost destroyed at 500 ºC.

To analyze longitudinal intensity oscillations, the 1D I vs $q_z$ plot (fig. 7.8b) is derived from longitudinal cuts (vertical red rectangle in fig. 7.8c) marked near the satellite peak. Though longitudinal intensity oscillations are not visible in the GISAXS patterns (fig. 7.7) as compared to another GISAXS report [22] still, the interference fringes are visible as damped intensity oscillations in fig. 7.8(b). The damping of the oscillations is a measure of the surface topography correlation. Slow damping corresponds to the high conformity of the film/multilayer. But in our case, high damping shows no correlation of the multilayer with the substrate because, in our case, the multilayer is a complex structure containing oblique-normal bilayers, and the Si substrate is not patterned as in the literature [22]. But we can see in fig. 7.8(b) that with increasing temperature up to 250 ºC, the presence of oscillations symmetry shows that the multilayer still resembles the modulation of pristine multilayer and this information matches with the column diffusion information obtained through the explanation of fig. 7.8(a) that up to the temperature of 250 °C, columns are not diffused.

## 7.5   Discussion: High UMA and its dependence on morphology and texturing

In oblique angle deposition, in the initial stage of film growth, due to the self-shadowing effect, the Volmer–Weber type of growth is expected [23]. In the present case, Co islands initially nucleate on the Si substrate at various nucleation centres. These islands, with further deposition, transform into elongated columns with different crystallographic orientations due to OAD. Here, the stress is found to be absent but mentioned in some of the studies in the literature [5,24], but in Co-related multilayers, i.e. Co/Cu and Co/Au multilayers, stress is not mentioned [11,12,13,14]. The possible reasons for the absence of stress in the present multilayer structure are the different angles (75°- higher deposition angle) used for OAD and the different thickness combinations. In general, OAD Co (hcp) film contains MCA due to texturing of the (002) peak along the column direction [9,10]. In our study, we also have to find the weak texturing of this c-axis along the column direction as confirmed by RHEED, GIXRD and GISAXS measurements, as the angle of texturing is the same as the angle of tilt β of columns from the substrate normal. Also, it is a well-established fact that OAD provides shape anisotropy whose strength depends on the angle of deposition [7,8,25,26,27,28,29] and film thickness [25,30]. Therefore, the observed magnetic anisotropy in the present case of pristine Co multilayer is mainly due to the shape anisotropy generated due to the OAD and weak texturing in the (002) peak in the structure of the film. According to the literature, The oblique deposition above α= 60° causes an easy axis along the in-plane direction of the OAD projection [7,8,25,26,27,28,29]. Hence, shape anisotropy is the main factor for magnetic anisotropy in the in-plane projection direction of OAD.

Thus, the overall minimization of the weak magneto-crystalline and strong shape anisotropy energy would result in the alignment of the corresponding easy axes in the in-plane projection of the OAD direction. As the sample is annealed at 350 °C, the observed good square loop (as compared to the hysteresis loop of the pristine sample) along the easy axis (fig. 7.2b) shows the significantly reduced pinning centres due to reduced porosity caused by diffusion as clear by GISAXS data for 350 °C (fig. 7.7d and 7.8a), structure becoming symmetric. The coercivity reduces up to 450 °C and increases at 500 °C (fig. 7.2d). This behavior of coercivity with temperature is the result of two competing processes: (i) $H_C$ reduces with reduced pinning centres (due to reduced porosity) with increasing temperature [31,32,33], and (ii) $H_C$ increases with increasing grain size/ particle size [34]. GISAXS data (fig. 7.7 and 7.8) shows significant diffusion up to 450 °C means pinning centres reduces significantly as the temperature reaches 450 °C; hence $H_C$ reduces. But GIXRD analysis and AFM data (fig. 7.6b and 7.6c) confirm that as the sample is heated to 500 °C, crystallite/grain size increases significantly, which will increase the $H_C$ at 500 °C. Also, the film becomes significantly rough, as shown by XRR data of 500 °C (fig. 7.6a), increasing pinning centres. Hence at 500 °C, the increased crystallite/grain size and roughness cause the increase in $H_C$. Finally, at 500 °C, texturing is significantly reduced as clear by GIXRD (fig. 7.5f) and RHEED (fig. 7.3e and 7.3f), which again has two causes: (i) fcc phase grows at 400 °C and starts dominating the hcp phase at 500 °C. Both phases coexist at 500 °C (ii) columns are diffused significantly, resulting in the realignment of the hcp structure (texturing). Also, at this temperature, the tilted columnar structure is almost diffused (fig. 7.7f and 7.8a).AFM measurements also confirm this information of diffusion as room temperature AFM image (fig. 7.6b) shows elongation along deposition, but when heated to 500 °C (fig. 7.6c), bigger grain size without any preferred elongation confirms the diffusion of columns. The removal of the Bragg peak in XRR (fig. 7.6a) also ensures the same. All this causes a significant reduction of UMA at 500 °C. It is to be noted that, in the case of fcc Co, [111] direction is the easy axis of magnetization [35]. At 400 °C, fcc (111) and hcp (002) peaks nearly overlap (fig. 7.5g). At this temperature, whether one or both peaks are textured, the texturing is too weak (as the ring is continuous), as clear from fig. 7.5(d) is similar to the weak texturing of the pristine sample (fig. 7.4b or 7.5b), hence contributing less to the UMA than to the pristine sample. Also, from GIXRD images, it is clear that, along the texturing direction, other peaks (therefore corresponding planes) also exist, which means not only the (002) hcp or (111) fcc peak but also the other peaks (or planes) exist along column direction throughout the heating which further decreases the contribution of MCA on overall UMA.

The reason behind the combined use of the oblique-normal deposition is that a cobalt film shows the UMA when deposited normally on a cobalt underlayer deposited at oblique incidence [8]. Moreover, not only in OAD Co underlayer but also when thin cobalt film is deposited at normal incidence on obliquely deposited non-magnetic Ta [36] and Pt [37] underlayers. In all these cases, UMA in the normally deposited magnetic layer is associated with elongated corrugations on the underlayer. Thus, the present work demonstrates the use of multiple obliquely deposited thin Co underlayers between normally deposited Co layers to induce strong magnetic anisotropy in the Co multilayer structure.

We know that in the obliquely deposited layer, magnetic anisotropy contains out-of-plane anisotropy contribution due to tilted columns with respect to the substrate, which means UMA is not exactly in-plane [7] though the thickness of OAD Co is kept as low as 6.0 nm [12]. In the present case, we have kept the individual low thickness of obliquely deposited Co layers (4.4 nm) throughout the multilayer stacks and found the in-plane UMA as found in our previous study of FeCo multilayer [5]. Moreover, separated columns in the OAD thin films produce more UMA than connected/collapsed columns at a larger thickness due to increased dipolar interaction in-between columns [5]. In many OAD Co studies, an out-of-plane contribution in magnetic anisotropy comes due to texturing of hcp (002) or fcc (111) planes which keep the easy axis away from the in-plane direction, and they used relative higher thicknesses (from tens to hundreds of nm) for deposition [9,10,11]. But in our case, OAD/NAD combination of Co ultra-thin film layers, both the shape and weak MCA, contribute to the effective in-plane UMA in Co multilayer. More importantly, the magnetic anisotropy in the present Co multilayer is significantly larger than the Co/Cu or Co/Au multilayers studied previously [11,12,13,14]. More importantly, in our case, high UMA is maintained up to 450 °C.

## 7.6    Conclusion

A multilayered columnar structure of Co material is prepared by depositing various Co ultra-thin layers by keeping the deposition in two directions (i) along the surface normal and (ii) at an angle of 75° from the surface normal to the substrate, alternatively. In this system, strong shape anisotropy and weak MCA resulted in a well-defined UMA with an easy axis of magnetization along the projection of the tilted nanocolumns in the film plane. This magnetic anisotropy is high enough compared to other magnetic Co layers deposited obliquely at various oblique angles. Nevertheless, the multilayer remains magnetic anisotropic up to 450 °C, which confirms good thermal stability. The possible reason for the absence of long-range stress in this multilayer is the relatively lower thickness of the individual layers and large pinning centres at OAD, causing short-range stress throughout the layers. Drastic reduction in the in-plane UMA after annealing at 500 °C is attributed to the diffusion and merging of columns and growth of the fcc phase with significant removal of texturing after the heat treatment. Thus, such OAD/NAD multilayers deposited at various oblique angles and using ultra-thin layer thickness can produce significantly higher controllable UMA up to high temperatures.

**CHAPTER 8**    Conclusions and Future Scope

**8.1    Conclusion**

**8.2    Future scope**

This chapter concludes the results obtained through experimental works discussed in chapters 4 to 7 of the present thesis work. Additionally, an overview of the work that can be done in the future, based on the results obtained in this thesis, has also been briefly discussed.

## 8.1 Conclusion

In the present thesis, our main motive was to find the correlation between magnetic anisotropy, stress, and columnar nanostructure in polycrystalline and amorphous thin films and multilayers. It is essentially required to achieve magnetic anisotropy in such films, which is otherwise not present due to the absence of long-range crystalline order in such nanostructures. For the comprehensive study of the understanding of magnetic anisotropy, an in-situ mini-vacuum chamber to study GI-SAXS and GIWAXS at a synchrotron radiation source has been developed. Due to its small size and weight, it is portable. Also, we have developed various sample heaters for temperature-dependent studies. Furthermore, a pyrolytic boron nitride thermal evaporator has been developed to deposit organic thin films to utilize the oblique angle deposition method to prepare organic spin valves.

In general, magnetic anisotropy in polycrystalline and amorphous thin films is attributed to the stress developed during growth. But direct evidence for the same is missing from the literature. Our study aims to determine the correlation between stress, nanocolumns, and magnetic anisotropy. Our study utilized this understanding to achieve magnetic polycrystalline films with strong tunable magnetic anisotropy. For this purpose, the present study selected Co and FeCo-based thin films and multilayer structures as model systems to demonstrate our findings. In-situ MOSS, MOKE, and GISAXS measurements are performed during growth and annealing to correlate stress with magnetic anisotropy. In addition, various other complementary measurements such as XRR, XRD AFM, TEM, and GIWAXS measurements are done to investigate further and confirm the results.

To reveal the origin of magnetic anisotropy in polycrystalline thin films and to accomplish the role of stress, an in-situ analysis of the growth of polycrystalline Co films has been conducted. It has been discovered that the anisotropic tensile stress in the films is generated during growth due to the island coalescence process. The competition between stress generation and stress relaxation is related to film thickness and leads to peak tensile stress at about 10 nm. The film displays a well-defined UMA, which varies in correlation with tensile stress in the film. Through the reduction of magnetoelastic and anisotropic energies, stress combines with non-zero magnetostriction in the film to produce the UMA. The findings are verified by depositing a small layer of Co on the Ag layer, as no magnetic anisotropy is seen because the Co layer is pinned to the Ag islands, which eliminates long-range stress. The origin of the UMA in polycrystalline thin films has been more clearly understood as a result of in-situ real-time examination.

**Manipulated magnetic anisotropy using external stress:** Previous work revealed that tunning magnetic anisotropy is possible by applying controlled stress to the film. E-beam deposition on bent Si (111) substrates is used to create polycrystalline Co films with varied long-range stress in the film. Tensile stress smoothens the film's surface, which is linked to the movement of the atoms involved in the applied stress. Both compressively and tensile stressed Co films exhibit uniaxial magnetic anisotropy with easy axis in different directions in the film plane. The direction of the easy axis is understood in terms of the minimization of magneto-elastic energy. It's significant to note that rising stress enhances magnetic anisotropy strength. Since crystalline anisotropy is missing in polycrystalline films, the current technique for introducing magnetic anisotropy could be important to tune magnetic anisotropy in the polycrystalline and amorphous films.

**Study of stress in OAD processes:** To overcome the limitation with the UMA in OAD deposited single layer thin films, soft magnetic FeCo layers are deposited in a multilayer columnar structure by combining normal angle deposition (NAD) and oblique angle deposition (OAD) processes. Strong shape anisotropy produced a well-defined UMA with the easy axis of magnetization along the projection of the slanted nanocolumns in the film plane, in contrast to the literature, which typically describes FeCo films as being magnetically isotropic in nature. The current work reveals a novel technique for fabricating a FeCo multilayer with a strong and regulated UMA (nearly five times higher than the OAD single layer FeCo film). Moreover, in contrast to the literature, compressive stress in the obliquely deposited layers minimizes magnetoelastic energy along the in-plane column direction. Thus, the shape and stress anisotropy associated with the ensemble of nanocolumns in the multilayer structure produces strong UMA. Heat treatments are also done in the multilayer to study temperature-dependent behaviour of UMA, morphology and stress. This report establishes that the current technique induces and controls UMA in the polycrystalline films/multilayers and offers a way to create strong in-plane UMA in thin films, which are typically magnetically isotropic. The research reported here opens a new path toward building multilayer structures out of a single material with density variation, which could profoundly affect the design of future technological innovations across various industries.

**Absence of stress at higher OAD:** In another OAD study, in comparison to OAD FeCo film/multilayer, Co multilayer shows the presence of MCA in total magnetic anisotropy. In literature, due to the presence of MCA, total magnetic anisotropy doesn't remain exactly in-plane. In our study, we have kept the total magnetic anisotropy in-plane. Here we have kept the OAD angle of 75° from the surface normal to the substrate, while it was 60° in the FeCo case. In this Co multilayer system, strong shape anisotropy and weak MCA produced a well-defined UMA with an easy axis of magnetization along the projection of the slanted nanocolumns in the film plane. Compared to other reported magnetic Co multilayers formed at various oblique angles, this magnetic anisotropy is sufficiently high. In contrast to the FeCo multilayer study, The comparatively thin individual layers and the significant pinning centres at higher OAD, which produce short-range stress throughout the layers, may cause the absence of long-range stress in this multilayer. The multilayer shows good thermal stability. The diffusion and merging of columns, the formation of the fcc phase, and the considerable elimination of texturing following the heat treatment are responsible for the drastic drop in the in-plane UMA.

**Overall role of stress in performed OAD studies:** As a result, such OAD/NAD multilayers can create much greater controlled UMA up to high temperatures when they are deposited at different oblique angles and employ ultra-thin layer thickness. Stress can be present in the OAD film/ multilayer and can be responsible for stress anisotropy depending upon the film/layer thickness and angle of OAD.

## 8.2   Future scope

(i) Various flexible substrates can be used to create higher magnetic anisotropy.

(ii) Flexible substrate + OAD technique can be used simultaneously to explore the possibility of drastically enhancing the strength of magnetic anisotropy.

(iii) In the case of OAD, normal deposition of the immiscible non-magnetic layer in between OAD magnetic layers is planned to maintain the magnetic anisotropy up to a very high temperature. We have already used Ag as a non-magnetic layer. As a result, a slight increase in magnetic anisotropy (5%) is observed, but the challenge is still to maintain the magnetic anisotropy at very high temperatures.

(iv) In the next session (appendix), we have shown the reduced diffusion of the top electrode inside the organic semiconductor (OSC) together with drastically increased coercivity. In future, the combination of various magnetic materials, their varying thickness and deposition angle is planned to introduce more effective organic spin valve (OSV) and organic giant magneto-resistance (OGMR) structures.

_Reduced diffusion and increased coercivity using single magnetic material as the top and bottom metallic layer_

A comparative study between two bilayers, namely: $Alq_3$(178nm)/$Co_{nor}$(65nm) and $Alq_3$(178nm)/ $Co_{obl}$75°(25nm) deposited on Si (100) substrate having a native oxide layer is performed,. The understanding developed during this work is used to prepare organic spin valve structure: $Co_{nor}$(20nm)/$Alq_3$(50nm)/$Co_{obl}$75°(10nm). XRR results show significantly reduced diffusion, while MOKE measurements show the produced magnetic anisotropy and increased coercivity for oblique bilayer and oblique trilayer. GISAXS analysis confirms the tilted columnar growth for oblique bilayer and oblique trilayer, which is responsible for the in-plane shape anisotropy. GIWAXS results confirm the role of the hcp (002) peak of the Co layer for the presence of weak magneto-crystalline anisotropy. Thus, the significantly reduced diffusion and highly increased coercivity make the OAD process a unique method for preparing spin valve structures using single magnetic material.

## A.1    Introduction

Over the past decades, organic semiconductors (OSCs)-based technology has advanced astronomically. In general, OSCs have numerous important benefits, such as being lightweight, low cost and offering mechanical flexibility compatible with plastic substrates. Also, it offers relatively simple molecular properties engineering large area coverage. Consumers can already purchase organic light-emitting diodes (OLEDs). In addition, many new applications for devices are about to be released, such as organic field effect transistors (OFETs), organic photovoltaic (OPV), and non-volatile data storage related to electric memories [1]. One of the most recent and intriguing developments in the thin film field is organic spintronics, which uses OSCs sandwiched between two ferromagnetic layers to achieve spin-polarized electron transport [2,3]. Since the discovery of organic spin valves (OSVs), considerable effort has been made to investigate spin injection, manipulation, and detection [4,5]. For the effective performance of OSC devices in applications in spintronics, the interfaces between ferromagnetic and organic material layers are crucial. In extreme instances, the top ferromagnetic layer's atoms might diffuse or penetrate inside the organic layer deeply, creating a conducting path to the bottom metallic layer. This drawback causes OSV failure [6]. Over time, it has become clear that the ferromagnetic-OSC structure's interfaces are the key to understanding the extraordinary characteristics of OSV devices. A lot of work is being done to study the magnetic and structural characteristics of the diffused layer at ferromagnetic-OSC interfaces. Still, it has yet to be experimentally realized how to effectively insect the spin into the OSV device [7].

In this study, we have mainly focused on reducing the penetration of the top metallic layer inside the OSC using oblique angle deposition (OAD). Otherwise, the growth of the metallic layer on the organic layer penetrates deep into the organic layer, which can cause the failure of the spin valve structure.   We have used Co as the top, and bottom metallic layers and Alq$_3$ as OSC sandwiched layer. Various magnetic, structural, and morphological measurements are performed, and results are correlated to gather the complete story of the study performed. It is found that the top cobalt layer is deposited obliquely, reducing diffusion inside the Alq3 material and increasing the coercivity drastically compared to the bottom Co layer. Strong shape anisotropy and weak magneto-crystalline anisotropy are observed in the obliquely deposited Co layer.

## A.2    Experimental details

We have deposited various bilayers and trilayers in the present study. To deposit organic material Alq3, we have used the thermal deposition method using a pyrolytic boron nitride heater, while the Co layer is deposited using the e-beam evaporation method. Both the depositions are done in a high vacuum chamber under a base pressure of $3 \times 10^{-7}$ mbar. Thickness is monitored using the calibrated quartz crystal thickness monitor. The rate of deposition for $Alq_3$ is kept at 2 Å/sec, while that for $Co_{nor}$ and $Co_{obl}$ are kept at 0.2 Å/sec. The bilayers $Alq_3$(178nm)/$Co_{nor}$(65nm) and $Alq_3$(178nm)/ $Co_{obl}$75°(25nm) have been deposited on Si (100) substrate having a native oxide layer on it. Here "nor" and "obl" in the subscript represent normal angle deposition and OAD, respectively, and we will use "normal bilayer" and "oblique bilayer" terms for the $Alq_3$(178nm)/$Co_{nor}$(65nm) and $Alq_3$(178nm)/$Co_{obl}$75°(25nm) bilayers respectively. These thicknesses have been confirmed using single layer $Alq_3$, Co and $Co_{obl}$ (75°) thin films, which are deposited together with these bilayers. Then we also deposited $Co_{nor}$(20nm)/$Alq_3$(50nm)/$Co_{obl}$75°(20nm) on native oxide containing Si(100) substrate, and we will use the term "oblique trilayer" for it. X-ray diffraction (XRD) and X-ray reflectivity (XRR) measurements are performed to get structure, thickness and density information. Magneto-optical Kerr effect (MOKE) is performed to analyze the magnetic behavior on normal and obliquely deposited Co. In-situ temperature-dependent MOKE study is performed in a UHV chamber with a base vacuum of $6 \times 10^{-10}$ mbar. For the morphological and structural (texturing) study, grazing incidence small angle x-ray scattering (GISAXS) and grazing incidence wide-angle x-ray scattering (GIWAXS) are performed at P03 (MiNaXS) beamline, PETRA III storage ring, DESY, Hamburg (Germany) having photon energy of E =13 keV.

## A.3    Characterizations

### A.3.1    XRD analysis: Structure of $Alq_3$ film

For the structural analysis of the base $Alq_3$ film, XRD is done using a Cu $K_\alpha$ beam and shown below in fig. A.1. Appearance of halos on the spectra and absence of sharp peaks suggests an amorphous-like structure; hence evaporated Alq3 films at room temperature are amorphous-like and composed of meridional isomers of Alq3 molecules [8].

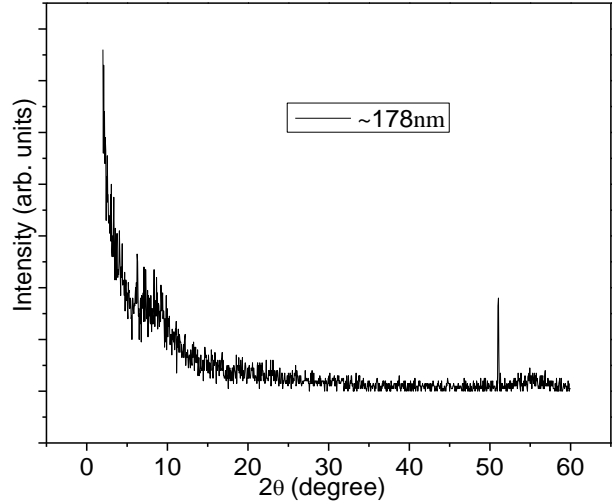

**Figure A.1:** shows the XRD spectra of the as-deposited Alq3 films.

## A.3.2    XRR: Thickness and density study

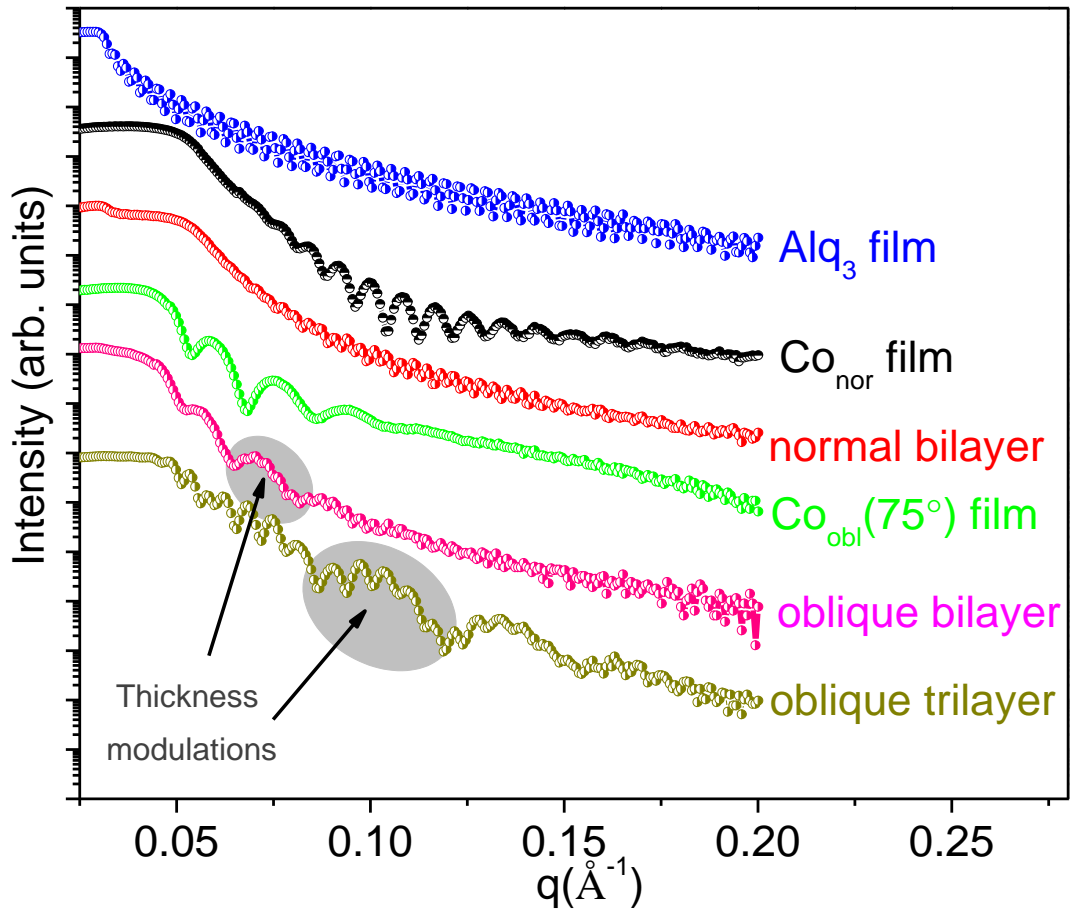

**Figure A.2:** XRR data of the various films mentioned alongside.

For the structural analysis of the films, XRR is performed on all single layers, bilayer and trilayer films, as shown in fig. A.2. Thickness of single layers $Alq_3$, $Co_{nor}$ and $Co_{obl}$ (75°) are found to be 177 nm, 64 nm and 25 nm, respectively, after fitting the respective XRR data. The electron scattering length density (ESLD) of OSC $Alq_3$ is significantly low as it is porous and hence shows a lower critical angle. On the other hand, the fitted XRR data of $Co_{nor}$ and $Co_{obl}$ films represents that the ESLD of $Co_{obl}$ is significantly low compared to $Co_{nor}$ because of the porous tilted columnar structure of $Co_{obl}$ film[9]. This is why, in our case, the roughness of the $Co_{obl}$ is found to be 2.7 nm compared to the roughness of $Co_{nor}$ film, which is 1.6 nm.

XRR data for the Bilayer and trilayer films are shown in the same fig. A.2. Very weak thickness modulations are present for the normal bilayer, showing significant diffusion of top Co layer inside the porous $Alq_3$ layer. Strong thickness modulations are observed in the case of oblique bilayer confirming significantly reduced diffusion of $Co_{obl}$ inside the porous $Alq_3$ layer. The critical angle for Co film and normal bilayer film matches each other, ensuring that the Co is not fully diffused inside the organic $Alq_3$ film. Proper thickness modulation for oblique trilayer up to higher q range again confirms the significantly reduced diffusion. Broad oscillations and narrow oscillations (inside broad oscillations) shows the top Co layer thickness and bilayer thickness respectivrly.

### A.3.3    MOKE study: pristine samples

MOKE measurements are performed to understand these bilayer and trilayer films' magnetic behaviour. For the normal bilayer, we don't observe any significant magnetic anisotropy (fig. A.3a). On the other hand, in the case of the oblique bilayer, in-plane projection of the OAD direction is found to be easy axis and marked as 0° (fig. A.3b) while in-plane direction perpendicular to it is the hard axis and marked as 90°. This anisotropic oblique bilayer is compared with the $Co_{obl}$ (75°) thin film. Figure A.3(c) shows the MOKE loops of $Co_{obl}$ (75°) thin film. Increased coercivity is observed as compared to the oblique bilayer. In the case of oblique trilayer (fig. A.3d), two-step loops are observed, corresponding to the different coercivities of top and bottom Co layers.

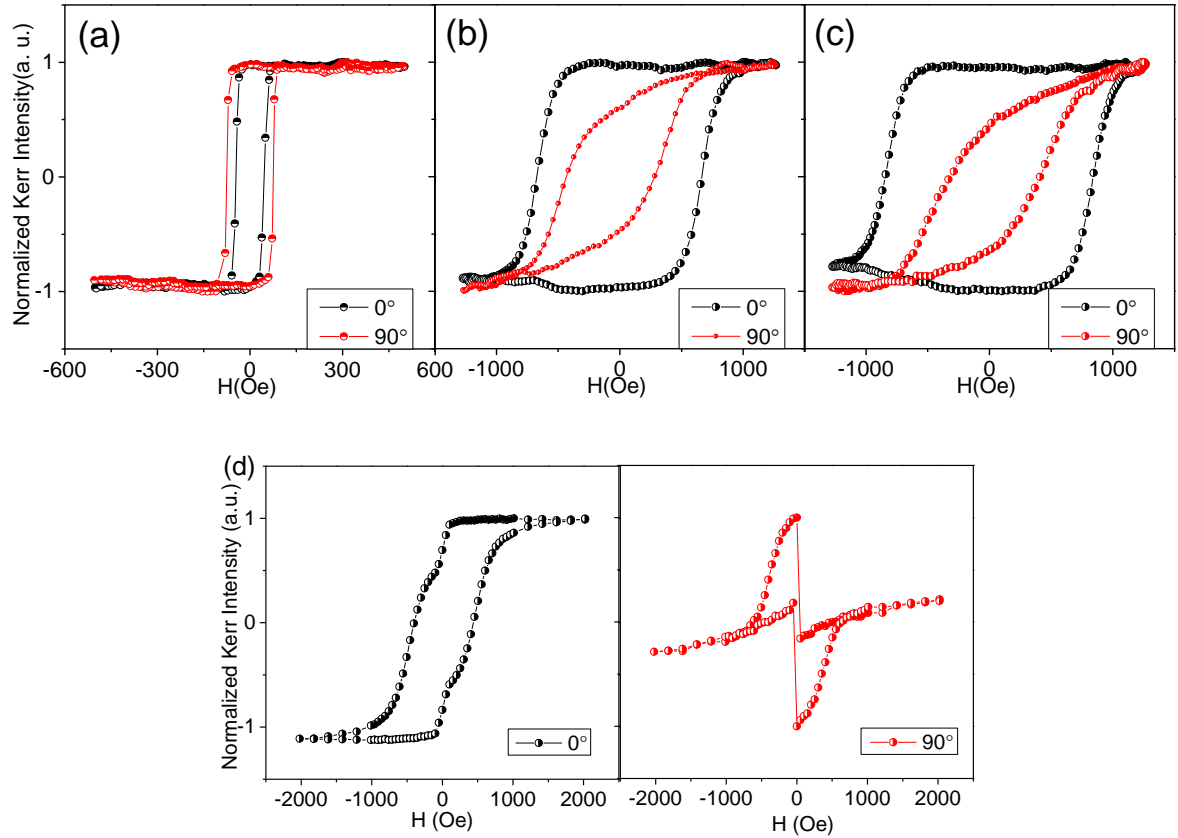

**Figure A.3:** MOKE loops along easy (0°) and hard (90°) direction of magnetization of (a) normal bilayer, (b) oblique bilayer, (c) Co75° film and (d) oblique trilayer on Si(100) having native oxide.

## A.3.4    In-situ MOKE study

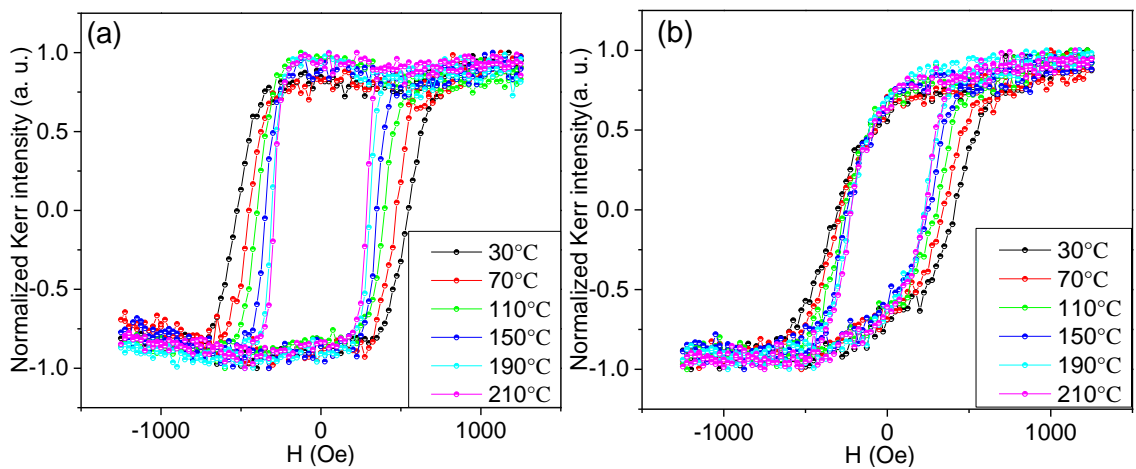

**Figure A.4:** MOKE loops along (a) easy (0°) and (b) hard (90°) axis oblique bilayer during in-situ heating.

In-situ heating is done for the oblique bilayer in the UHV chamber up to 210 °C. MOKE loops taken along the easy and hard axis are shown in fig. A.4. It is clear from the MOKE loops that coercivity is decreasing continuously, and loops are becoming square along both axis with increasing temperature, denoting that magnetic anisotropy is continuously decreasing with increasing temperature. At 230 °C, the film is peeled off from the substrate.

### A.3.5 GISAXS analysis

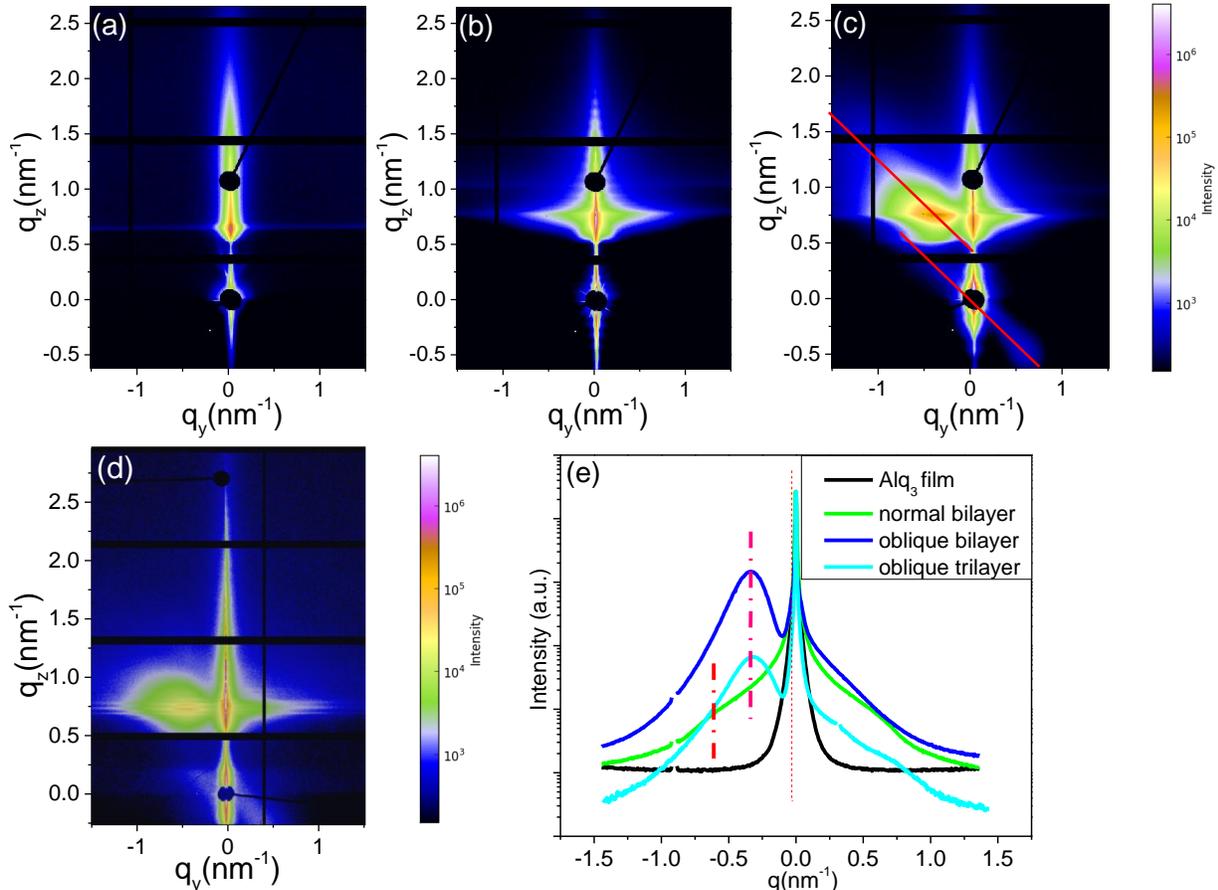

**Figure A.5:** GISAXS images taken for (a) Alq₃, (b) normal bilayer (c) oblique bilayer and (d) oblique trilayer sample while (e) is the 1D intensity data vs $q_y$ profiles at the Yoneda wing region (at $q_z = 0.75$ nm⁻¹) for all these films.

Details about the buried nanostructures have been obtained using GISAXS measurements for the pristine samples. All the details about the measurements' geometry have already been mentioned in the previous chapters 6 and 7. Obtained GISAXS images are shown in fig. A.5. Both for Alq₃ (fig. A.5a) and normal bilayer (fig. A.5b) film, the GISAXS spectra is symmetric on both side of $q_y = 0$ nm⁻¹ while it is asymmetric for the oblique bilayer (fig. A.5c) as expected. In the case of the oblique bilayer, it confirms that, due to OAD, tilted Co columns are formed, similar to previous studies. This asymmetric intensity distribution along the red line (fig. A.5c) makes an angle of 45° with respect to the horizontal direction. It means the tilted columnar structure makes an angle β= 45° from the substrate normal while deposition direction is kept at α = 75°. A similar asymmetric intensity distribution is observed for the oblique trilayer (fig. A.5d) as for the oblique bilayer. 1D data is also extracted from these images and shown in fig. A.5(e). The shifting of satellite peak to lower $q_y$ value (peak shifting from dashed red line to dashed pink line) for oblique bilayer as compared to normal bilayer confirms the increased lateral spacing, which arises due to columnar structure. The satellite peak position for the oblique trilayer is at a slightly lesser $q_y$ value than the oblique bilayer, indicating the increased separation between columns.

## A.3.6    GIWAXS analysis

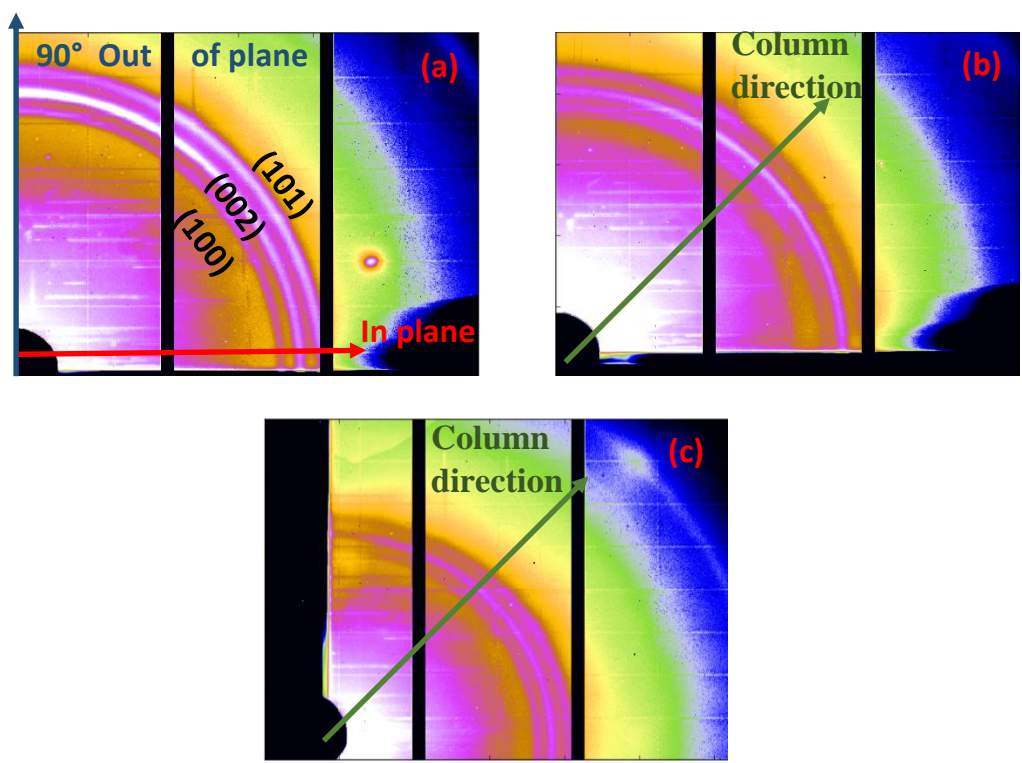

**Figure A.6:** GIWAXS images taken for (a) Normal bilayer, (b) oblique bilayer (c) oblique trilayer. In-plane (or 0° out of plane), 90° out of plane and average column directions are shown using arrows.

In parallel to GISAXS measurements, for the structural analysis, GIWAXS measurements are performed. Figure A.6 shows the GIWAXS images of normal (fig. A.6a) and oblique bilayers (fig. A.6b) together with oblique trilayer (fig. A.6c). Various peaks correspond to the Co hcp phase. For the oblique bilayer and oblique trilayer, GIWAXS measurements are done by keeping the column's direction perpendicular to the beam direction. In the case of a normal bilayer (fig. A.6a), the hcp (002) peak is textured along out of plane direction. For the oblique bilayer, this hcp (002) peak is more textured along average column direction rather than in-plane and 90° out of plane directions, which is mentioned in literature[10,11]. The texturing for the oblique trilayer does not seem that clear.

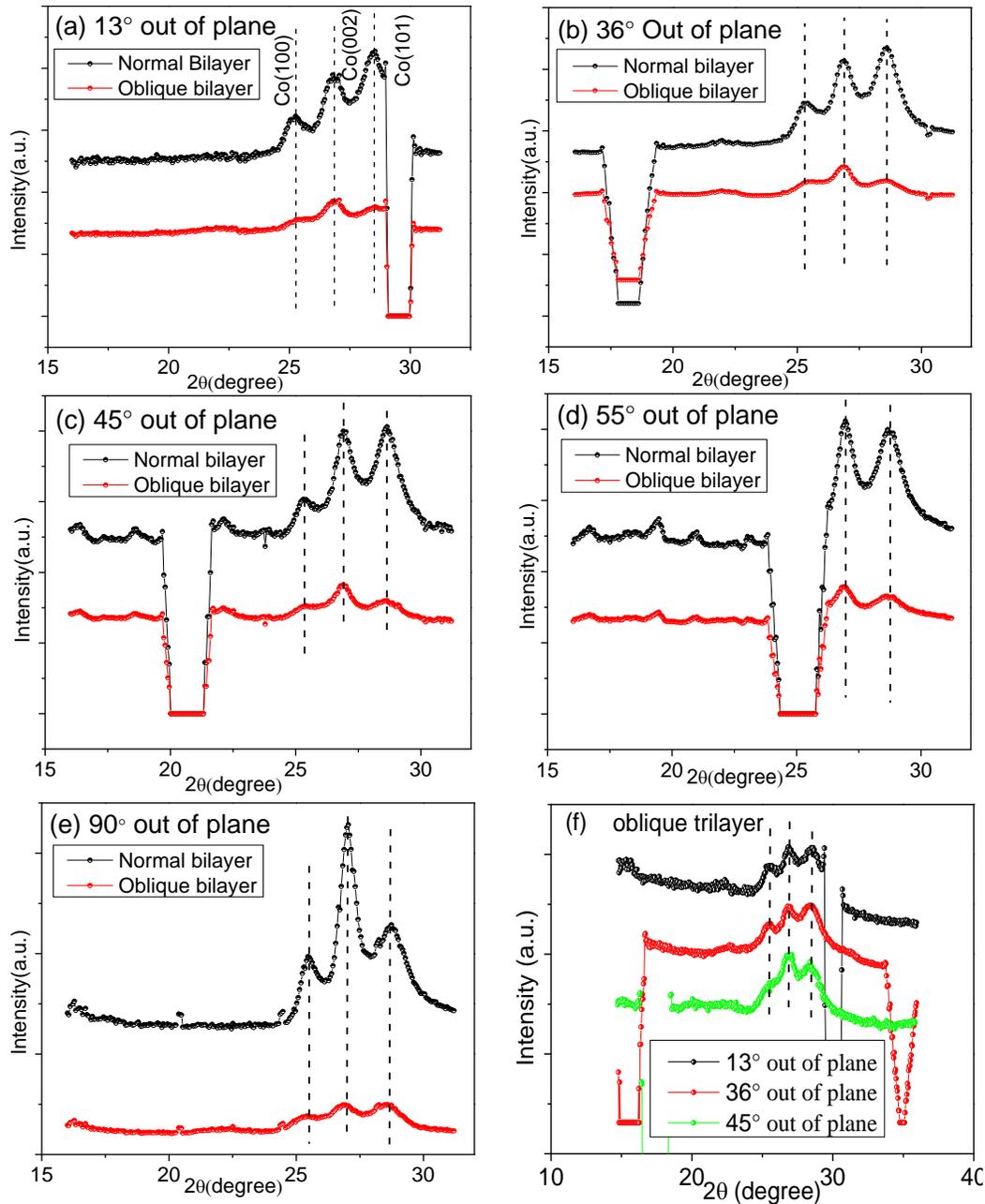

**Figure A.7:** GIWAXS 1D plots taken along in-plane (0° out of plane) to various out-of-plane directions (mentioned in the respective plots) for both normal and oblique bilayers (a-e), while (f) shows the same for oblique trilayer. Various hcp peaks are denoted in (a).

1D plots are also collected using GIWAXS images along various in-plane and out-of-plane directions for the in-depth analysis of peak texturing and stress and arranged in fig. A.7. It is again clear from these plots that for normal bilayer (black plots; fig. A.7a to e), when we move from in-plane to out-of-plane directions, the relative peak intensity of hcp (002) peak increases as compared to other hcp peaks. This means the hcp (002) peak is becoming more textured toward the out-of-plane direction. In contrast, for the oblique bilayer (red plots; fig. A.7a to e), this hcp (002) peak is more textured along 45° out of the plane direction, which is the average column direction confirmed by GISAXAS analysis. No peak shift is found for both normal and oblique bilayers, which ensures the absence of stress in the bilayer. In the case of the oblique trilayer (fig. A.7f), texturing of the hcp (002) peak is also along 45° out of the plane as for the oblique bilayer.

## A.4    Discussion

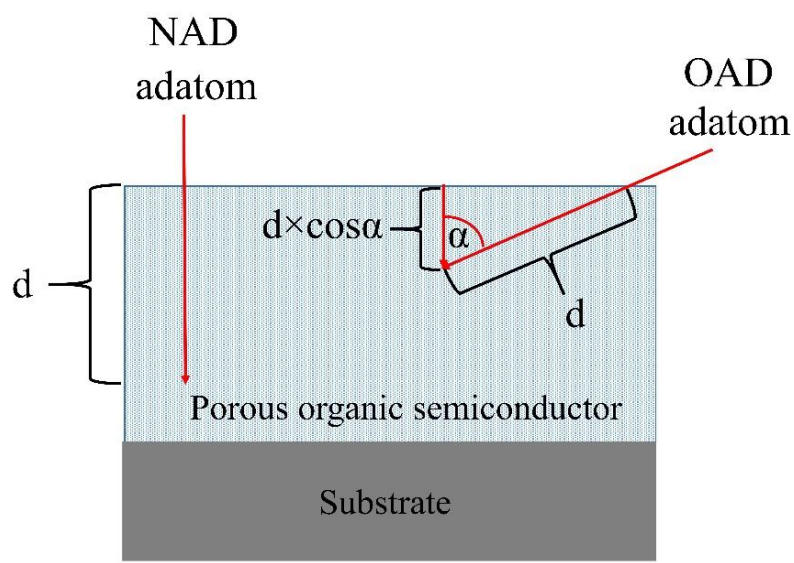

**Figure A.8:** Diffusion scenario in NAD and OAD cases.

In the case of normal deposition, the diffusion of the top magnetic metallic layer is significantly high, up to hundreds of nm, inside the OSC Alq$_3$ because Alq$_3$ has a porous structure [6]. But in the case of OAD of metallic layer on porous OSC, the distance covered by the magnetic layer inside the porous organic layer will be the same if the incoming metallic atom sees the same density of Alq$_3$ in the OAD direction as it sees in the normal deposition direction. Due to the same velocity (i.e. energy) of adatom (metallic atom) in both the cases

NAD or OAD, adatom covers the same diffusion distance "d" inside the porous OSC, but in different directions, as shown in fig. A.8. Here, α is the deposition angle of incoming adatoms from the substrate's normal direction during OAD. Hence the effective distance covered by the incoming adatom inside the porous OSC in the vertical direction is "d×cosα". The higher the deposition angle α, the lower will be the effective diffusion. We have used this theory for the deposition to meet the solution to the challenge mentioned in the literature [6].

As expected, Alq$_3$ is found to be in amorphous form. The density of the Co layer in a normal bilayer or Co$_{nor}$ film on Si substrate is the same as confirmed by the XRR measurements (fig. A.2) because the critical angle is the same. The same applies to the obliquely deposited Co layer in the oblique bilayer and Co$_{obl}$ (75°) on the Si substrate. But in the case of the normal bilayer, thickness modulations are not present in XRR, confirming rough interface and high diffusion, while in the case of oblique bilayer and oblique trilayer, thickness modulation ensures the relative sharp interface as compared to the normal bilayer hence less diffusion. Even for the oblique trilayer, thickness modulations are present up to higher q values means lesser roughness than the oblique bilayer.

MOKE data shows significantly good magnetic anisotropy in oblique bilayer as compared to normal bilayer (fig. A.3), which is expected because normal deposition on Si substrate shows inherent magnetic anisotropy due to long-range stress [12] but Co deposition on Alq$_3$ (normal bilayer) shows no anisotropy because of the absence of long-range stress (fig. A.7) due to significant rough interface [12] (no modulation in XRR; fig. A.2) which creates pinning centres that relax the long-range stress in the film. The direction of the easy magnetic axis is along the plane containing the deposition direction and perpendicular to the substrate plane, while the direction perpendicular to the deposition plane is the hard axis. Magnetic anisotropy is not perfect in-plane because the hysteresis loop along the in-plane projection of columns is not a perfect square, but it is near a square loop; hence the easy axis is almost in-plane. GISAXS measurements confirm that in the case of the oblique bilayer (fig. A.5c), the tilted columnar structure makes an average tilt angle of 45° from the substrate normal, which matches well with the texturing direction of the hcp (002) peak (fig. A.6b). OAD deposition creates in-plane shape anisotropy [13] while texturing of hcp (002) peak along column direction creates the easy axis of magneto-crystalline anisotropy along column's direction [10,11]. GIWAXS measurements (fig. A.7) confirm the absence of stress in our study. Thus, the resultant magnetic anisotropy in the present case is a combination of in-plane shape anisotropy and 45° out-of-plane magneto-crystalline anisotropy; hence the resulting easy axis of total magnetic anisotropy will not be perfect in-plane direction. But almost square loop in MOKE data (fig. A.3b) shows that the contribution of magneto-crystalline anisotropy is significantly less than shape anisotropy in total magnetic anisotropy as in the previous study, which causes due to weak texturing. Hence we can say that the resultant magnetic anisotropy in the present case is in-plane uniaxial magnetic anisotropy (UMA). As we move from normal to the oblique bilayer, magneto-crystalline anisotropy contributes in the out-of-plane direction, which increases the saturation field along the in-plane direction (fig. A.3b) and also the increased pinning centres due to porous OAD film (fig. A.5d) increases $H_c$ which in turn increase saturation field. This coercivity increase is nearly six times even if we compare the coercivity of a normal bilayer to the coercivity along the hard axis of the oblique bilayer. The increased coercivity in the case of $Co_{obl}(75°)$ film on Si substrate (fig. A.3c) as compared to oblique bilayer (fig. A.3b) is due to increased pinning centres which is possibly due to the increased porosity in $Co_{obl}(75°)$ film. The slightly decreased anisotropy in the oblique bilayer (on Alq3) as compared to $Co_{obl}(75°)$ film on Si substrate is possibly due to the contributions in the different magnitude of shape and magneto-crystalline anisotropies due to different base layers used (soft $Alq_3$ layer in the bilayer and Si in single layer film). Hysteresis loops become more square as we increase the temperature for the oblique bilayer (fig. A.4), which is due to the further decreased contribution of magneto-crystalline anisotropy. Decreasing coercivity with heating is due to reducing pinning centres with heating. For the oblique trilayer, one step loop is observed along the easy axis (fig. A.3d), while the hard axis shows butterfly structure. Each loop shows two different coercivities. Lower coercivity corresponds to the bottom Co layer deposited normally, while higher coercivity corresponds to the top obliquely deposited Co layer. Thus, even if we have used the same magnetic material Co for top and bottom magnetic layers, Different coercivities of top and bottom magnetic layers make this trilayer structure a potential spin valve structure as the diffusion is significantly reduced due to OAD through the thickness of the sandwiched OSC layer $Alq_3$ is kept very low. Hence such structures could be potentially excellent spin valve structures of the future. Here, in this oblique trilayer case, GISAXS measurements (fig. A.5d and A.5e) confirm the tilted columnar structure of the top Co layer, and GIWAXS results (fig. A.6c and A.7f) confirm the weak texturing of hcp (002) peak along the same 45°out of plane direction as in case of the oblique bilayer. Thus in the case of oblique trilayer, magnetic anisotropy is in-plane as the hysteresis loop looks almost perfect square (fig. A.3d) along the easy axis, and this in-plane magnetic anisotropy is mainly due to shape anisotropy and having the weak contribution of magneto-crystalline anisotropy.

## A.5    Conclusion:

In this study, we have compared two bilayers mainly: (i) Co deposited normal to Alq$_3$ (normal bilayer), (ii) Co deposited obliquely (75°) to Alq$_3$ (oblique bilayer) and used the knowledge to deposit oblique trilayer ( spin-valve). GISAXS measurements have confirmed tilted growth of obliquely deposited Co on Si and Alq3 layers. XRR data have demonstrated that the diffusion of Co layer inside OSC Alq$_3$ is significantly reduced when deposited obliquely compared to normal deposition. At the same time, the coercivity is drastically increased for the oblique Co layer compared to the normal deposited Co layer. Also, the oblique bilayer became significantly magnetic anisotropic, while the normal bilayer remained isotropic. Finally, a two-step loop is observed for the oblique trilayer together with considerably reduced diffusion even if the thickness used for sandwiched Alq$_3$ layer is only 50nm. It is a significant outcome of the present study, where the diffusion of the top magnetic layer inside the OSC reduces significantly. At the same time, OAD tunes the film's coercivity, which provides the capability to prepare a spin valve with the same magnetic material. It allows for making effective spin valve structures by eliminating the problem of huge diffusion up to hundreds of nm inside the OSC material.

# *List of figures*

## *Chapter 3:*

**Fig. 3.1:** (a) 2D AUTOCAD design of the xrd- mini chamber (b) top view of the mini chamber showing the angles covered by kapton windows.

**Fig. 3.2:** (a) Actual photograph of the chamber and its various components (b) Inside heater assembly of the chamber.

**Fig. 3.3:** PID controller to control the heater's temperature in the mini chamber.

**Fig. 3.4:** Geometry for GISAXS measurements.

**Fig. 3.5:** (a) GISAXS patterns of $Co_{obl}/Ag_{nor}$ multilayer at room temperature (R.T.) and (b) GISAXS 1-D profiles with increasing temperatures. GISAXS 1-D profiles are taken at the red marked area in fig. 3 (a) corresponding to the Yoneda wings region.

**Fig. 3.6:** (a) GIWAXS patterns of $Co_{obl}/Ag_{nor}$ multilayer at room temperature (R.T.) and (b) GIWAXS 1-D profiles with increasing temperatures. The yellow areas in GIWAXS 1-D profiles correspond to the detector's intermodular gap (IDG), where no photons can be detected.

**Fig. 3.7:** Top Kapton window developed for the Fluorescence measurements.

**Fig. 3.8:** Schematic design of (a) pyrolytic boron nitride thermal evaporator (b) base of the heater.

**Fig. 3.9:** (a) Schematic diagram of thermal evaporator (b) actual image of evaporator.

**Fig. 3.10:** (a) Main deposition chamber, (b) attached thermal evaporator inside the chamber (c) controller assembly.

**Fig. 3.11:** XRR data of 170nm $Alq_3$ deposited using a thermal evaporator.

## *Chapter 4:*

**Figure 4.1.** (a) Original and (b) Schematic view of the MOSS chamber with MOSS components (laser and detector).

**Figure 4.2.** (a) Original and (b) Schematic view of the sample holder with connections for Resistivity measurement. Laser is a part of MOSS components. The details about sample placement in the holder with the connections for resistivity and laser array on the sample are also included in the figure. (c) top view of the sample with connections. Two perpendicular stresses, present in the thin film, are also shown.

**Figure 4.3.** In-situ real-time evolution of resistance $R_{Co}$ (right axis, blue plot) and stress-thickness product- $\sigma_x \times d_f$ (left axis, red plot) as a function of the nominal film thickness $d_f$. Inset gives $\sigma_y \times d_f$ vs $d_f$ plot.

**Figure 4.4.** Hysteresis loops along x (hard axis of magnetization; $\theta=90_0$) and y (easy axis of magnetization; $\theta=0_0$) directions for Co film at various thicknesses marked.

**Figure 4.5.** Azimuthal angle dependence of coercivity ($H_c$) is plotted in the polar plots for the corresponding thickness marked.

**Figure 4.6.** Azimuthal angle dependence of the $H_c$ fitted with $H_a$ and $H_{cw}$ as a fitting parameter for film thickness $d_{Co}=15.2$ nm.

**Figure 4.7.** Thickness-dependent variation of (a) $H_a$ and (b) $H_{cw}$.

## *Chapter 6:*

the vertical green rectangle marked in fig. 9(c). (Normalized with exposure time). β in fig. (b) and (c) is the angle between the columns and the substrate normal.

**Figure 6.11:** (a) XRR pattern of oblique angle (α=60º) deposited FeCo single layer film (b) Schematic of the OAD deposition and MOKE measurement geometry of obliquely deposited single FeCo film. θ is the azimuthal angle in the film plane, where θ=0º is the direction along the in-plane direction of OAD.

**Figure 6.12:** MOKE loops for Single layer FeCo thin films of ~36 nm thickness deposited at (a) OAD=0° - normal to the substrate (b) OAD=45º (c) OAD=60º, and (d) OAD= 80º.

**Figure 6.13:** MOKE loop along easy (0°) and hard (90°) directions for as prepared FeCo multilayer sample.

## *Chapter 7:*

**Figure 7.1.** (a) Schematic diagram of the deposition geometry for both normal and oblique deposition of Co layers, green rod shows the axis about which the substrate is rotated (b) final multilayer having 10 bilayers stack (Co multilayer) on Si (100) substrate containing native oxide layer.

**Figure 7.2**: Hysteresis loops along the easy and hard axis at (a) room temperature (pristine), (b) 350 °C, (c) 450 °C and (d) 500 °C, while (e) shows the area difference between the easy and hard axis and Hc along the easy axis at various temperatures.

**Figure 7.3:** RHEED images along 0° and 90° at (a), (b) room temperature, (c), (d) 400 °C and (e),(f) 500 °C.

**Figure 7.4:** GIXRD image (for x-ray λ =0.505917 Å) of Co multilayer along (a) 0º and (b) 90º directions (c) corresponding 1-D plots.

**Figure 7.5:** Temperature dependent GIXRD (for x-ray λ =0.78445 Å) images at (a-b) room temperature (pristine sample), (c-d) 400 °C and (e-f) 500 °C for x-ray along (a-c-e) 0º and (b-d-f) 90º to the columns (in-plane projection), (g) fitting of the 1D data shows phase change.

**Figure 7.6:** (a) XRR plots of pristine and annealed samples of Co multilayer. The vertical green line shows the shift in the critical angle of the multilayer system. AFM images of pristine (b) and annealed samples (c) at 500 °C of Co multilayer. The arrow shows the in-plane direction of the oblique angle deposition.

**Figure 7.7:** (a) GISAXS geometry, 2-D GISAXS images at (b) RT, (c) 250 °C, (d) 350 °C, (e) 450 °C and (f) 500 °C (not in-situ heating during GISAXS measurements, these are GISAXS measurements for already heated samples). (g) Both the faces of the oblique columns scatter the x-ray in respective directions (the direction of the incident x-ray is perpendicular to the plane containing the columns).

**Figure 7.8:** (a) 1-D GISAXS line plot for respective temperatures at Yoneda wing region (b) vertical 1-D GISAXS line plot for respective temperatures for respective vertical red line in (c).

## *Appendix A:*

**Figure A.1:** shows the XRD spectra of the as-deposited Alq3 films.

**Figure A.2:** XRR data of the various films mentioned alongside.

# *List of tables*

# List of abbreviations

Atomic force microscopy (AFM)

Charged coupled device (CCD)

Chemical vapor deposition (CVD)

dc magnetron sputtering (dcMS)

Easy axis magnetization (EAM)

Electron scattering length density (ESLD)

Exchange bias (EB)

Face-centred cubic (fcc)

Face-centred tetragonal (fct)

Four-probe resistivity (FPR)

Giant magneto-resistance (GMR)

Glancing angle deposition (GLAD)

Grazing-incidence small-angle x-ray scattering (GISAXS)

Grazing incidence wide-angle x-ray scattering (GIWAXS)

Grazing incidence x-ray diffraction (GIXRD)

Hard disk drives (HDDs)

Helmholtz coils (HCs)

High power impulse magnetron sputtering (HiPIMS)

High vacuum (HV)

In-plane magnetic anisotropy (IMA)

Ion-beam sputtering (IBS)

Lateral Force Microscopy (LFM)

Left circularly polarized (LCP)

Magnetic anisotropy (MA)

Magnetic tunnel junction (MTJ)

Magneto-crystalline anisotropy (MCA)

Magneto-elastic anisotropy (MEA)

Magneto-optical Kerr effect (MOKE)

Multi-beam optical stress sensor (MOSS)

Normal angle deposition (NAD)

Oblique angle deposition (OAD)

Organic field effect transistor (OFET)

Organic light-emitting diode (OLED)

Organic photovoltaic (OPV)

Organic semiconductor (OSC)

Organic spin valve (OSV)

Oxygen-free high thermal conductivity (OFHC)

Perpendicular magnetic anisotropy (PMA)

Physical vapor deposition (PVD)

Photoelastic modulator (PEM)

Reflection high energy electron diffraction (RHEED)

Right circularly polarized (RCP)

Room temperature (RT)

Scanning probe microscopy (SPM)

Self-shadowing effect (SSE)

Spin-orbit interaction (SOI)

Stainless steel (S.S.)

Stoner-Wohlfarth (SW)

Transmission electron microscopy (TEM)

Tunnel magneto-resistance (TMR)

Turbo molecular pump (TMP)

Ultra high vacuum (UHV)

Uniaxial magnetic anisotropy (UMA)

Vibrating Sample Magnetometer (VSM)

Voltage-controlled magnetic anisotropy (VCMA)

X-ray diffraction (XRD)

X-ray reflectivity (XRR)